\pdfoutput=1

\documentclass[final]{vutinfth} 
\usepackage{fixltx2e}  
\usepackage{amsmath}   
\usepackage{amssymb}   
\usepackage{mathtools} 
\usepackage{microtype} 
\usepackage{enumitem}  
\usepackage{multirow}  
\usepackage{booktabs}  
\usepackage{nag}       
\usepackage{hyperref}  
\usepackage[acronym,toc]{glossaries} 

\setsecnumdepth{subsection} 

\usepackage{gensymb}
\usepackage{epsfig}
\usepackage{color}
\usepackage{acronym}
\usepackage{epsfig}
\usepackage{epstopdf}
\usepackage{float}
\usepackage{graphicx}
\usepackage{wrapfig}
\usepackage{lscape}
\usepackage{rotating}
\usepackage{epstopdf}

\usepackage[table]{xcolor}    
\usepackage{algpseudocode}
\usepackage{amsfonts}
\usepackage{tabularx}
\usepackage{color}
\usepackage{tabu}
\usepackage{url}
\usepackage{tablefootnote}
\usepackage[gen]{eurosym}
\usepackage{float}
\restylefloat{table}
\usepackage{color}
\usepackage{listings}
\usepackage{longtable}
\usepackage[numbers]{natbib}
\usepackage{subcaption}
\usepackage{wrapfig}
\usepackage{textcomp}
\usepackage{filecontents}
\usepackage{array}
\usepackage{cleveref}
\crefname{equation}{Equation}{Equations}
\crefname{figure}{Figure}{Figures}
\crefname{table}{Table}{Tables}
\crefname{section}{Section}{Sections}
\crefname{chapter}{Chapter}{Chapters}

\usepackage[T1]{fontenc}
\usepackage[utf8]{inputenc}
\usepackage{algpseudocode}
\usepackage{amsmath}
\usepackage{float}
\usepackage{amssymb}
\usepackage{glossaries}
\usepackage[super]{nth}
\usepackage{mathptmx}
\usepackage{todonotes}
\usepackage{algpseudocode}
\usepackage{algorithm}
\usepackage{varwidth}
\usepackage{relsize}
\usepackage{lipsum}
\usepackage{dcolumn}
\usepackage{snapshot}
\usepackage{makeidx}
\usepackage{amsthm} 
\usepackage{etoolbox,changepage,lipsum}
\usepackage{thmtools}
\usepackage{geometry}
\usepackage[all]{nowidow}
\DeclarePairedDelimiter{\ceil}{\lceil}{\rceil}

\newcommand{\eqtopmargin}{-0.1cm}
\newcommand{\eqbottommargin}{-0.3cm}
\newcommand{\eqtopmarginbig}{-0.4cm}
\newcommand{\eqbottommarginbig}{-0.4cm}
\newcommand{\tabletopmargin}{-0cm}
\newcommand{\tablecaptionmargin}{-0cm}
\newcommand{\tablebottommargin}{-0cm}
\newcommand{\figtopmargin}{-0cm}
\newcommand{\figbottommargin}{-0cm}
\newcommand{\figcaptionmargin}{-0cm}
\newcommand{\sectionendmargin}{-0cm}
\newcommand{\sectiontitlemargin}{-0cm}
\newcommand{\subsectionendmargin}{-0cm}
\newcommand{\subsectiontitlemargin}{-0cm}

\newtheoremstyle{rqstyle}
    {\topsep}                    
    {\topsep}                    
    {\itshape}                   
    {}                           
    {\scshape}                   
    {.}                         
    {.5em}                       
    {}  

\theoremstyle{rqstyle}
\newtheorem{researchquestion}{Research Question}
\newtheorem{scicontrib}{Scientific Contribution}
\newacronym{rq}{RQ}{research question}
\newacronym{sc}{SC}{scientific contribution}

\newcommand{\coe}{$CO_{2}e$}

\newacronym{coe}{\coe}{$CO_{2}$ equivalent}
\newacronym{ewma}{EWMA}{exponentially weighted moving average}
\newacronym{dvfs}{DVFS}{dynamic voltage \& frequency scaling}
\newacronym{vm}{VM}{virtual machine}
\newacronym{era}{ERA}{energy reduction assets}
\newacronym{api}{API}{application programming interface}
\newacronym{os}{OS}{operating system}
\newacronym{rtp}{RTP}{real-time pricing}
\newacronym{qos}{QoS}{quality of service}
\newacronym{sla}{SLA}{service level agreement}
\newacronym{rtep}{RTEP}{real-time electricity pricing}
\newacronym{iaas}{IaaS}{infrastructure as a service}
\newacronym{paas}{PaaS}{platform as a service}
\newacronym{saas}{SaaS}{software as a service}
\newacronym{pm}{PM}{physical machine}
\newacronym{pue}{PUE}{power usage efficiency}
\newacronym{cue}{CUE}{carbon usage effectiveness}
\newacronym{cef}{CEF}{carbon emission factor}
\newacronym{hpc}{HPC}{high-performance computing}
\newacronym{db}{DB}{database}
\newacronym{dc}{DC}{data center}
\newacronym{oltp}{OLTP}{online transaction processing}
\newacronym{mse}{MSE}{mean squared error}
\newacronym{ga}{GA}{genetic algorithm}
\newacronym{tc}{TC}{treatment category}
\newacronym{wtp}{WTP}{willingness to pay}
\newacronym{bcf}{BCF}{best cost fit}
\newacronym{bfd}{BFD}{best fit decreasing}
\newacronym{iot}{IoT}{Internet of Things}
\newacronym{bcffs}{BCFFS}{Best Cost Fit Frequency Scaling}
\newacronym{fps}{FPS}{frames per second}
\newacronym{hevc}{HEVC}{high efficiency video coding}
\newacronym{arima}{ARIMA}{Autoregressive Integrated Moving Average}
\newacronym{arma}{ARMA}{Autoregressive Moving Average}
\newacronym{aam}{AAM}{Automatic ARIMA Modeling}
\newacronym{ses}{SES}{Simple Exponential Smoothing}
\newacronym{mape}{MAPE}{Mean Absolute Percentage Error}
\newacronym{it}{IT}{information technology}

\def\name{Dražen Lučanin}

\newtoggle{full}
\newtoggle{publicationURLs}

\newcommand{\gaschedulerEnSavingsNoGeotemp}{$28.6\%$}
\newcommand{\gaschedulerVMNumSimulation}{10k}

\newcommand{\VMPricingTdown}{400 s}
\newcommand{\VMPricingRevenueIncrease}{$43\%$}
\newcommand{\VMPricingCustomerIncrease}{$5.89\times$}
\newcommand{\VMPricingEnSavings}{$39\%$}

\newcommand{\VMPricingslanum}{eight}

\newcommand{\VMPricingslanummult}{1--60}
\newcommand{\VMPricingslasall}{\gls{sla} 1-8}

\newcommand{\VMPricinggadown}{27 minutes}
\newcommand{\VMPricinggaav}{98.12\%}

\newcommand{\VMPricingDRmin}{12.5\%}
\newcommand{\VMPricingDRmax}{66.67\%}

\newcommand{\Multicoreschedulerbase}{\gls{bfd}}
\newcommand{\Multicoreschedulerfreq}{\gls{bcffs}}
\newcommand{\Multicoreschedulermigr}{\gls{bcf}}

\newcommand{\Multicoreensavingsmax}{$14.57\%$}
\newcommand{\Multicoreensavingsfreq}{$14.13\%$}

\newcommand{\Multicoreensavingsmaxintel}{$10.06\%$}
\newcommand{\Multicoreensavingsfreqintel}{$9.86\%$}

\newcommand{\Multicorevmnumsimulationmulticore}{2k}
\newcommand{\Multicorepmnumsimulation}{2k}

\graphicspath{{./img/}{./img/thesis/}{./img/gascheduler/}{./img/volatility/}
{./img/vmpricing/}{./img/powermodel/}{./img/multicore/}{./img/equilibrium/}
{./img/forecasting/}}
\DeclareGraphicsExtensions{.pdf,.png,.jpg}

\makeindex
\makeglossaries
\glstocfalse 

\setauthor{}{Dražen Lučanin}{}{male}

\setadvisor{Univ. Prof. Dr.}{Ivona Brandić}{}{female}

\setfirstreviewer{Univ. Prof. Dr.}{Univ. Prof. Dr. Ivona Brandić}{}{female}
\setsecondreviewer{Univ. Prof. Dr.}{Univ. Prof. Dr. Helmut Hlavacs}{}{male}

\setaddress{Bürgerspitalgasse 22/15, 1060 Wien}
\setregnumber{0848810}
\setdate{21}{4}{2016}

\settitle{Energy Efficient Cloud Control and Pricing in Geographically Distributed Data Centers}{Energy Efficient Cloud Control and Pricing in Geographically Distributed Data Centers}

\setthesis{doctor}
\setdoctordegree{techn.}
\begin{document}

\addtitlepage{naustrian} 
\addtitlepage{english} 

\frontmatter 

\addstatementpage

\begin{acknowledgements*}
\vspace{-0.95cm}

The research leading to this thesis has received funding from the following
projects: (1) the ``Holistic Energy Efficient Approach
for the Management of Hybrid Clouds''
(HALEY) project,
granted under the Vienna University of Technology research award;
(2) several international meeting and training school grants by
\href{http://www.cost804.org/static/index.html}{COST Action IC804\
 ``Energy efficiency in large scale distributed systems''};
(3) the short term scientific mission (STSM) \ 
grant by \href{http://www.nesus.eu/}{COST Action IC1305\
  ``Network for Sustainable Ultrascale Computing'' (NESUS)}.

I would like to thank my supervisor Univ. Prof. Dr. Ivona Brandić
for her support and mentoring during my research
at the Vienna University of Technology.
I would also like to thank Univ. Prof. Dr. Helmut Hlavacs
for providing feedback as an external reviewer of this thesis.
Additionally, I would like to thank Senior Lecturer Dr. Rizos Sakellariou
for our research collaboration and for hosting my research visit
at the University of Manchester.
I am also grateful to my research collaborators
Ilia Pietri, Simon Holmbacka and Foued Jrad
for their help in our joint work.

Big thanks go to my close colleagues and collaborators Soodeh Farkohi,
Toni Mastelić, Ivan Brešković, Michael Maurer and Vincent Chimaobi Emeakaroha
for the discussions, talks and fun we had during our period of working together.
Additionally, I would like to thank the other members of the research groups
I have had the privilege of collaborating with -- the Distributed Systems Group
and the Electronic Commerce Group at the Vienna University of Technology
and the Division of Electronics at the Ruđer Bošković Institute in Zagreb,
Croatia.

Finally, I would like to thank my partner Sara, my family and my friends
who were always there for me when I needed advice or encouragement.
I also have to thank my favorite musicians and bands who created
the many great songs that were a musical inspiration during my work.

\begin{flushright}
Dražen Lučanin\\
April 2016
\end{flushright}






\end{acknowledgements*}


\begin{abstract}

The fast pace of progress in the domain of applications
provided over the Internet has created a need for computational resources
delivered as an on-demand utility, without having to manually manage
computer hardware. Cloud computing has emerged as a very popular paradigm where
resources such as virtual machines are provided as a scalable, pay-as-you-go
service, catering to applications in a multitude of fields.

On the other hand, the rapid cloud computing growth has turned
the energy consumption of data centers hosting
the cloud's hardware infrastructure
into a global environmental problem and a major cost factor.
It is estimated that data centers constitute
1.5\% of global electricity usage.
At the same time, to serve increasing user requirements,
modern cloud providers are operating multiple geographically distributed
data centers.
Distributed data center infrastructure changes\
the rules of cloud control, as\
energy costs depend on current regional electricity prices and temperatures
that we call geotemporal inputs.\
Furthermore, pricing policies at which cloud providers can offer
computational resources depend on the \gls{qos}.
With such pricing schemes
and the increasing energy costs in data centres, balancing energy savings
with performance and revenue losses
is a challenging problem for cloud providers.
%
Existing cloud control methods are
suitable only for a single data center
or do not consider all the available cloud control actions that
can reduce energy costs in geographically distributed data centers.

In this thesis, we propose a pervasive cloud control approach
consisting of multiple methods
for dynamic resource reallocation and hardware configuration adapted to
volatile geotemporal inputs.
The proposed methods consider the \gls{qos} impact
of cloud control actions
and the data quality limits of time series forecasting methods.
We offer a cloud controller design that supports future extensions when
new decision support components need to be added.
We also propose novel pricing schemes which account for the
computational resource availability and costs
that arise from our cloud control approach
to enable both flexible, energy-aware and high performance cloud computing.

We evaluate our cloud control methods empirically and in a number of simulations
using historical traces
of electricity prices, temperatures, workloads and other data.
We estimate the potential energy cost savings by
comparing our methods to state-of-the-art baseline methods.
We explore a variety of input parameters to provide
a range of guidelines for practical application of our methods in cloud systems.
Our results show that significant energy cost savings are possible without
harming the \gls{qos} or service revenue
in geographically distributed cloud computing.


\end{abstract}

\cleardoublepage

\begin{kurzfassung}

Der schnelle Fortschritt im Bereich von Internet-Anwendungen hat es
erforderlich gemacht, Rechenressourcen on-demand zur Verfügung zu stellen
ohne Computer-Hardware manuell managen zu müssen. Cloud Computing hat sich
als sehr beliebtes Paradigma herauskristallisiert, um Ressourcen wie virtuelle
Maschinen mannigfaltigen Anwendungen skalierbar
in Form eines "Pay-as-you-go"-Services zugänglich zu machen.

Andererseits hat das rapide Wachstum von Cloud Computing den Energieverbrauch
von Datenzentren, die die Cloud-Hardware betreiben, zu einem globalen
Umweltproblem und einem gewichtigen Kostenfaktor gemacht. Man schätzt,
dass Datenzentren 1,5\% des globalen Elektrizitätsverbrauchs ausmachen.
Gleichzeitig betreiben moderne Cloud-Provider mehrere geographisch verteilte
Datenzentren, um steigenden Nutzeranforderungen gerecht zu werden.
Eine verteilte Infrastruktur von Datenzentren verändert
die Cloud-Kontrollregeln, da Energiekosten von den aktuellen regionalen
Elektrizitätspreisen und Temperaturen (sog. geozeitliche Einflussfaktoren)
abhängen. Zusätzlich hängt die Preispolitik, über die Cloud-Provider
Rechenressourcen anbieten, vom Quality-of-Service (QoS) ab.
Auf Grund dieser Preisschemata und der steigenden Energiekosten in Datenzentren,
ist das Austarieren zwischen Energieersparnissen und damit einhergehenden
Verlusten an Leistung und Gewinnen ein herausforderndes
Problem für Cloud-Provider. Bestehende Cloud-Kontrollmethoden existieren
nur für einzelne Datenzentren oder berücksichtigen die verfügbaren
Cloud-Kontrollaktionen nicht, die zur Reduzierung von Energiekosten
in geographisch verteilten Datenzentren führen können.

In dieser Dissertation schlagen wir eine durchdringende Cloud-Kontrollmethode
vor, die aus mehreren Methoden für dynamische Resourcereallokation und
Hardwarekonfiguration basierend auf volatilen geozeitlichen
Einflussfaktoren besteht. Die vorgestellten Methoden betrachten den QoS-Einfluss
von Cloud-Kontrollaktionen, sowie Limits für die Datenqualität
von Vorhersagemethoden von Zeitreihen.
Wir präsentieren einen Cloud-Kontrollmechanismus, der zukünftige Erweiterungen,
falls neue Komponenten zur Entscheidungsfindung hinzugefügt werden müssen,
ermöglicht. Außerdem stellen wir neuartige Preisschemata vor,
die die Verfügbarkeit der Rechenressourcen, sowie die Kosten,
die durch unsere Cloud-Kontrollmethode entstehen, in Betracht ziehen,
um flexibles, energiebewusstes und hochperformantes Cloud-Computing
zu ermöglichen.

Die Cloud-Kontrollmethoden werden empirisch und durch Simulationen
evaluiert, die u.a. auf historischen Daten von Elektrizitätspreisen,
Temperaturen und Auslastung beruhen. Die potentiellen Energiekostenersparnisse
werden geschätzt, indem unsere Methoden mit Methoden,
die Stand der Technik sind, verglichen werden. Wir untersuchen
eine Vielzahl an Einflussparametern, um eine Reihe von Empfehlungen
für die praktische Verwendung unserer Methoden in Cloud-Systemen abzugeben.
Die Resultate zeigen, dass signifikante Energiekostenersparnisse möglich sind,
ohne das QoS oder die Gewinne von Services im geographisch verteilten
Cloud-Computing zu beeinträchtigen.


\end{kurzfassung}

\selectlanguage{english}

\setcounter{tocdepth}{1}
\setcounter{secnumdepth}{2}
\tableofcontents 
\newpage
\listoffigures 
\newpage
\listoftables 
\listofalgorithms
\addcontentsline{toc}{chapter}{List of Algorithms}

\chapter*{Previous Publications}
\label{ch:publications}
This thesis is based on work published in peer-reviewed scientific journals,
conferences and workshops. Here we list the six core papers used
as the foundation of this thesis.
Parts of these papers are contained in verbatim without
explicitly being referenced.
Additionally, we also list other publications in the domain of cloud computing
and other fields of computer science
that were not included in this thesis,
but where the author of this thesis is a co-author.

\subsection*{Refereed Publications in Journals}

\begin{enumerate}
\item \textbf{Dražen Lučanin}, Ivona Brandić.\
  \emph{Pervasive Cloud Controller for Geotemporal Inputs.}\
  IEEE Transactions on Cloud Computing, 2016.
  \href{http://dx.doi.org/10.1109/TCC.2015.2464794}{\path{doi:10.1109/TCC.2015.2464794}}
\item \textbf{Dražen Lučanin}*, Ilia Pietri*, Simon Holmbacka*, Ivona Brandic, Johan Lilius, Rizos Sakellariou.\
  \emph{Performance-Based Pricing in Multi-Core Geo-Distributed Cloud Computing.}\
  IEEE Transactions on Cloud Computing (under review, *equal contribution).
\end{enumerate}

\subsection*{Refereed Publications in Conference Proceedings}

\begin{enumerate}\setcounter{enumi}{2}
  \item \textbf{Dražen Lučanin}, Ilia Pietri, Ivona Brandić, Rizos Sakellariou.
    \emph{A Cloud Controller for Performance-Based Pricing.}
    \nth{8} IEEE International Conference on Cloud Computing (CLOUD 2015),\
    27 June -- 2 July, 2015, New York, USA.
    \href{http://dx.doi.org/10.1109/CLOUD.2015.30}{\path{doi:10.1109/CLOUD.2015.30}}
  \item \textbf{Dražen Lučanin}, Foued Jrad, Ivona Brandić, and Achim Streit.
    \emph{Energy-Aware Cloud Management through Progressive SLA Specification.}
    \nth{11} International Conference on
    Economics of Grids, Clouds, Systems, and Services
    (GECON 2014).
    16–18 September, 2014, Cardiff, UK.
    \href{http://dx.doi.org/10.1007/978-3-319-14609-6_6}{\path{doi:10.1007/978-3-319-14609-6_6}}
  \item \textbf{Dražen Lučanin}, Ivona Brandić.
    \emph{Take a break: cloud scheduling optimized for real-time electricity pricing.}
    Proceedings of the \nth{3} International Conference on Cloud and Green Computing (CGC),
    30 September -- 2 October, 2013, Karlsruhe, Germany.
    \href{http://dx.doi.org/10.1109/CGC.2013.25}{\path{doi:10.1109/CGC.2013.25}}
\end{enumerate}

\subsection*{Refereed Publications in Workshop Proceedings}
\begin{enumerate}\setcounter{enumi}{5}
  \item \textbf{Dražen Lučanin}, Michael Maurer, Toni Mastelić, Ivona Brandić.
    \emph{Energy Efficient Service Delivery in Clouds in Compliance with the Kyoto Protocol.}
    \nth{1} International Workshop on Energy-Efficient Data Centers,
    8th May, 2012, Madrid, Spain.
    \href{http://dx.doi.org/10.1007/978-3-642-33645-4_9}{\path{doi:10.1007/978-3-642-33645-4_9}}
\end{enumerate}

\subsection*{Other Refereed Publications}
\begin{enumerate}\setcounter{enumi}{6}
   \item Soodeh Farokhi, Pooyan Jamshidi, \textbf{Dražen Lučanin}, Ivona Brandić.\
     \emph{Performance-based Vertical Memory Elasticity.}\
     \nth{12} IEEE International Conference on Autonomic Computing (ICAC 2015),\
     7--10 July, 2015, Gronoble, France.
     \iftoggle{publicationURLs}{(\href{http://www.infosys.tuwien.ac.at/staff/sfarokhi/soodeh/papers/Soodeh-Farokhi_CameraReady_ICAC-2015.pdf}{preprint})}
     \href{http://dx.doi.org/10.1109/ICAC.2015.51}{\path{doi:10.1109/ICAC.2015.51}}

   \item Toni Mastelić, \textbf{Dražen Lučanin}, Andreas Ipp, Ivona Brandić.\
     \emph{Methodology for trade-off analysis when moving scientific\
       applications to the Cloud.} CloudCom 2012, \nth{4} IEEE International\
     Conference on Cloud Computing Technology and Science.\
     3--6 December, 2012, Tapei, Taiwan.
     \iftoggle{publicationURLs}{(\href{http://ieeexplore.ieee.org/xpl/articleDetails.jsp?arnumber=6427575}{IEEE Xplore})}
     \href{http://dx.doi.org/10.1109/CloudCom.2012.6427575}{\path{doi:10.1109/CloudCom.2012.6427575}}
\end{enumerate}

\begin{enumerate}\setcounter{enumi}{8}
      \item Dragan Gamberer, \textbf{Dražen Lučanin}, Tomislav Šmuc. \emph{Descriptive\
	modeling of systemic banking crises.} The \nth{15} International\
	Conference on Discovery Science (DS 2012), 29--31 October, 2012, Lyon,\
	France. \iftoggle{publicationURLs}{(\href{http://lis.irb.hr/~gambi/DS2012paper/ds2012-for-web.pdf}{preprint},\
	\href{http://bib.irb.hr/prikazi-rad?&lang=EN&rad=601608}{CROSBI})}
        \href{http://dx.doi.org/10.1007/978-3-642-33492-4_8}{\path{doi:10.1007/978-3-642-33492-4_8}}

      \item Matija Gulić, \textbf{Dražen Lučanin}, Nina Skorin-Kapov. \emph{A Two-Phase\
	Vehicle based Decomposition Algorithm for Large-Scale Capacitated Vehicle\
	Routing with Time Windows.} Proceedings of the \nth{35} International\
	Convention on Information and Communication Technology, Electronics\
	and Microelectronics -- MIPRO, 21--25 May, 2012, Opatija, Croatia.\
	\iftoggle{publicationURLs}{(\href{http://ieeexplore.ieee.org/xpl/articleDetails.jsp?tp=&arnumber=6240808}{IEEE Xplore})}

       \item \textbf{Dražen Lučanin}, Ivan Fabek, Domagoj Jakobović.\
	\emph{A visual programming language for drawing and executing flowcharts.}\
	Proceedings of the \nth{34} International Convention on Information\
	and Communication Technology, Electronics and Microelectronics -- MIPRO.\
	23--27 May, 2011, Opatija, Croatia. \textbf{Best student paper award.}\
	\iftoggle{publicationURLs}{(\href{http://arxiv.org/abs/1202.2284}{arXiv},\
	\href{http://ieeexplore.ieee.org/search/srchabstract.jsp?tp=&arnumber=5967331}{IEEE Xplore})}

      \item Matija Gulić, \textbf{Dražen Lučanin}, Ante Šimić, Šandor Dembitz.\
	\emph{A digit and spelling speech recognition system for the Croatian language.}\
	Proceedings of the \nth{34} International Convention on Information and\
	Communication Technology, Electronics and Microelectronics -- MIPRO. 23--27\
	May, 2011, Opatija, Croatia.\
	\iftoggle{publicationURLs}{(\href{http://ieeexplore.ieee.org/search/srchabstract.jsp?tp=&arnumber=5967330}{IEEE Xplore})}
\end{enumerate}


\addcontentsline{toc}{chapter}{Selected Publications}

\mainmatter
\chapter{Introduction}
\label{ch:introduction}

Cloud computing is becoming the prevalent way to deliver modern software. 
The formal definition of cloud computing was given in \cite{foster_cloud_2008}
as \emph{a large-scale distributed computing paradigm that is
driven by economies of scale, in which a pool of
abstracted, virtualized, dynamically-scalable, managed
computing power, storage, platforms, and services are
delivered on demand to external customers over the
Internet.}
In essence, computational resources are provided to users remotely
over the Internet.


The worldwide public cloud services market is projected to reach \$204 billion
according to a 2016 Gartner report \cite{_gartner_2016}.
A reason for the success of cloud computing is that it enables users to
get access to computational resources in a black box manner,
without necessarily having expert knowledge
required to set up and maintain all the underlying technology.
They can instead focus on their own domain of work.
This separation of concerns fuels the modern \gls{it} sector,
enabling numerous research institutions, existing companies as well as emerging
startups to innovate and address problems in all areas of human work.





In this thesis we focus on \gls{iaas} and public clouds.
It is the fastest growing segment of the cloud services market
for several years already \cite{_gartner_2016,_gartner_2013}.
Examples of public \gls{iaas} clouds are the Amazon Elastic Compute Cloud
(Amazon EC2)~\cite{_amazon_2016}
and the Google Compute Engine~\cite{_google_2016}.
Public clouds are a popular cloud computing model where access to
computational resources is provided to customers as a business service
on the open market. Parts of the described work can be applied to other models
as well, however. For example, the \gls{paas} model also relies
on an underlying infrastructure layer,
large private clouds might also require energy efficient management
similar to public clouds etc. We approach our research from
the perspective of a cloud provider, presenting methods mainly meant for
cloud providers to implement.


To describe the \gls{iaas} public cloud model, it is useful to look
at the individual components necessary for such a service. The service provider,
i.e. the cloud provider owns a number of physical machines
located in one or more data centers (warehouse-like buildings equipped
for housing, powering and cooling physical computer infrastructure).
Using virtualisation technology, each \gls{pm}
can host multiple virtual machines, isolated from each other and having access
to a certain amount of physical resources (CPU cores, RAM, storage,
network bandwidth). The \gls{vm}, an \gls{iaas} product,
is offered as a computational resource to customers for usage over the Internet.
The \gls{vm}'s usage is monitored to calculate the total price charged
to the user for the service.
The \gls{vm} pricing, e.g. hourly price, and \gls{qos}, e.g. the \gls{vm}'s
technical specifications and availability, are defined in a \gls{sla},
a contract between the cloud provider and the customer.
We define cloud control as methods for choosing the actions
a cloud provider can apply to \gls{vm}s and underlying \gls{pm}s
that affect computational resource usage and the \gls{qos}.

The high level goal of the research described in this thesis is to
explore and evaluate new
cloud control and \gls{vm} pricing techniques
that improve energy efficiency in \gls{iaas} public clouds.
More specifically,
the focus is on the challenges present in cloud systems consisting of multiple
geographically distributed data centers
where environmental conditions such as electricity prices
and cooling efficiency change over time between data center locations
and thus require new
cloud control and \gls{sla} specification approaches.

To introduce the work presented in this thesis, in the following section
we describe the problems motivating our research.
In Section~\ref{ch:intro:sec:rq} we list the individual research questions
we address. In Section~\ref{ch:intro:sec:contributions} we summarily present the
contributions of our thesis corresponding to the research questions.
In Section~\ref{ch:intro:sec:methodology} we give a general methodological
framework used throughout the thesis. Section~\ref{ch:intro:sec:organisation}
presents the organisation of the remainder of the thesis.


\section{Problem Statement and Research Goals}
\label{ch:intro:sec:problem}





To satisfy growing cloud computing demands, data centers are consuming more and
more energy, accounting for 1.5\%
of global electricity usage~\cite{jonathan_koomey_growth_2011}\
and annual electricity bills of over \$40M
for large cloud providers~\cite{qureshi_cutting_2009}.
In fact, the ICT sector's 2\% global $CO_{2}$ emissions have surpassed
those of aviation \cite{_gartner_2007}, making energy efficiency of data centers
a major environmental issue.
At the same time, a trend of more geographically distributed data centers
can also be seen, e.g. Google has twelve data centers across four continents.
As new paradigms develop, such as smart buildings with
integrated data centers \cite{privat_smart_2013}, computation is more and more
shaping as a distributed utility.
Such cloud deployments result in dynamically changing energy cost conditions
and require new approaches to cloud control.

Assorted technological innovations have brought forth the optimisation
of several independent systems that affect cloud operation, creating
a heterogeneous and dynamically variable environment.
The technologies of the next-generation electricity grid\
known as the “smart grid”, distributed power generation,\
microgrids and deregulated electricity markets have lead to\
demand response and \gls{rtep} options\
where electricity prices change hourly or even by the minute\
\cite{yang_integrating_2013,weron_modeling_2006}.\
Additionally, new solutions for cooling data centers\
(an energy overhead reported to range from 15\% to 45\%\
of a data center's power consumption \cite{barroso_datacenter_2009})\
based on outside air economizer technology result\
in cooling efficiency depending on local weather conditions.\
We call such time- and location-dependent factors \emph{geotemporal inputs}.\
Geotemporal inputs may also include renewable energy availability\
\cite{goiri_parasol_2013},\
peak load electricity pricing \cite{le_reducing_2011}\
or demand response \cite{liu_data_2013,berl_modelling_2013}\
as they constitute time- and location-dependent factors that impact\
the final energy costs as well.\
To reduce energy costs,\
new energy-aware cloud control methods are needed that can leverage
geographical data center distribution together with geotemoral inputs such as
electricity prices \cite{weron_modeling_2006}\
and cooling efficiency \cite{zhou_optimization_2012}.

Furthermore, as IT-based optimisation solutions enter more and more domains,\
we may expect the emergence of new geotemporal inputs in the future.\
Examples include more options for precisely calibrating electricity usage\
and pricing in smart grids, local renewable energy generation,\
further geographical distribution and bringing data centers closer to users\
through smart buildings \cite{privat_smart_2013}.\
As a result, more advanced metering infrastructure\
for quantifying cloud service demand and usage through smart cities,\
smart homes, mobile technology or more generally the\
\gls{iot} will capture more data, allowing cloud providers
to make better informed decisions.\ 
Hence, to account for cloud environment evolution,\
it has to be possible to extend
the cloud controller 
with yet-to-be-realised geotemporal inputs and other factors.




A side-effect of energy-aware cloud control is that it may impact the \gls{qos}.
For example, \gls{vm} availability can be reduced,\
because certain management actions like \gls{vm} migrations\
cause temporary \gls{vm} downtimes \cite{liu_performance_2011}.\
As long as the resulting availability is higher than the value\
guaranteed in the \gls{sla}, cloud providers\
can benefit from the cost savings. However, current cloud providers only offer\
high availability \gls{sla}s, e.g. 99.95\% in case of
Google~\cite{_google_2016} and Amazon~\cite{_amazon_2016}.\
Such \gls{sla}s do not leave enough flexibility\
to apply energy-aware cloud management or result in \gls{sla} violations.
Therefore, new \gls{sla} specification models are needed that facilitate
energy-aware cloud control.



As \gls{iaas} clouds are the fastest growing segment of the cloud
services market~\cite{_gartner_2016}, their business models
are continuously evolving. New \textit{performance-based pricing} pricing models
are being introduced by cloud providers such as ElasticHosts~\cite{elastichosts}
and CloudSigma~\cite{cloudsigma} that radically change the cloud computing
revenue models and require new cloud control approaches.
In performance-based pricing, the cost of \gls{vm} provisioning is based on the
selected CPU frequency along with the allocated amount of RAM and the use of
other resources.
The \gls{vm}'s performance can be modified by choosing from a range of CPU
frequencies and matching prices -- even at runtime.
%
%
A power management action commonly used to reduce energy costs is CPU frequency
scaling \cite{miyoshi2002critical}. Potential energy savings can be estimated
based on geotemporal inputs. CPU frequency scaling actions, however, also cause
service revenue losses in performance-based pricing.
Hence, 
cloud control approaches that balance energy savings and service revenue
losses caused by CPU frequency scaling in performance-based pricing are needed.

\section{Research Questions}
\label{ch:intro:sec:rq}


After giving the detailed problem description in the previous section as a
context for the issues that motivate our research, we describe the
open research questions that we addressed in this thesis.
The first two research questions focus on energy efficiency in cloud control,
while the next two consider the effects of energy-aware cloud
control on \gls{qos}, \gls{sla} specification and pricing. 


\begin{researchquestion}
  How can we empirically reduce energy consumption in \gls{iaas} clouds
  based on real-time electricity prices?
  \label{rq:volatility}
\end{researchquestion}

Existing \gls{vm} scheduling methods that can be applied empirically
in \gls{iaas} clouds typically\
monitor resource usage and performance, adhering to the agreed constraints\
and improving energy efficiency in a best-effort manner by throttling\
any surplus resources \cite{maurer_enacting_2011,beloglazov_energy-aware_2012}.\
What such methods usually neglect is the complexity\
of the electrical grid that powers data centers and that not all\
energy has the same environmental impact -- it differs between\
generators, depends on demand, grid congestion, time of day, weather conditions\
and many other factors.\
An important aspect of current electrical grids is that due to huge demand\
peaks during midday, electricity often gets generated in more environmentally\
unsustainable ways such as by\
diesel generators \cite{andrews_potential_2008}, which is reflected\
in higher real-time electricity prices \cite{klingert_sustainable_2012}.
An open problem, therefore, is to integrate such real-time factors into
existing cloud systems, such as the open source OpenStack cloud manager,
distribute computational resource usage to reduce energy costs
and measure the potential savings.




\begin{researchquestion}
  How can we dynamically control cloud resources based on multiple
  geotemporal inputs to improve energy efficiency and preserve \gls{qos}?
  \label{rq:pervasive}
\end{researchquestion}

With multiple geotemporal inputs and \gls{qos} constraints,
new challenges emerge that
existing cloud control approaches do not wholly address.
Initial \gls{vm} placement based on geotemporal inputs researched
in grid computing~\cite{xu_temperature_2013}
or network request routing~\cite{qureshi_cutting_2009}
never reallocates a running \gls{vm}.
Dynamic \gls{vm} consolidation methods using live \gls{vm}
migrations~\cite{feller_snooze:_2012,beloglazov_managing_2013,
maurer_enacting_2011} are focused on a model suitable
for a single data center, where no cost heterogeneity
inherent to geotemporal inputs is considered.
We explore methods that address all these issues,
so-called pervasive control, i.e.
controlling the cloud's resource allocation dynamically to both\
consolidate resources and adapt to\
multiple geotemporal inputs by utilising cost-efficient data centers
through long-term planning facilitated by time series forecasting.
A challenge in pervasive cloud control
is the \gls{vm} migration action overhead
of the \gls{vm} downtime during the transfer~\cite{liu_performance_2011}.
Forecasting of geotemporal inputs is necessary
to find the optimal balance between energy cost saving
and \gls{vm} migration overhead trade-offs that minimise \gls{qos} degradation.
With time series forecasting that enables long-term planning,
the issue of data quality also
has to be considered to account for the forecasting accuracy and reach.
Additionally, designing a controller to enable
adding new geotemporal inputs is necessary 
to integrate the solution into diverse cloud deployments.






\begin{researchquestion}
  How can we define \gls{sla}s for energy-aware cloud services adapted
  to geotemporal inputs that affect the \gls{qos}?
  \label{rq:sla}
\end{researchquestion}


The challenges of pervasive cloud control and \gls{vm} scheduling\
according to volatile geotemporal inputs
are that too frequent \gls{vm} pausing or migrations to reallocate
the cloud’s resource consumption cause downtimes~\cite{liu_performance_2011}
which harm the \gls{qos} and breach \gls{vm} availability terms
specified in the \gls{sla}.
Various \gls{sla} approaches
alternative to the typical fixed-price \gls{vm}s exist,
such as auction-based price negotiation
in Amazon spot instances \cite{chen_tradeoffs_2011}\
or calculating costs per resource utilisation \cite{berndt_towards_2013}.\
However, the state of the art is still offering only high availability
\gls{vm}s, without enough flexibility to apply energy-aware cloud control.
Hence, estimating availability and price values that can be guaranteed\
in \gls{sla}s for \gls{vm}s managed based on geotemporal inputs\
is still an open research issue. This problem is challenging,\
because exact \gls{vm} availability and energy costs depend\
on electricity markets, weather conditions, application memory access patterns\
and other volatile factors.
Methods for analysing the effects of cloud controllers
to determine the availability and price terms
that can be specified
in \gls{sla}s are necessary to support both energy-aware cloud control
and a predictable \gls{qos}.



\begin{researchquestion}
  How can we adapt cloud control to performance-based pricing,
  modern CPU architectures and variable application workloads?
  \label{rq:pricing}
\end{researchquestion}


Another approach to reduce energy consumption in data centers is by applying
CPU frequency scaling, however this reduces service revenue under
the emerging performance-based \gls{vm} pricing schemes.
Cloud control solutions relying on CPU frequency scaling\
exist, e.g. \cite{von2009power,shi2011towards}, but only consider
fixed \gls{vm} pricing where the trade-offs of energy savings and
performance-based \gls{vm} pricing are not considered.\
The currently existing clock frequency governors available at the operating
system level that adjust CPU clock frequency according to workload changes have
proven to be inefficient in responding to the required \gls{vm} performance
level \cite{holmbacka2015energy,holmbacka2014energy}.
%
Other open challenges are that modern physical machines have multiple
CPU cores with complex utilisation-to-power-dissipation models.
Additionally, besides the traditional CPU architecture like Intel's,
smartphone-technology-based ARM CPU architectures are emerging in
large scale cloud platforms
\cite{rajovic2013supercomputing,francesquini2015benchmark} with significantly
different power models. Finally,
the performance impact of the clock frequency may also vary between different
workloads \cite{spiliopoulos2014power}.\
For example, CPU-bound workloads are more sensitive to the reduction
in frequency, while the performance of I/O-bound workloads is less affected.
The sensitivity of the workload to CPU frequency reduction is called
the workload's CPU-boundedness $\beta$ following the approach
in~\cite{etinski2010optimizing}.\
A cloud control solution has to model and consider
an environment with performance-based pricing, multi-core CPUs,
different CPU architectures and variable workloads
to be of practical relevance.

\section{Scientific Contributions}
\label{ch:intro:sec:contributions}

\begin{figure}
\centering
\includegraphics[width=1.0\columnwidth]{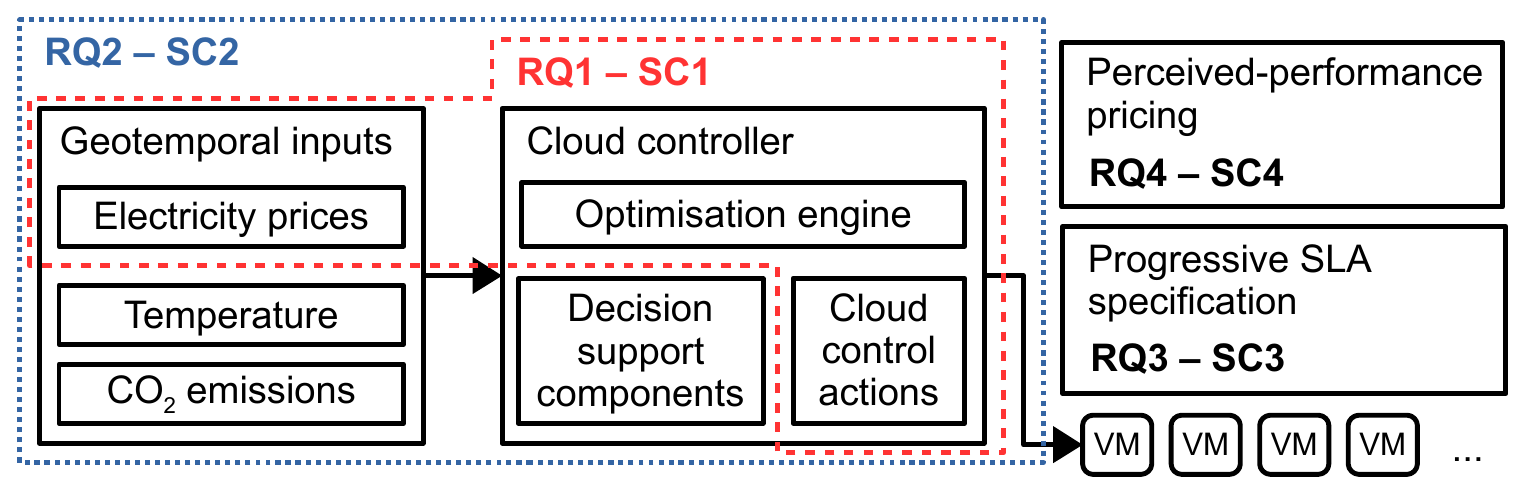}
\vspace{\figcaptionmargin}
\caption{Organisation of scientific contributions.}
\label{ch:intro:fig:organisation}
\vspace{\figbottommargin}
\end{figure}

Based on the research questions identified in the previous section, we now
list the individual contributions presented in this thesis. Each of the
contributions was previously published as a peer-reviewed paper that is
referenced with the corresponding contribution in this section. A diagram
showing the relation between the different research questions and scientific
contributions in a geographically distributed cloud system is shown
in Fig~\ref{ch:intro:fig:organisation}.

\begin{scicontrib}
Experimental evaluation of a \gls{vm} scheduling method for reducing
energy costs under real-time electricity pricing.
\label{sc:peak_pauser}
\end{scicontrib}

We introduce the grid-conscious cloud model\
which relies on\
computation elasticity to optimise energy usage
under dynamic electricity prices.\
Building on this principle, to validate its feasibility in
real-life cloud systems, we present a scheduling method\
that analyses historical electricity price data~\cite{_ameren_2016}\
to determine the most probable peak hours\
and then pauses the managed \gls{vm}s.\
We implemented this scheduling method -- the peak pauser cloud controller
on the industry-standard OpenStack\
cloud manager \cite{ken_deploying_2011}\
to empirically measure its effectiveness.\
We calculated further savings projections\
based on the available assessments of hardware\
used in Google's production environment \cite{fan_power_2007}\
and obtained encouraging results.\
Since pausing \gls{vm}s is an invasive action\
towards the user, we present \emph{green instances}, a business option\
similar to Amazon's spot instances \cite{agmon_ben-yehuda_deconstructing_2011}\
where the user is offered\
a better service price in exchange for reduced \gls{vm} availability.
This contribution addresses \gls{rq}~\ref{rq:volatility}. It was previously
published in~\cite{lucanin_take_2013-1} and is presented
in Chapter~\ref{ch:volatility} of this thesis.




\begin{scicontrib}
  A pervasive cloud controller 
  considering multiple geotemporal inputs
  that unifies \gls{sla} compliance,
  \gls{vm} and \gls{pm} per-resource constraint checking,
  live \gls{vm} migration overhead,
  as well as data center energy costs.
  \label{sc:pervasive_controller}
\end{scicontrib}


In this thesis we propose a novel pervasive cloud controller\
designed for\ 
resource allocation optimisation\
that efficiently utilises cloud infrastructure,\
accounting for geographical data center distribution\
under geotemporal inputs.\
%
%
Our optimisation engine\ 
supports components for\ 
costs based on geotemporal inputs, \gls{vm} migration overheads,\
\gls{qos} requirements and other inputs to be\ 
composed in a unified optimisation problem specification.\
A schedule of \gls{vm} migrations is planned ahead of time\
in a forecast window.\ 
This allows the controller to minimise energy costs by planning\
over a long-term period such as hours or days,\
while retaining the required \gls{qos},\
i.e. not incurring too frequent VM migrations.\
To assess the application of our\ 
pervasive cloud controller\
in diverse cloud deployments,\
we present a number of guidelines showing\ 
how the effectiveness changes under different geotemporal input patterns,\
geographical distributions and forecast data quality.

As a proof of concept evaluation\ 
we present\ 
an implementation of the pervasive cloud controller\
based on a hybrid genetic algorithm\
for optimising the schedule of \gls{vm} migrations.\
A time-series-based schedule representation is developed for integration\
with geotemporal inputs\
to facilitate\
long-term planning.\
A realistic duration-agnostic model,\
with no a priori \gls{vm} lease duration knowledge assumption,\ 
improves the compatibility with real cloud deployments,\
such as Amazon EC2, Google Compute Engine or private\
OpenStack clouds.\ 
Multiple decision support components including energy cost,\
\gls{qos},\ 
migration overhead and capacity constraints\
are combined into an extensible fitness function.

We evaluate the pervasive cloud controller\ 
in a large-scale simulation consisting of \gaschedulerVMNumSimulation{} \gls{vm}s\ 
using historical electricity price\
and temperature traces\ 
to show the resulting energy cost savings and \gls{qos} impact.\
Based on our simulation results, energy cost savings\
can be increased up to\ 
\gaschedulerEnSavingsNoGeotemp{} compared to a baseline scheduling algorithm\
\cite{beloglazov_openstack_2014}\
with dynamic \gls{vm} consolidation.\
Furthermore, we expand the evaluation\
to provide guidelines for cloud providers\
in terms of how different geotemporal input value ranges and\
geographical data center distributions\
affect the method's effectiveness.\
We provide a data quality analysis by evaluating\
the controller under different forecasting errors.\
Finally, we validate the architecture’s extensibility\
by performing the simulation with different subsets\
of decision support components.
This contribution addresses \gls{rq}~\ref{rq:pervasive}. It was previously
published in~\cite{lucanin_pervasive_2016} and is presented
in Chapter~\ref{ch:gascheduler} of this thesis.

\begin{scicontrib}
  A progressive \gls{sla} specification based on \gls{vm} availability and cost
  analysis from historical cloud control actions based on geotemporal inputs.
  \label{sc:sla_specification}
\end{scicontrib}

In this thesis, we propose a novel approach for estimating\
the optimal number of \gls{sla}s, as well as their\
availability and price values under energy-aware cloud management.\
Specifically, we present a method to analyse past traces\
of dynamic cloud management actions based on geotemporal inputs\
to estimate \gls{vm} availability and price values\
that can be guaranteed in an \gls{sla}.\
Furthermore, we propose a progressive \gls{sla} specification\
where a \gls{vm} can belong to one of multiple treatment categories,\
where a treatment category defines the type of energy-aware management actions\
that can be applied. An \gls{sla} is generated for each treatment category\
using our availability and price estimation method.

We evaluate our method by estimating availability and price values\
for \gls{sla}s of \gls{vm}s managed by\
two energy-aware cloud controllers --\
the pervasive cloud controller utilising live \gls{vm} migration
from \gls{sc}~\ref{sc:pervasive_controller}
and the peak pauser cloud controller from \gls{sc}~\ref{sc:peak_pauser}.
We evaluate the SLA specification in a user \gls{sla} selection simulation
based on multi-auction theory~\cite{jrad2013}
using Wikipedia and Grid’5000 workloads to represent multiple user types.
Our results show that more users with different requirements
and payment willingness can find a matching \gls{sla} using our specification,
compared to existing high availability \gls{sla}s.
Average energy savings of \VMPricingEnSavings{} per \gls{vm} can be achieved
due to the extra flexibility of lower availability \gls{sla}s.
Furthermore, we determine the optimal number of offered \gls{sla}s
based on customer conversion.
This contribution addresses \gls{rq}~\ref{rq:sla}. It was previously
published in~\cite{drazen_lucanin_energy-aware_2014} and is detailed
in Chapter~\ref{ch:vmpricing}.



\begin{scicontrib}
  A cloud control method for allocating \gls{vm}s and scaling CPU frequencies
with perceived-performance \gls{vm} pricing and non-linear power modelling.
  \label{sc:freq_scaling}
\end{scicontrib}

In this thesis, we introduce a compound cloud control model which considers
multiple factors representative of modern cloud systems. We combine:
(1) Realistic power modelling accounting for multi-core, Intel and ARM architectures;
(2) Energy cost calculation based on geotemporal inputs;
(3) Performance-based \gls{vm} pricing;
(4) Variable \gls{vm} CPU-boundedness that determines the performance impact of CPU frequency scaling.



To describe real-world power dissipation behaviour, we developed a
non-linear power model based on real experiments performed on multi-core Intel
and ARM CPU architectures representative of modern data center
infrastructures~\cite{rajovic2013supercomputing,francesquini2015benchmark}.
As we show, on such power models, traditional
race-to-idle approaches~\cite{sasaki2013model,seeker2014energy} are no longer
valid, which also motivates the cloud control method we introduce.

To tackle varying \gls{vm} workloads, we propose a novel
perceived-performance pricing scheme for determining
the \gls{vm} price based on the application-level performance.\
This scheme allows energy-aware cloud control that treats
\glspl{vm} differently based on the actual impact that CPU frequency
scaling will have on their workload performance,
considering their measured CPU-boundedness.

To address the data center energy consumption and performance-based \gls{vm}
revenue trade-offs,\
we introduce the \Multicoreschedulerfreq{} cloud controller.\
The controller we propose uses our multi-core power model for ARM and Intel CPU
architectures in an energy calculation method that factors in geotemporal inputs
from multiple geo-distributed data centers.\
To account for performance-based \gls{vm} pricing,
ElasticHosts \cite{elastichosts} and CloudSigma \cite{cloudsigma} price data
was used to model their behaviour and precisely compute the effects
of each CPU frequency level.\
The \Multicoreschedulerfreq{} cloud controller then combines both models in a
two-phase algorithm, where firstly \glspl{vm} are allocated between
geo-distributed data centers and subsequently CPU frequencies are set for each
\gls{pm} where energy savings exceed service revenue losses.\

The controller and the models were mapped onto the Philharmonic simulator
\cite{lucanin2014energy}. Simulations with a wide range of scenarios are used
to estimate the energy savings and service revenue stemming from our
cloud control approach.\ 
The results obtained by the \Multicoreschedulerfreq{} cloud controller are
compared and evaluated using
two baseline controllers~\cite{beloglazov_energy-aware_2012} and historical
traces of real-time electricity prices~\cite{alfeld_toward_2012}
and temperatures~\cite{lucanin2014energy}.
The \gls{vm} CPU-boundedness values used in the simulation are distributed
according to the PlanetLab~\cite{planetlab} dataset of \gls{vm} CPU usage.
The results indicate that energy savings up to \Multicoreensavingsmax{} without
significant service revenue reductions can be achieved using the
\Multicoreschedulerfreq{} cloud controller.
This scientific contribution addresses \gls{rq}~\ref{rq:pricing}. It includes
combined work published in~\cite{lucanin_cloud_2015}
and~\cite{lucanin_performance-based_2016}. The details
are presented in Chapter~\ref{ch:multicore}.



\section{Methodology}
\label{ch:intro:sec:methodology}

\begin{figure}
\centering
\includegraphics[width=1.0\columnwidth]{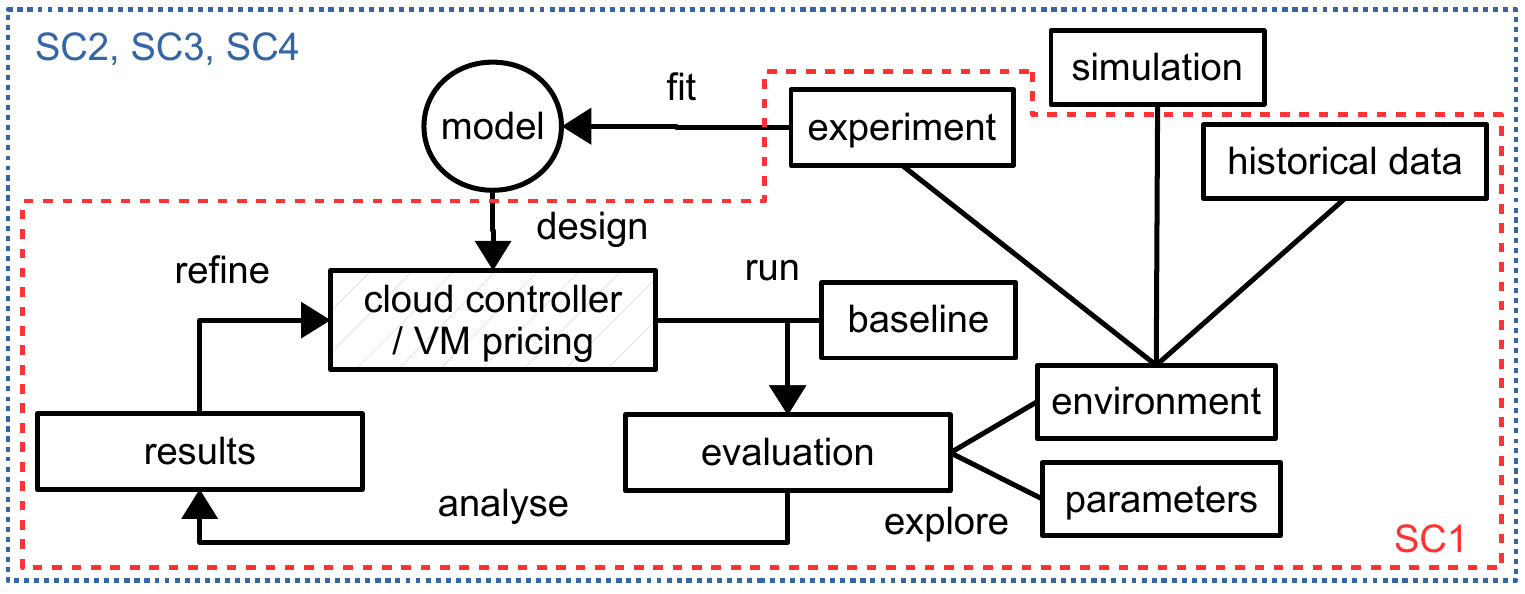}
\vspace{\figcaptionmargin}
\caption{Research methodology.}
\label{ch:intro:fig:methodology}
\vspace{\figbottommargin}
\end{figure}

After an overview of the thesis contributions, in this section we describe the
general methodology applied in each of the contributions to
evaluate their effectiveness towards addressing the
corresponding research questions. Generally speaking, our work combines aspects
of several paradigms:
(1) empirical research where mathematical models are obtained from real-world
data collected in practical experiments;
(2) software simulation, an approach often
used in computer science, where real-world behaviour is approximately imitated
(i.e. simulated) in a computer program developed for this purpose;
(3) data analysis or data science, where statistical methods
are applied to process the data collected from experiments or simulations
and extract conclusions.

To analyse the results, we relied on the data analysis tools and libraries
available in the Python programming language ecosystem.
Most importantly, we used
the Jupyter (formerly IPython) interactive notebook~\cite{perez_ipython:_2007},
the Pandas~\cite{mckinney_data_2010} library for tabular data manipulation
and time series analysis and
the Matplotlib~\cite{hunter_matplotlib:_2007} plotting library
to generate most of the figures in this thesis.

The general applied methodological approach is illustrated
in Fig.~\ref{ch:intro:fig:methodology}. The steps are:

\begin{enumerate}
  \item Model the properties of cloud systems and the effects various
    cloud control various actions have on them. The models are obtained by
    performing real-world experiments, making a hypothesis of the mathematical
    form that could describe the behaviour and fitting the model's parameters
    using methods such as linear regression.
  \item Design a cloud control algorithm based on the modelled
    cloud properties and actions to improve the output metrics, such as energy
    consumption, cost and \gls{qos}.
    Alternatively, in some of the contributions, primarily
    a \gls{vm} pricing scheme is designed.
  \item Define an environment for evaluating the cloud controller as
    a real-world experiment or as a simulation resembling real-world behaviour
    using the available cloud model and historical data traces.
  \item Use said evaluation environment to evaluate the cloud controller and
    possible existing cloud controllers as a baseline to compare the results.
    Based on the results, refine the design of the controller
    and repeat the process. Explore different controller and input
    data parameters in the process as well.
\end{enumerate}

The methodology outlined above with subtle variations is applied
to all our scientific contributions.
We will shortly explain how each of the scientific contributions
adheres to this methodology,
as we also illustrate in Fig.~\ref{ch:intro:fig:methodology}.

In \gls{sc}~\ref{sc:peak_pauser} we evaluate the peak pauser cloud controller
compared to applying no actions as a baseline in an experimental deployment
on the OpenStack cloud manager~\cite{ken_deploying_2011}. We analyse
results collected from a wattmeter to calculate the energy costs
for both scenario based on historical real-time electricity price data
and compare the energy and cost savings.

\gls{sc}~\ref{sc:pervasive_controller} includes several models fitted from
experimental data such as the \gls{vm} migration energy overhead and downtime,
the cooling model and add our own model for combining the cost components.
We use the Philharmonic simulator~\cite{drazen_lucanin_philharmonic_2014}
we developed to evaluate
the proof-of-concept pervasive cloud controller
and compare it to two baseline methods~\cite{beloglazov_energy-aware_2012}.
We analyse the effects of an assortment
of parameters, e.g. the effects of different time series forecasting errors.
Historical electricity price~\cite{alfeld_toward_2012}
and temperature~\cite{_forecast_2015} datasets for US-only and worldwide
data center locations were used in the simulations.

\gls{sc}~\ref{sc:sla_specification} evaluates a progressive \gls{sla}
specification approach using a dataset of historical energy-aware cloud control
traces (obtained in turn using past geotemporal inputs).
The \gls{sla} specification method was evaluated in a user behaviour simulation
based on multi-auction theory~\cite{jrad2013} where the users'
requirements were modelled from historical Wikipedia and Grid’5000 workloads.

Finally, \gls{sc}~\ref{sc:freq_scaling} evaluates a cloud controller based on
\gls{vm} placement and CPU frequency scaling, using a methodology similar
to \gls{sc}~\ref{sc:pervasive_controller} using the Philharmonic simulator
and historical geotemporal inputs. Additional modelling was applied
to make the simulations more practically relevant,
including workloads with different CPU boundedness properties,
a power models for multi-core CPUs and both Intel and ARM architectures.
A variety of parameters were analysed in the results, such as
how different pricing schemes, cloud providers or CPU architectures
affect the potential energy savings.

\section{Thesis Organisation}
\label{ch:intro:sec:organisation}

The remainder of the thesis is organised as follows:

\begin{itemize}
  \item Chapter~\ref{ch:relatedwork} systematically examines the various
    research branches grouping work related to ours. We cover the
    state-of-the-art methods in each of these branches and highlight some of the
    open challenges that we address in our work.
  \item Chapter~\ref{ch:background} gives some more background on geotemporal
    inputs
    and methods to forecast their future behaviour.
    We also enter into the details
    of \gls{coe} trading where \gls{coe} emissions become a cost component
    similar to electricity prices,
    a model we previously
    published in~\cite{lucanin_energy_2012}. 
  \item Chapter~\ref{ch:volatility} offers an empirical approach to designing
    a cloud control method for reducing energy costs in the context of real-time
    electricity pricing. The focus is mostly set on integration with real-life
    systems such as OpenStack and measuring the outcome using wattmeters
    to validate the method's feasibility.
  \item Chapter~\ref{ch:gascheduler} presents
    a pervasive cloud controller that accounts for multiple geotemporal inputs,
    including real-time electricity pricing
    and temperature-dependent cooling efficiency and defines a more general
    approach for managing an extensible set
    of geotemporal inputs and decision criteria.
    We propose a proof-of-concept genetic algorithm
    for controlling \gls{vm} migration in a geographically distributed cloud
    and show substantial energy cost savings possible using the method.
  \item Chapter~\ref{ch:vmpricing} focuses on the \gls{qos} effects energy-aware
    methods from the previous chapters pose. We present a method for progressive
    \gls{sla} specification where \gls{vm}s are priced and controlled
    differently based on the selected \gls{sla}. A customer behaviour simulation
    methodology is presented to evaluate the economical outcome of
    this progressive \gls{sla} system.
  \item Chapter~\ref{ch:multicore} analyses the emergence of performance-based
    pricing schemes in \gls{iaas} markets where \gls{vm} properties such as CPU
    frequency can be changed at runtime and affect the hourly \gls{vm} price.
    We present a detailed multi-core power consumption model based on different
    CPU frequencies. We then propose a novel perceived-performance pricing model
    which also considers the CPU-boundedness properties
    of the \gls{vm} workload. Finally, we propose a cloud control method that
    balances the revenue and energy cost impact to determine optimal
    CPU frequency scaling actions across a geographically distributed cloud.
  \item Chapter~\ref{ch:conclusion} summarises the thesis, gives the concluding
    remarks and outlines limitations and possible future work directions.
\end{itemize}


\chapter{Related Work}
\label{ch:relatedwork}

In the domain of energy efficiency in cloud computing, there is a large body of
related work that has been published. We structure related work into the
following list of categories:
(1) cloud control methods -- cloud resource allocation techniques and
cloud management actions which aim to improve energy efficiency
in cloud computing;
(2) energy efficiency legislation -- existing rules, regulations
and best practice recommendations from optimising the whole
data center efficiency towards optimising its constituting parts;
(3) \gls{sla} management methods -- existing and emerging \gls{vm} pricing
models suited for energy-aware cloud management;
(4) Power modelling -- methods to mathematically describe the power consumption
behaviour of cloud systems and cloud control actions that can be applied to it.
We will examine each of these two groups separately now.

\section{Cloud Control}

Energy efficiency is often considered in cloud control -- a survey is
available in \cite{berl_energy-efficient_2010}).
We categorises cloud control approaches relevant to the work presented
in this thesis into methods for:
(1) scaling CPU frequencies in \gls{pm}s hosting \gls{vm}s,
(2) initial \gls{vm} placement, (3) subsequent dynamic \gls{vm} reallocation
using live migration and (4) the optimisation theory used to find the best
allocation strategy.


\subsection{CPU Frequency Scaling}

Frequency scaling is the focus in many studies with the aim to reduce
power consumption by decreasing the CPU frequency
\cite{etinski2010optimizing,von2009power,wu2014green}. The cloud scheduler
in \cite{wu2014green} sorts and allocates the incoming jobs to \gls{vm}s based
on the user \gls{sla}s. The minimum resource requirements are allocated
to each \gls{vm} and the CPU frequencies of the \gls{pm}s with low load are
reduced so that resource wastage is minimised without affecting the performance
of the executing jobs. In \cite{von2009power}, the proposed scheduler allocates
the queued \gls{vm}s to \gls{pm}s, while reducing the CPU frequencies at runtime
so that \gls{vm} performance requirements can be met, preferring \gls{pm}s that
operate at lower frequencies.
As opposed to related work, the controller we propose in
Chapter~\ref{ch:multicore} scales the CPU frequencies
taking into account the workload CPU-boundedness while controlling
the impact of frequency reduction on the provider's profit.
The impact of frequency scaling on workload performance has been
investigated in many studies
\cite{hsu2003design,etinski2010optimizing,miyoshi2002critical,freeh2007analyzing}. 
The authors in \cite{hsu2003design} introduce a metric to model workload
CPU-boundedness based on the observation that frequency scaling potentials
depend on the workload CPU-boundedness, with less CPU-bound applications being
benefited more.
In \cite{miyoshi2002critical}, the authors investigate the power-performance
characteristics of systems with frequency scaling capabilities and introduce
a metric to determine energy-efficient performance points to operate the system.
This is also the focus in \cite{freeh2007analyzing}, investigating the impact
of frequency scaling on workload performance for different HPC workloads
in order to achieve energy-performance trade-offs.

Although \gls{dvfs} cloud controllers have been proposed before
\cite{hsuan2013cloud,zhuo2014cloud,ioannou2011cloud,alnowiser2014cloud},\
our adaptive approach scales the operating frequencies based
on the \gls{vm} CPU-boundedness and the impact frequency reduction has
on the provider's gross profit under performance-based pricing.
Also, in most papers multi-core modelling is not considered or
is simplified as a linear combination of the number of cores used.
The clock frequency of the systems used in cloud platforms are usually modelled
according to its dynamic power as a product of the clock frequency and the core
voltage. In contrast to such systems, we focus on adopting a more accurate
multi-core power model; still simple enough to be integrated in real systems.
The model accounts for real-world influences more accurately such as the heat
dissipation influencing the static power significantly \cite{hallis2013power}.


Besides for dynamically scaling CPU, other hardware-related approaches include
completely suspending \gls{pm}s~\cite{da_costa_green-net_2009}, which we
analysed in the context of geotemporal inputs in Chapter~\ref{ch:volatility}.
Insights about choosing hardware based on power consumption\
are given in \cite{beloglazov_energy-aware_2012,beloglazov_energy_2010}.\

\subsection{Initial VM Placement}

Looking at the group of methods that only perform initial\
placement and consider geotemporal inputs during host selection,\
the approach was pioneered by Quereshi et al.~\cite{qureshi_cutting_2009}.\
Their work shows the potential of optimising\
distributed systems (adaptive network routing in content delivery networks)\
for \gls{rtep}, estimating savings up to 40\% of the full electricity cost.\
Similar\ 
routing approaches are explored\
in~\cite{rao_minimizing_2010,lin_online_2012,li_towards_2012}\
and considering both electricity prices\
and $CO_2$ emissions in~\cite{doyle_stratus:_2013}.\
Initial placement is also researched\
in the context of map-reduce jobs \cite{buchbinder_online_2011}\
and based both on \gls{rtep} and cooling in computational grids
in~\cite{guler_cutting_2013,liu_renewable_2012}.\
Load balancing where \gls{rtep} are considered
is studied in~\cite{luo_spatio-temporal_2015}.
Both web request routing and map-reduce jobs are different\
than the IaaS \gls{vm} hosting model we are researching in that\
they require only initial placement.\
Methods presented in \cite{guler_cutting_2013,liu_renewable_2012}\
take into account cooling and \gls{rtep} to place jobs in a grid system,
but apply them on a grid system\
where jobs need to be scheduled only initially with no reallocation.\
A theoretical analysis of placing grid jobs\
with regards to electricity prices, job\
queue lengths and server availability\
is given in \cite{ren_provably-efficient_2012}, but without considering a\
realistic model of \gls{vm} migration overhead and \gls{qos} impact.\
The research was extended with a cooling model\
in~\cite{polverini_thermal-aware_2013}.
A power-aware job scheduler with no rescheduling\
and assuming a priori job duration knowledge\
is presented in \cite{yang_integrating_2013}.
Job scheduling that considers an application's data access requirements,
network capacity limits etc. in geo-distributed data centers
is presented in~\cite{yin_joint_2015}.
In \cite{coutinho_workflow_2011} the HGreen heuristic is proposed to schedule
batch jobs on the greenest resource first, based on prior energy efficiency
benchmarking of all the nodes, but not how to optimize a job once it is
allocated to a node - how much of its resources is it allowed to consume. A
similar multiple-node-oriented scheduling algorithm is presented in
\cite{wang_energy-efficient_2011}.
A survey of \gls{vm} placement methods regarding energy efficiency,
\gls{sla}s and\gls{qos} is given in~\cite{pires_virtual_2015}.
Multi-objective approaches \cite{fard_multi-objective_2012,yu_multi-objective_2007}
are also becoming popular to initial allocation,
where it is seen that the scheduler's goal is to satisfy economic cost,
energy consumption as well as reliability constraints, which is nearer to our\
approach of weighing service availability and energy cost and efficiency.

A job scheduling algorithm for geographically distributed data centers with\
temperature-dependent cooling efficiency\
is given in \cite{xu_temperature_2013}, but only for initial job allocation\
and no subsequent migrations.\
A method for using migrations across geographically distributed data centers\
based on cooling efficiency is shown in \cite{le_reducing_2011},\
but for a grid system, where\
job durations are declared by users in advance, unlike cloud systems\
where \gls{vm} deletion is entirely up to the user, in a pay-as-you go scheme.\
Also, no \gls{qos} aspects are considered in the optimisation, which can lead\
to too many migrations per \gls{vm}.

\subsection{Dynamic VM Consolidation}

The dynamic \gls{vm} consolidation group includes methods targeted
at modern \gls{iaas} clouds where live \gls{vm} migration is
used to dynamically reallocate resources and reduce energy consumption.
The topic of using live \gls{vm} migrations to reduce energy consumption in
cloud computing has been widely researched~\cite{feller_snooze:_2012,
beloglazov_managing_2013,maurer_enacting_2011,
beloglazov_energy_2010,
clark_live_2005,chabarek_power_2008,
li_modeling_2011,beloglazov_managing_2013}.
All the methods we found, however, value energy the same, no matter
the time or location, and therefore overlook
the additional challenges and optimisation potential of geotemporal inputs.
Feller et al. proposed a distributed scheduling algorithm for dynamic
\gls{vm} consolidation using live migrations, based on
hierarchical group management~\cite{feller_snooze:_2012}.
Beloglazov et al.~\cite{beloglazov_managing_2013} introduced
a \gls{vm} consolidation method that minimises the migration frequency
in an online controller,
taking future workload predictions into consideration.
A rule-based \gls{vm} consolidation approach
is developed in \cite{maurer_enacting_2011}.
Practical cloud control utilising \gls{vm} migrations
with a focus on high scalability in production VMware systems
is researched in \cite{kesavan_practical_2013}.
A \gls{vm} allocation method for gaming clouds that considers \gls{qos}
and revenue aspects is presented in~\cite{hong_placing_2015}.
Deshpande et al.~\cite{deshpande_traffic-sensitive_2015} present a method for
coordinating live \gls{vm} migration considering network traffic usage
to not affect application network requirements.
Consolidation based on RAM and CPU usage is
researched and evaluated on a real data center
in \cite{mastroianni_probabilistic_2013}.
Aikema et al. \cite{aikema_energy-cost-aware_2011} show a method of rescheduling
low-priority jobs in clusters to improve the data center's
carbon footprint, which goes in the direction
of our green instances in the context of computational grids.
A more adaptive method for detecting changes in \gls{qos} requirements
and subsequent calibration of virtualised resources
is presented in~\cite{liu_aggressive_2015}.
Low-level \gls{vm} migration time, energy and bandwidth metrics are studied in\
\cite{liu_performance_2011,aikema_green_2012-1,nguyen_improvement_2014}.

\subsection{Pervasive Control}
In a group we named pervasive cloud, \gls{vm}s are\
dynamically migrated to adapt both to user requests and\
changes in geotemporal inputs that enable energy cost savings,\
while considering forecast data quality and \gls{qos} requirements.\
As already stated, to the best of our knowledge, no cloud control method\
solves this problem completely. There have been advances in this direction,\
however.\
There have been initial advances in this direction.\
Cauwer et al. \cite{cauwer_study_2013} presented a method\
of applying time series forecasting of electricity prices\
to detect how a data center's resource consumption should be controlled in a\
geographically distributed cloud, but for a simplified model with no\
concrete actions that should be applied on \gls{vm}s.\ 
Determining exact per-\gls{vm} actions is a challenging trade-off problem,\
between closely following volatile geotemporal input changes and\
minimising the number of migrations to retain high \gls{qos},\
as we will examine in Chapter~\ref{ch:gascheduler}.\
Time series forecasting is used 
in the cloud computing domain in~\cite{calheiros_workload_2015} to predict
application workload in a \gls{saas} cloud to better utilise computational
resources.
Abbasi et al. \cite{abbasi_dynamic_2011} started researching migrating\
\gls{vm}s based on \gls{rtep}, but for a limited\
workload distribution scenario where a \gls{pm} hosts only a single \gls{vm}.\
Additionally, temperature-dependent cooling energy overhead and\
forecasting errors are not considered.\
Our work addresses these challenges through a holistic model supporting\
multiple decision support components and\
long-term planning facilitated by forecasting\
and data quality assertion.\



\subsection{Optimisation Algorithms}

Finally, we look at work related to ours from an algorithmic perspective.
Approaches for schedule optimisation based on a forecast horizon are
explored as rolling-horizon real-time task scheduling in
\cite{zhu_real-time_2014} and a genetic algorithm for this purpose
was used for multi-airport capacity management
with receding horizon control in \cite{hu_multiairport_2007}.
A genetic algorithm for optimising energy consumption in computational grids
using \gls{dvfs} is presented in \cite{kolodziej_hierarchical_2013}.
Ant colony optimisation, also an evolutionary optimisation algorithm is
used in~\cite{tiwari_improved_2016} to schedule tasks in grid computing.
Tabu search, a related meta-heuristic optimisation method,
is used for static data center location
and capacity planning
with a focus on network traffic in \cite{larumbe_tabu_2013}.
Another multi-objective evolutionary method is biogeography-based optimisation,
mimicking the distribution of biological species through time and space,
which was used in~\cite{zheng_multi-objective_2015} for \gls{vm} placement
that minimises power consumption, migration time and network traffic.
Other approaches use deterministic approaches, such as constrained
optimisation problem solving which is used in~\cite{chiang_efficient_2015}
to determine power saving policies that reduce idle \gls{pm} power.
However, multi-objective evolutionary algorithms are still an active
research topic~\cite{qiu_adaptive_2016,li_stable_2014,ma_multiobjective_2016},
so it is likely that they will find more usage in real-time cloud control
scenarios with time.

\section{Energy Efficiency Legislation}



Measures of controlling energy efficiency in data centers do exist -- metrics
such as power usage efficiency (PUE)~\cite{_green_2012-2}, carbon usage
efficiency (CUE), water usage efficiency (WUE)~\cite{_green_2012} and others
have basically become the industry standards through the joint efforts of policy
makers and cloud providers gathered behind
The Green Grid consortium~\cite{_green_2012-1}.
The problem with these metrics, though, is that they only
focus on the infrastructure efficiency -- turn as much energy as possible into
computing inside the IT equipment. Once the power gets to the IT equipment,
though, all formal energy efficiency regulation stops, making it more of a
black-box approach. For this reason, an attempt is made in our work to bring
energy efficiency control to the interior operation of clouds -- resource
scheduling.

So far, the measurement and control of even such a basic metric as PUE is not
mandatory. It is considered a best practice, though, and agencies such as the
U.S. Environmental Protection Agency (EPA) encourage data centers to measure it
by rewarding the best data centers with the Energy Star award
\cite{_energy_2016}.

A good overview of cloud computing and sustainability is given in
\cite{garg_environment-conscious_2011}, with explanations of where cloud
computing stands in regard to \coe\ emissions. Green policies for scheduling are
proposed that, if accepted by the user, could greatly increase the efficiency of
cloud computing and reduce \coe\ emissions. Reducing emissions is not treated as
a source of profit and a possible way to balance SLA violations, though, but
more of a general guideline for running the data center to stay below a certain
threshold. We discuss cloud computing in the context of the Kyoto protocol
and its cap-and-trade system in Chapter~\ref{ch:background}.


\section{SLA Management Methods}

The disadvantages of current \gls{vm} pricing models\
relying on constant rates\
have been shown by Berndt et al. \cite{berndt_towards_2013}
using a game theory approach.\
They propose a method for variable \gls{vm} pricing,\
based on the actual \gls{vm} utilisation.\ 
A new charging model for \gls{paas} providers, where variable-time requests\
can be specified by the users,\
is developed in \cite{vieira_towards_2013}.\
A fine-grained pricing scheme for \gls{iaas} providers was proposed
in~\cite{jin_towards_2015} accounting for overhead costs such as
the \gls{vm} startup time.
Ibrahim et al. applied machine learning\ 
to compensate for interference\
between \gls{vm}s in a pay-as-you-go\
pricing scheme \cite{ibrahim_towards_2011}.\
Though related, Amazon spot instances \cite{chen_tradeoffs_2011}\
permanently terminate\
\gls{vm}s that get outbidden, hence requiring fault-resilient\
application architectures.\
Aside from this, they\ 
perform exactly like other Amazon instances, \ 
again not allowing temporary downtimes necessary every day\
for energy-aware cloud management.\
Alternative auction-based pricing systems are explored
in~\cite{mashayekhy_physical_2015,bonacquisto_procurement_2015}.
Amazon spot instances and similar auction-based schemes do show that
end users are willing to accept\
a more complex pricing model to lower their costs for certain applications,\
indicating the feasibility of such approaches in real cloud deployments.\
Zhang et al. \cite{zhang_dynamic_2011} propose model predictive control\
as a method for maximizing\
the provider's revenue by matching customer demand\
in terms of supply and prices.\
Toosi et al.~\cite{toosi_revenue_2015} present a method for managing
a diverse set of pricing options, such as on-demand or
spot instances, and solving the optimisation problem of
allocating data center capacity to each data plan.
Cloud market mechanics are analysed in \cite{breskovic_creating_2013}, where\
service standardisation and automatic adaptation to user requirements\
using clustering algorithms and monitor is explored\
and in \cite{breskovic_maximizing_2013} where\
a modified cloud market liquidity metric is proposed\
to monitor market efficiency.\
A market-based system based on bidding and adapting to user needs
is discussed in \cite{chase_managing_2001}.

But for many of these energy management techniques to have a significant impact,
computation has to be flexibly offset, at which point \gls{qos} concerns become
an issue and it becomes problematic to choose the users most appropriate
for such green services. To resolve this issue, the authors in \cite{Buyya2012}
proposed a green cloud architecture, whereby ``green broker'' middleware manages
the selection of the ``greenest'' cloud provider for users.
The green cloud broker depends on a Carbon Emission Directory (CED)
that conserves all the data related to energy efficiency of the clouds
and a Green Offer Directory (GOD). In contrast, the \gls{sla} management
approach we present in Chapter~\ref{ch:vmpricing} relies on
\gls{qos} requirements and pricing information, letting a service broker
select the best match from a competitive market dominated by energy-conscious
clouds. Another expansion of the typical \gls{sla} negotiation process
is presented in \cite{klingert_sustainable_2012}, where GreenSLAs are used to
define energy-related cloud provider properties.
None of the mentioned pricing approaches consider\ 
energy-aware cloud management based on geotemporal inputs or\
the accompanying \gls{qos} and energy cost uncertainty,\
which is the focus of our work.

The basis of our service broker's economic model in Chapter~\ref{ch:vmpricing}
is the utility-based algorithm, which is adopted from multi-attribute
auction theory \cite{asker08}, to describe user preferences as a function
of weighted SLA attributes taking into account the user's payment willingness.
This economic model is well-established in literature in the context of
web services for the optimisation
of service value networks \cite{Lamparter2007b} and for the service portfolio
design \cite{Knapper2011}. It has also been applied in the context of
smart grids \cite{flath2013flexible} for the evaluation of variable electricity
rates in order to achieve efficient matching of electricity consumer needs
and generator capabilities.


\section{Power Modelling}



The literature has shown many power models and approaches to model
the power dissipation in computer systems
\cite{martinez2010model,rauber2012model,tudor2012model,sasaki2013model,cupertino2014model,shao2013model}.
Most models are constructed bottom-up from physical characteristics
on top of which practical aspects such as frequency scaling is applied.
The dynamic power dissipation is expressed in many examples
\cite{cho2010model,shen2012model,bharathwaj2013model} as the relation
$f \cdot v^\alpha$ where $f$ is the clock frequency, $v$ is the core voltage
and $\alpha$ is a constant used to comply with the real platform
as close as possible.

As the dynamic power dissipation can be expressed
with this simple bottom-up formula,
the ever growing static power proves more difficult.
The leakage current causing the static power is expressed in~\cite{kim2003power}
as a relationship between transistor gate width, thermal voltage
and architectural parameters such as the insulation material.
Moreover, leakage is also caused by electron tunnelling through the insulator.
This means that a bottom-up modelling of static power
is significantly more difficult. We instead used a top-down view
of the power model, purely based on real-world experiments,
which provides a more realistic view of the complete system including cores,
buses, memories, temperature, operating system influence and other software.



\chapter{Background on Geographically Distributed Cloud Computing}
\label{ch:background}



As detailed in the previous chapter, cloud computing is revolutionizing the
ICT landscape by providing scalable and efficient computational resources
on demand. However, the data centers hosting these computational resources
are responsible for considerable energy consumption and subsequent $CO_2$
emissions. In this chapter, we provide background on the energy consumption
in the context of geographically distributed clouds.
%
%
An opportunity to optimise computational resource management lies in the fact
that more and more cloud providers own geographically-distributed data centers,
with temporal variation of energy and cost efficiency at different
locations -- different cooling efficiency factors and real-time electricity
prices. To tap into such opportunities, there is a need for automatic,
fast and precise analysis of dynamically-changing variables, providing
insights into their underlying characteristics and future behaviour.
Time series analysis and forecasting is a statistical tool that provides
such capabilities. In this chapter we give a comparative description
and evaluation of two methods for time series forecasting. Automatic
Autoregressive Integrated Moving Average Modelling is an expert system
that performs adaptive data preprocessing and model fitting resulting
in a stationary process representing the series. The Theta method is
based on exponential smoothing of the series with random drift.
Both methods participated in the M3 competition where they were evaluated
on 3003 time series datasets. Additionally, we performed our own evaluation
on real-time electricity price and temperature datasets, important
in cloud computing. We give an overview of the results of the M3 competition
and our simulation in the context of cloud computing.
%

Another geotemporal input is the \gls{coe} cost of the energy used by the
data centers.
It is possible that cloud providers might one day be faced with legislative
restrictions, such as the Kyoto protocol,
defining \gls{coe} caps.
A lot has been done around
energy efficient data centers, yet there is very little work done in
defining flexible models considering $CO_2$.
In this chapter we present a first attempt of modelling data centers in compliance
with legislation such as the Kyoto protocol.
We discuss a novel approach for trading credits for
emission reductions across data centers to comply with their constraints.
$CO_2$ caps can be integrated with Service Level Agreements and juxtaposed
to other computing commodities (e.g. computational power, storage),
setting a foundation for implementing next-generation schedulers
and pricing models that support Kyoto-compliant
$CO_2$ trading schemes.


This chapter is structured as follows --
forecasting of geotemporal inputs
is analysed in Section~\ref{ch:bg:sec:forecasting}.
In Section \ref{ch:bg:sec:methods}, we first examine
each method for time series forecasting individually -- \gls{aam}
and the Theta method.
We examine their underlying theoretical mechanisms (autoregressive versus
exponentially-weighted moving averages) and the impact they impose.
We follow by describing the empirical benchmarks of both methods as part of the
M3 competition and in our own simulation where we evaluate them on cloud-related
data in Section~\ref{ch:bg:sec:evaluation}. We give an overview
of the results in Section~\ref{ch:bg:sec:results}.
In Section \ref{ch:bg:sec:equilibrium}, we detail the \gls{coe} emission
legislation's implications to cloud computing.
Section \ref{ch:bg:sec:kyoto} gives some background as to why cloud computing
might become subject to the Kyoto protocol.
Section \ref{ch:bg:sec:model} presents our
model in a general \coe-trading cloud scenario, we then go on to define a formal
model of individual costs to find a theoretical balance and discuss the
usefulness of such a model as a scheduling heuristic.
Section \ref{ch:bg:sec:conclusion} gives a summary of the chapter.




\section{Forecasting Geotemporal Inputs}







\label{ch:bg:sec:forecasting}

\begin{figure}
\centering
\includegraphics[width=0.7\columnwidth]{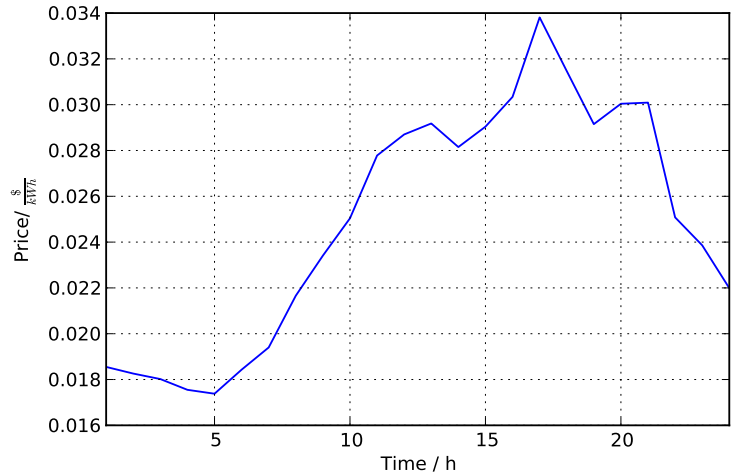}
\caption{A time series of hourly electricity prices in Illinois,
USA within a single day (visualisation of the
data available in~\cite{_ameren_2016}).}
\label{fig:illinois-1day}
\end{figure}


To fully optimise energy consumption in cloud computing, it is necessary
to consider
temporal and geographical variations in electricity prices
and colling efficiency.\
They offer an opportunity to e.g. migrate \gls{vm}s to regions with
optimal dynamic conditions.\
This way we prioritize \gls{vm} execution in energy- and cost-efficient\
space-time windows.
Two example geotemporal inputs that affect geographically distributed
data centers are:
\
\begin{enumerate}
\item \emph{Real-time electricity prices.} In regions with deregulated\
electricity markets, electricity is available at prices that change\
hourly with generation, distribution and demand conditions\
\cite{weron_modeling_2006}.\
We call this purchasing option \gls{rtep}.
Such markets exhibit a price volatility up to two orders of magnitude\
higher than that of other commodities \cite{weron_modeling_2006}.\
For the market we observed~\cite{_ameren_2016}, hourly electricity price peaks\
are more than double the low value within the same day on the average\
(visible in Fig.~\ref{fig:illinois-1day}).
\item \emph{Cooling efficiency.} To cool the data center,\
additional energy is necessary and the exact amount depends on the outside\
air temperature -- cheaper cooling at low temperatures and\
more expensive cooling at high temperatures \cite{guler_cutting_2013}.\
Air temperature also changes dynamically, depending on weather conditions.
\end{enumerate}
\
To be able to make informed decisions in cloud management based on these
dynamic, time-dependent variables, knowledge of their future behaviour
is necessary. Time series forecasting, a large sub-field of statistics
is concerned with exactly such problems \cite{brockwell_introduction_2002}.
Time series forecasting is used across many domains such as finance,
macro-economy, weather forecasting, signal processing etc. and it is a technique
to quickly and automatically predict future variable values. In this section the
aim is to give a survey of the state-of-the-art techniques for time series
forecasting, their precision and suitability in the context of data important
for resource management in cloud computing.

To collect information on the quality of available forecasting methods,
the M3 competition \cite{makridakis_m3-competition:_2000} was held in 2000.
It evaluated the available methods on the same dataset encompassing 3003
time series from different domains. In this section we will examine and compare
two interesting methods that took part in this competition.
The first method offers a way to automatically build \gls{arima} models and is
therefore known as \gls{aam}. The \gls{aam} method is presented in
\cite{melard_automatic_2000}. It predicts future behaviour of a time series
based on the statistical non-stationary data properties.
The Theta method presented in \cite{hyndman_unmasking_2003}
is an empirically-successful time series forecasting method (winner of the
M3 competition \cite{makridakis_m3-competition:_2000}), shown to be equivalent
to \gls{ses} -- an ad-hoc data smoothing procedure.

After defining the methods, we analyse their effectiveness
in the M3 competition, restating some of the findings important
for our survey from \cite{makridakis_m3-competition:_2000}. The Theta method
proved to be the most successful general-purpose forecasting method,
while the \gls{aam} method achieved better results when there was a longer
history available.


\subsection{Methods for Time Series Forecasting}
\label{ch:bg:sec:methods}

We define a time series as observations ordered in time:

\begin{equation}
x_t\ ;\ t=1,\ldots , T ,\, x_t \in \mathbb{R}^n
\end{equation}

For $n=1$ we have a time seris of scalar values -- e.g. hourly observations of the electricity price in Illinois illustrated in Fig. \ref{fig:illinois-1day}, while for a larger $n$ we have multi-dimensional data for every point in time.
As a simple example of a time series that will be used to illustrate forecasting, daily temperatures in Kraków for four days are shown in Table \ref{tab:krakow}.

\begin{table}[h]
\centering
\caption{Daily temperatures in Kraków~\cite{_accuweather.com_2016}}
   \begin{tabular}{l | *{4}{l}}
   \noalign{\bigskip}
	Date (June, 2013) & 18 & 19 & 20 & 21 \\
	\hline
	Temperature (C) & 31 & 32 & 26 & 26 \\
   \end{tabular}
\label{tab:krakow}
\end{table}

\begin{figure}
\centering
\includegraphics[width=0.65\columnwidth]{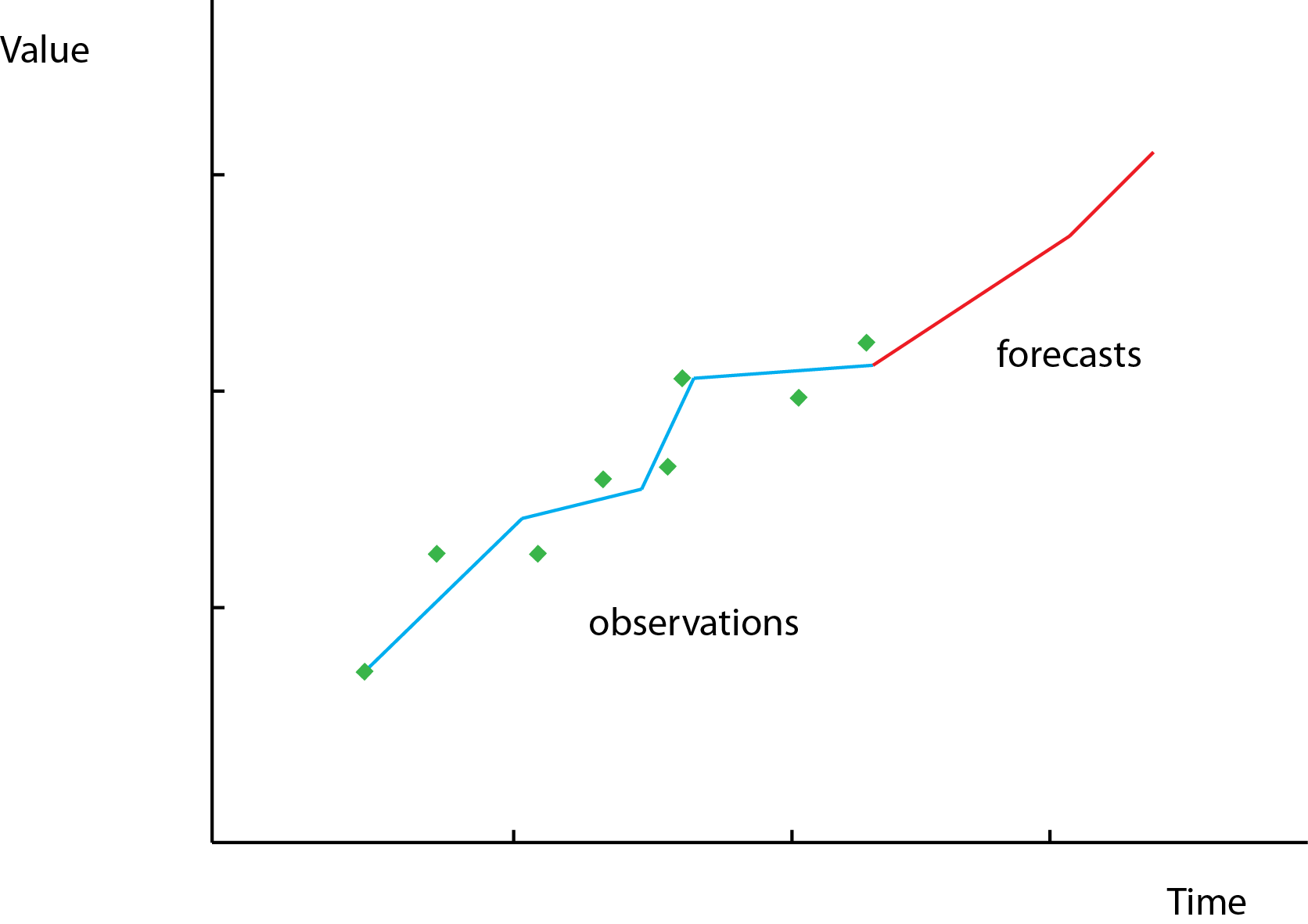}
\caption{An illustration of time series forecasting.}
\label{ch:bg:fig:forecasting}
\end{figure}

The goal of a time series forecasting method is to take as input a series $x_t$
of time-stamped values over a period in history $t=1,\ldots T$ and produce
as output the predicted values for the predetermined number of future
time stamps $x_t,\ t=T+1,\ldots T+m$. An illustration of this process
is shown in Fig.~\ref{ch:bg:fig:forecasting}.
A statistical model (the full line) is built based on known past observations
(green dots) to predict future behaviour (the red part of the line).
In the remainder of this section, we will show how this is performed
by two methods -- \gls{aam} and the Theta method.

\subsubsection{Automatic ARIMA Modeling}
\label{sec:AAM}

An \gls{arima} process is a statistical model used to describe the behaviour of a time series. It is a generalisation of the \gls{arma} model. First, an initial step is performed to detect and remove non-stationarity\footnote{Stationary processes are stochastic processes whose joint probability distribution does not change when shifted in time.} from the data. After this, the time series $X_t$ is represented by an $ARIMA(p',q)$ model:

\begin{equation}
\left(
  1 - \sum_{i=1}^{p'} \alpha_i L^i
\right) X_t
=
\left(
  1 + \sum_{i=1}^q \theta_i L^i
\right) \varepsilon_t \,
\end{equation}

where $L$ stands for the lag function, $\alpha_i$ and $\theta_i$ parameters have to be fitted on the training data and $(p',q)$ parameters define the order of the autoregressive and moving average parts.
Since fitting an ARIMA process to match the empirical data requires statistical and domain-specific knowledge, Mélard and Pasteels built \gls{aam}, an expert system \cite{melard_automatic_2000} that performs automatic, self-calibrating \gls{arima} parameter fitting. Some of the features offered by the \gls{aam} expert system are:

\begin{itemize}
	\item Keeping the user informed of the intermediate and final results.
	\item The model can be modified afterwards.
	\item Adjusting the level of automation according to the expertise of the user.
	\item Regarding only the most "substantive models" for lag structures.
	\item Working efficiently in case of a large number time series dataset.
	\item Intervention analysis for the appropriate treatment of outliers
\end{itemize}

This way, complex statistical models of a time series can be built without having to be an expert in the field.
The automatic steps applied by the expert system are:

\begin{enumerate}
\item Determining the seasonal difference -- the Kruskal and Wallis test of seasonal differences is performed, while the autocorrelation function is used to check if the lag coefficient stays normal after the elimination of the seasonal component.
\item Choosing the transformation -- the Tau test of rank correlation coefficients is used to find an appropriate transformation of the data to build the \gls{arima} model.
\item Intervention analysis -- a step that automatically handles the outliers in the processed series (e.g. removes them).
\end{enumerate}

The resulting model is checked by analysing the convergence of the optimisation process and afterwards for stationarity and invertibility of the autoregressive and moving average polynomials.

\subsubsection{The Theta Method}
\label{sec:Theta}

Assimakopoulos and Nikolopoulos first introduced the Theta method in 2000 \cite{assimakopoulos_theta_2000}, later than most of the classical forecasting methods still used today. Since it proved to be the the best general-purpose method on the M3 competition \cite{makridakis_m3-competition:_2000} and contained complex algebraic operations, Hyndman and Billah analysed it in detail and presented a clearer, simpler form of the method in \cite{hyndman_unmasking_2003}, equivalent to the original.

\begin{figure}
\centering
\includegraphics[width=0.7\columnwidth]{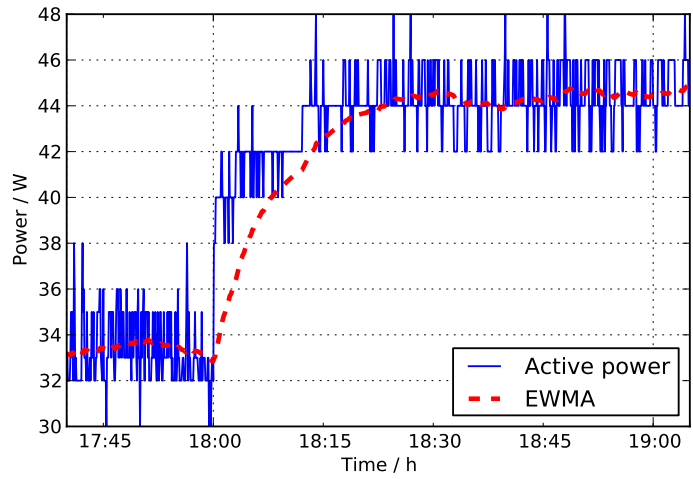}
\caption{An illustration of simple exponential smoothing on a time series of server power consumption samples (obtained from the author's experimental testbed).}
\label{fig:pause_power}
\end{figure}

The Theta method is unveiled in \cite{hyndman_unmasking_2003} to be a special case of \gls{ses}, a method for smoothing data for presentation or forecasting. It models the point as the average of a historical window where each point is multiplied by an exponentially decreasing parameter. A graph of two time series -- the original and its exponential smoothing is shown in Fig. \ref{fig:pause_power}. For a detailed description of \gls{ses} see e.g. \cite{brockwell_introduction_2002}.
The \gls{ses} represents a series $X_t$ as:

\begin{align}
s_1& = x_0 \label{eq:ses-start} \\
s_t& = \alpha x_{t-1} + (1-\alpha)s_{t-1},\ t>1 \label{eq:ses}
\end{align}

where $0<\alpha<1$ determines how much older values influence future values and can be fitted based on the training dataset.
This substitution of the original series produces an exponentially-weighted moving average, which can be seen by expressing $s_t$ using only the original series elements:

\begin{align}
s_t& = \alpha \sum_{i=1}^{t-1}\left[(1-\alpha)^{i-1} x_{t-i}\right]
+ (1-\alpha)^{t-1} x_0.
\end{align}

From this expression it can be seen that the values closer to the currently estimated point carry a higher weight parameter, therefore having a stronger impact. The farther back in past we go, the less the values impact the estimation.
To forecast unknown values using \gls{ses}, for the first point $T+1$ the expression \eqref{eq:ses} can be used, where $s_{t+1}$ is the predicted value for $x_{t+1}$
For subsequent values, though, there is are no more observations of $x_t$,
so a modified expression -- \emph{bootstrapping of forecasts} is applied
where the last known observation $x_T$ is always used:

\begin{align}
s_{t+1}& = \alpha x_{T} + (1-\alpha)s_{t},\ t>T
\end{align}

\begin{figure}
\centering
\includegraphics[width=0.65\columnwidth]{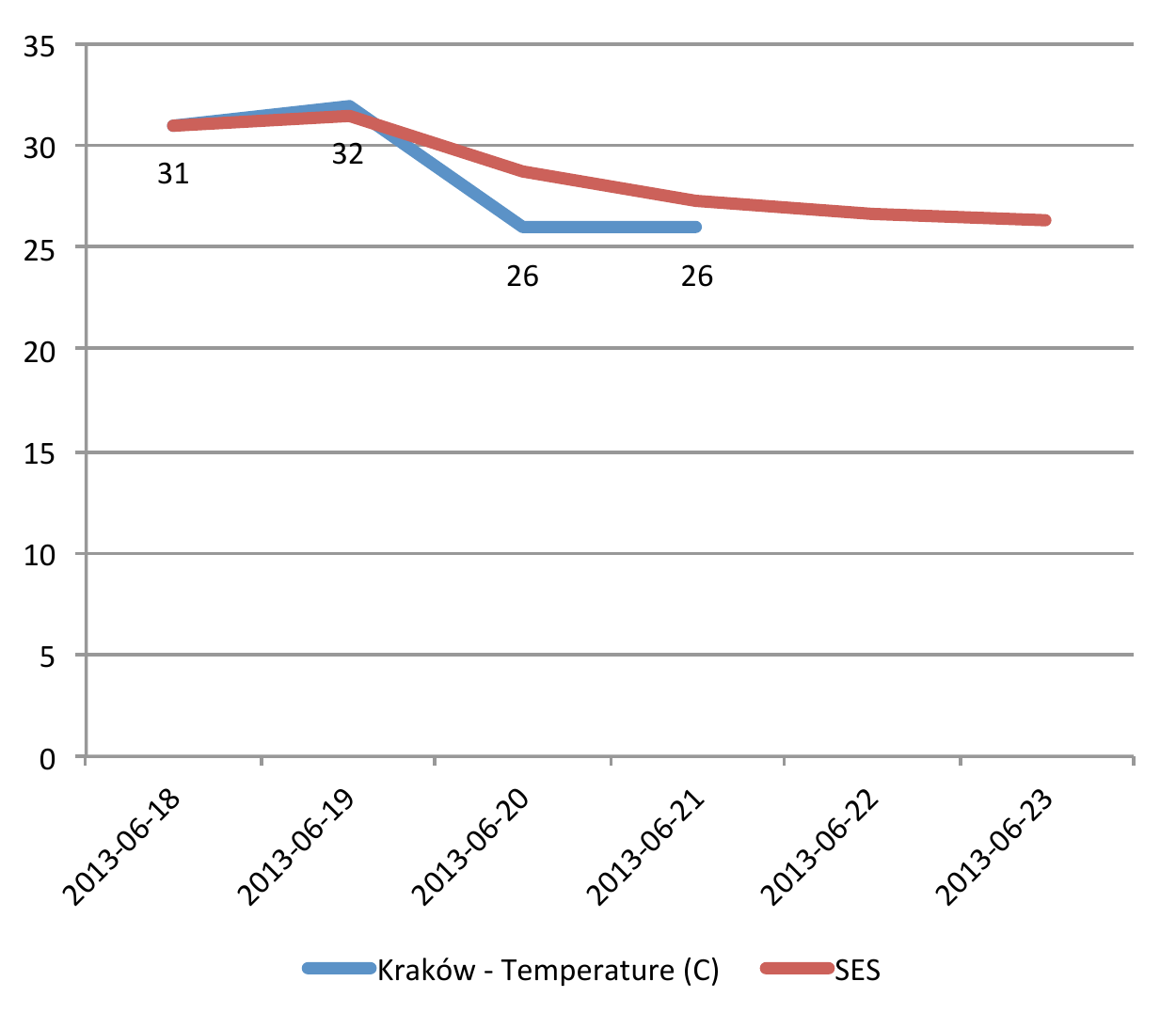}
\caption{Simple exponential smoothing applied on the example from Table \ref{tab:krakow}.}
\label{fig:krakow}
\end{figure}

Coming back to the example of daily temperature values in Kraków from Table \ref{tab:krakow}, we can apply the formula for \gls{ses} to calculate the model. According to \eqref{eq:ses-start} $s_0 = x_0 = 31$. For subsequent cases, expanding \eqref{eq:ses} we get $s_1=31.5$, $s_2=31.5$ etc. By applying bootstrapping of forecasts, we can start predicting unknown values (taking the last known observation $26\ C$): $s_5=26.7$, $s_6=26.3$\ldots This example is shown in Fig. \ref{fig:krakow}.
Additionally, as determined by \cite{hyndman_unmasking_2003}, the Theta method adds a random drift of the form:

\begin{align}
s'_t = s_t + e_t
\end{align}

where s(t) is the deterministic \gls{ses} model and $e_t$ is a zero-long-run-mean stationary random variable.

\subsection{Evaluation}
\label{ch:bg:sec:evaluation}

In this section we will give a short description of the procedures applied to
evaluate and compare different forecasting methods
in the M3 competition~\cite{makridakis_m3-competition:_2000} and in
our own simulation where we used data important for managing cloud resources.

\subsubsection{The M3 Competition}

The 3003 series of the M3-Competition were selected on a quota basis to include various types of time series data (micro, industry, macro, \ldots) and different time intervals between successive observations (yearly, quarterly, \ldots). A minimum number of observations for each series was set to 14 observations for yearly series, 16 for quarterly, 48 for monthly and 60 for other series -- this measure ensured that there was enough data to develop each forecasting model. Table \ref{tab:domains} lists the number of series per domain.

\begin{table*}
\centering
\caption{The classification of the 3003 time series used in the M3-Competition (source \cite{makridakis_m3-competition:_2000})}
   \begin{tabular}{*{8}{l}}
	\noalign{\bigskip}
	\hline\noalign{\smallskip}
	\multirow{2}{*}{Time interval} & \multicolumn{7}{l}{Types of time series data}  \\
	\cline{2-8}\noalign{\smallskip}
	& Micro & Industry & Macro & Finance & Demographic & Other & Total \\
	\hline\noalign{\smallskip}
	Yearly & 146 & 102 & 83 & 58 & 245 & 11 & 645\\
	Quarterly & 204 & 83 & 336 & 76 & 57 &  & 756\\
	Monthly & 474 & 334 & 312 & 145 & 111 & 52 & 1428\\
	Other & 4 &  &  & 29 & & 141 & 174\\
	\noalign{\bigskip}
	Total & 828 & 519 & 731  & 308 & 413 & 204 & 3003\\
	\hline
   \end{tabular}
\label{tab:domains}
\end{table*}


To compare the performance of each method to some well-known method, the Naïve2 method was chosen as the benchmark method. It is a random walk model that is applied to seasonally adjusted data by assuming that seasonality is known.
The other benchmark method to compare the results against (see Fig. \ref{fig:m3}) was the Dampen Trend Exponential Smoothing \cite{gardner_forecasting_1985}.

\subsubsection{Forecasting Cloud-related Data}

To test the performance of the described forecasting methods in the context of cloud computing, we built two time series datasets. The first dataset consisted of hourly electricity prices available as a \gls{rtep} option through Ameren in Illinois, USA. They also provide historical data which we used for our evaluation~\cite{_ameren_2016}. The dataset consists of hourly prices for a period of 122 days, starting at 2012-05-06 00:00.
The temperature data was obtained from the National Climatic Data Center \cite{ncdc_climate_2013}, spanning the same time period as the electricity price data. Chicago, Illinois, USA was chosen as the location.

\subsubsection{Measuring the Error}

The methods evaluated in the M3 competition were used to make
the following number of forecasts beyond the available data available
to them: six for yearly, eight for
quarterly, 18 for monthly and eight for other data.
For cloud-related data, we did not perform out-of-sample forecasting
to measure the accuracy, but instead evaluated the forecast values
against the observations on the the whole dataset.
Time series forecasting methods are evaluated by comparing
the $n$ forecast values $(F_i)$ obtained from the fitted model
with the actual values that continue the series $(A_i)$,
where $i \in \{1\ldots n\}$. From these values we can calculate the \gls{mape}:

\begin{equation}
M = \frac{1}{n} \sum_{t=1}^n\left|\frac{A_t - F_t}{A_t}\right|
\end{equation}

In an ideal case of correctly forecasting all the values,
we would have a MAPE of 0. As the number and amount of errors increase,
\gls{mape} becomes larger.

\subsection{Results}
\label{ch:bg:sec:results}

We will now present the results of the M3 competition
and of our own simulation of forecasting cloud-related time series.

\subsubsection{M3 Comparison}

\begin{table}
\centering
\caption{Average symmetric MAPE: Naïve2 (benchmark), Theta and AAM methods (source \cite{makridakis_m3-competition:_2000})}
   \begin{tabular}{l||l|l}

	Method & Average MAPE & \# obs \\
	\hline
	Naïve2 & 15.47 & 3003 \\
	Theta & 13.01 & 3003 \\
	AAM & 14.63 & 2184 \\
	\hline
   \end{tabular}
\label{tab:mape}
\end{table}

Table \ref{tab:mape} lists the average \gls{mape} values obtained after
evaluating the Theta and AAM method
on the M3 competition dataset~\cite{makridakis_m3-competition:_2000}.
The benchmark method Naïve2 is included to give some perspective
to the difference in results.
We can see from the results that the Theta method gives substantially
better results on the large scale of several thousands time series
included in the M3 dataset. Note that the \gls{aam} method is evaluated
on only 2184 series -- the reason for this is that yearly series
do not have a history long-enough to be modelled by \gls{arima},
so they would corrupt the results as explained in~\cite{melard_automatic_2000}.

\begin{figure*}
\centering
\includegraphics[width=0.99\textwidth]{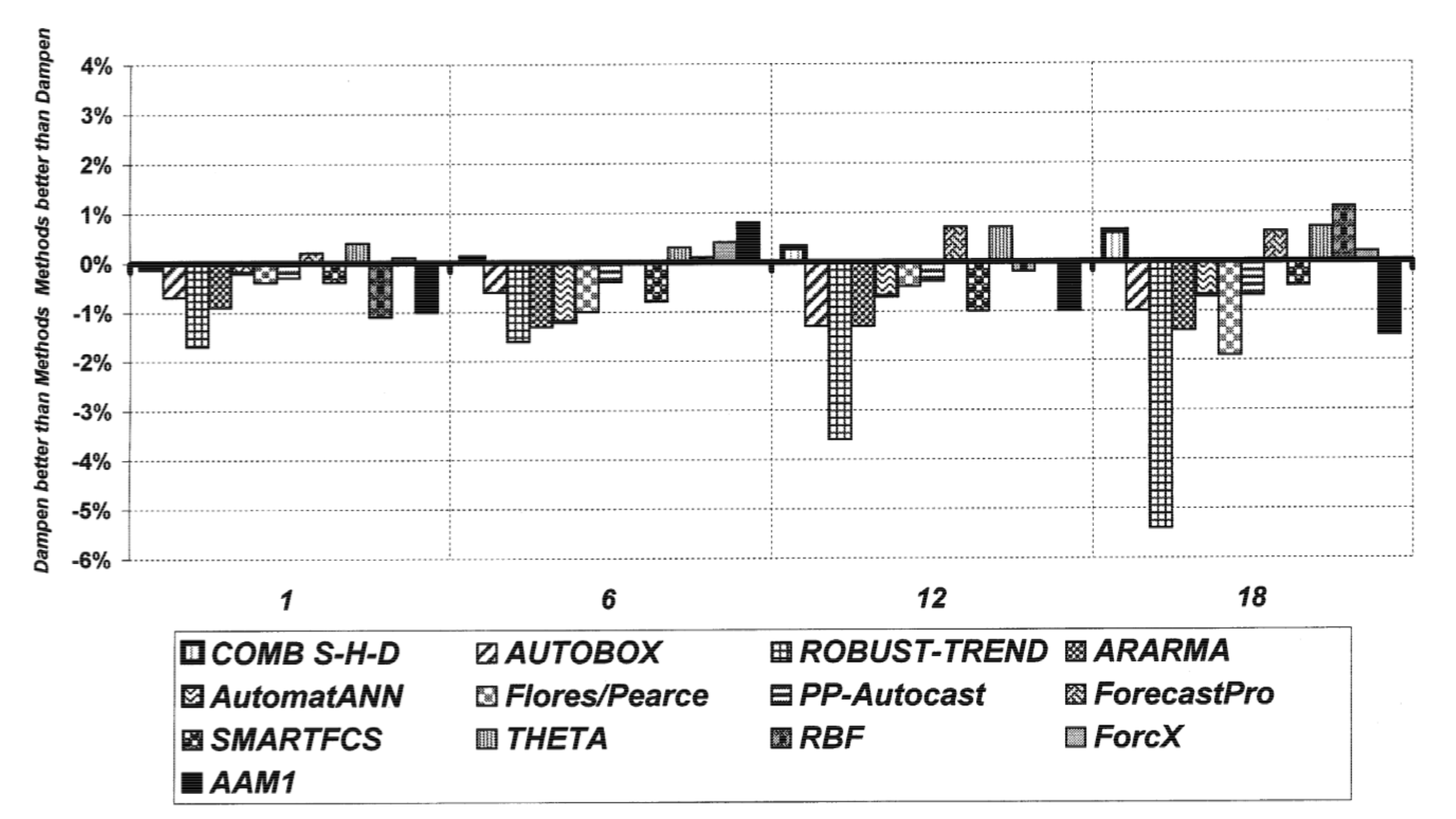}
\caption{Average symmetric MAPE compared to the Dampen benchmark method (source \cite{makridakis_m3-competition:_2000})}
\label{fig:m3}
\end{figure*}

Fig. \ref{fig:m3} shows the average symmetric \gls{mape} compared to the Dampen benchmark method for \emph{all methods} that took part in the M3 competition. Different horizons (types of series) are grouped under labels 1, 6, 12 and 18. Taking a finer-grained look at the results, the AAM method outperformed the Theta method in horizon 6, which consists of financial monthly data. This kind of high-frequency data usually offers a larger history to learn from, which is beneficial for fitting complex models, such as \gls{arima}. This shows that for certain kinds of data it still makes sense to consider the \gls{aam} method.

\subsubsection{Cloud-related Data Forecasting Precision}

\begin{figure}
\centering
\includegraphics[width=0.7\columnwidth]{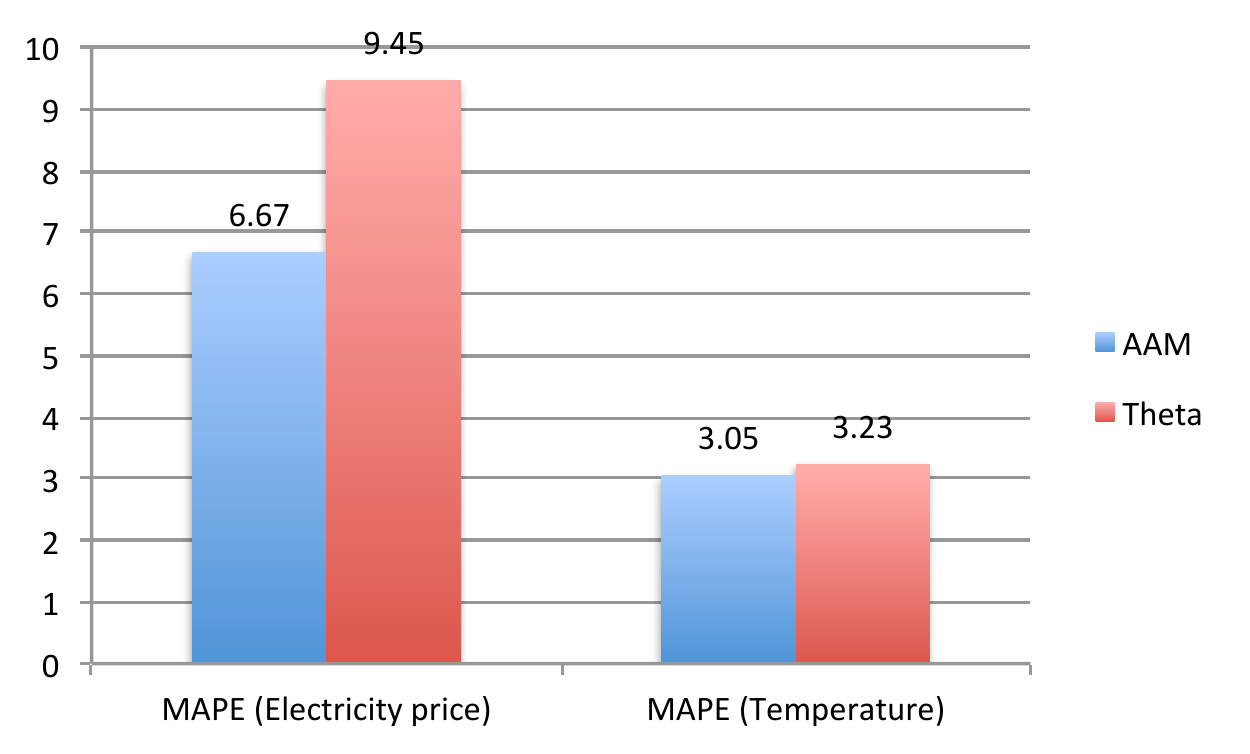}
\caption{MAPE values of forecasting electricity prices and air temperature time series using AAM and the Theta method.}
\label{fig:results-cloud}
\end{figure}

\begin{figure*}
\centering
\includegraphics[width=0.75\textwidth]{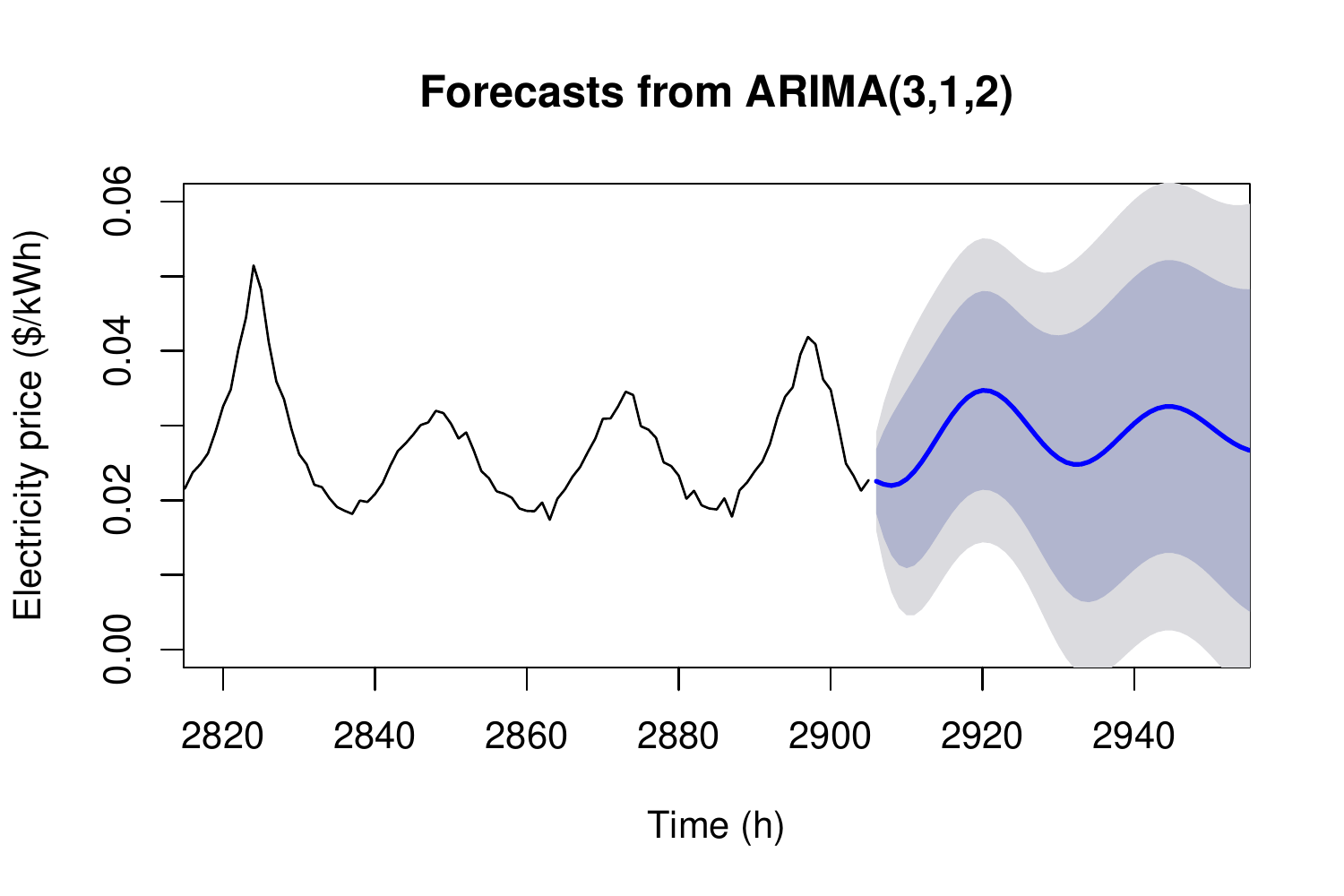}
\caption{Forecasts of electricity prices with confidence intervals using AAM.}
\label{fig:forecasting-cloud-el}
\end{figure*}

\begin{figure*}
\centering
\includegraphics[width=0.75\textwidth]{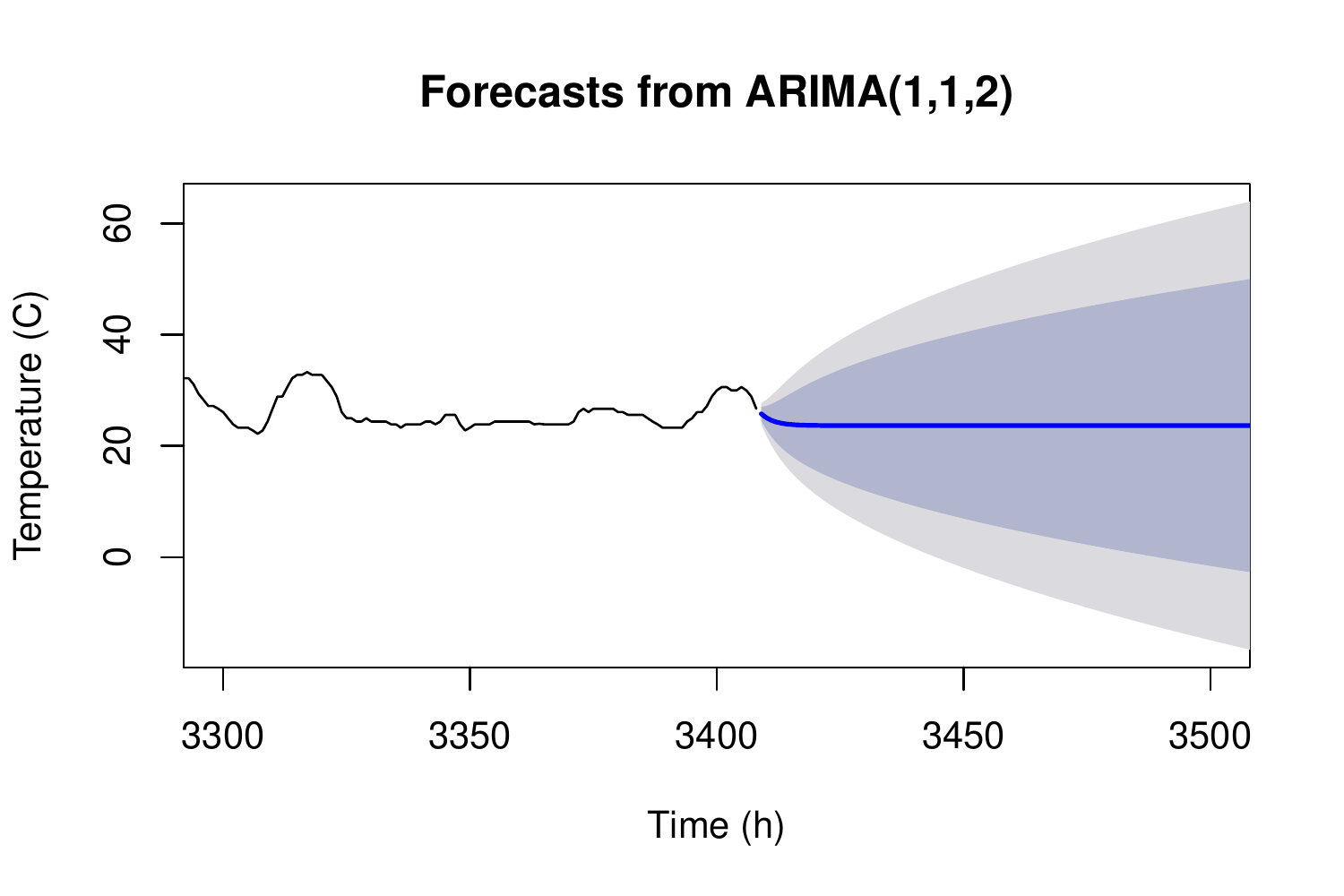}
\caption{Forecasts of air temperature with confidence intervals using AAM.}
\label{fig:forecasting-cloud-temp}
\end{figure*}

The \gls{mape} values for applying \gls{aam} and the Theta method on electricity
prices and temperature values are shown in Fig.~\ref{fig:results-cloud}.
Unlike the M3 competition, the AAM method proved to be more precise on both
datasets. It resulted in a significantly lower \gls{mape} value for the
electricity price time series, meaning that it was able to better fit its
observations. The absolute values are somewhat better than that of the M3
competition, but that can be attributed to not using out-of-sample testing,
but instead having a unified testing and training dataset.
A visualisation of the forecasts of electricity price and temperature data using
the more successful \gls{aam} method with its confidence intervals are shown
in Fig.~\ref{fig:forecasting-cloud-el} and Fig.~\ref{fig:forecasting-cloud-temp}
respectively.

\section{Emission Trading Implications on Cloud Computing}






\label{ch:bg:sec:equilibrium}

The energy produced to power the ICT industry
(and data centers constitute a major energy consumption portion)
is responsible for 2\% of all the carbon dioxide equivalent (\coe\ -- greenhouse
gases normalized to carbon dioxide by their environmental impact)
emissions~\cite{_gartner_2007},
thus accelerating global warming~\cite{_ipcc_2012}.

Cloud computing facilitates users to buy computing resources from a cloud
provider and specify the exact amount of each resource (such as storage space,
number of cores etc.) that they expect through a Service Level Agreement
(SLA)\footnote{We consider the traditional business model where the desired
specifications are set in advance, as is still the case in most
\gls{iaas} clouds.}. The cloud provider then honors this
agreement by providing the promised resources to avoid agreement violation
penalties (and to keep the customer satisfied to continue doing business).
However, cloud providers are usually faced with the challenge of satisfying
promised SLAs and at the same time not wasting their resources as a user very
rarely utilizes computing resources to the maximum
\cite{beauvisage_computer_2009}.

In order to fight global warming, the Kyoto protocol was established by the
United Nations Framework Convention on Climate Change (UNFCCC or FCCC). The goal
is to achieve global stabilisation of greenhouse gas concentrations in the
atmosphere at a level that would prevent dangerous anthropogenic interference
with the climate system~\cite{_kyoto_2012}.
The protocol defines control mechanisms to reduce \coe\
emissions by basically setting a market price for such emissions.
Currently, flexible models
for \coe\ trading are developed at different organizational and political level
as for example at the level of a country, industry branch, or a company.
As a result, keeping track of and reducing \coe\ emissions is becoming more and
more relevant after the ratification of the Kyoto protocol.


In this section we propose a \coe-trading model for transparent scheduling of
resources in cloud computing adhering to the Kyoto protocol guidelines
\cite{ellerman_european_2012}, we present a conceptual model for \coe\
trading compliant to the Kyoto protocol's emission trading scheme. We consider
an \emph{emission trading market (ETM)} where \emph{credits for emission
reduction (CERs)} are traded between data centers. Based on the positive or
negative \emph{CERs} of the data center, a cost is set for the environmental
impact of the energy used by applications.
Thereby, a successful application scheduling decision can
be made after considering the (i) energy costs,
(ii) \coe\ costs and (iii) SLA violation costs. Second, we propose a
\emph{wastage-penalty} model that can be used as a basis for the
implementation of Kyoto protocol-compliant scheduling and pricing models.
Finally, we discuss potential uses of the model as an optimisation heuristic in
the resource scheduler.

\subsection{Applying the Kyoto Protocol to Clouds}
\label{ch:bg:sec:kyoto}

The Kyoto protocol \cite{grubb_kyoto_1999} commits involved countries to
stabilize their greenhouse gas (GHG) emissions by adhering to the measures
developed by the United Nations Framework Convention
on Climate Change (UNFCCC)~\cite{_kyoto_2012}.
These measures are commonly known as \emph{the
cap-and-trade system}. It is based on setting national emission boundaries --
caps, and establishing international emission markets for trading emission
surpluses and emission deficits. This is known as \emph{certified emission
reductions} or \emph{credits for emission reduction} (CERs). Such a trading
system rewards countries which succeeded in reaching their goal with profits
from selling CERs and forces those who did not to make up for it financially by
buying CERs. The European Union Emission Trading System (EU ETS) is an example
implementation of an emission trading market~\cite{_eu_2012}. Through such
markets, CERs converge towards a relatively constant market price, same as all
the other tradable goods.

Individual countries control emissions among their own large polluters
(individual companies such as power generation facilities, factories\ldots) by
distributing the available caps among them. In the current implementation,
though, emission caps are only set for entities which are responsible for more than 25
Mt\coe/year \cite{_environment_2012}. This excludes individual data centers
which have a carbon footprint in the kt\coe/year range \cite{_data_2016}.

It is highly possible, though, that the Kyoto protocol will expand to smaller
entities such as cloud providers to cover a larger percentage of polluters and
to increase the chance of global improvement. One such reason is that currently
energy producers take most of the weight of the protocol as they cannot pass the
responsibilities on to their clients (some of which are quite large, such as
data centers). In 2009, three companies in the EU ETS with the largest shortage
of carbon allowances were electricity producers \cite{_eu_2012-1}. Another
indicator of the justification of this forecast is that some cloud providers,
such as Google already participate in emission trading markets to achieve carbon
neutrality \cite{_googles_2012}.

For this reason, we hypothesize in this section that cloud providers are indeed
part of an emission trading scheme and that \coe\ emissions have a market price.

\subsection{Wastage-Penalty Balance in a Kyoto-Compliant Cloud}
\label{ch:bg:sec:model}
In this section we present our \coe-trading model that is to be integrated with
cloud computing. We show how an economical balance can be found in it. Lastly,
we give some discussion as to how such information might be integrated into a
scheduler to make it more energy and cost efficient.

\subsubsection{The \coe-Trading Model}
The goal of our model is to integrate the Kyoto protocol's \coe\ trading
mechanism with the existing cloud computing service-oriented paradigm. At the
same time we want to use these two aspects of cloud computing to express an
economical balance function that can help us make better decisions in the
scheduling process.

\begin{figure}[h!]
\centering
\includegraphics[width=0.7\textwidth]{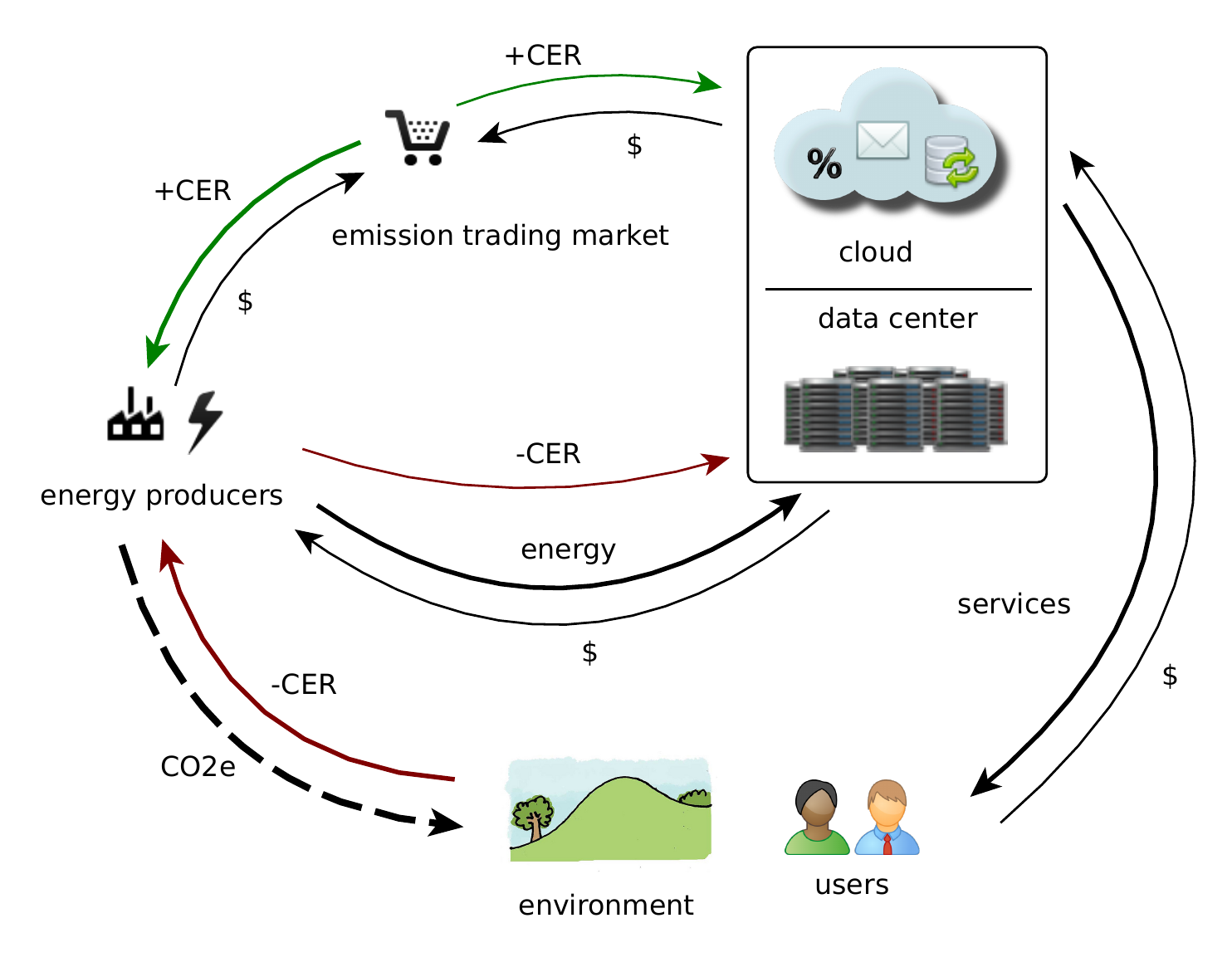}
\caption{Cloud computing integrated with the Kyoto protocol's emission trading scheme}
\label{fig:model}
\end{figure}

The model diagram in Fig. \ref{fig:model} shows the entities in our model and
their relations. A cloud offers some computing resources as services to its
users and they in turn pay the cloud provider for these services. Now, a cloud
is basically some software running on machines in a data center. To operate, a
data center uses electrical energy that is bought from an energy producer. The
energy producers are polluters as they emit \coe\ into the atmosphere. As
previously explained, to mitigate this pollution, energy producers are bound by
the Kyoto protocol to keep their \coe\ emissions bellow a certain threshold and
buy CERs for all the excess emissions from other entities that did not reach
their caps yet over the emission trading market (ETM). This is illustrated by
getting negative CERs (-CERs) for \coe\ responsibilities and having to buy the
same amount of positive CERs (+CERs) over the ETM. It does not make any real
difference for our model if an entity reaches its cap or not, as it can sell the
remaining \coe\ allowance as CERs to someone else over the ETM. Most
importantly, this means that \emph{\coe\ emissions an entity is responsible for
have a price}.

The other important thing to state in our model is that
\emph{\coe\ emission responsibilities for the energy that was bought is
transferred from the energy producer to the cloud provider}. This is shown in
Fig. \ref{fig:model} by energy producers passing some amount of -CERs to the
cloud provider along with the energy that was bought. The cloud provider then
has to buy the same amount of +CERs via the ETM (or he will be able to sell them
if he does not surpass his cap making them equally valuable).

The consequences of introducing this model are that three prices influence the
cloud provider: (1) energy cost; (2) \coe\ cost; (3) service cost. To maximize
profit, the cloud provider is motivated to decrease energy and \coe\ costs and
maximize earnings from selling his service. Since the former is achieved by
minimizing resource usage to save energy and the latter by having enough
resources to satisfy the users' needs, they are conflicting constraints. Therefore, an
economical balance is needed to find exactly how much resources to provide.

The service costs are much bigger than both of the other two combined (that is
the current market state at least, otherwise cloud providers would not operate),
so they cannot be directly compared. There are different ways a service can
be delivered, though, depending on how the cloud schedules resources. The aim of a
profit-seeking cloud provider is to deliver just enough resources to the user so
that his needs are fullfilled and that the energy wastage stays minimal. If a
user happens to be tricked out of too much of the resources initially sold to
him, a service violation occurs and the cloud provider has to pay a penalty
price. This means that we are comparing the energy wastage price with the
occasional violation penalty. This comparison is the core of our wastage-penalty
model and we will now explain how can a wastage-penalty economical balance be
calculated.

\subsubsection{The Wastage-Penalty Model for Resource Balancing}

As was briefly sketched in the introduction, the main idea is to push cloud
providers to follow their users' demands more closely, avoiding too much
resource over-provisioning, thus saving energy. We do this by introducing additional cost
factors that the cloud provider has to pay if he wastes too much resources  --
the energy and \coe\ costs shown in Fig. \ref{fig:model}, encouraging him to
breach the agreed service agreements and only provide what is actually needed.
Of course, the cloud provider will not breach the agreement too much, as that
could cause too many violation detections (by a user demanding what cannot be
provided at the moment) and causing penalty costs. We will now expand our model
with some formal definitions in the cloud-user interface from Fig.
\ref{fig:model} to be able to explicitly express the wastage-penalty balance in
it.

\begin{figure}
\centering
\vspace{0cm}
\includegraphics[width=0.75\textwidth]{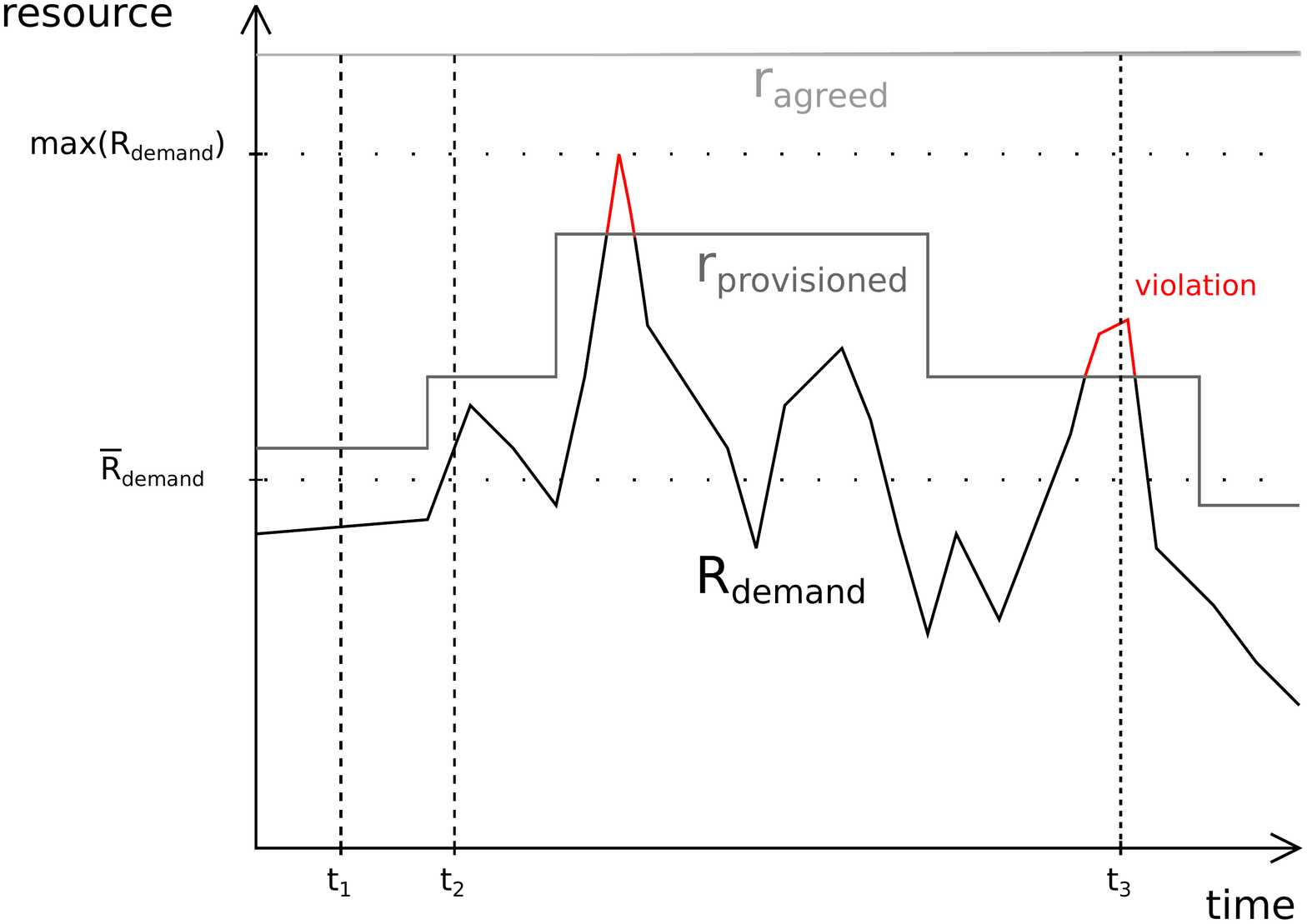}
\caption{Changes in the \emph{provisioned} and \emph{demand} resource
amounts over time}
\label{fig:agreed-provisioned}
\end{figure}

We assume a situation with one cloud provider
and one cloud user. The cloud provides the user with a single, abstract resource
that constitutes its service (it can be the amount of available data storage
expressed in GB, for example). To provide a certain amount of this resource to
the user in a unit of time, a proportional amount of energy
is consumed and indirectly a
proportional amount of \coe\ is emitted. An example resource scheduling scenario
is shown in Fig. \ref{fig:agreed-provisioned}. An SLA was signed that binds the
cloud provider to provide the user a constant resource amount, $r_{agreed}$. The
cloud provider was paid for this service in advance. A user uses different
resource amounts over time. At any moment the $R_{demand}$ variable is the
amount required by the user. To avoid over-provisioning the provider
does not actually provision the promised resource amount all the time,
but instead adapts this value dynamically,
$r_{provisioned}$ is the resource amount allocated to the
user in a time unit. This can be seen in Fig. \ref{fig:agreed-provisioned} as
$r_{provisioned}$ increases from $t_1$ to $t_2$
to adapt to a sudden rise in $R_{demand}$.

As we can not know how the user's demand changes over time, we will think of
$R_{demand}$ as a random variable. To express $R_{demand}$ in an explicit way,
some statistical method would be required and research of users' behaviour
similar to that in \cite{beauvisage_computer_2009} to gather real-life data
regarding cloud computing resource demand. To stay on a high level of
abstraction, though, we assume that it conforms to some statistical distribution
and that we can calculate its mean $\overline{R}_{demand}$ and its maximum
$max(R_{demand})$. To use this solution in the real world, an appropriate
distribution should be input (or better yet -- one of several possible
distributions should be chosen at runtime that corresponds to the current user
or application profile). We know the random variable's expected value E and
variance V for typical statistical distributions and we can express
$\overline{R}_{demand}$ as the expected value $E(R)$ and $max(R_{demand})$ as
the sum of $E(R)+V(R)$ with a limited error.


Let us see how these variables can be used to model resource wastage costs. We
denote the energy price to provision the whole $r_{agreed}$ resource amount per time unit $c_{en}$ and similarly the \coe\ price $c_{co_2}$. By only using
the infrastructure to provision an amount that is estimated the user will
require, not the whole amount, we save energy that would have otherwise been
wasted and we denote this evaded wastage cost $c_{wastage}$. Since $c_{wastage}$
is a fraction of $c_{en}+c_{co_2}$, we can use a percentage $w$ to state the
percentage that is wasted:

\begin{equation}
\label{eq:cw}
c_{wastage} = w * (c_{en}+c_{co_2})
\end{equation}

We know what the extreme cases for $w$ should be -- $0\%$ for provisioning
approximately what is needed, $\overline{R}_{demand}$; and the percentage
equivalent to the ratio of the distance between $\overline{R}_{demand}$ and
$r_{agreed}$ to the total amount $r_{agreed}$ if we provision $r_{agreed}$:

\begin{equation}
w =
\begin{cases}
1-\frac{\overline{R}_{demand}}{r_{agreed}} , & \textrm{if } r_{provisioned}=r_{agreed} \\
0, & \textrm{if } r_{provisioned}=\overline{R}_{demand}
\end{cases}
\end{equation}

We model the distribution of $w$ between these extreme values using linear
interpolation: 
average resource utilization - a ratio of the average provisioned resource
amount ($\overline{r}_{provisioned}$) and the promised resource amount
($r_{promise}$):

\begin{equation}
\label{eq:w}
w = \frac{r_{provisioned} - \overline{R}_{demand}}{r_{agreed}}
\end{equation}

If we apply \ref{eq:w} to \ref{eq:cw} we get an expression for the wastage cost.:

\begin{equation}
\label{eq:wastage}
c_{wastage} = \frac{r_{provisioned} - \overline{R}_{demand}}{r_{agreed}} * (c_{en}+c_{co_2})
\end{equation}


Let us now use a similar approach to model penalty costs. If a user demands more
resources than the provider has provisioned, an SLA violation occurs. The user gets only the provisioned amount of resources in this
case and the provider has to pay the penalty cost $C_{penal}$. While $c_{en}$
and $c_{co_2}$ can be considered constant for our needs, $C_{penal}$ is a random
variable, because it depends on the user's behaviour which we can not predict
with 100\% accuracy, so we will be working with $E(C_{penal})$, its expected
value.



$E(C_{penal})$, the expected value of $C_{penal}$ can be calculated as:

\begin{equation}
\label{eq:penalty}
E(C_{penal}) = p_{viol} * c_{viol}
\end{equation}

where $c_{viol}$ is the constant cost of a single violation (although in reality
probably not all kinds of violations would be priced the same) and $p_{viol}$ is
the probability of a violation occurring. This probability can be expressed as a
function of $r_{provisioned}$, $r_{agreed}$ and $R_{demand}$, the random
variable representing the user's behaviour:

\begin{equation}
p_{viol} = f(\overline{r}_{provisioned}, r_{promise}, R_{demand})
\end{equation}

Again, same as for $c_{wastage}$, we know the extreme values we want for
$p_{viol}$. If 0 is provisioned, we have 100\% violations and if
$max(R_{demand})$ is provisioned, we have 0\% violations:

\begin{equation}
p_{viol} =
\begin{cases}
100\%, & \textrm{if } r_{provisioned}=0 \\
0\%, & \text{if } r_{provisioned}=max(R_{demand})
\end{cases}
\end{equation}

and if we assume a linear distribution in between we get an
expression for the probability of violations occuring, which is needed for
calculating the penalty costs:

\begin{equation}
\label{eq:violation}
p_{viol}= 1 - \frac{r_{provisioned}}{max(R_{demand})}
\end{equation}


Now that we have identified the individual costs, we can state our goal
function. If the cloud provider provisions too much resources the $c_{wastage}$
wastage cost is too high. If on the other hand he provisions too little
resources, tightens the grip on the user too much, the $E(C_{penal})$ penalty
cost will be too high. The economical balance occurs when the penalty and
wastage costs are equal - it is profitable for the cloud provider to breach the
SLA only up to the point where penalty costs exceed wastage savings. We can
express this economical balance with the following equation:

\begin{equation}
c_{wastage} = E(C_{penal}) + \text{[customer satisfaction factor]}
\end{equation}

The \emph{[customer satisfaction factor]} could be used to model how our
promised-provisioned manipulations affect the user's happiness with the quality
of service and would be dependant of the service cost (because it might
influence if the user would be willing to pay for it again in the future). For simplicity's sake we
will say that this factor equals 0, getting:

\begin{equation}
\label{eq:costs}
c_{wastage} = E(C_{penal})
\end{equation}

Now, we can combine equations \ref{eq:wastage}, \ref{eq:costs}, \ref{eq:penalty}
and \ref{eq:violation} to get a final expression for $r_{provisioned}$:

\begin{equation}
\label{eq:equilibrium}
r_{provisioned} =\frac{max(R_{demand})*\left[\overline{R}_{demand}*(c_{en}+c_{co_2}) + r_{agreed}*c_{viol}\right]}{max(R_{demand})*(c_{en}+c_{co_2})+r_{agreed}*c_{viol}}
\end{equation}

This formula is basically \emph{the economical wastage-penalty balance}. All the
parameters it depends on are constant as long as the demand statistic stays the
same. It shows how much on average should a cloud provider breach the
promised resource amounts when provisioning resources to users so that the
statistically expected costs for SLA violation penalties do not surpass the
gains from energy savings. Vice versa also holds -- if a cloud provider
provisions more resources than this wastage-penalty balance, he pays more for
the energy wastage (energy and \coe\ price), than what he saves on SLA violations.

\subsubsection{Heuristics for Scheduling Optimisation with Integrated Emission
Management}

In this section we discuss a possible application of our wastage-penalty model
for the implementation of a future-generation data center. Knowing the
economical wastage-penalty balance, heuristic functions can be used to optimize
resource allocation to maximize the cloud provider's profit by integrating both
service and violation penalty prices and energy and \coe\ costs. This is useful,
because it helps in the decision-making process when there are so many
contradicting costs and constraints involved.

A heuristic might state: ``try not to provision more than $\pm x\%$ resources
than the economical wastage-penalty balance''. This heuristic could easily be
integrated into existing scheduling algorithms, such as
\cite{maurer_enacting_2011,maurer_simulating_2010} so that the cloud provider
does not stray too far away from the statistically profitable zone without
deeper knowledge about resource demand profiles. The benefits of using our
wastage-penalty model are:
\begin{itemize}
  \item a new, expanded cost model covers all of the influences from Fig. \ref{fig:model}
  \item \coe-trading schema-readiness makes it easier to take part in emission trading
  \item a Kyoto-compliant scheduler module can be adapted for use in resource
  scheduling and allocation solutions
  \item the model is valid even without Kyoto-compliance by setting the \coe\
  price $c_{co_2}$ to 0, meaning it can be used in traditional ways by weighing
  only the energy wastage costs against service violation penalties.
\end{itemize}

The wastage-penalty balance in \ref{eq:equilibrium} is a function of significant
costs and the demand profile's statistical properties:

\begin{equation}
r_{provisioned} =g(max(R_{demand}),\overline{R}_{demand},r_{agreed},c_{en},c_{co_2},c_{viol})
\end{equation}

This function enables the input of various statistical models for user or
application demand profiles ($max(R_{demand})$ and $\overline{R}_{demand}$) and
energy ($c_{en}$), \coe\ ($c_{co_2}$) and SLA violation market prices
($c_{viol}$). With different input parameters, output results such as energy
savings, environmental impact and SLA violation frequency can be compared. This
would allow cloud providers and governing decision-makers to simulate the
effects of different scenarios and measure the influence of individual
parameters, helping them choose the right strategy.





\section{Summary}
\label{ch:bg:sec:conclusion}


In this chapter we offered some insights into the dynamic variables
affecting geographically distributed cloud
systems and analysed time series forecasting
as a tool for predicting their future behaviour.
We compared two time series forecasting methods -- the statistically
complex \gls{aam} technique consisting of an expert system that performs various
data transformations to build a stationary process representing the series'
temporal behaviour; and the Theta method, based on \gls{ses} with random drift,
relying on simpler algebraic expressions to build a geometrically-intuitive
smoothed model. An empirical evaluation of both methods on a large dataset of
thousands of different kinds of time series was performed to find the best
forecasting methods in \cite{makridakis_m3-competition:_2000}, which
allowed us to see how both methods perform in the same situations.
Additionally, we evaluated the methods on datasets relevant to data centers --
real time electricity prices~\cite{_ameren_2016} and outside air temperature
values \cite{ncdc_climate_2013}.

The Theta method had the smallest average \gls{mape}, therefore "winning" the M3
competition. The results of this competition have shown that simple and
straightforward methods often outperform complex, theoretically advanced models
on the average. This means that simple methods, such as the Theta method,
represent good general-purpose tools that can be used if we have little
knowledge of the domain or the art and theory of forecasting. Still, in areas
where high forecasting precision is important, it would be good to
collect a larger training dataset and apply more complex methods,
such as \gls{aam}, to test if they would maybe perform better.

In our evaluation on cloud-related data, the \gls{aam} method proved to be
better in forecasting both electricity prices and temperatures. It also proved
to be better in the M3 competition when there is a long history available,
as is the case with financial monthly data.

An interesting family of methods which also performed well in the M3 competition
were those that combined several models to automatically choose the best one
for the data at hand. These hybrid methods that take the best of both worlds are
probably the future of forecasting in our more and more data-driven world.

We have shown that time series forecasting can indeed give insights about
the future behaviour of dynamic variables. Such methods could find many
uses in improving the management of computational resources in geographically
distributed data centers forming the backbone of cloud computing.

In this chapter we also presented an approach for Kyoto protocol-compliant
modelling of data centers where \gls{coe} also become a cost component,
similar to electricity prices.
We presented a conceptual model for \coe\
trading compliant with the Kyoto protocol's emission trading scheme. We
consider an \emph{emission trading market (ETM)} where \coe\ obligations are
forwarded to data centers, involving them in the trade of credits for emission
reduction (CERs). Such measures would ensure a \coe\ equilibrium and encourage
more careful resource allocation inside data centers.
To aid decision making inside this \coe-trading system, we proposed a
\emph{wastage-penalty} model that can be used as a basis for the
implementation of Kyoto protocol-compliant scheduling and pricing models.


\chapter{Experimental VM Scheduling for Real-Time electricity Pricing}
\label{ch:volatility}
In this chapter we set the foundations of a grid-conscious cloud,
taking a more informed view of the electrical
grid by analysing real-time electricity prices.
We propose a scheduling algorithm that predicts
electricity price peaks and throttles energy consumption
by pausing virtual machines.
We empirically evaluate the approach on the
OpenStack cloud manager in an experimental implementation
and show reductions in energy consumption and costs.
Finally, we define green instances in which cloud providers can offer
such services to their customers under better pricing options.
We first give some insights into seasonal patterns in real-time pricing options
in Section~\ref{sec:electricity}.
In Section~\ref{sec:scheduler} we define green instances and state the
peak pauser scheduling algorithm, with a rundown of its evaluation procedure
in Section~\ref{sec:methodology} and the actual results are
presented in Section~\ref{sec:results}.

\begin{figure*}
\centering
\includegraphics[width=1.\textwidth]{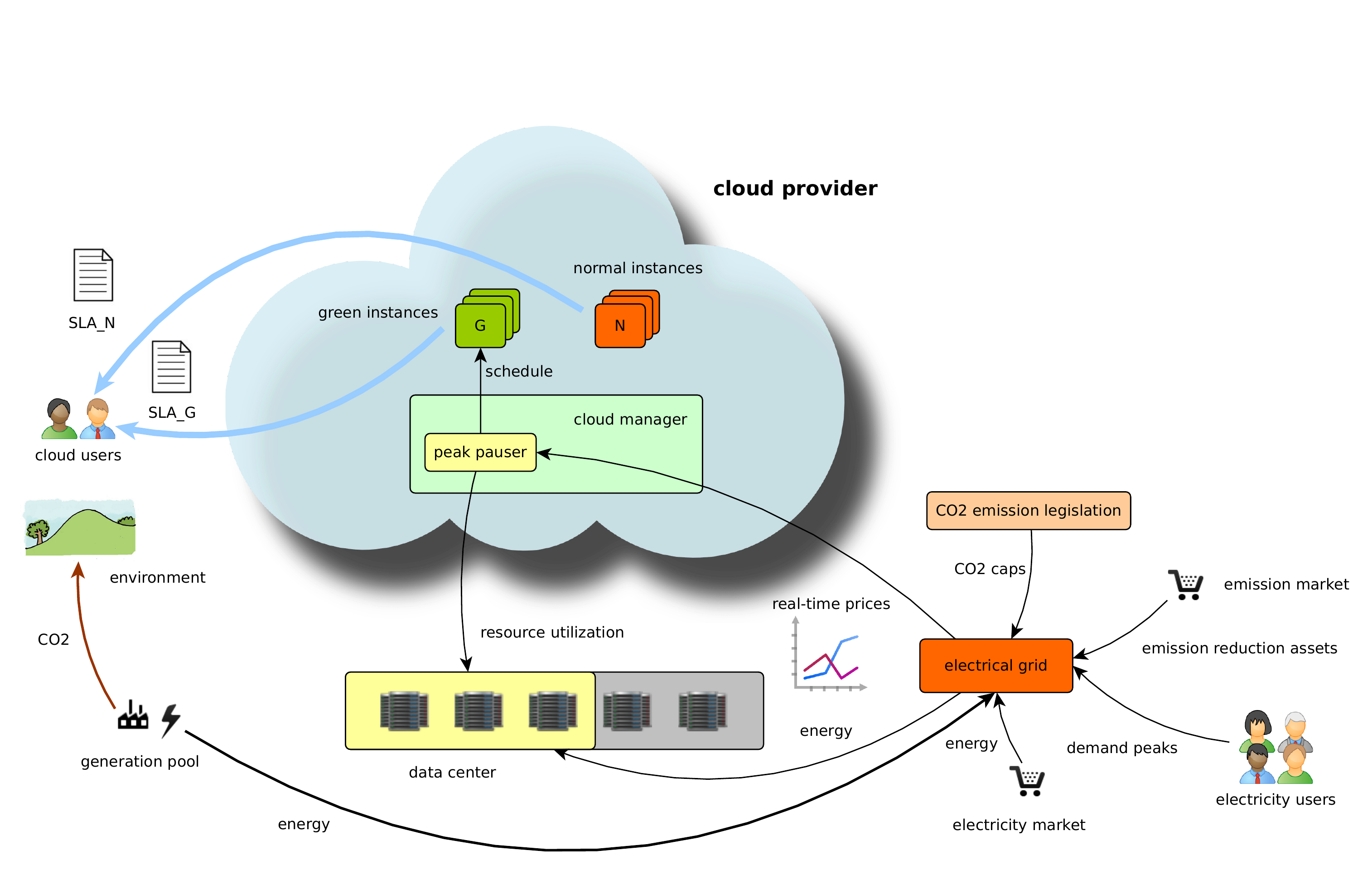}
\caption{The grid-conscious cloud model.}
\label{fig:grid-conscious_cloud}
\end{figure*}

\section{Foundations of the Grid-conscious Cloud}



The grid-conscious cloud model is illustrated in\
Fig \ref{fig:grid-conscious_cloud}.\
Assuming a real-time electricity pricing model,\
the price of electricity offered by the utility changes hourly.\
Various factors such as active generation pools, electricity markets,\
volatile demand \cite{weber_uncertainty_2004,weron_modeling_2006},\
and \gls{coe} emission legislation and markets from Chapter~\ref{ch:background}
that influence the electrical grid are condensed in this price signal.\
Combined with the elasticity of cloud computing, this creates an\
opportunity to optimize the conversion of energy to computation\
that would benefit both sides.\
%

The grid-conscious cloud is based on the idea that \emph{computation is more\
elastic than energy.} Moving energy around, storing it or building infrastructure\
to satisfy demand can be very expensive and environmentally inefficient\
\cite{weber_uncertainty_2004}.\
Computation, on the other hand -- especially using modern paradigms such as\
virtualisation and cloud computing, can be scaled outwards or inwards,\
moved to other locations and postponed for a later time, depending on the\
needs of users and resource providers.

The provider of a grid-conscious cloud is\
motivated to weigh the electricity price parameter in its scheduling and\
resource utilization process. It needs a scheduler\
that can respond to changing electricity prices,\
altering the cloud's physical resource usage and energy demand.\
Changing resource usage potentially affects end users so it\
might be necessary to open a new business interface towards the end users.\
Taking all this into account, the grid-conscious cloud needs\
to satisfy three major requirements which we examine in the following subsections:\
(1) a real-time pricing option of buying electricity, (2) a resource scheduler\
that considers electricity prices to throttle energy demand when\
they are high and (3) a business model to offer this kind of service to end users\
with a justification for potential performance degradation.

\subsection{Patterns in Real-Time Electricity Prices}
\label{sec:electricity}


\begin{figure}
\centering
\begin{subfigure}{0.48\textwidth}
  \centering\includegraphics[width=\textwidth]{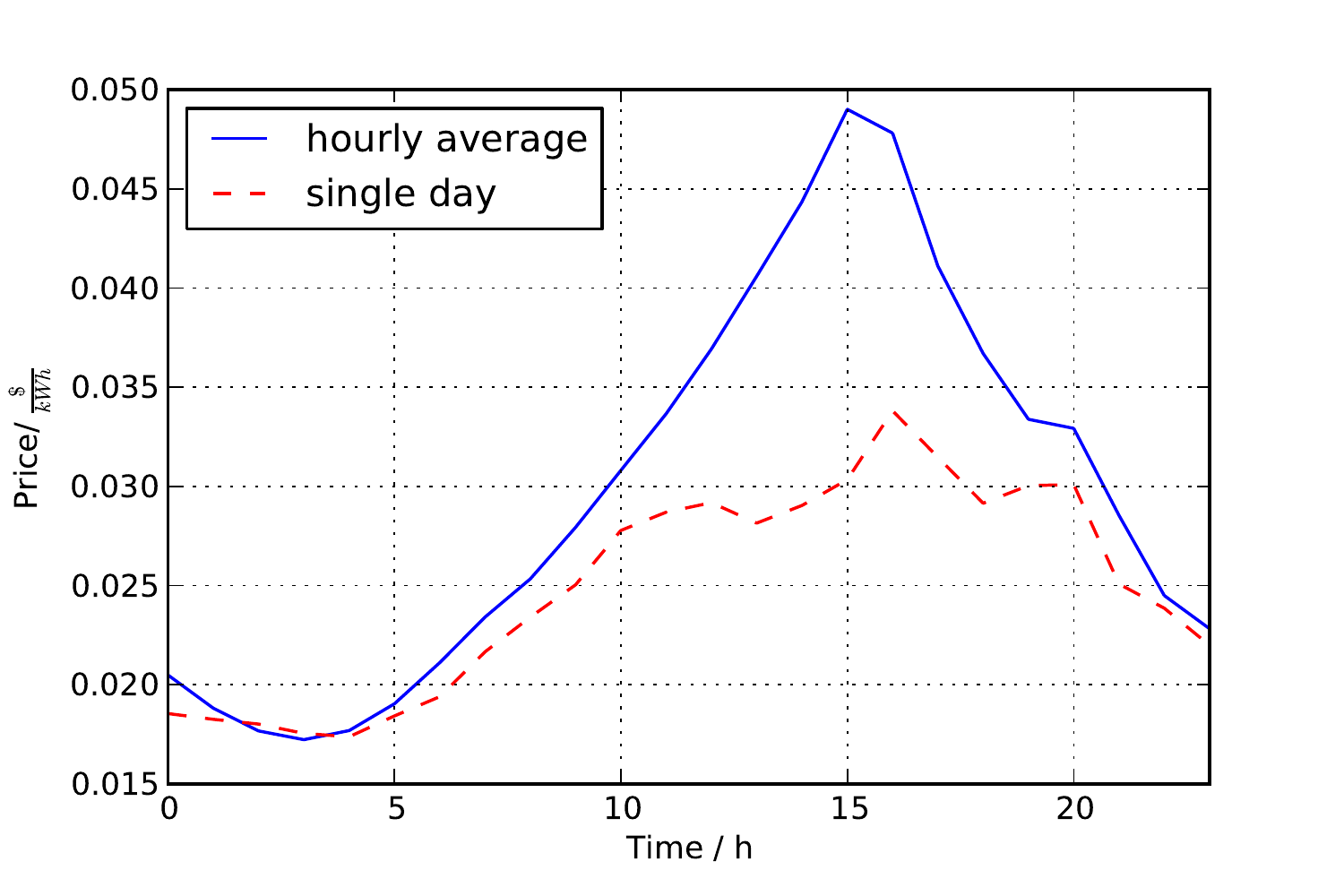}
  \caption{Peak hour distribution}\label{fig:prices-avg-single}
\end{subfigure}
\begin{subfigure}{0.48\textwidth}
  \centering\includegraphics[width=\textwidth]{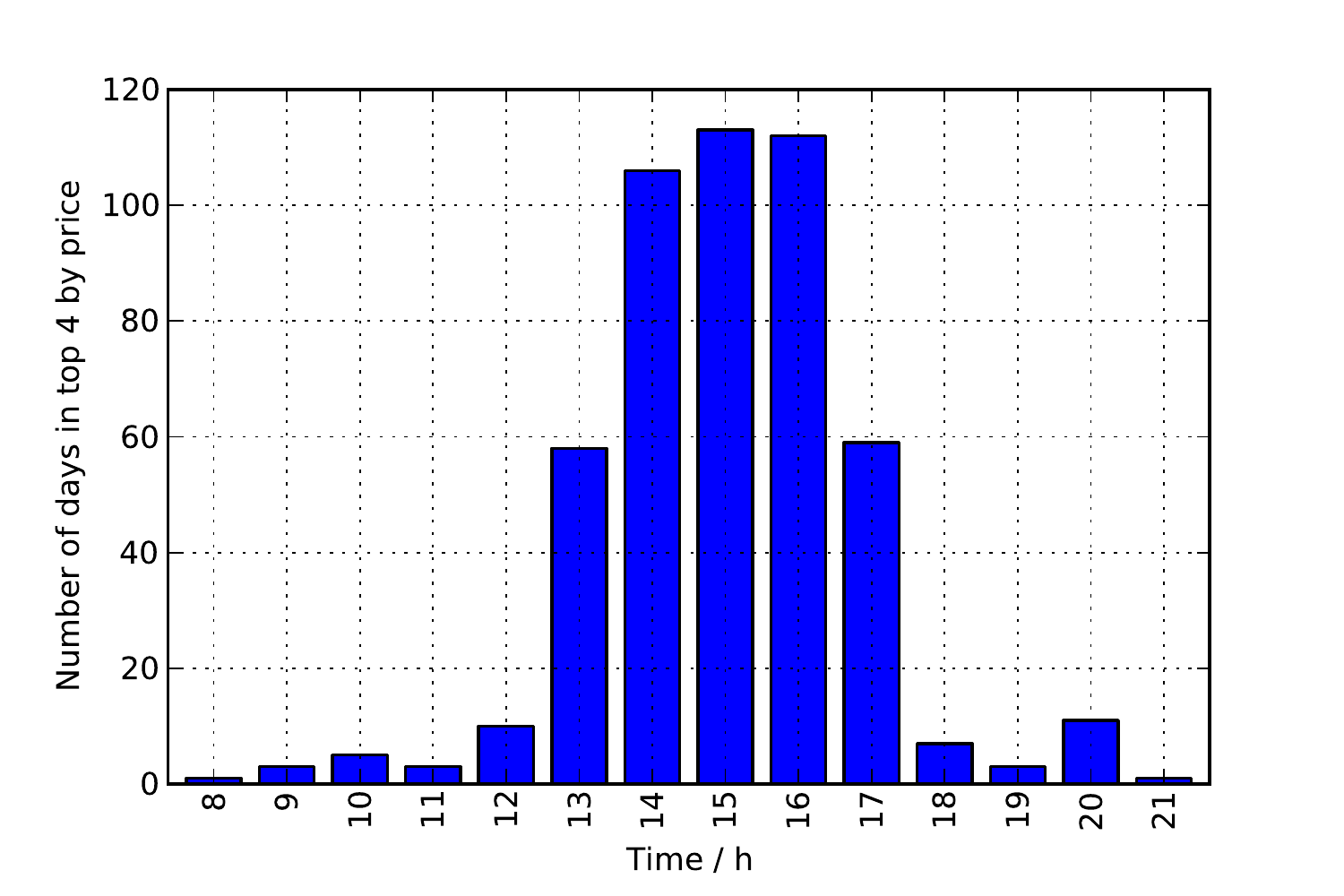}
  \caption{Peak electricity price hours distribution}\label{fig:expensive_histogram}
\end{subfigure}
\caption{Historical electricity prices (data taken from~\cite{_ameren_2016})}
\end{figure}

Peak-load pricing is a technique where demands for a public good\
in different periods of the day, month or year are considered\
to find the optimal capacity and the optimal peak-load prices.\
The method was being developed\
as far back as 1949 \cite{boiteux_peak-load_1960}.\
This is the basis of electricity spot markets,\
where electricity can be bought\
at real-time prices \cite{schweppe_spot_1988}.

As discussed in Chapter~\ref{ch:background},
many utilities (such as Ameren~\cite{_ameren_2016}) offer\
a \gls{rtp} option\
where electricity prices change hourly,\
based on market supply and demand.
Generally speaking,\
market prices are highest during times of peak demand\
(during daytime, usually peaking at 15:00), as\
shown in Fig. \ref{fig:prices-avg-single}. In \gls{rtp},\
customers are given price signals\
to guide their\
energy use and ease the burden on utilities during peak hours.\
Such customers are rewarded by potentially saving money\
when compared to the standard rate.\
There exist other pricing options, such as \emph{critical peak pricing}\
and \emph{time of use pricing} which could be suitable for\
our grid-conscious cloud model. In this study, however, we focus only\
on \gls{rtp} -- considered to be the most direct and efficient\
demand-response program \cite{albadi_demand_2007}.

An important point to consider that was analysed in\
\cite{klingert_sustainable_2012}\
is that in some cases utility companies have to resort to\
environmentally more harmful energy generation methods to handle peak demand.\
This is true in the UK, for example, where the National Grid can call\
upon about 2 GW of diesel-generated power to meet demand spikes,\
resulting in hundreds of hours of expensive and harmful fuel usage\
annually \cite{andrews_potential_2008}.\
The result is that a high electricity price also\
indicates a high rate of \gls{coe} emissions.\
This proportionality is further fortified under the Kyoto protocol's\
cap-and-trade model, where\
the \coe\ emissions themselves carry a market price as we analysed
in Chapter~\ref{ch:background}. This additional \coe\ emission price\
increases the price of energy coming from environmentally harmful sources\
even more.\
These specifics vary between different locations, however, which underlines\
the importance of having access to information grid through\
smart grids to make more informed, efficient decisions.

Another viewpoint of real-time prices is that it encourages polite behaviour,\
complying with the utilities' needs. This enables\
utilities to be more efficient (in terms of cost efficiency, reliability and\
environmental impact) and is said to create \emph{energy\
reduction assets} \cite{_understanding_2012}.\
Such assets can then be exchanged for certain other\
benefits, such as lower energy prices, meaning that they\
benefit both sides.\
These power-reduction assets can be achieved more directly by integrating\
our grid-conscious cloud model with a direct demand/response control\
mechanism available in modern smart grids \cite{hassan_survey_2010}.

These examples show that there exist both monetary and ecological\
incentives for reducing energy consumption during hours of electricity\
price peaks -- especially in computing clouds, which are substantial\
energy consumers \cite{koomey_worldwide_2008}.


%
%

\subsection{Optimizing the Scheduler: the Peak Pauser Algorithm}
\label{sec:scheduler}

The aim of this algorithm is to curtail energy consumption of a cloud\
during the hours of the day with the highest electricity price.\
Its pseudo-code is shown in Alg. \ref{alg:pp}.

The algorithm first\
finds $n$ hours in a day that are statistically most probable to be\
the peak-price hours.\
This number is defined using the parameter $downtime\_ratio$ given to the algorithm\
and is defined as
\
\begin{equation}
downtime\_ratio = \frac{\text{n}}{24}
\end{equation}
\
We then define the set of \emph{expensive hours} by calculating
hourly price averages over the provided historical dataset,
sorting them descending
and selecting the first $n$ hours in the $find\_expensive\_hours$ function
that requires historical electricity price data such as~\cite{_ameren_2016}.

%

We verified this function by analysing how often certain hours of the\
day appear among the top 4 by price (Fig. \ref{fig:expensive_histogram}\
shows the regular cyclic nature with peaks in the afternoon)\
and found it to result in only a negligible error\
\footnote{The root-mean-square error of the sum of expensive hours in a day\
determined by our function compared to an optimal daily-changing set\
that assumes a priori price knowledge\
equals $0.0058\ \$/kWh$ or  $\approx 3\% $ of the absolute amounts.}
compared to an\
ideal scenario where we know the prices in advance, sufficiently good\
for the purposes of our discussion.

Knowing how to determine expensive hours, we can define\
the scheduling algorithm as an endless loop\
(the $peak\_pauser$ function) that checks if the electricity is predicted\
to be expensive at the moment (in the $is\_expensive$ function that\
returns a boolean value depending on the current time's membership in the\
$expensive\_hours$ set).\
If the price is expensive, the scheduler pauses the set $G$ of instances\
it controls (green instances, which we will further discuss in the next section)\
and if it is not expensive, instances in $G$ are unpaused (or left running if\
they were not paused before). The scheduler can then remain idle for the remainder\
of the hour, so as not to waste resources.

\begin{algorithm}
\caption{The peak pauser algorithm.}
\label{alg:pp}
\begin{algorithmic}

\Function{find\_expensive\_hours}{$downtime\_ratio$}
\State $prices\gets\text{historical hourly prices}$
\State $avg\_prices\gets$ group $prices$ by hour and calculate mean
\State sort $avg\_prices$ descending by price
\State $n\gets ceil(downtime\_ratio*24)$\Comment{$ceil$: find first larger integer}
\State $expensive\_hours\gets$ {first $n$ elements of $avg\_prices$}
\State \Return $expensive\_hours$
\EndFunction
\item[]

\State  \begin{varwidth}[t]{\linewidth}
	$expensive\_hours\gets find\_expensive\_hours($\par
		\hskip\algorithmicindent $downtime\_ratio)$
\end{varwidth}

\item[]
\Function{is\_expensive}{}
	\State $time\gets\text{current time of day}$
	\State\Return $time.hour\in expensive\_hours$
\EndFunction
\item[]

\Function{peak\_pauser}{$G$}
\While{$True$}
	\If{$is\_expensive()$}
		\State pause $\forall instance \in G$
	\Else
		\State unpause $\forall \text{paused instance} \in G$
	\EndIf
	\State idle for the remainder of the hour
\EndWhile
\EndFunction

\end{algorithmic}
\end{algorithm}

A possible alternative to pausing would be to switch to\
battery power supply \cite{palasamudram_using_2012,bianchini_parasol:_2012}\
during expensive hours.\
Additional logic can be added to the algorithm by dynamically\
determining duration of the pause interval (the parameter $downtime\_ratio$),\
based on e.g. considering the\
current day's deviation from monthly or annual averages. This\
way we could have longer pause periods during unusually ``expensive'' days\
and close-to-normal operation on ``cheaper'' days. For the purposes\
of our evaluation, we chose a predefined value $downtime\_ratio=0.16$,\
yielding in 4 paused hours.

\subsection{The Green Instance Model}

To justify \gls{vm} pausing to users, we propose \emph{green instances}\
as an option\
in the manner of Amazon's \emph{spot instances}\
\cite{agmon_ben-yehuda_deconstructing_2011}.\
Spot instances\
have a disadvantage -- they will only be running when\
there are free computing resources; and an advantage -- a\
more affordable price. Expanding on this philosophy of acceptable\
trade-offs, we devised the green instance model.\
We envision it as an option where\
(1) compute instances are offered at a reduced availability time,\
but (2) at a lower price and (3) with environmental metrics\
presented to the user.\
From the cloud provider's perspective,\
this allows for more flexibility\
while scheduling to reduce energy costs and lessen the environmental\
impact which is good for public relations.

This would be an opt-in model, where users would be offered an additional\
\gls{sla} -- $SLA_G$ to rent green instances or choose normal instances\
($SLA_N$). The set $G$ of instances managed\
by the peak pauser would be restricted to green instances only.\

Only users who applied for green instances and\
therefore accept occasional downtimes (at relatively\
predictable times of day) would feel the consequences of the peak pauser\
and create\
\gls{era}s. \gls{era}s basically mean that the cloud provider\
gets better electricity prices in return for\
helping the electrical grid's efficient operation.\
As a result, the cloud provider\
can offer lower prices to the end user, to compensate for the reduced\
\gls{vm} availability. So, where in spot instances\
users sacrifice performance for profitability, in green instances they\
sacrifice availability for profitability.


As a further reinforcing factor in favour of green instances,\
users could be presented with the approximated energy savings arising\
from choosing green over normal instances.\
Curry et al. \cite{curry_environmental_2012} denote this type of information\
\emph{environmental charge-backs} ($EC$). Using \gls{cef} and \gls{pue}\
defined in \cite{belady_carbon_2010} we could express it as:
\
\begin{equation}
EC = CEF * PUE * (\text{VM energy consumption})
\end{equation}
\
VMeter developed in \cite{bohra_vmeter:_2010} might be used to determine\
the energy consumption of a user's \gls{vm}s or the approach from\
\cite{alonso_tools_2012} for parallel applications.

Green instances would target the same user group as Amazon's spot instances do\
and the sole fact that Amazon, a commercial corporation, still offers this\
service shows that there are interested parties.\
As a potential use-case,\
many automated background processes would be fit for running on green instances.\
Examples of such processes are nightly builds of software (frequent\
and long-lasting due to compile-time optimization for various architectures),\
automated testing, web crawling, offline data mining etc.\
However, for applications that require\
real-time human interaction, access to data in a storage-as-a-service manner\
\cite{livenson_towards_2011}, normal instances would be preferred.


\section{Evaluation Methodology}
\label{sec:methodology}

The goal of the empirical testing environment is to show how the peak pauser\
scheduler influences energy efficiency and application execution time. \ 
To measure the effects of applying the peak pauser scheduler in practice,\
it was implemented on top of the open source cloud manager OpenStack \cite{ken_deploying_2011},\
hosting a \gls{vm}.\
The server running the experiment was connected to a wattmeter that provides\
us with precise information about power consumption.\
Additionally, to evaluate our scheduler under more realistic conditions than our\
prototype environment, we simulated the scheduler's effects\
in production environments, such as that of Google \cite{fan_power_2007}.

We will now go through the details of our evaluation setup and show how\
we measured the effectiveness of the peak pauser algorithm empirically and\
by simulating production systems.

\subsection{Empirical Testing}

\begin{figure}
\centering
\includegraphics[width=0.75\columnwidth]{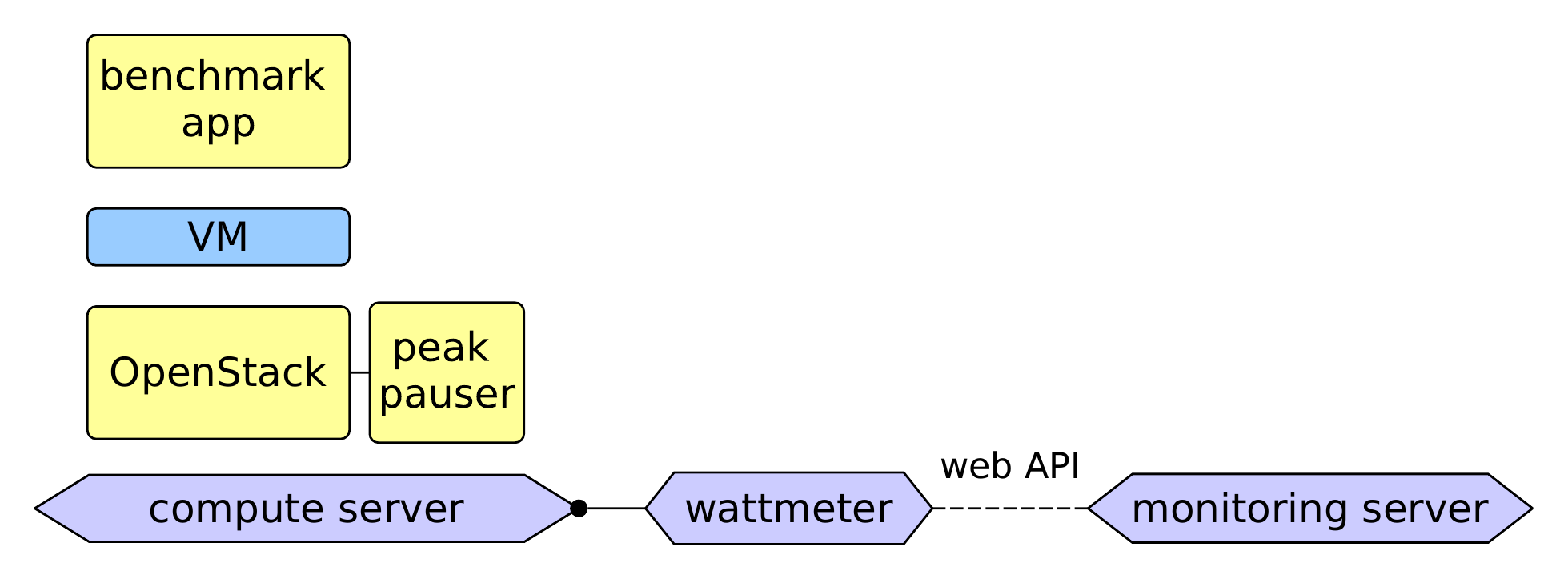}
\caption{Experiment deployment}
\label{fig:deployment}
\end{figure}

We ran a predictable\
synthetic benchmark application inside a \gls{vm} controlled by the peak pauser\
for approximately 24 hours.\
Exactly the same experiment was performed once again, but\
\emph{without a scheduler} to obtain comparison results.\
Two assumptions were made due to electrical price data availability -- that\
the data center is located in Illinois, USA\
which offers a real-time pricing option~\cite{_ameren_2016} and that\
the experiment occurred a couple of weeks in the past.

The parameters of the peak pauser were set to pausing for a total\
of 4 hours in a day ($downtime\_ratio=0.16$). The statistically most probable peak hours\
were determined according to 3 months
of historical electricity prices~\cite{_ameren_2016}\
before (non-inclusive)\
the day the experiment was assumed to be running on. 

As a benchmark application we used Berserk~\cite{_berserk_2015}, our framework
for running CPU-intensive tasks -- repeated\
recursive calculation of Fibonacci numbers.\
The benchmark application was run inside a \gls{vm}\
deployed on an OpenStack \cite{ken_deploying_2011} compute server\
\footnote{AMD Opteron 4130 2.6 GHz CPU, 8 GB RAM, Ubuntu 12.04\
server 64-bit GNU/Linux OS with OpenStack Essex}\
whose power consumption was measured\
%
%
as illustrated in Fig. \ref{fig:deployment}.\
The peak pauser scheduler was run on the same physical server as OpenStack and\
the Philharmonic~\cite{drazen_lucanin_philharmonic_2014} framework we developed
used its API to control the managed \gls{vm}.\



We measured \emph{active power} consumed by the compute server using a wattmeter\
\footnote{EATON ePDU PW104MA0UC34}\
offering a web interface for collecting data.\

\subsection{Estimating Savings in Production Systems}
\label{sec:synthetic}

To broaden our evaluation by\
estimating savings in\
today's production systems, we relied on a study by Google\
\cite{fan_power_2007} to gain insight into their power consumption during peak and low demand.\
According to their study, peak power consumption of a\
server ranges from 100 to 250 W and\
idle power (the power consumed when nothing is executed on this server) can be\
as low as 50-65\% of this amount. This ratio of idle and peak power is referred to\
as the \emph{idle ratio}. It represents the energy elasticity of a server and\
turned out to be very important in our study.\

In addition to this, existing mechanisms such as suspend and wake-on-lan are\
already available for completely turning off under-utilized\
components. In fact the topic of energy-proportional servers is\
currently being studied extensively\
\cite{meisner_powernap:_2009}\
and it is likely that energy elasticity will\
improve in the future, aiming for an ideal idle ratio of 0.

\begin{figure}
\centering
\includegraphics[width=0.75\columnwidth]{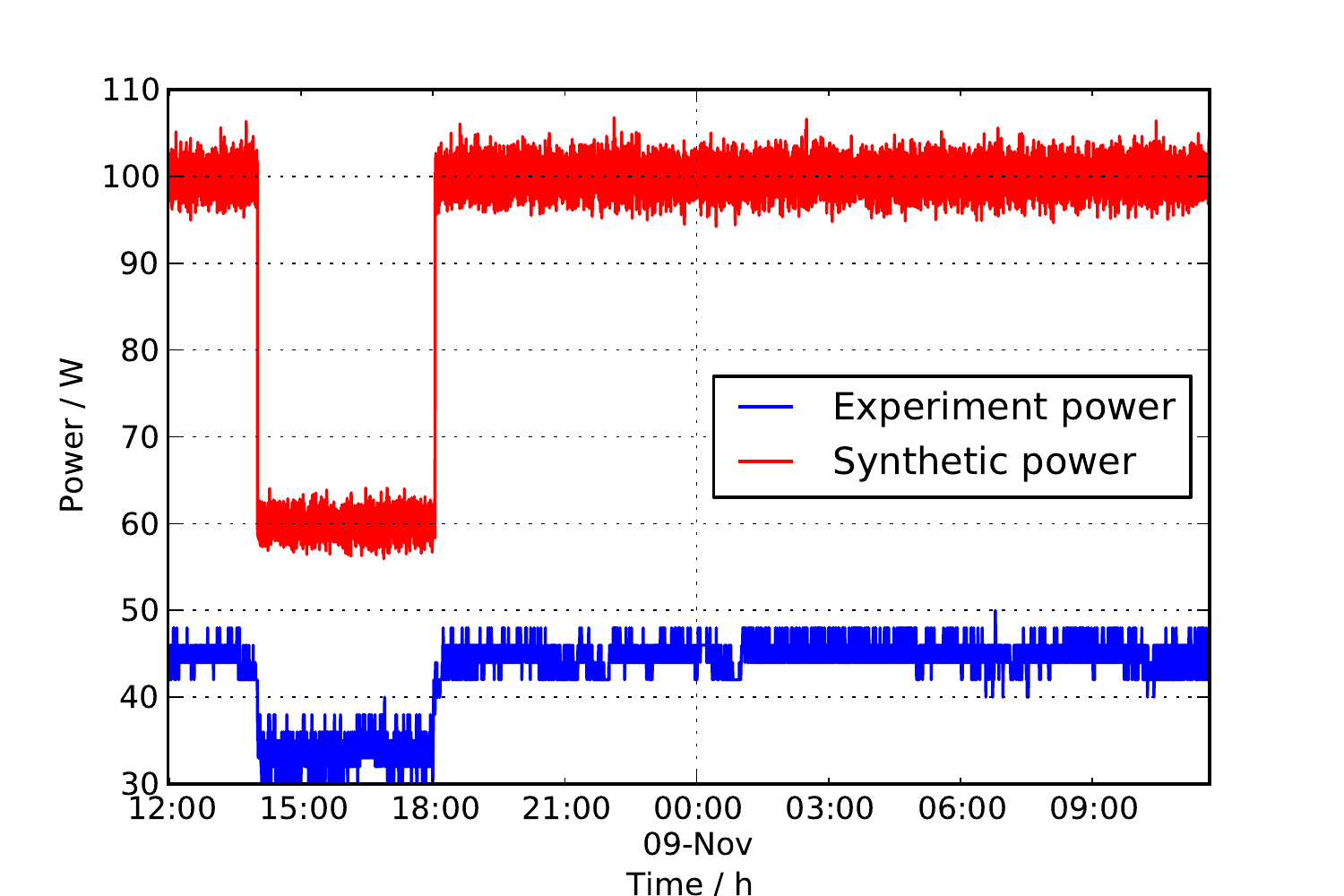}
\caption{Power consumption in our experiment and a derived synthetic signal.}
\label{fig:synthetic}
\end{figure}

A synthetic power time series was generated\
according to a simple model of\
our empirically collected data and scaled to match production-quality\
parameters from \cite{fan_power_2007}.\
We assumed normally distributed\
oscillation around the peak (during \gls{vm} execution) and idle\
(during a pause event) power values with a variance matching our experiment.\
The empirical and synthetic power signals\
can be seen in Fig. \ref{fig:synthetic}.
Power is centered around 100 W peak power during \gls{vm} execution
and 60 W idle power while the \gls{vm} is paused.
The synthetic signal was used for estimating\
savings under different energy elasticity parameters.

\subsection{Calculating Electricity Costs}

We previously established electricity costs to be more proportional to the\
actual environmental impact than the bare energy consumption,\
which we measure. To get as close as we can to knowing the\
impact our system has on the environment, we therefore\
want to calculate the total monetary expense. As we only have\
access to real low-level metrics obtained using the wattmeter\
to monitor server behaviour in an empirical experiment, we
have to calculate these monetary expenses on our own. To achieve\
this, we tried to mimic the pricing system actually used by the utilities\
in a real-time electricity price business model and we will explain\
these methods in detail here.

We sampled active power measurements of the compute server every 5 seconds.\
From this, we can calculate\
energy consumption and, according to real-time\
electricity prices~\cite{_ameren_2016} corresponding to\
the appropriate time intervals, derive a total energy price. $S_{total}$,\
the total electricity price in a time interval $T=\overline{t_0t_N}$\
is calculated from a numerical integral (we used the basic rectangle rule):

\begin{equation}
S_{total} =\sum_{t=t_0}^{t_N-1} \frac{t_N-t_0}{N}*P_t*C_t
\end{equation}

where $P_t$ and $C_t$ stand for power and electricity price\
in moment $t$, respectively and $N$ is the number of samples.

\section{Results}
\label{sec:results}

We will now present the results obtained after running the empirical\
experiment, followed by our estimation\
of savings in a production environment based on a synthetic power signal.\
Finally, we assemble the results into a green instance \gls{sla}\
that could be offered to users.

\subsection{Empirical Savings}

\begin{figure}
\centering
\begin{subfigure}{0.48\textwidth}
  \centering\includegraphics[width=\textwidth]{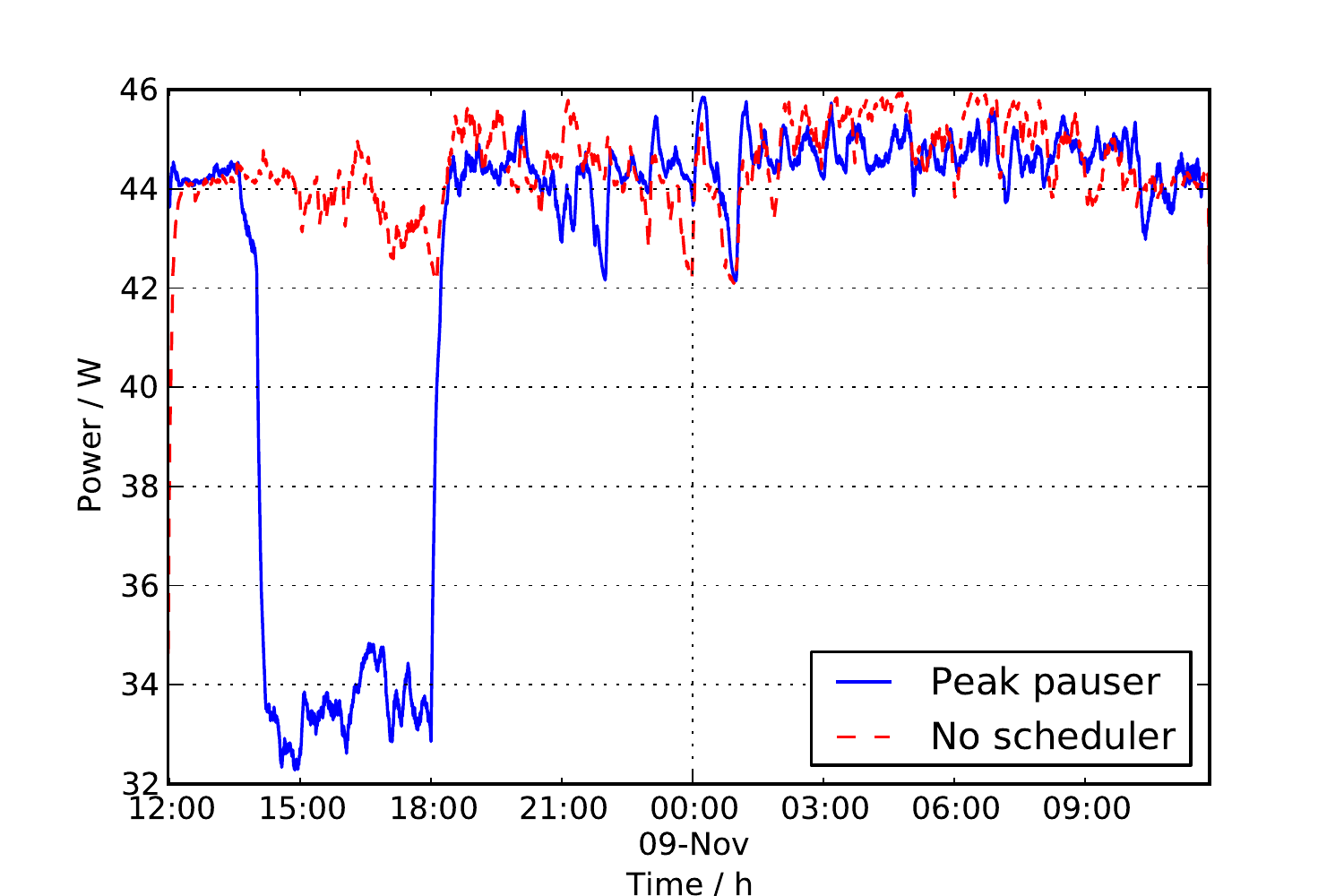}
  \caption{Dynamic behaviour}\label{fig:dynamic_comparison}
\end{subfigure}
\begin{subfigure}{0.48\textwidth}
  \centering\includegraphics[width=\textwidth]{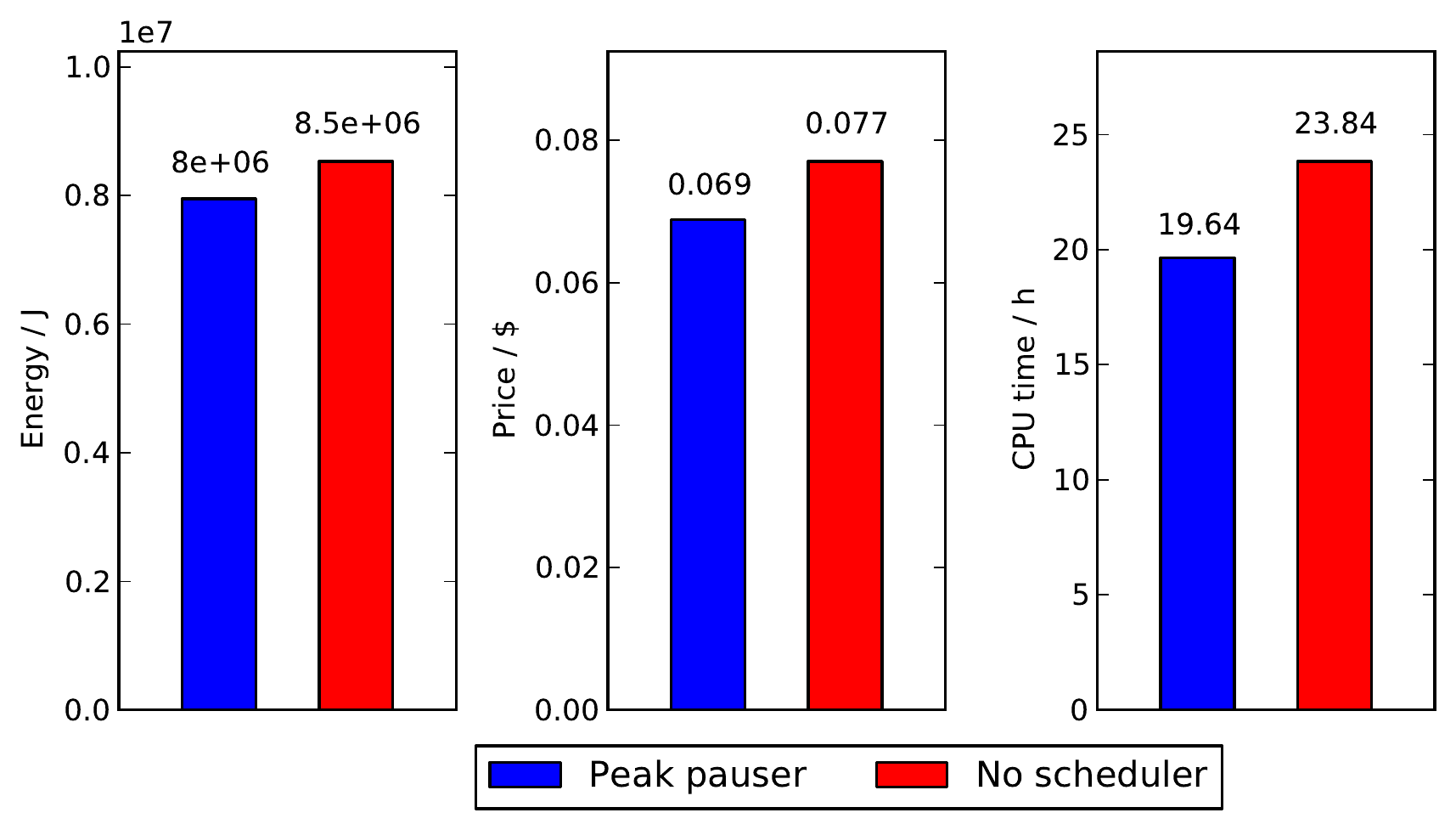}
  \caption{Aggregated results}\label{fig:aggregated_comparison}
\end{subfigure}
\caption{Empirical evaluation results}
\end{figure}

We are interested in a comparison of running the experiment for 24 hours\
with and without the peak pauser scheduler with regards to:
\begin{itemize}
  \item total consumed energy (also considering the energy consumed\
  while the \gls{vm} is paused)
  \item total electricity price based on dynamic, hourly charging
  \item benchmark CPU time
\end{itemize}

Runtime results can be seen in Fig. \ref{fig:dynamic_comparison}. The\
curves are smoothed using\
the \gls{ewma} method \cite{roberts_control_1959}. They\
show how the compute server's power consumption changes throughout\
the whole experiment. The drop in power from cca. 44 to 34 W\
($\approx 23\%$ or an idle ratio of $\approx 77\%$) can clearly be\
seen from 14 to 18 h. The \gls{vm}\
is paused during this time interval.

The results of aggregating measurements\
over the entire 24 hours of the experiment\
are shown in Fig. \ref{fig:aggregated_comparison}.\
The total energy consumption is $\approx 5.3\%$ lower\
than the amount consumed without\
a scheduler, due to the the reduced power consumption during \gls{vm} pausing.\
The difference in the electricity price is even\
larger, since the peak pauser only excludes the most expensive\
(financially and environmentally) hours of the day, the amount spent\
during a single day is $\approx 6.9\%$ lower when using the peak pauser.

The CPU time the benchmark application received in the experiment is\
4 hours less when using the peak pauser scheduler. This\
amounts to $\approx 17.6\%$ fewer actual calculations. This is a considerable\
performance deterioration,\
however it would only affect green instances whose owners agreed on\
fewer computing resources to increase energy efficiency and get\
a better price. An important thing to note here is\
that the current benchmark implementation\
only works in a single thread, not being able to consume all the CPU cores.\
In reality, where many processes are running, consuming many CPU cores,\
the ratio of energy and price savings to CPU time wastage would be\
greater, which we examine next.

\subsection{Projected Savings}
\label{sec:results-projected}

\begin{figure}
\centering
\includegraphics[width=0.70\columnwidth]{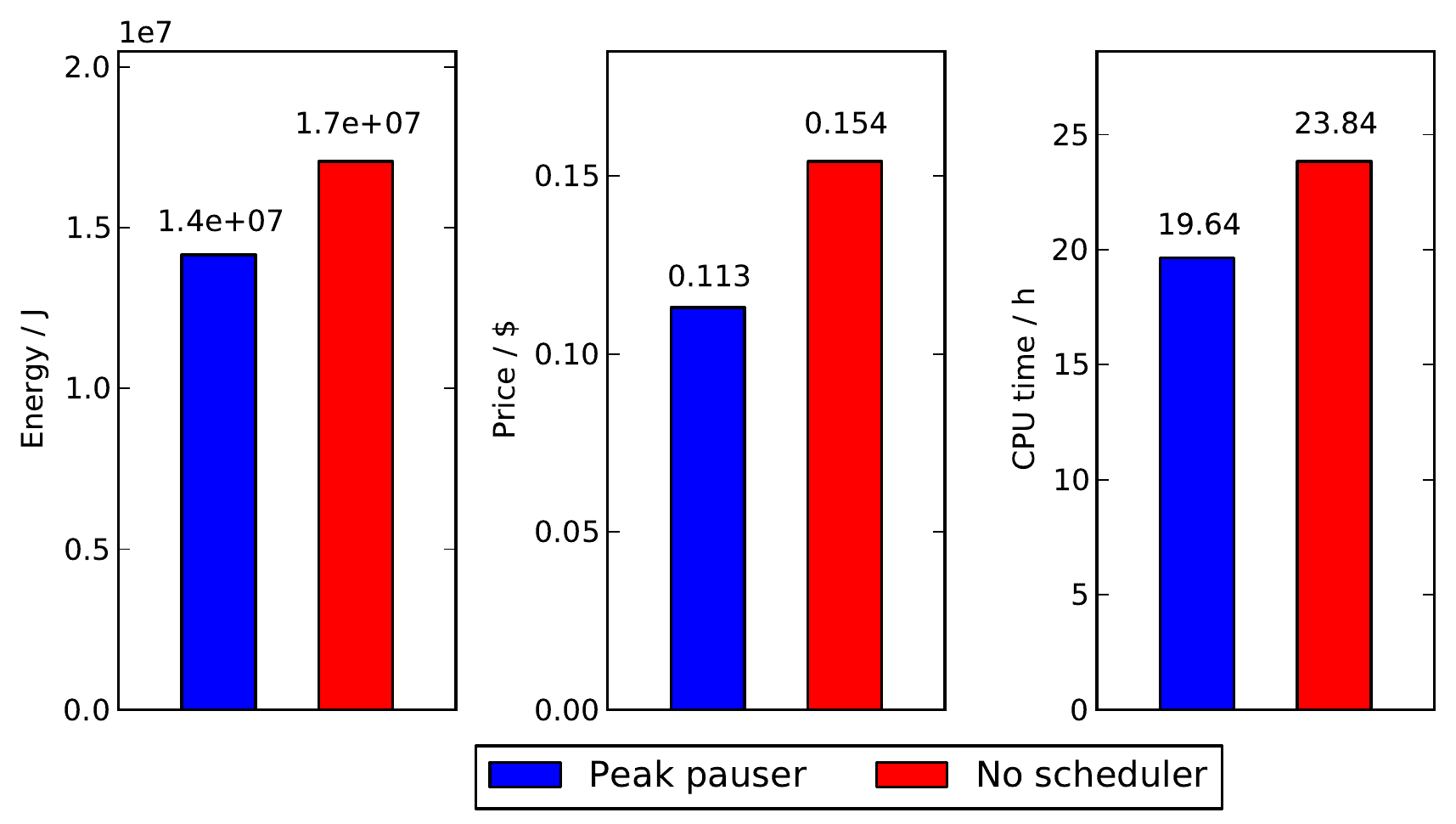}
\caption{Synthetic data results}
\label{fig:aggregated_comparison-synthetic}
\end{figure}

After applying the same energy and price calculation methods on synthetic data
described in Section~\ref{sec:synthetic}, much better savings can be achieved.
Fig.~\ref{fig:aggregated_comparison-synthetic} shows the aggregated results on
a server with a power of 200 W during peak load and 0 W while being idle. This
scenario assumes an ideally energy proportional server or the utilization of a
suspend and wake-on-LAN mechanism on the physical machine
(albeit a bit simplified
as no delay is added to account for these actions).
The raw energy savings are $\approx 17.1\%$, roughly equal to the
drop in availability. The price savings exceed both of these values and
amount to $\approx 26.63\%$ which is a considerable improvement.

Savings based on more combinations of peak power and idle ratio
(the ratio of idle and peak power) parameters
are given in Table~\ref{tab:savings}. It is interesting to note how peak power
does not seem to influence savings much -- the difference between 100 and 200 W
accounts for less than 1\%.
The idle ratio, on the other hand, has a high impact,
underlining the importance of dynamically adjustable,
power-proportional servers.


\begin{table}
\centering
\caption{Estimated energy and price savings}
   \begin{tabular}{l||l|l|l|l|}
   \cline{2-5}
&\multicolumn{2}{ c| }{Energy savings} &\multicolumn{2}{ c| }{Price savings}\\\cline{2-5}

	\hline
$idle\ ratio$  |  $P_{peak}$ & 100 W & 200 W & 100 W & 200 W\\
	\hline
	\hline
0\% & 16.96\% & 17.01\% & 26.56\% & 26.63\%\\
30\% & 11.93\% & 11.94\% & 18.68\% & 18.69\%\\
60\% & 6.82\% & 6.82\% & 10.67\% & 10.67\%\\

	\hline
   \end{tabular}
\label{tab:savings}
\end{table}


Based on the estimated annual electricity costs for a company as large as Google\
conservatively placed at \$38M \cite{qureshi_cutting_2009}, even with\
a realistic power usage efficiency of\
1.3 the above savings amass to very large numbers and show considerable impact -- both\
economically and environmentally.

\subsection{Resulting SLA}


Given a 4-hour daily pause, instance availability is 83.3\%.\
If we consider the 0--200 W scenario from\
Fig. \ref{fig:aggregated_comparison-synthetic},\
a \gls{pue} of 1.3 and a \gls{cef} of 1537.82 lb/MWh measured\
in \cite{_united_2008} for Illinois, we can calculate the annual environmental charge-back\
for green instances to be 1300 kg\gls{coe}. This is 300 kg less than a normal instance\
would produce (equivalent to driving an average car for 811 km). This is an approximation,\
but it gives an idea of the order of magnitude and as such might be presented\
to users as an additional advantage of green instances.

If we assume a normal instance cost of \$0.060 per hour,\
the 26.6\% savings in electricity costs would mean that green instances could\
be offered at \$0.044 (disregarding many other factors such as equipment\
amortisation and maintenance costs which are out our our work's scope).

\section{Summary}

In this chapter we have shown a practical way of utilizing information about\
the electrical grid in the domain of cloud computing\
to visibly reduce energy consumption and costs.

We presented the \emph{peak pauser} scheduling algorithm that offers a clean way of\
managing virtual machines in a computing cloud. It pauses\
computation during hours when the electricity\
price is statistically most probable to peak.\
This mechanism reduces energy costs\
through controllable availability reduction.\
Offering this kind of service as \emph{green instances}\
under special \gls{sla}s to willing\
users only, would ensure that no harm is done from the user point of view.\
Furthermore, this represents environmentally friendly behaviour, because\
energy production applies most stress to the environment\
during times of peak demand,\
when it has to resort to faster and more inefficient generation methods.

Our prototype implementation and experimental evaluation show that\
savings are indeed possible in real-life systems.\
Results stemming from our synthetically-scaled projections of different\
parameters give a sketch of potentially even higher gains if the methods\
were to be used in production systems of today's leading cloud providers.






\chapter{Pervasive Cloud Control for Geotemporal Inputs}
\label{ch:gascheduler}
In this chapter, we expand the grid-conscious cloud model from the previous
chapter for geographically distributed data centers and additional
geotemporal inputs like temperatures affecting cooling efficiency.
Distributed data center infrastructure changes\
the rules of cloud control, as\
energy costs depend on current regional electricity prices and temperatures.\
We denote geotemporal inputs, cloud requirements, regulations and other\
factors that guide the cloud provider's actions\
as \emph{decision support components}.
We define \emph{forward compatiblility} as being able to cope with\
additional decision support components without drastic changes of\
the core architecture.\
Hence, to account for emerging technologies surrounding\
the cloud ecosystem, 
a maintainable control solution needs to be forward-compatible
with new decision support components.\
Existing cloud controllers are focused on \gls{vm} consolidation methods\
suitable only for a single data center\
or consider migration just in case of workload peaks,\
not accounting for all the aspects of geographically distributed data centers.\
In this chapter, we propose a pervasive cloud controller\
for dynamic resource reallocation adapting to\
volatile time-dependent and location-dependent factors, \
while considering the \gls{qos} impact of too frequent migrations\
and the data quality limits of time series forecasting methods,
such as the methods analysed in Chapter~\ref{ch:background}.\
The controller is designed with extensible decision support components.\
We evaluate it in a simulation\
using historical traces of electricity prices and temperatures.\
By optimising for these additional factors,\
we estimate \gaschedulerEnSavingsNoGeotemp{} energy cost savings\
compared to baseline dynamic \gls{vm} consolidation.\ 
We provide a range of guidelines for cloud providers,\
showing the environment conditions necessary to achieve\
significant cost savings and we validate the controller’s extensibility.\




In the remainder of this chapter,
Section~\ref{ch:gascheduler:sec:approach} explains the research problem intuitively\
on a real example of geotemporal inputs\
and provides a high-level description\
of our pervasive cloud controller.\
The formal specification of the plug-and-play decision support components and\
the optimisation problem specification\
is presented in Section~\ref{ch:gascheduler:sec:model}.\
The proof-of-concept implementation of the forward-compatible\
optimisation engine of the pervasive cloud controller we developed\
is explained in Section~\ref{ch:gascheduler:sec:gascheduler} and\
in Section~\ref{ch:gascheduler:sec:evaluation} we describe the evaluation methodology and\
discuss the results.


\begin{figure*}
\centering
\includegraphics[width=0.99\columnwidth]{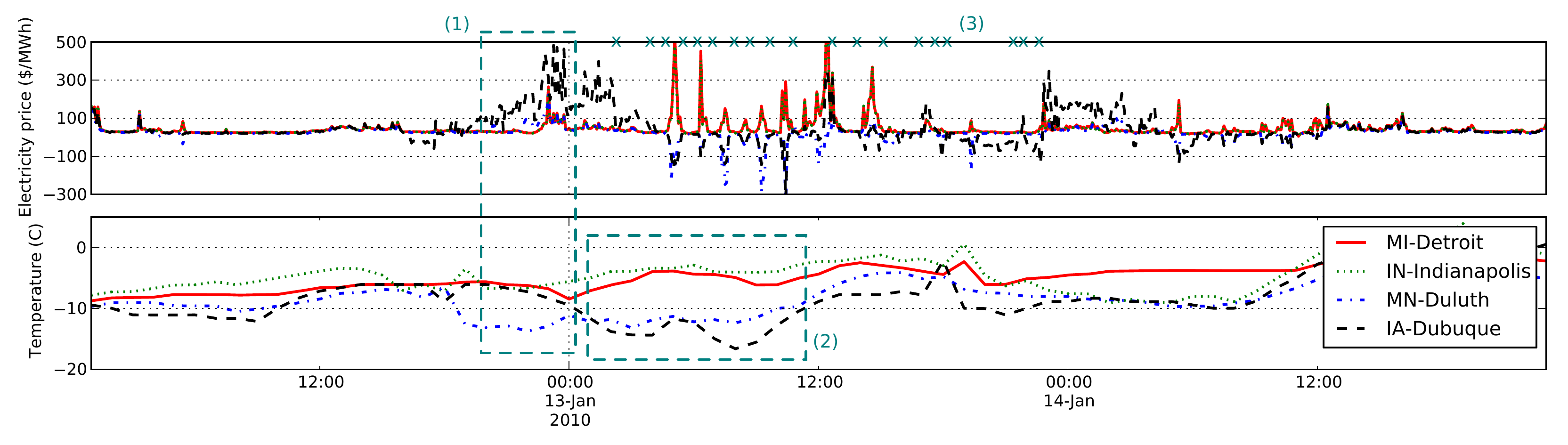}
\caption{Geotemporal inputs (real-time electricity prices and temperatures)\
at four locations in the USA during a period\
of three days.}
\label{fig:geotemporal-inputs}
\end{figure*}

\section{Geotemporal Cloud Environments}
\label{ch:gascheduler:sec:approach}

%

We now explain the geotemporal environment surrounding\
geographically distributed clouds on a real example of electricity\
prices and temperatures. On this use case, we will give an overview\
of the problem on an intuitive level and give a high-level description\
of our pervasive cloud controller, before detailing the\
formal specification of the model in the following section.

\subsection{Problem Overview}

Example geotemporal inputs\
for four US cities\
are shown in Fig.~\ref{fig:geotemporal-inputs}.\
Temperature values within a single day change up to 15 degrees
between peaks and lows
and even larger relative differences can be observed
in the volatile electricity prices.
Electricity price data was obtained from \cite{alfeld_toward_2012}
and temperatures from \cite{_forecast_2015}.
%
%
Rapid changes in geotemporal inputs can occur dynamically.\
The peak that can be seen in Dubuque\
on January 12th from 21:00 to 23:00 (1),\
results in five or more times the average prices.\
It can be observed that temperature peaks occur towards the end of the day,\
while lows occur during nights.\
Even though electricity price is more volatile,\
%
partial dependence on previous data points can be seen.\
This means that it is possible to model their behaviour to\
forecast probable future values, and in fact is done\
in practice \cite{weron_modeling_2006,_forecast_2015}.

To explain the potential and challenges of geotemporal inputs\
in the context of cloud computing, let us assume that there are\
two data centers -- one in Detroit and\
another one in Dubuque.\
For the data shown in\
Fig.~\ref{fig:geotemporal-inputs} during the period (1),\
it makes sense to run more \gls{vm}s\
in Detroit when temperatures are the same, because electricity\
is more expensive in Dubuque (constantly over 100 \$/MWh, reaching 500 \$/MWh)\
than in Detroit\
(less than 100 \$/MWh).\
However, when it gets 10 C colder in Dubuque three hours later (2)\
and electricity prices become lower than in Detroit, less\
energy would be consumed on cooling there, resulting in lower energy costs,\
so it is better to migrate\
a number of \gls{vm}s from Detroit to Dubuque\
and shift computational load this way.\

A challenge in adapting cloud control for geotemporal inputs is that\
the cloud provider cannot migrate \gls{vm}s\
between different locations too rapidly,\
as this wastes bandwidth, incurs an energy overhead and impacts \gls{qos}.\
This is underlined even more by the volatile variable behaviour\
observable in electricity prices.\
In Fig.~\ref{fig:geotemporal-inputs},\
we marked by crosses (3)\
all the moments throughout January 13th\
when ratios between electricity prices\
in Detroit and Dubuque change significantly,\
offering an opportunity to save on energy costs by reallocating\
\gls{vm}s using live migrations.\
We can see that 19 migrations would be performed this way.\
If we assume a downtime caused by a live \gls{vm} migration to last for\
one minute, which is possible\
based on the model presented in \cite{liu_performance_2011},\
this would result in a \gls{vm} availability of 98.68\%.\
This availability\
is considerably lower than the 99.95\% availability rates\
advertised by Amazon and Google in their \gls{sla}\ 
and incurs extra data transfer costs.\ 
The challenges arising from this are:\
(1) To profit from geotemporal inputs in cloud computing,\
the trade-offs of the energy savings of geotemporal inputs,\
the migration overheads and impact on \gls{qos}, as well as the data accuracy\
provided by the forecasting methods for future geotemporal inputs all have\
to be considered and reconciled in a long-term plan.\
(2) In reality, the problem has to be\
solved on a much larger scale with more data centers and\
thousands of \gls{vm}s.\
New ways of controlling \gls{vm}s\
across geographically distributed data centers\
have to be developed to address these challenges.\ 



\subsection{Pervasive Cloud Controller} 

We now present our pervasive cloud controller approach by explaining\
the identified requirements, defining the architecture of the\
solution and giving a high-level overview of its workflow.\ 

\subsubsection{Requirements}
In our approach, we consider\
an \gls{iaas} cloud provider hosted on multiple\
geographically distributed data centers.\
The cloud is assumed to be operating in an environment comprising\
geotemporal inputs such as \gls{rtep} and\
temperature-dependent cooling efficiency\
(other inputs can be added as well).\
The cloud is governed by a controller system\
that manages virtual and physical machines\
in all data centers and can issue actions, such as migrating\
a \gls{vm} from one \gls{pm} to another, suspending or resuming a \gls{pm}.

\vspace{\tabletopmargin}
\begin{table}[H]
\centering
\caption{Example SLA.}
\vspace{\tablecaptionmargin}
\label{tab:sla}
\begin{tabular}{lllll}
\toprule
 \#CPUs & RAM & storage & availability & price\\
\midrule
4 & 15 GB & 80 GB & $99.95\%$ & \$0.28/hour\\
\bottomrule
\end{tabular}
\end{table}
\vspace{\tablebottommargin}

The first input are user goals represented by \gls{sla}s, specifying the\
number of requested \gls{vm}s,\
their resource and \gls{qos} requirements.\ 
An example of user requirements for an Amazon \emph{m3.xlarge} \gls{vm}\
specified in an \gls{sla} is shown in Table~\ref{tab:sla}.\
The second input are the geotemporal inputs, providing time series\
metrics describing each of the data center locations, such as\
electricity price and temperature data,\ 
example values of which are shown in Fig.~\ref{fig:geotemporal-inputs}.\
Each geotemporal input is a time series of past and current\
values and, using time series forecasting, it is possible to predict\
future values and the accompanying  data quality\
(reach and the most likely error rate).\
The controller's task is to output a schedule that determines\
where each \gls{vm} is deployed\
and for each \gls{pm} if it is running or suspended at any point in time.\

\subsubsection{Architecture}
\begin{figure}
\centering
\includegraphics[width=0.85\columnwidth]{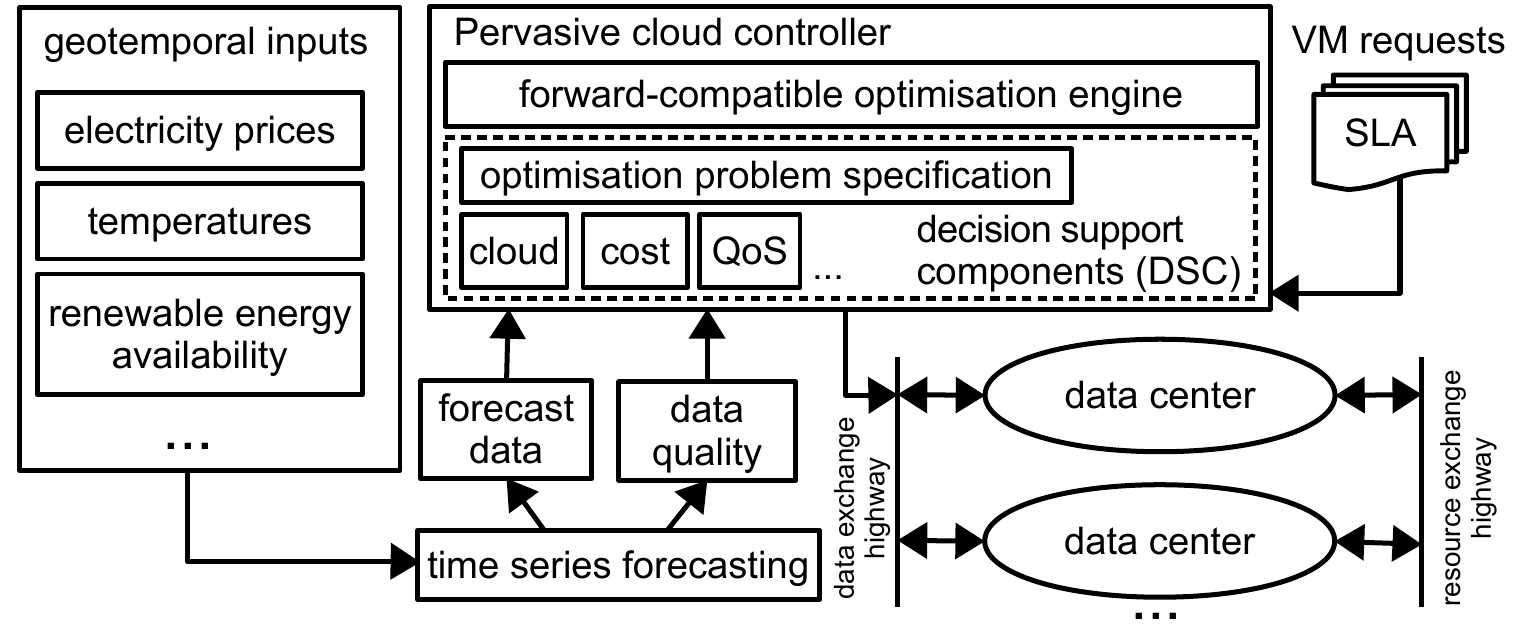}
\caption{Pervasive cloud controller architecture.}
\label{fig:gascheduler-overview}
\end{figure}

Fig.~\ref{fig:gascheduler-overview} shows the architecture of\
our proposed pervasive cloud controller for managing\
a geographically distributed cloud based on geotemporal inputs.\
On a high level, geotemporal inputs are used to obtain forecast\
data and the corresponding quality.\
This input is, together with the \gls{sla}s, provided to the pervasive\
cloud controller, which generates\
a long-term schedule of control actions to apply\
to the geographically distributed data centers.\

Looking at the architecture details,\
geotemporal input forecasts are converted by the controller into\
values meaningful to the cloud provider, e.g. data center energy costs\
that combine \gls{rtep} with the cooling overhead, environmental impact etc.\
These measures along with other internal measures like cloud capacity and\
the \gls{qos} stemming\
from actions planned for \gls{vm}s are all combined\
into an optimisation problem specification as\
decision support components.\
To support new geotemporal inputs, \gls{sla} metrics or cloud regulations,\
it is important for the decision support components to be extensible\
in a plug-and-play manner,\
i.e. without requiring architectural changes.\
We formally present the decision support component model\
in the following section.\
The role of ensuring decision support component extensibility\
lies in a\
forward-compatible optimisation engine.\
It considers the decision support components as criteria\
to plan and optimise\
a schedule of control actions for a future period.\
The challenging part of ensuring forward compatibility\
with new decision support components is that\
there has to be a separation of\
the schedule evaluation logic and the optimisation logic.\
The schedule is evaluated using geotemporal inputs\
and the time-based allocation of \gls{vm}s to \gls{pm}s,\
to estimate the actions' outcome in terms of costs,\
\gls{qos} and any other decision support components.\
This evaluation is then used by the controller in a black-box manner\
to explore the search space of possible actions\
using its custom optimisation logic and the high-level information\
about each schedule returned by the evaluation logic.\
The selected schedule is applied over time\
to physical and virtual machines in the cloud\
(forwarding control actions through a data exchange highway)\
by utilising live \gls{vm} migrations as a resource exchange highway\
to redistribute computational load between the data centers.\



%

\subsubsection{Workflow}
\label{sec:pcc:workflow}

\begin{figure}
\centering
\includegraphics[width=0.6\columnwidth]{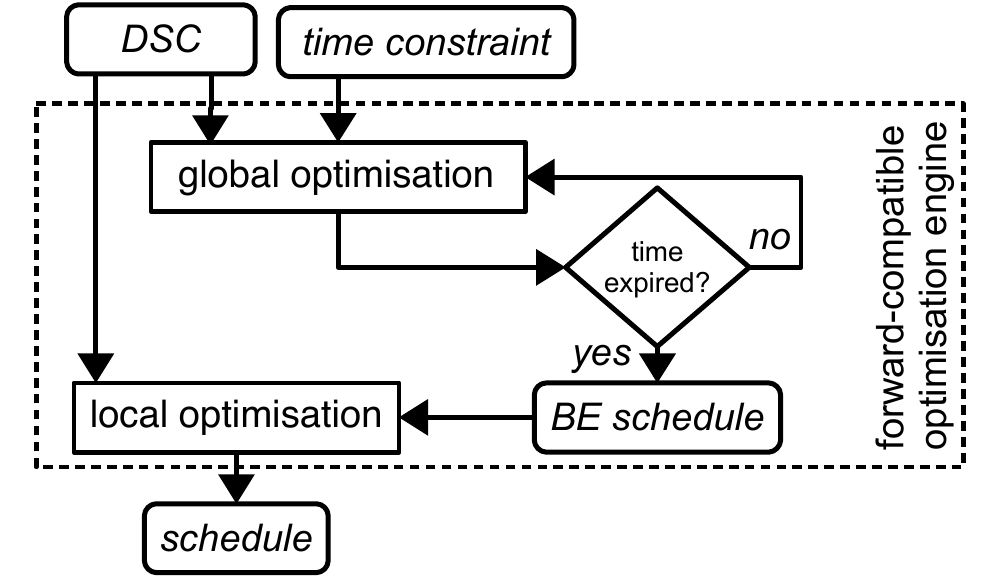}
\vspace{-0.1cm}
\caption{Optimisation engine workflow.}
\label{fig:pcc}
\vspace{\figbottommargin}
\end{figure}

The forward-compatible optimisation engine worfklow\
is illustrated in Fig.~\ref{fig:pcc}.\
It collects the decision support components,\
related together as an optimisation problem specification\
and produces a schedule of\
control actions to apply in the cloud.\
Given that bin packing of allocating \gls{vm}s to \gls{pm}s\
is an NP-hard problem and\
in our case we add to it the dimension of time and time-related\
\gls{qos} requirements (e.g. the \gls{vm} migration frequency),\
an optimal solution can not be found at runtime for an arbitrary\
problem size. To overcome this,\
we propose a two-stage optimisation process.
The first stage is a global optimisation\
method that sweeps the whole search space looking for a global optimum\
within a time constraint (provided as an additional parameter\
that the cloud provider specifies).\
We then take the best-effort schedule this method was able to find\
(BE schedule) and pass it to a second stage local optimisation method that\
continues to improve it with a primary goal\
of satisfying all the hard constraints\
among the decision support components\
that the global optimisation failed to satisfy.\
%
In Section~\ref{ch:gascheduler:sec:gascheduler} we\
present our proof-of-concept implementation\
for the global and local optimisation methods\
based on a hybrid genetic\
algorithm with greedy local constraint satisfaction.\
We later use these methods to evaluate the pervasive cloud controller.\

\section{Decision Support Components}
\label{ch:gascheduler:sec:model}

In this section we formally specify the\ 
cloud, \gls{qos} and cost decision support components\
and show how they are related together\
into an optimisation problem specification\
of optimisation goals and constraints.\
These decision support components are then used by the optimisation\
engine in the schedule generation. 

\subsection{Cloud and \gls{qos} Components}
\label{ch:gascheduler:sec:cloud_components}
We consider a single \gls{iaas} cloud provider that is\
represented by $DCs$, a set of $d$ geographically distributed data centers\
and $PMs$, a set of $p$ physical machines it operates.\
We define each physical machine's location at one of the data centers.

\vspace{\eqtopmarginbig}
\begin{align}
\forall pm \in PMs,\ loc(pm) \in DCs
\end{align}
\vspace{\eqbottommarginbig}

As we are modelling dynamic system behaviour, we define a time period\
in the range from $t_{0}$ to $t_{N}$ (denoted $[t_{0}, t_{N}]$),\
of $N$ discrete (arbitrarily small) periods.
\


User requirements are defined by $vmreqs_t$, a set of virtual machine\
requests at a moment $t$, each of which can either ask for a new \gls{vm}\
to be booted or an existing\
one to be deleted. These events are controlled by end users and we assume\
no prior knowledge of the users' requests\
(as is the case in \gls{iaas} clouds like Amazon EC2).\
Based on the past and current $vmreqs_t$, we define $VMs_t$,\
a set of VMs provided to the users at time $t$.

The cloud provider defines an extensible set of $r$ resource types\ 
that have to be specified through an \gls{sla},\
e.g. number of CPUs and amount of RAM.\
The exact resources for a single \gls{vm} are\
an ordered $r$-tuple of values, defining the \gls{vm}'s $spec$:

\vspace{\eqtopmarginbig}
\begin{align}
spec_{vm} = \left(res_{vm,i},\, \forall i \in \{1, \ldots r\}\right),\ \forall vm \in VMs_t
\end{align}
\vspace{\eqbottommarginbig}

where $res_{vm,i}$ is the $i$-th resource's value.\
For the example shown in Table~\ref{tab:sla}, there are three quantitative\
resource types\
(number of CPUs, amount of RAM and amount of storage)\
that are provided on the infrastructure level, so $r=3$.\
Given the concrete values for an \emph{m3.xlarge} $vm$ instance,\
we have $spec_{vm} = (4, 15, 80)$.\
Similarly, the capacity of a \gls{pm} is defined as an r-tuple\
of the resource amounts it has:

\vspace{\eqtopmarginbig}
\begin{align}
spec_{pm} = \left(cap_{pm,i},\, \forall i \in \{1, \ldots r\}\right),\, \forall pm \in PMs
\end{align}
\vspace{\eqbottommarginbig}

We define the \gls{vm} allocation\ 
at moment $t$ as:

\vspace{\eqtopmarginbig}
\begin{align}\label{eq:alloc}
\forall pm \in PMs,\ alloc_t(pm) \subseteq VMs_t,
\text{ s.t. } \forall vm \in alloc_t(pm) \text{ is hosted on $pm$ at moment $t$}
\end{align}
\vspace{\eqbottommargin}

Effectively, $alloc_t,\ \forall pm \in PMs$\
is the cloud state at moment $t$.\
For any two subsequent moments $t_i$ and\
$t_{i+1},\ t_i \in [t_0, t_{N-1}]$,\
a $vm$ is considered \textit{migrated} if\
$\exists pm_j, pm_k \in PMs$ s.t.\
$vm \in alloc_{t_i}(pm_j)$ and $vm \in alloc_{t_{i+1}}(pm_k)$.\
The number of such migrations for a $vm$ in some relevant period\
specified by the cloud provider\ 
(e.g. an hour) is denoted $R_{mig}(vm)$\ 
and represents the rate of migrations.

\subsection{Cost Components}
\label{ch:gascheduler:sec:cost_components}

Progress has been made in\
modelling various aspects of cloud energy costs and\
we shortly outline the relevant findings\ 
of the existing energy-aware cost model\
using our notation.\
We then proceed with presenting our own\
pervasive cost model\
for expressing energy costs of \gls{iaas} clouds\
based on geotemporal inputs.\

\subsubsection{Energy-Aware Cost Model}

Power consumption $P_t(pm)$ of a $pm \in PMs$\
is modelled in \cite{guler_cutting_2013}\ 
as a function of utilisation $util_t(pm)$ at time $t$,\
with $P_{peak}$ and $P_{idle}$ standing for the\
server's power consumption during peak and idle load, respectively.

\vspace{\eqtopmarginbig}
\begin{align}\label{eq:power_base}
P_t(pm) = pow(util_t, pm) = P_{idle} + util_t(pm) \cdot (P_{peak} - P_{idle})
\end{align}
\vspace{\eqbottommarginbig}

The impact of time series forecasting errors\
is modelled in \cite{cauwer_study_2013},\
where the predicted value $\hat{x_t}$ of a real value $x_t$\
at time $t$ is:\

\vspace{\eqtopmarginbig}
\begin{align}\label{eq:forecasting_error}
\hat{x}_t = \mathcal{N}(x_t, \sigma_{pred}^2),\ \forall t \in [t_{0}, t_{N}]
\end{align}
\vspace{\eqbottommarginbig}

where 
$\mathcal{N}(x_t, \sigma_{pred}^2)$ is
a Gaussian distribution\
with mean $x_t$ and standard deviation $\sigma_{pred}$.\

Temperature-dependent cooling efficiency resulting from\
computer room air conditioning using outside air economizers\
is modelled in \cite{xu_temperature_2013}.\
Cooling efficiency is expressed as partial \gls{pue} $pPUE_{dc,t}$\
at data center $dc$ at time $t$,\
which affects the power overhead based on the following formula:

\vspace{\eqtopmargin}
\begin{align}\label{eq:P_tot}
pPUE_{dc,t} = \frac{P_t(pm) + P_{cool,t}(pm)}{P_t(pm)} = \frac{P_{tot,t}(pm)}{P_t(pm)}
\end{align}
\vspace{\eqbottommargin}

where $P_{cool,t}(pm)$ is the power necessary to cool $pm$, and\
$P_{tot,t}(pm)$ stands for the combined cooling and\
computation power.\
The dynamic value of $pPUE_{dc,t}$ is modelled as a function of\
temperature $T$ to match hardware specifics as:\

\vspace{\eqtopmarginbig}
\begin{align}\label{eq:pPUE}
pPUE_{dc,t} = 7.1705 \cdot 10^{-5}T_{dc,t}^2 + 0.0041T_{dc,t} + 1.0743
\end{align}
\vspace{\eqbottommargin}
%

Based on the migration model developed in \cite{liu_performance_2011},\
the combined energy consumption overhead\
of the source and destination hosts $E_{mig}$\
for a single migration can be calculated as a function of\
the migrated \gls{vm}'s memory $V_{mem}$, data transmission rate $R$,\
memory dirtying rate $D$ and a pre-copying termination threshold $V_{thd}$.

\vspace{\eqtopmarginbig}
\begin{eqnarray}\label{eq:E_mig}
E_{mig} = f(V_{mem}, D, R, V_{thd})
\end{eqnarray}
\vspace{\eqbottommarginbig}

%
%

\subsubsection{Pervasive Cost Model}

Based on our extensible resource types,\
we define a generic model of server utilisation\
$util_t(pm)$ at time $t$ of a $pm \in PMs$\
as a weighted sum\
of the individual resource type utilisations:

\vspace{\eqtopmargin}
\begin{align}
util_t(pm) = \mathlarger{\sum}_{\forall i \in \{1, \ldots r\}} w_i \cdot \frac{\mathlarger{\sum}_{\forall vm \in alloc_t(pm)}res_{vm, i}}{cap_{pm, i}}
\label{eq:util}
\end{align}
\vspace{\eqbottommargin}

where $w_i,\ \forall i \in \{1, \ldots r\}$ is a value in $[0,1]$ describing\
the weight resource type $i$ has on\
the physical machine's power consumption\
(exact amounts depend on hardware specifics; the values we used\
are discussed in Section~\ref{ch:gascheduler:sec:evaluation}).\ 
Variables $cap_{pm, i}$ and $res_{vm, i}$ are the amounts of that resource\
available or requested by the $pm$ or $vm$, respectively.

We model the power consumption $P_t(pm)$ of a $pm \in PMs$\
using the basic approach from Eq.~\ref{eq:power_base}, but\
we extended it to model fast suspension of empty hosts\
(a technology explained in \cite{meisner_powernap:_2009}).\
Also, to model additional load variation,\
we define $P_{peak}$ and $P_{idle}$ as time series of\
a server's power consumption during peak and idle load\
depending on the time $t$, instead of being constant.

\vspace{\eqtopmargin}
\begin{align}\label{eq:power}
P_t(pm) =
\left\{
	\begin{array}{ll}
	    0 & \hspace{-0.5cm} \mbox{if } util_t(pm) = 0\\
		pow(util_t, pm)  & \mbox{otherwise.}
	\end{array}
\right.
\end{align}
\vspace{\eqbottommargin}

%
We use a common time series notation $\{x_t:\ t \in T\}$,\
where $T$ is the index set and $\forall t \in T$, $x_t$\
is the time series value at time stamp $t$.\
For each data center location $dc$ in $DCs$,\
there is a time series of real-time electricity\
prices $\{e_{dc,t}:\ t \in [t_{0}, t_{N}]\}$.\
Similarly, at each location there is a time series of temperature values\
$\{T_{dc,t}:\ t \in [t_{0}, t_{N}]\}$.\
To analyse forecasting errors,\
on both electricity and temperature time series,\
we apply Eq.~\ref{eq:forecasting_error}.\
To explore its impact in the evaluation,\
we vary $\sigma_{pred}$, which determines the accuracy\
of the forecast.\
%
We assumed the temperature-dependent\
cooling efficiency model from Eq.~\ref{eq:P_tot}\ 
to express $P_{tot,t}$
and kept the polynomial model and the fitted\
factors from Eq.~\ref{eq:pPUE}\
where $pPUE$ ranges from 1.02 for -25 C to 1.3 for 35 C.\

%

Combining all the equations so far, the cloud's energy cost $C$ can be\
approximated using the rectangle integration method:

\vspace{-0.2cm}
\vspace{\eqtopmargin}
\begin{align}\label{eq:price}
C = \mathlarger{\sum}_{pm \in PMs} \frac{t_N-t_0}{N} \mathlarger{\sum}_{t=t_0}^{t_N-1} P_{tot,t}(pm)e_{loc(pm),t}
\end{align}
\vspace{\eqbottommargin}

Similarly, by omitting the electricity price component $e_{loc(pm),t}$\
we calculate the cloud's energy consumption $E$.\
%
For adding the migration overhead,\
we considered the model\ 
from Eq.~\ref{eq:E_mig}\
and converted it to a cost using a mean electricity price\ 
between the\ 
locations at the time of the migration.\
Bandwidth costs were not considered,\
as the necessary business agreement details are not public\ 
-- e.g. Google leases optical fiber cables,\
instead of paying for traffic.\
The final cost of all the migrations was added to the total energy consumption\
$E$ and total energy cost $C$.

\subsection{Optimisation Problem Specification}

Based on the\ 
decision support components\
we can define the optimisation problem specification as\ 
$\big\{C_i: i \in \{1, \ldots prob\_con\}\big\} \cup \
\big\{G_j: j \in \{1, \ldots prob\_goal\}\big\}$,\
which are the sets of $prob\_con$ constraints and\
$prob\_goal$ optimisation goals composed\
of decision support components.\
This optimisation problem specification can be extended with arbitrary\
requirements.\
We now\ 
state the optimisation problem specification\
with two constraints and two goals\
that we\
use in our evaluation.\

In every moment, every VM has to be allocated to one server\
(belong to its $alloc$ set) that acts as its host.\
This is the \textit{allocation constraint} ($C_1$):

\vspace{\eqtopmargin}
\begin{align}\label{eq:c1}
\forall t \in [t_0, t_N],\ \forall vm \in VMs_t,\
 \exists_{=1} pm \in PMs,\text{ s.t. } vm \in alloc_t(pm)
\end{align}
\vspace{\eqbottommargin}

The \textit{capacity constraint} ($C_2$) states that at any given time a server cannot host VMs\
that require more resources in sum than it can provide.\

\vspace{\eqtopmargin}
\begin{align}\label{eq:c2}
\mathlarger{\sum}_{\forall vm \in alloc_t(pm)} res_{vm,i} < cap_{pm,i},\
\forall pm \in PMs,\ \forall i \in \{1, \ldots r\},\ \forall t \in [t_0, t_N]
\end{align}
\vspace{\eqbottommargin}

The \textit{cost goal} ($G_1$) is to minimise the cloud's electricity cost $C$\
expressed in Eq.~\ref{eq:price}, stemming from \gls{pm} utilisation,\
cooling efficiency, electricity prices and migration overhead.

The \textit{\gls{qos} goal} ($G_2$) is to minimise the rate of migrations\
$R_{mig}(vm)$ in a designated interval,\
$\forall vm \in VMs_t,\ \forall t \in [t_0, t_N]$.\
In the following section we present an optimisation engine\
for dealing with such a problem specification.\


\section{Forward-compatible Optimisation Engine}
\label{ch:gascheduler:sec:gascheduler}


In this section we show concrete implementations of the\
optimisation engine workflow from Fig.~\ref{fig:pcc}.\
As already stated in Section~\ref{sec:pcc:workflow},\
to tackle the NP-hard scheduling problem,\
we use a two-stage approach, with best-effort global optimisation\
and a deterministic local optimisation for hard constraint satisfaction.\
For the first stage global optimisation,\
we propose a genetic algorithm \cite{goldberg_genetic_1989} where\
a population of potential solutions\ 
is evolved using\
genetic operators (crossover and mutation).\
For the second stage local optimisation,\ 
we propose a deterministic greedy local search where the best solution\
obtained by the genetic algorithim within the given time limit is\
further improved. The algorithm's main goal is to satisfy the hard\
capacity constraints, in case they were not already satisfied by the\
genetic algorithm, but it also considers the decision support components\
to reduce energy costs based on geotemporal inputs.

\subsection{Algorithm Selection Justification}


The reason the genetic algorithm was chosen for global optimisation\
(in the workflow from Fig.~\ref{fig:pcc}) is that\
using a fitness function for schedule selection\ 
matches the requirement of separated optimisation\
and solution evaluation logic.\
Furthermore, it satisfies\ 
the decision support component extensibility requirement through\
multiple fitness components with associated weights.\ 
There is also a benefit in keeping a population of solutions and not just\
a single best one, as is the case in deterministic optimisation techniques.\
Inputs change over time -- requests to boot new or delete\
old \gls{vm}s arrive,\
temperature or electricity price forecasts change.\
Upon such a change, our genetic algorithm\
propagates\ 
a part of the old\
population to the new environment and there is a higher chance\
that some solutions\ 
will still be fit\
(or\ 
a good evolution basis).\ 

Greedy approaches are often used in deterministic local optimisation,
e.g. in~\cite{beloglazov_energy-aware_2012}.\
For the purpose of improving an existing\
schedule to satisfy primarily the hard constraints,\
without considering the full multi-objective trade-offs,\
it proved as a good addition to the genetic algorithm in our experiments.\

\subsection{Forecast-based Planning}

It is possible to forecast future values of geotemporal inputs\
to a certain extent \cite{weron_modeling_2006,_forecast_2015}.\
This facilitates planning of more efficient cloud management actions.\
For example, knowing whether a shift in electricity prices between two\
data center locations is the result of a temporary spike or\
a longer trend, enables more cost-efficient scheduling choices.\

Time series forecasting is possible in a\
domain-agnostic manner, by dynamically fitting\
auto-regressive integrated moving average models \cite{melard_automatic_2000}.\
As we are dealing with temperatures and electricity prices, widely used data,\
we assume domain-specific forecast information sources,\
such as the announcement of electricity prices\
(e.g. on day-ahead markets~\cite{weron_modeling_2006}\
and a weather forecast web service~\cite{_forecast_2015}).


\subsection{Cloud Control Schedule}

At the current moment $t_c$, we have information about future values\
for the geotemporal inputs for a period of time\
we call a \textit{forecast window} that ends at $t_f$:\
$fw = [t_c, t_f]$.\
The size of the forecast window is determined by the\
available forecast data and the desired accuracy level.\
Given the current cloud configuration,\ 
we are able to estimate the effects\
of any cloud control actions in terms of the optimisation problem\
inside the forecast window\
by applying the cost model from Eq.~\ref{eq:price}.\
We represent a cloud control schedule as a\
time series $\{action_t:\ t \in fw\}$\
of planned cloud control actions in the forecast window\
.\
In this chapter we consider \gls{vm}\
live migration actions \cite{liu_performance_2011}\
and suspension of empty \gls{pm}s that reduces\
idle power consumption \cite{meisner_powernap:_2009}.\
A control action $action_t$\
is described as an ordered pair $(vm,pm)$, specifying\
which $vm$ migrates to which $pm$.\
Migrations determine \gls{vm} allocation over time\
(Eq.~\ref{eq:alloc})\
and implicitly \gls{pm} suspension (Eq.~\ref{eq:power}).


\begin{figure}
\centering
\vspace{\figtopmargin}
\includegraphics[width=0.8\columnwidth]{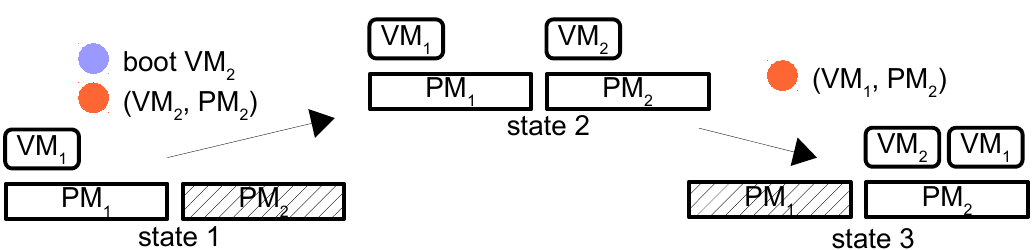}
\vspace{\figcaptionmargin}
\caption{Example of state transitions based on control actions.}
\label{fig:states}
\end{figure}

The representation of the cloud as a sequence of transitions between\
states (as defined through $alloc_t$ in Eq.~\ref{eq:alloc}),\
triggered by migration actions is illustrated in Fig.~\ref{fig:states}.\
Initially, $vm_1$ is hosted on $PM_1$. $PM_2$ is suspended, as it is empty.\
A migration of $VM_2$ to $PM_2$\
transitions the cloud to a new state\
where $PM_2$ is awoken from suspension and hosting $VM_2$.\
An incoming request for \gls{vm} booting\
can be represented as a migration with no source \gls{pm},\
like in the first transition.\
Next, an action migrates $VM_1$ to $PM_2$, after which $PM_2$ is hosting\
both \gls{vm}s and $PM_1$ is suspended.

\begin{figure}[H]
\centering
\vspace{\figtopmargin}
\includegraphics[width=0.9\columnwidth]{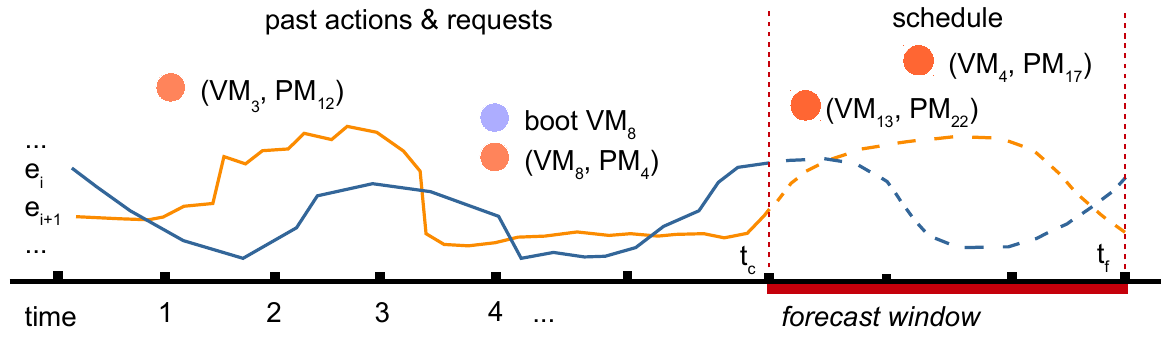}
\vspace{\figcaptionmargin}
\caption{Schedule optimisation inside a forecast window.}
\label{fig:forecasting}
\end{figure}

The time-based aspect of a schedule\
is illustrated in Fig.~\ref{fig:forecasting}.\
Past events that occurred before the current moment $t_c$,\
such as a \gls{vm} migration at $time=1$ or a new user request\
to boot a \gls{vm} at $time=4$, determine the current cloud state.\
Based on the past values of different geotemporal variables, such\
as electricity prices $e_i$ and $e_{i+1}$ at different data center\
locations, we are able to get their value forecasts in the\
forecast window.\
Different control actions of a schedule inside the forecast window\
can then be tried out\
at any moment between $t_c$ and $t_f$.\
The control actions can be evaluated\
to determine the resulting \gls{vm} locations and estimate costs with\
regard to the different geotemporal input forecasts and other optimisation\
problem aspects, such as constraints or \gls{sla} violations.\
When a schedule has been selected for execution, any immediate actions\
are applied and the forecast window moves as time passes,\
new requests arrive and geotemporal inputs change.\

\subsection{Hybrid Genetic Algorithm Implementation}



We now present\ 
the hybrid genetic algorithm.\ 
The most challenging parts in its design were\ 
the genetic operators that had to semantically match\
our optimisation problem domain,\
the fitness function that combines the decision support components\
and the hybrid part of the algorithm, i.e. the greedy\
local improvement.\ 

\subsubsection{Schedule Fitness}

In a genetic algorithm, we keep not only one, but a population of multiple problem\
solutions -- multiple cloud control schedules, in our case.\
An essential part of the algorithm is to evaluate the fitness\ 
of each of these schedules.\
The fitness function we derived for this purpose is adapted from the\
decision support components, but constrained only to the forecast window\
(as we cannot affect actions that were already executed or plan further\
ahead than the available forecasts).\ 
Additional decision support components can be treated in a similar manner.\

For a schedule $s$, we calculate the capacity and\
allocation constraint fitness, as a measure of how well\
the \gls{vm}s are allocated and the \gls{pm}s are within their capacity,\
using the time-weighted function:

\vspace{\eqtopmargin}
\begin{align}
constraint(s) = \mathlarger{\sum}_{\forall t \in fw}\frac{w_{alloc} \cdot R_{unalloc, t} + w_{cap} \cdot R_{overcap, t}}{|fw|}
\end{align}
\vspace{\eqbottommargin}

where $R_{unalloc, t}$ and $R_{overcap, t}$ are ratios of\
$vm, \forall vm \in VMs_t$ for which $C_1$ (Eq.~\ref{eq:c1}) does not hold and\
$pm, \forall pm \in PMs$ for which $C_2$ (Eq.~\ref{eq:c2})\
does not hold at moment $t$, respectively.\
$|fw|$ is the cardinality of $fw$\
(the number of time periods in the interval).\
Constants $w_{alloc}$ and $w_{cap}$ in $[0, 1]$ define the components' weights.

The acceptable migration rate $R_{mig\_min}$ and the maximum allowed\
migration rate $R_{mig\_max}$ can be specified in the \gls{sla},\
or determined by the cloud provider's bandwidth expenses.\
Their values are used to calculate the \gls{qos} component\
from the actual migration rate $R_{mig}$ over the period of $fw$.

\vspace{\eqtopmargin}
\begin{align}
qos\_pen(vm) =
\left\{
	\begin{array}{ll}
	    0 & \mbox{if } R_{mig} < R_{mig\_min}\\
   	    1 & \mbox{if } R_{mig} > R_{mig\_max}\\
	    \frac{R_{mig} - R_{mig\_min}}{R_{mig\_max} - R_{mig\_min}} & \mbox{otherwise.}
	\end{array}
\right.
\end{align}
\vspace{\eqbottommargin}

The $qos\_pen$ expression gives a penalty for\
too frequent migrations per \gls{vm}s\
that grows linearly from 0 to 1 for the migration rate $R_{mig}$\
from $R_{mig\_min}$ to $R_{mig\_max}$. We then calculate the average\
\gls{qos} penalty over all the \gls{vm}s.

\vspace{\eqtopmargin}
\begin{align}
qos(s) = \frac{\mathlarger{\sum}_{\forall vm \in VMs_{t_c}}qos\_pen(vm)}{|VMs_{t_c}|}
\end{align}
\vspace{\eqbottommargin}

As a cost estimation, we use a simplified expression -- $utilprice$.\
First, we calculate the average value of utilisation\
multiplied by $pPUE$ and $e$\
over all \gls{pm}s in the forecast window.\
This is a heuristic of the total energy costs from Eq.~\ref{eq:price}.

\vspace{\eqtopmargin}
\begin{align}
up\_avg(s) = \frac{\mathlarger{\sum}_{\forall pm \in PMs}\mathlarger{\sum}_{\forall t \in fw}util_t(pm) \cdot pPUE_{loc(pm),t} \cdot e_{loc(pm),t}}{|PMs|\cdot|fw|}
\end{align}
\vspace{\eqbottommargin}

Then we normalise it to a $[0,1]$ interval\
by dividing it with $up\_worst$,\
which is the same as $up\_avg$, but calculated\
for a constant maximum utilisation $util_t(pm)=1$.

\vspace{\eqtopmargin}
\begin{align}
utilprice(s) = \frac{up\_avg(s)}{up\_worst}
\end{align}
\vspace{\eqbottommargin}

To measure how tightly \gls{vm}s are packed among the available \gls{pm}s,\
we estimate the consolidation quality $consolid$\ 
by considering only positive utilisation time series elements denoted as\
$\{ pos\_util_t(pm): t \in fw,\
\text{ s.t. } util_t(pm)>0 \}$.

\begin{align}
consolid(s) = 1 - \frac{1}{|PMs|} \mathlarger{\sum}_{\forall pm \in PMs}\frac{\mathlarger{\sum}_{\forall t \in fw}pos\_util_t(pm)}{|pos\_util(pm)|}
\end{align}
\vspace{\eqbottommargin}


Finally, we can calculate the fitness as:

\vspace{\eqtopmarginbig}
\begin{align}
\label{eq:fitness}
\begin{split}
fitness(s) = w_{ct} \cdot constraint(s) + w_q \cdot qos(s) +\\
+ w_{up} \cdot utilprice(s) + w_{cd} \cdot consolid(s)
\end{split}
\end{align}

with $w_{ct}$, $w_q$, $w_{up}$ and $w_{cd}$ in $[0, 1]$ determining\
each component's impact.\ 
The optimal schedule converges towards a fitness of 0 and the\
worst schedule towards 1.\
This solution evaluation form suits the forward compatibility requirement,\
as it can easily be extended with new decision support components by\
weighing them into the total fitness summation.\

\subsubsection{Genetic Operators}

The \textit{creation} procedure creates a random schedule as\
a time series of actions $(vm,pm),\ vm \in VMs_{t_c}, pm \in PMs$\
at random times $t_r\ \in fw$. The number of migrations is uniformly\
distributed between the parameters $min\_migrations$ and $max\_migrations$.\

Given schedules $s_1$ and $s_2$, the \textit{crossover} operator creates\
a child schedule $s_3$ by choosing a random moment\
$t_r\ \in [t_c, t_f]$:

\vspace{\eqtopmargin}
\begin{align}\label{eq:crossover}
s_3 =
\left\{
	\begin{array}{ll}
	    s_{1,t}: & t \in [t_c, t_r],\\
   	    s_{2,t}: & t \in [t_r, t_f]
	\end{array}
\right\}
\end{align}
\vspace{\eqbottommargin}


The \textit{mutation} operator applied to a schedule $s$ removes a random\
action $a$\ 
and inserts\ 
one at a random moment $t_r\ \in fw$:

\vspace{\eqtopmarginbig}
\begin{align}\label{eq:mutation}
s = s \setminus a \cup \{t_r : (vm,pm)\},\\
a \in s, vm \in VMs_{t_c}, pm \in PMs
\end{align}
\vspace{\eqbottommarginbig}

\subsubsection{Algorithm}

The core \gls{ga} is listed in Alg.~\ref{alg:ga}.\
Among other parameters not introduced so far, it also receives\
the population size $pop$, crossover, mutation and random creation rates\
$cross$, $mut$ and $rand$ in $[0,1]$ and the\
maximum number of generations $gen$.

\begin{algorithm}
\caption{Core genetic algorithm.}
\label{alg:ga}
\begin{algorithmic}[1]
\State $schedules = create\_population(pop, rand, [existing])$ \label{alg:ga:creation}
\State $i = 0$
\While{$True$}
	\State $calculate\_fitness(schedules)$ \label{alg:ga:fitness} \Comment{Eq.~\ref{eq:fitness}}
	\State sort $schedules$ by fitness, best first
	\If{$i = gen$} \label{alg:ga:termination}
		\State break loop
	\EndIf
	\State $parents = roulette\_wheel(schedules, cross)$
	\State $children = crossover(parents)$  \Comment{Eq.~\ref{eq:crossover}}
	\State $schedules = schedules[0:pop\cdot(1-cross)] + children$\label{alg:ga:new}
	\State $mutation(schedules, mut)$\label{alg:ga:mutation}  \Comment{Eq.~\ref{eq:mutation}}
	\State $i = i + 1$
\EndWhile
\State $existing = schedules$
\State \Return $schedules[0]$
\end{algorithmic}
\end{algorithm}
%


After creating the population, every schedule is evaluated using\
the fitness function (Eq.~\ref{eq:fitness}) in line~\ref{alg:ga:fitness}.\
Schedules are selected for crossover using the \textit{roulette wheel}
selection method with chances for selection\
proportional to a schedule's fitness.\
A new generation is created by replacing the worst schedules with\
the newly created children (line~\ref{alg:ga:new}).\
A random segment of the population proportional to $mut$ is\
changed by applying the mutation operator (line~\ref{alg:ga:mutation}).\
The termination condition is fulfilled after $gen$ generations\
have passed (line~\ref{alg:ga:termination}) and the best schedule\
is returned.\
To cope with forecast-based planning,\
an existing population is partially propagated to the\
next schedule reevaluation, after the forecast window moves (line~\ref{alg:ga:creation}).\


The \gls{ga} is not guaranteed to satisfy all the constraints\
within a limited time period. To remedy this, we expand the optimisation with\
a greedy constraint satisfaction algorithm that is applied to the best\
schedule selected by the genetic algorithm to reallocate any offending\
\gls{vm}s using a \gls{bcf} heuristic we developed\
that also considers geotemporal inputs.\

The greedy constraint satisfaction pseudo-code is listed in Alg.~\ref{alg:bcf}.\
The algorithm receives as input $schedule$, the output of the \gls{ga}.\
We begin by marking for reallocation all \gls{vm}s which cause\
any hard constraint violations ($C_1$ or $C_2$) in line~\ref{alg:bcf:nonalloc}\
and additionally all \gls{vm}s from underutilised hosts\
in line~\ref{alg:bcf:underutil}.\ 
The \gls{vm}s will then be reallocated\
in the outermost loop (line~\ref{alg:bcf:vmloop})\
starting with larger \gls{vm}s first\
whose placement is more constrained (line~\ref{alg:bcf:vmsort}).\
Available \gls{pm}s are split into $active$ and $nonactive$ lists,\
depending on whether they are suspended or not.\
We sort $inactive$ in line~\ref{alg:bcf:sortinactive}\
such that larger \gls{pm}s come first for activation\
(preferable to more smaller machines, because of the idle power overhead)\
and data centers with lower combined electricity price\
and cooling overhead cost are preferred.\
The target \gls{pm} to host $vm$ is selected in the inner loop\
in line~\ref{alg:bcf:mappedloop} by first sorting $active$\
to try and fill out almost full \gls{pm}s first and prefer\
lower-cost location in case of ties (line~\ref{alg:bcf:sortactive})\
and activating the next \gls{pm} from $inactive$ if $vm$ does not\
fit any of the $active$ \gls{pm}s (line~\ref{alg:bcf:popinactive}).\
When a fitting \gls{pm} is found, the action is added\
to $schedule$ and the algorithm continues with the next \gls{vm}.

\begin{algorithm}
\caption{Greedy constraint satisfaction -- BCF heuristic.}
\label{alg:bcf}
\begin{algorithmic}[1]
\State $to\_alloc \gets$ empty list
\State append all constraint-violating \gls{vm}s to $to\_alloc$ \label{alg:bcf:nonalloc}
\State append \gls{vm}s from all underutilised \gls{pm}s to $to\_alloc$ \label{alg:bcf:underutil}
\State sort $to\_alloc$ by resource requirements decreasing \label{alg:bcf:vmsort}
\For{$vm \in to\_alloc$} \label{alg:bcf:vmloop}
    \State $active \gets $ all PMs where at least one VM is allocated 
    \State $inactive \gets $ all PMs where no VMs are allocated
    \State sort $inactive$ by capacity decreasing, cost increasing\
    \label{alg:bcf:sortinactive}
    \State $mapped \gets False$
    \While{not $mapped$} \label{alg:bcf:mappedloop}
        \State sort $active$ by free capacity, cost increasing \label{alg:bcf:sortactive}
        \For {$pm \in active$}\label{alg:bcf:pmloop}
            \If {$vm$ fits $pm$}\label{alg:bcf:pmfits}
                \State $mapped \gets True$
                \State break loop
            \EndIf
        \EndFor
        \If {not $mapped$}
            \State pop $inactive[0]$ and append it to $active$ \label{alg:bcf:popinactive}
        \EndIf
    \EndWhile
    \State modify $schedule$ by adding a migration $(vm, pm)$
\EndFor
\State \Return $schedule$
\end{algorithmic}
\end{algorithm}

\section{Evaluation}
\label{ch:gascheduler:sec:evaluation}


We evaluated the proposed progressive cloud controller in a large-scale simulation of \gaschedulerVMNumSimulation{} \gls{vm}s based\
on real traces of geotemporal inputs.\
To be able to simulate cloud behaviour under geotemporal inputs,\
we developed our own open source simulator\
Philharmonic.\ 
%
The goals of the evaluation are:\
(1) Estimate energy and cost savings, as well as the \gls{qos}\
attainable using the pervasive cloud controller;\
(2) Analyse the impact of various inputs, such as data center geography,\
different geotemporal inputs or controller parameters;\
(3) Validate the pervasive cloud controller extensibility by\
running the simulator with different decision support component subsets;\
(4) Define cloud provider guidelines, such as how temperature variation\
or forecast data quality affect the energy savings and illustrate their\
usage in a case study.

\subsection{Evaluation Methodology}
\label{ch:gascheduler:sec:evaluation_methodology}

In this part we give an overview of the simulator's implementation\
and proceed with explaining all the simulation details,\
such as the datasets, parameters and the baseline controller.

\subsubsection{Philharmonic Simulator}
\label{sec:philharmonic}

\begin{figure}
\vspace{\figtopmargin}
\centering
\includegraphics[width=0.5\columnwidth]{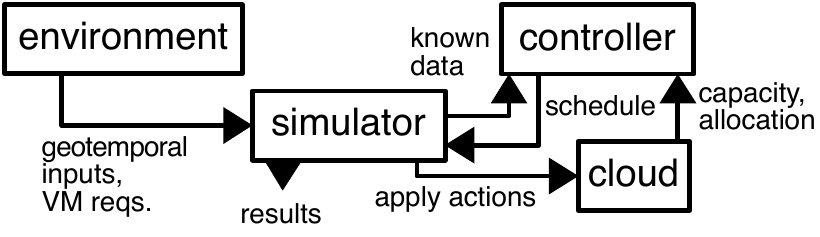}
\vspace{-0.2cm}
\caption{Philharmonic simulator overview.}
\label{fig:philharmonic}
\end{figure}

A high-level overview of\ 
the Philharmonic simulator~\cite{drazen_lucanin_philharmonic_2014}
that we developed\
is shown in Fig.~\ref{fig:philharmonic}.\
A simulation in Philharmonic\ 
consists of iterating through\
a series of equally-spaced time periods,\
collecting the currently available electricity price\
and temperature forecasts, as well as the incoming \gls{vm} requests\
from the environment component.\
The controller is called with the data known at that moment\
about the environment and the cloud\
to reevaluate the schedule for the forecast window and potentially\
schedule new or different actions.\
The simulator applies any actions scheduled for\
the current moment on the cloud model\
and continues with the next time step, repeating\
the procedure.\
The applied actions are used to\
calculate the resulting energy consumption and electricity\
costs, using the model from Section~\ref{ch:gascheduler:sec:model}.

\subsubsection{Simulation Details}
\label{ch:gascheduler:simulation_details}

\begin{figure}
\vspace{\figtopmargin}
\centering
\includegraphics[width=1.0\columnwidth]{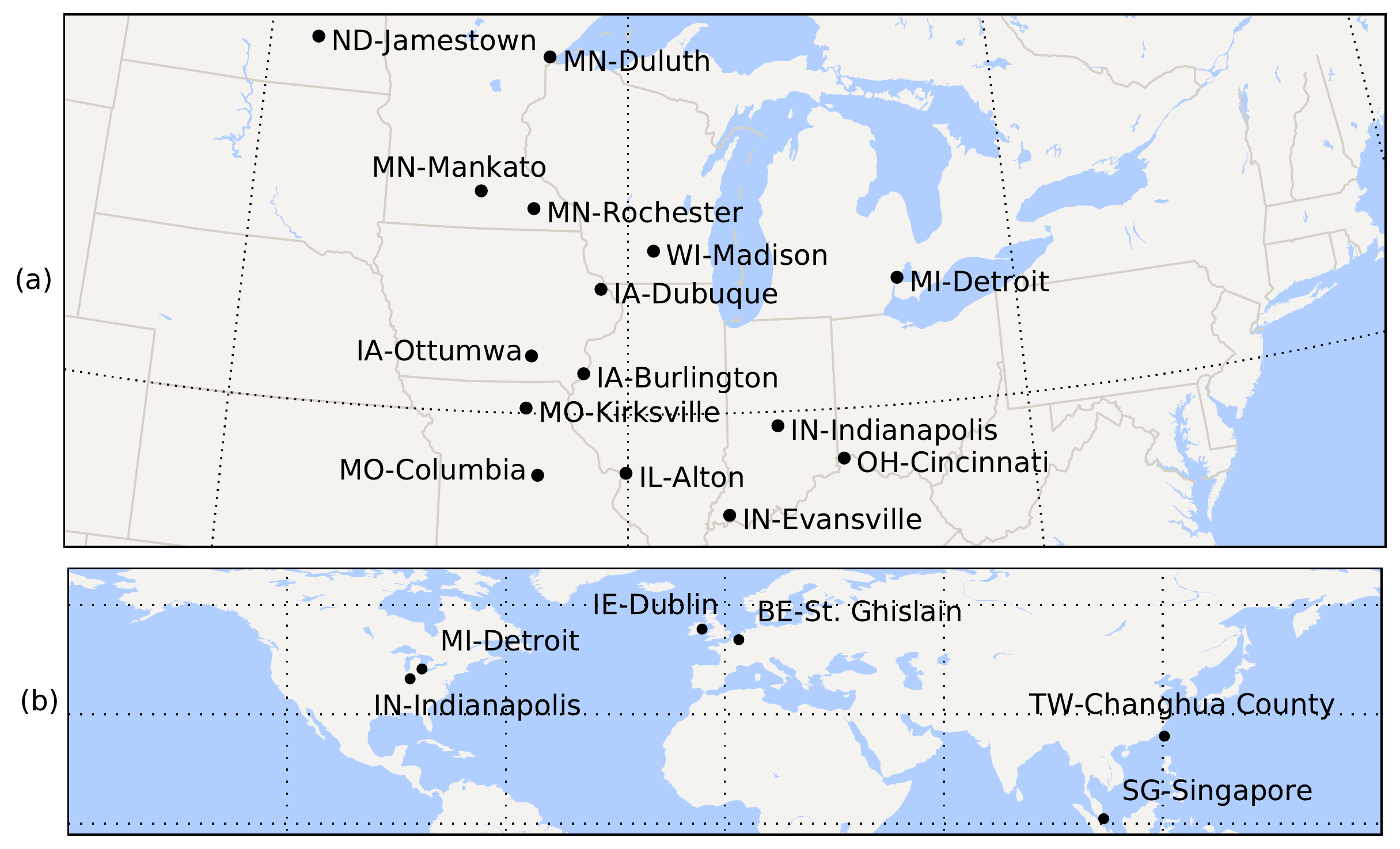}
\vspace{\figcaptionmargin}
\caption{Cities used as data center locations in the simulation based on a:\
(a) US dataset
(b) world-wide dataset.}
\label{ch:gascheduler:fig:cities}
\vspace{\figbottommargin}
\end{figure}


Real historical traces for electricity prices and temperatures\
were used in the simulation for 15 cities in the USA\
shown in Fig.~\ref{ch:gascheduler:fig:cities} (a).\
The electricity price dataset described in~\cite{alfeld_toward_2012} was used.\
The temperatures were obtained from\
the Forecast web service~\cite{_forecast_2015}.\
To evaluate less correlated geotemporal inputs,\
we used a world-wide dataset for six cities accross three continents\
shown in Fig.~\ref{ch:gascheduler:fig:cities} (b),\
choosing non-US cities to match\
the locations of Google's data centers.\ 
Temperatures were again obtained\
from the Forecast API \cite{_forecast_2015}.\
Due to lack of \gls{rtep} data for the four non-US cities,\
we artificially generated these traces from the data\
known for other US cities.\
We shifted the time series based on the time zone offsets\
and added a difference in annual mean values to resemble\
local electricity prices.

\begin{figure}
\centering
\vspace{\figtopmargin}
\includegraphics[width=0.9\columnwidth]{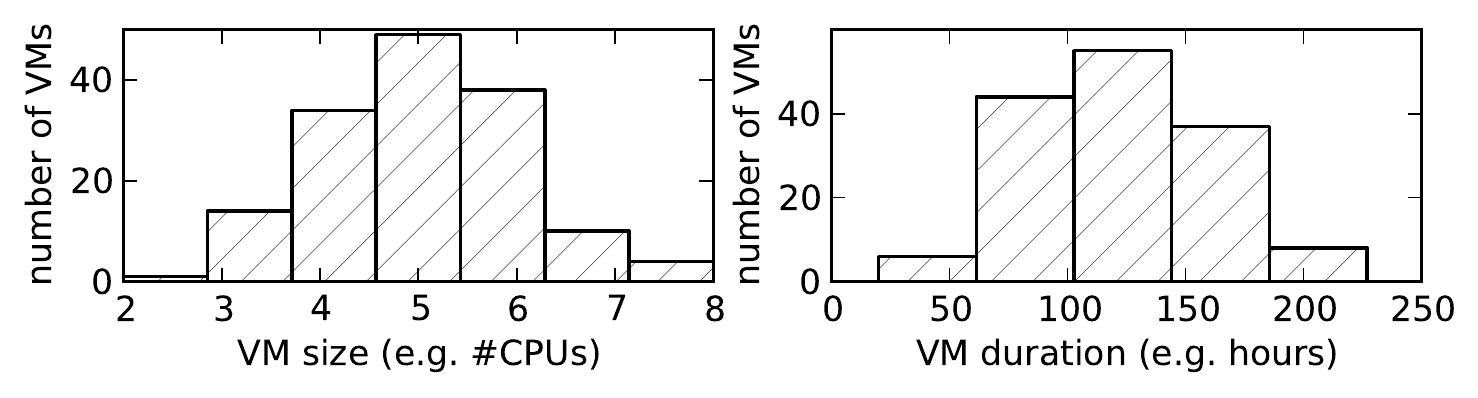}
\caption{VM request resource and duration histogram.}
\label{fig:vmreqs}
\vspace{\figbottommargin}
\end{figure}

User requests for \gls{vm}s were generated randomly by\
uniformly distributing the creation time and duration.\
The specifications of the
requested \gls{vm}s were modelled\
by normally distributing each resource type.\
An example \gls{vm} request distribution is illustrated\
in Fig.~\ref{fig:vmreqs}.\
%
%
Available cloud infrastructure was generated by uniformly\
distributing physical machines among different data centers.\
Capacities for each each machine's resources were generated\ 
in the same manner as the \gls{vm} requests.

\begin{table*}[ht]
\vspace{\tabletopmargin}
\centering
\caption{Simulation parameters}
\vspace{\tablecaptionmargin}
\label{tab:parameters-simulation}
\begin{tabular}{ccccccc}
\toprule
 period & duration & d & p (CPU; RAM)    & v (CPU; RAM) & $P_{peak}$ & $P_{peak}$ \\
\midrule
  1 h   & 2 weeks  & 6 & 2,000 (8-16; 16-32) & 10,000 (1-2; 2-4) & 200 & 100 \\
\bottomrule
\toprule
$(\mu_P, \sigma_P^2)$ & $R_{mig\_min}$ &   $R_{mig\_max}$ & $R$ & $D$ & $V_{thd}$ \\
\midrule
(0, 25) & 1/4 & 1 & 1 Gb/s & 0.3 Gb/s & 0.1 Gb/s \\
\bottomrule
\end{tabular}
\vspace{\tablebottommargin}
\end{table*}

The exact simulation parameters used in the evaluation\
are listed in Table~\ref{tab:parameters-simulation}.\
The time is defined by its period (simulation step size) and\
the total duration that determines the number of steps.\
To define the cloud, the number of data centers is given as $d$\
(for the world-wide scenario, for the US scenario we consider $d=15$),\
and for $PMs$ and $VMs$ their number $p$ and $v$\
(of boot requests in case of \gls{vm}s --\
there were about 50\% as much delete events as well)\
with minimum and maximum resource\
values in parentheses for the resources\
we assumed in this simulation -- number of CPUs\
and the amount of RAM in GB.\
Besides running the simulation for the large-scale scenario with\
\gaschedulerVMNumSimulation{} \gls{vm}s, we also simulated\
100 and 1k \gls{vm}s, but as the difference in normalised savings\
was only marginal, we only include the large-scale results.\
Uniform resource weights were used in the $util$ function (Eq.~\ref{eq:util}).\
We used $P_{peak}$ with $P_{idle}$ values $50\%$ of the peak,\
as reported in \cite{fan_power_2007}, and added normal random noise\
of the form $\mathcal{N}\ (\mu_P, \sigma_P^2)$ to account for load variation.\
$R_{mig}$ was calculated hourly\
and constant migration model parameters ($R$, $D$, $V_{thd}$) were used.\
We used these settings in all the simulations, unless otherwise specified.\



As a baseline controller for results comparison, we implemented a method\
for \gls{vm} consolidation dynamically adapting to user requests using a\
\gls{bfd} placement heuristic developed\
by Beloglazov et al. in \cite{beloglazov_energy-aware_2012}.\
We implemented the updated version of the controller that is currently\
developed for inclusion in\
the OpenStack open source cloud manager\ 
as project Neat \cite{beloglazov_openstack_2014}.\
Based on our classification of related work, this is\
a level two cloud management\
method that dynamically reallocates \gls{vm}s,\
treating all energy uniformly.\

\begin{table*}[ht]
\vspace{\tabletopmargin}
\centering
\caption{Optimisation engine parameters}
\vspace{\tablecaptionmargin}
\label{tab:parameters-scheduler}
\begin{tabular}{cccccccc}
\toprule
 fw &   pop &   gen &   cross &   mut &   rand   & $min\_migr$ & $max\_migr$ \\
\midrule
 12 h &   100 &   100 &    0.15 &  0.05 &    0.3  & 0 & $fw |VMs_{t_c}|/3$ \\
\bottomrule

\toprule
$w_{up}$ & $w_{ct}$ & $w_{alloc}$ & $w_{cap}$ & $w_q$ & $w_{cd}$ \\
\midrule
0.4 & 0.1 & 0.4 & 0.6 & 0.4 & 0.1 \\
\bottomrule
\end{tabular}
\vspace{\tablebottommargin}
\end{table*}


The optimisation engine's algorithm parameters
are listed in Table~\ref{tab:parameters-scheduler}.
All the weights used in the fitness function (Eq.~\ref{eq:fitness})
were 
systematically calibrated using automated parameter exploration,
which we cover later.
The values listed in the table show the parameter combination that achieved
the highest energy savings with the least number of constraint violations.






\subsection{Dynamic Controller Analysis}

This part of the analysis aims to compare our pervasive cloud controller\
with the baseline controller by\
visualising individual actions.\
The use case is the scenario with world-wide data centers\
and other parameters we already described\
in Table~\ref{tab:parameters-simulation}.\

\begin{figure*}
\includegraphics[width=1.0\textwidth]{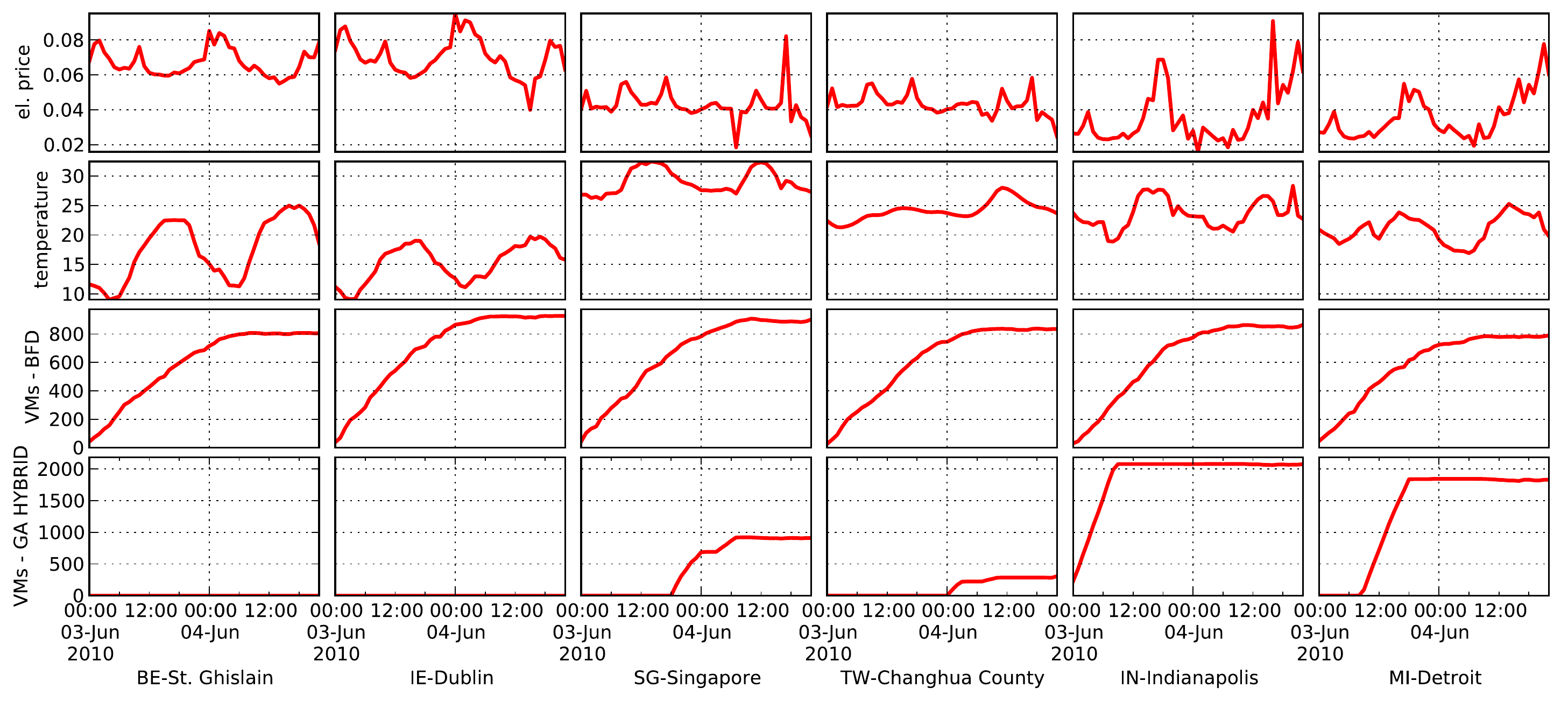}
\vspace{\figcaptionmargin}
\caption{Dynamic VM management comparison of the controllers.}
\label{fig:results-dynamic}
\vspace{\figbottommargin}
\end{figure*}

In Fig.~\ref{fig:results-dynamic} we see 6x4 graphs, where the columns
represent data centers. First two rows show geotemporal inputs --\
electricity price and temperatures.\
The next two row shows the dynamic number of active \gls{vm}s\
at the corresponding data center.\
The third row shows the behaviour of the baseline controller\
and the fourth row of the pervasive cloud controller.\
The x-axes of all the graphs cover the same time span and the graphs\
in the same column are aligned to the same x-axis, shown in the bottom.\
Similarly, the graphs in the same row share the same y-axis.\

The electricity prices are lowest in the USA\
(with an increase towards the end of the day),\
followed by Asian locations and the European locations have significantly\
higher prices.\
Temperatures start the lowest in Europe, but then approach 20 C.\
The other locations oscillate around 20 C, except for the peaks in Asian\
locations where 30 C are approached.\
We can see that the baseline method roughly uniformly distributes\
the \gls{vm}s across all the available data centers,\
disregarding the geotemporal inputs.\
The pervasive cloud controller allocates \gls{vm}s in the first 18 hours\
filling out the US capacities, targeting lower electricity prices.\
No \gls{vm}s are allocated in the European locations during this period,\
due to high electricity prices and enough capacity at other locations.\
The Asian locations are initially empty,\
but after the temperature peak is over,\
\gls{vm}s are migrated to the Singapore data center\
and at the end of the first day\
(when Asian electricity prices start to decrease even bellow the US values),\
the Taiwan data center as well.\
This shows us the desired behaviour of the pervasive cloud controller\
where geotemporal inputs are monitored and adapted to by reallocating\
load to the most cost-efficient data center location.\

\subsection{Aggregated Simulation Results}

To give an estimation of the benefits of using our pervasive cloud controller\
in a large-scale scenario described in Table~\ref{tab:parameters-simulation}\
and analyse various environmental parameters,\
we collected the performed actions during the whole simulation and calculated the\
aggregated energy consumption and costs.\ 




\subsubsection{Cost Savings}

\begin{figure}
\vspace{\figtopmargin}
\centering
\includegraphics[width=0.75\columnwidth]{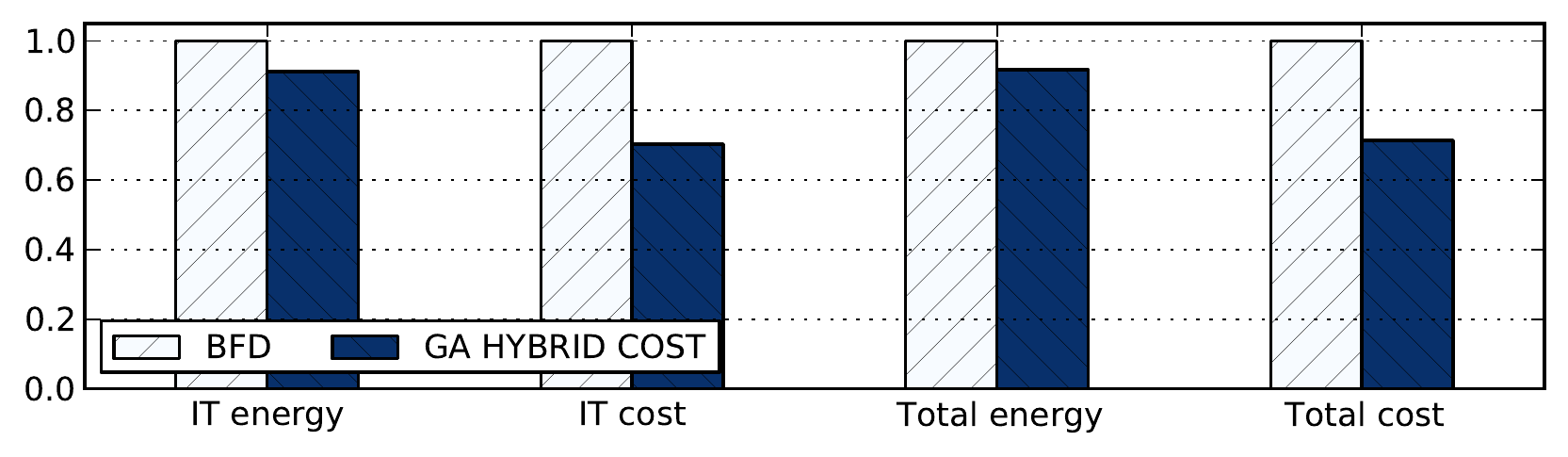}
\vspace{\figcaptionmargin}
\caption{Normalised energy and costs of the pervasive cloud controller\
compared to the baseline method in a simulation of \gaschedulerVMNumSimulation{} VMs.}
\label{fig:results-decisions}
\vspace{\figbottommargin}
\end{figure}

\begin{table}
\centering
\caption{Absolute energy consumption and costs}
\vspace{\tablecaptionmargin}
\label{tab:results-decisions}
\begin{tabular}{lrr}
\toprule
{} &        BFD &  GA HYBRID COST \\
\midrule
IT energy (kWh)    &    6226.00 &         5673.63 \\
IT cost (\$)        &     309.40 &          217.64 \\
Total energy (kWh) &    7488.01 &         6869.55 \\
Total cost (\$)     &     370.20 &          264.39 \\
\bottomrule
\end{tabular}
\end{table}

The normalised results of the simulation based on the world-wide dataset\
with \gaschedulerVMNumSimulation{} \gls{vm}s\
are shown in Fig.~\ref{fig:results-decisions}.\
A group of columns is shown for each of the examined quality metrics -- IT energy,\
IT cost, total energy and total cost.\
Inside each group, there is a column for both of the scheduling algorithms:\
the baseline algorithm (BFD)\
and the pervasive cloud controller (GA HYBRID COST).\
The values are normalised as a relative value of\
the baseline algorithm's results.\
The absolute values are listed in Table~\ref{tab:results-decisions}.\
The pervasive cloud controller achieves savings\
of \gaschedulerEnSavingsNoGeotemp{} in total energy cost compared to the baseline.\
We can see that significant savings can be achieved\
using our pervasive cloud controller,\
which is especially relevant for large cloud providers\
such as Google or Microsoft that spend over \$40M annually on data center\
electricity costs \cite{qureshi_cutting_2009}.\



\subsubsection{Decision Support Component Variation}

\newcommand{\ganesavings}{13\%}
\newcommand{\gantsavings}{7.5\%}

To validate the controller's extensibility and show that it can work\
with different decision support components, we performed\
the same \gaschedulerVMNumSimulation{} \gls{vm} simulation\
with different subsets of the decision support components\
considered by the optimisation engine.\
This analysis also gives an overview of the impact individual geotemporal\
inputs have in the total achieved energy savings.

\begin{figure}
\centering
\includegraphics[width=0.75\columnwidth]{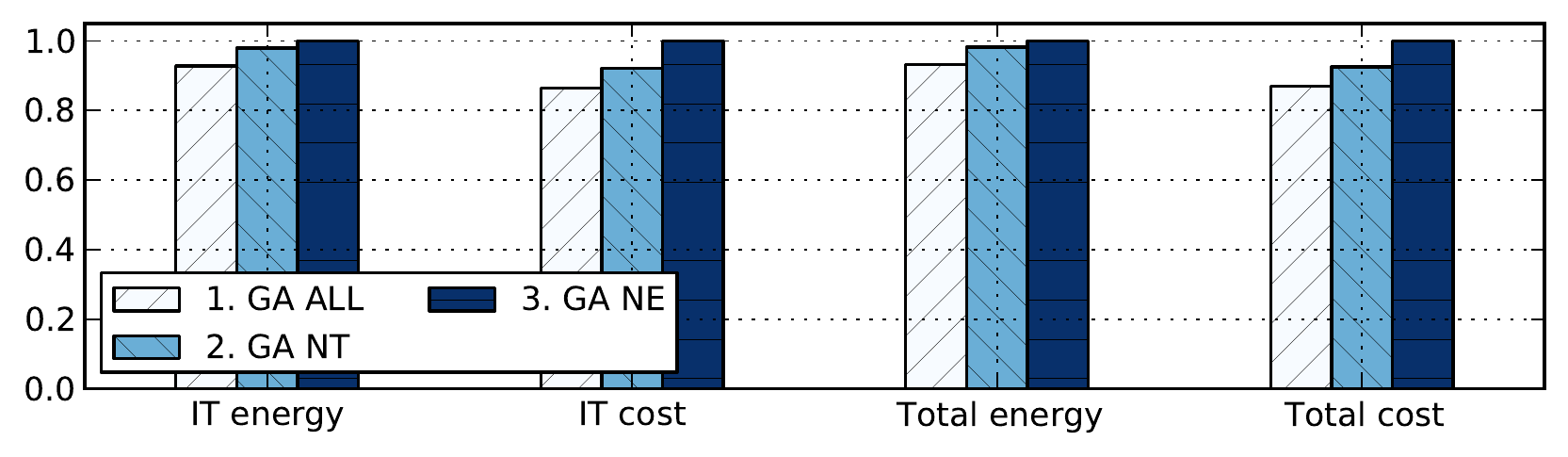}
\vspace{\figcaptionmargin}
\caption{Normalised energy and costs of the pervasive cloud controller\ 
with various decision support components:\ 
temperature and electricity price (1), no temperature (2)\
and no temperature or electricity cost (3).}
\label{fig:extensibility}
\end{figure}

The results are shown in Fig.~\ref{fig:extensibility}.\
Each column stands for one of the simulation scenarios\
-- both temperatures and electricity price components (GA ALL),\
electricity price component, but no temperature (GA NT) and\
no electricity price or temperature components (GA NE).\
Total cost savings of \gantsavings{} are achieved when both\
components are considered compared to not considering temperatures.\
Savings are \ganesavings{} when compared to not considering both components,\
which is a significant difference.\
In the same manner we turn on or off certain decision support components\
as a configuration option in the controller's implementation,\
new geotemporal inputs and rules can be added in the future when necessary.\

\subsubsection{Geography Variation} 

Different cloud providers will have different data center locations and\
geographical distributions.\
To estimate the impact of this geographical distribution of data centers\
on the possible cost savings achievable using our scheduling method,\
we simulate its effects on two such scenarios -- the world-wide scenario\
and the USA scenario,\
as described in Section~\ref{ch:gascheduler:sec:evaluation_methodology}.\
Furthermore, as the USA dataset of geotemporal inputs consists of real\
historical electricity price traces (which we did not have to artificially\
adapt to different time zones and local averages) it further testifies to the\
validity of our approach.\
Lastly, even though current cloud providers have incentives to spread\
their data centers further apart to bring services closer\
to a world-wide user base, with the advent of smart buildings\
\cite{privat_smart_2013}\
we might see more localised data center distributions based on\
neighborhood, city or region organisations.

\begin{figure}
\centering
\includegraphics[width=0.75\columnwidth]{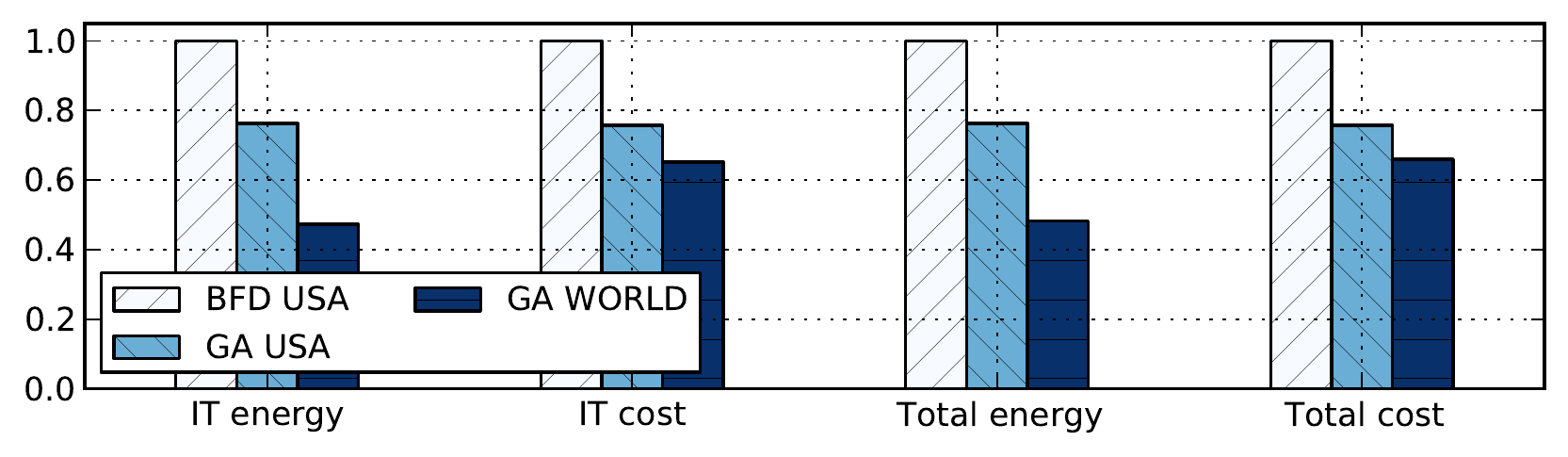}
\vspace{\figcaptionmargin}
\caption{Normalised energy and costs of the pervasive cloud controller\
compared to the baseline method for the USA and world-wide datasets.}
\label{fig:usa-world}
\vspace{\figbottommargin}
\end{figure}

The results of the simulation for the USA and world-wide dataset\
for \gaschedulerVMNumSimulation{} \gls{vm}s are shown in Fig.~\ref{fig:usa-world}.\
The simulation settings were the same\
we explained in Section~\ref{ch:gascheduler:sec:evaluation_methodology},\
except for the locations of the physical machines.\
The baseline controller was run for the USA dataset (BFD USA),\
and we can see the normalised results compared to this baseline\
for the pervasive cloud controller simulated on the USA dataset (GA USA)\
and the world-wide dataset (GA WORLD).\
It can be seen that  significant cost savings of 24\% are achieved\
even for the USA-only scenario.\
The pervasive cloud controller's energy consumption and costs\
are lower in the world-wide scenario than for the US-only data centers,\
though -- a further 10\% decrease is possible.\



\subsubsection{\gls{qos} Analysis}
\label{ch:gascheduler:sec:qos_analysis}

\newcommand{\onemigrationpercentage}{$20\%$}
\newcommand{\aggregatedmigrationsmin}{one}
\newcommand{\aggregatedmigrationslow}{two}
\newcommand{\aggregatedmigrationsmax}{three}
\newcommand{\aggregatedmigrationsci}{1.26--1.5}

\begin{figure}
\vspace{\figtopmargin}
\centering
\includegraphics[width=0.75\columnwidth]{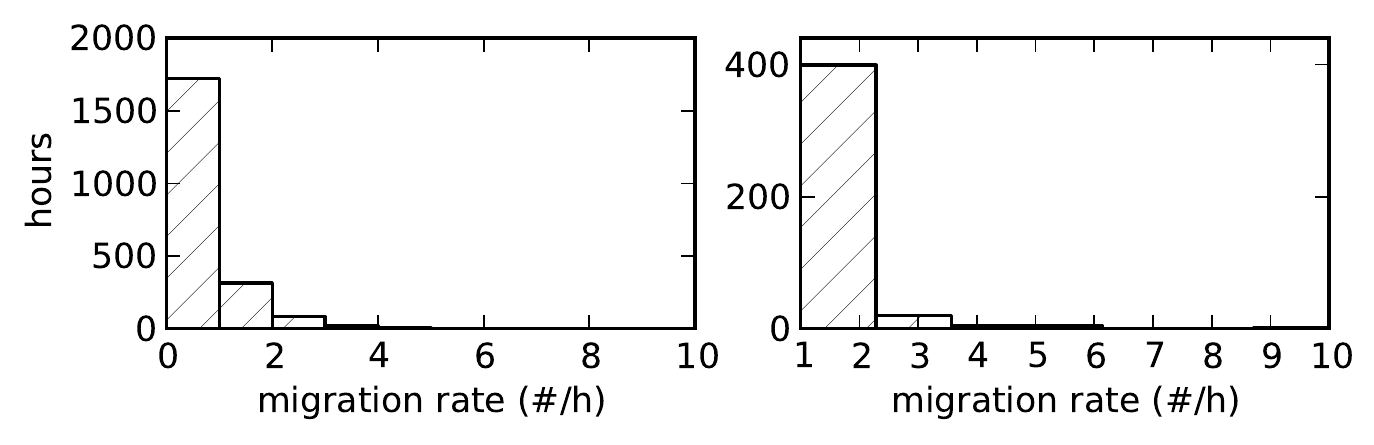}
\vspace{\figcaptionmargin}
\caption{Hourly migration rate histogram (two zoom levels).}
\label{fig:migrations-rate}
\end{figure}

Aside from understanding the cost savings\
from the cloud provider's perspective,\
we also have to analyse the \gls{qos}, i.e.\
how the controller affects end users of \gls{vm}s.\
To measure this, we count how often the migration actions\
occur, i.e. the migration rate.\
To get more data,\
we ran the simulation to cover three months.\
A histogram of hourly migration rates of all the \gls{vm}s\
obtained from the simulation of the pervasive cloud controller\
can be seen in Fig.~\ref{fig:migrations-rate}.\
The two plots show different zoom levels, as there are progressively less\ 
hours with higher migration rates.\
Most of the time, no migrations are scheduled,\
with one migration per hour happening\
about \onemigrationpercentage{} of the time.\


\begin{figure}
\vspace{\figtopmargin}
\centering
\includegraphics[width=0.75\columnwidth]{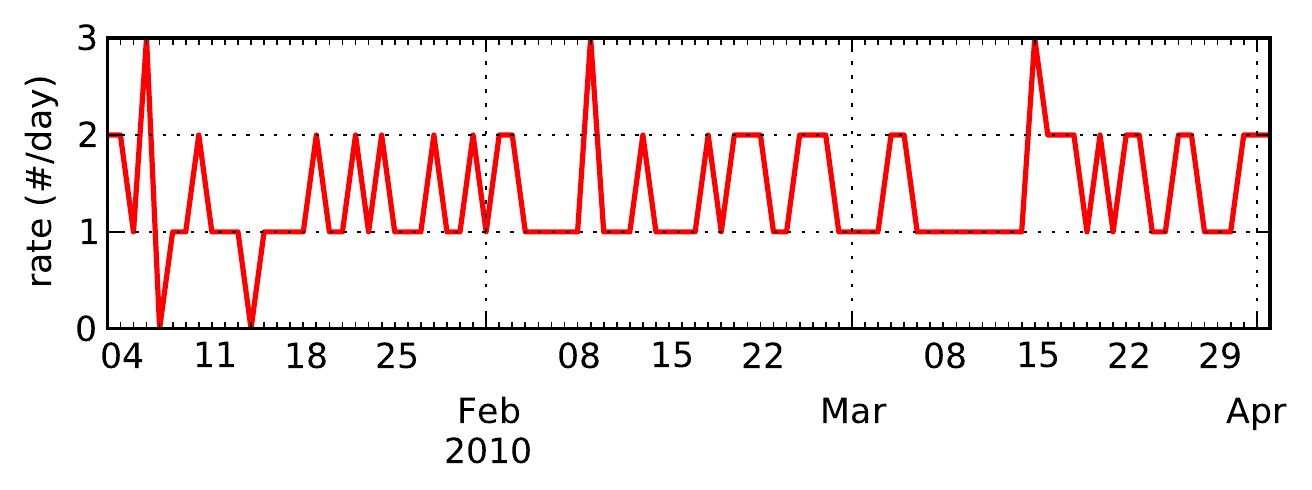}
\vspace{\figcaptionmargin}
\caption{Aggregated worst-case per-VM migration rate.}
\label{fig:migrations-aggregated}
\vspace{\figbottommargin}
\end{figure}

To get a \gls{qos} metric meaningful to the user,\
we group migrations per \gls{vm}\
(as users are only interested in migrations of\
their own \gls{vm}s)\
and process them in a\
function that aggregates migrations over a daily interval.\
We then define the \textit{aggregated worst-case} metric by\ 
counting the migrations per \gls{vm} per day and selecting\
the highest migration count among all the \gls{vm}s in every interval.\
Such a metric could be useful e.g. in defining the lower bound for the\
availability rate in an \gls{sla}.\
The aggregated worst-case migration rate for the simulation\
of the pervasive cloud controller\
is shown in Fig.~\ref{fig:migrations-aggregated}.\
There are \aggregatedmigrationsmin{} or \aggregatedmigrationslow{} migrations\
per \gls{vm} per day most of the time,\
with an occasional case with a higher rate such as the peaks with\
\aggregatedmigrationsmax{} migrations.\

Given that this data is highly dependent of the\
scheduling algorithm parameters used\
and the actual environmental parameters for a specific cloud deployment,\
fitting one specific statistical distribution to the data to get the desired\
percentile value that can be guaranteed in an \gls{sla} would be\
hard to generalise for different use cases and might require manual modelling.\
Instead, we propose applying the\
distribution-independent\
bootstrap confidence interval method \cite{efron_introduction_1994}\
to estimate the aggregated migration rate.\ 
In our simulation,\ 
the 95\% confidence interval for the mean daily per-\gls{vm} migration rate is\ 
\aggregatedmigrationsci{} migrations per day.\


\subsubsection{Genetic Algorithm Parameter Exploration}

To explore how the optimisation engine behaves under different \gls{ga}\
parameters, we ran the simulation with different parameter values and\
compared the resulting energy costs.\
We explored the weights of the different\
fitness function components in (Eq.~\ref{eq:fitness}).\
We developed a method for automatically running the simulation with\
different parameter combinations in the Philharmonic simulator.\ 
We covered a set $\{0, 0.1, 0.2, \ldots\ 1.\}$\
for each of the four weight parameters,\
exploring all the combinations with a constraint that their sum equals 1\
(as only weight ratios make a difference in the \gls{ga},\
not their absolute values),\
resulting in 285 combinations.\
This method can be used to calibrate the controller for different environments\
by finding the parameter combination that achieves\
the highest energy savings or the best \gls{qos},\
similar to how different objectives are optimised\
in a Pareto frontier.\

\begin{figure}
\includegraphics[width=0.7\columnwidth]{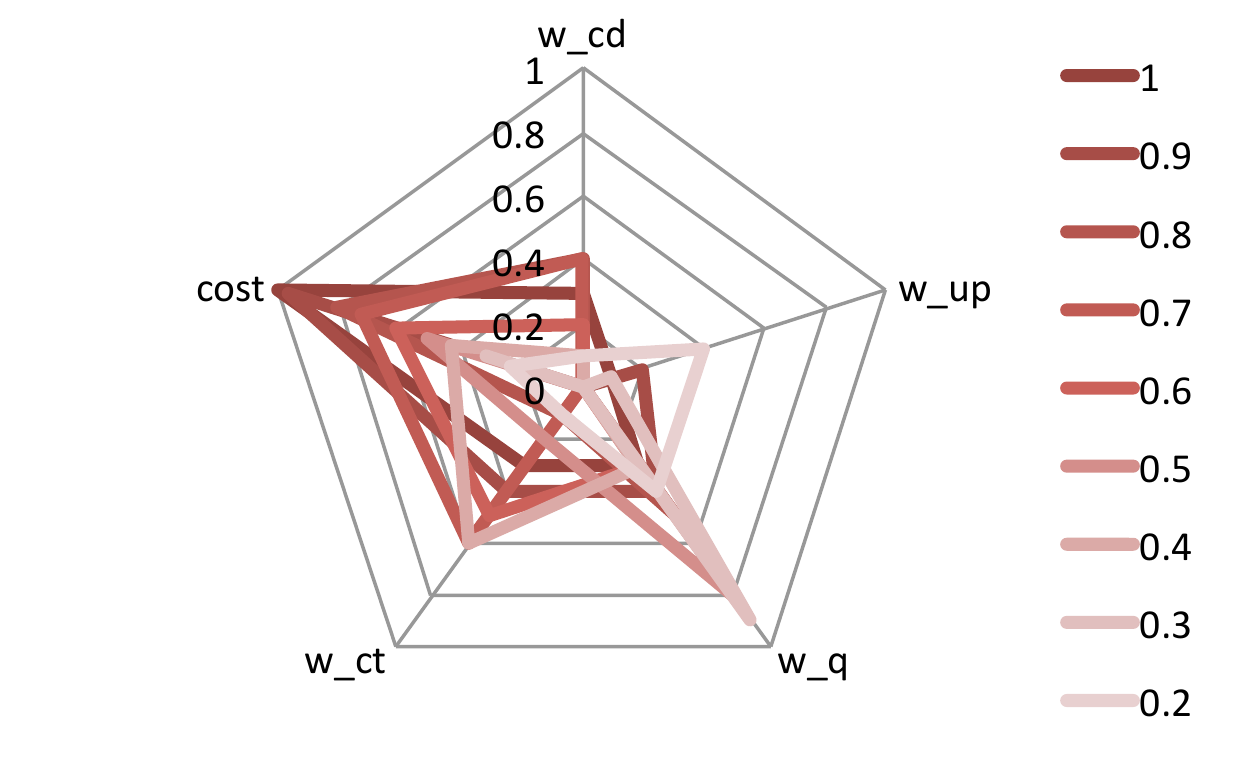}
\vspace{\figcaptionmargin}
\caption{Relative energy costs for different GA parameters.}
\label{fig:parameter-exploration}
\end{figure}

A radio chart with the results\
is shown in Fig.~\ref{fig:parameter-exploration}.\
The five axes show the four fitness component weights with the fifth axis\
as the energy cost expressed\
relative to the worst-case combination. One combination is shown as a\
pentagon of the same colour, indicating the 5-tuple\
$(w_{ct}, w_q, w_{cd}, w_{up}, cost)$.\
Combination colour is sorted by relative cost as well, with darker\
colours having higher and lighter colours lower energy costs.
For clarity, we only show a subset of the combinations with the\
rounded relative cost closest to a step of 0.1.\
We can see that the lowest energy costs are achieved for\
high $w_{up}$ and $w_q$ weights, meaning that energy cost\
and a low number of migrations is prioritized.\
Higher energy costs were obtained when constraint satisfaction ($w_{ct}$)\
and \gls{vm} consolidation ($w_{cd}$) is prioritized,\
as neither of these components includes geotemporal inputs.

\subsubsection{Temperature Range Variation} 


\newcommand{\temperaturerangersquared}{0.31}

As different cloud providers have data centers at various locations,\
where temperature ranges can be very different, we analyse the impact of\
temperature range variation on pervasive cloud control effectiveness.\
Temperature variation is affected by the time of the year and the range\
of daily temperatures will vary over time.\
For this reason, we performed the simulation with different starting times\
throughout the year, which resulted in different temperature ranges\
for different simulation runs.

\begin{figure}
\includegraphics[width=1.0\columnwidth]{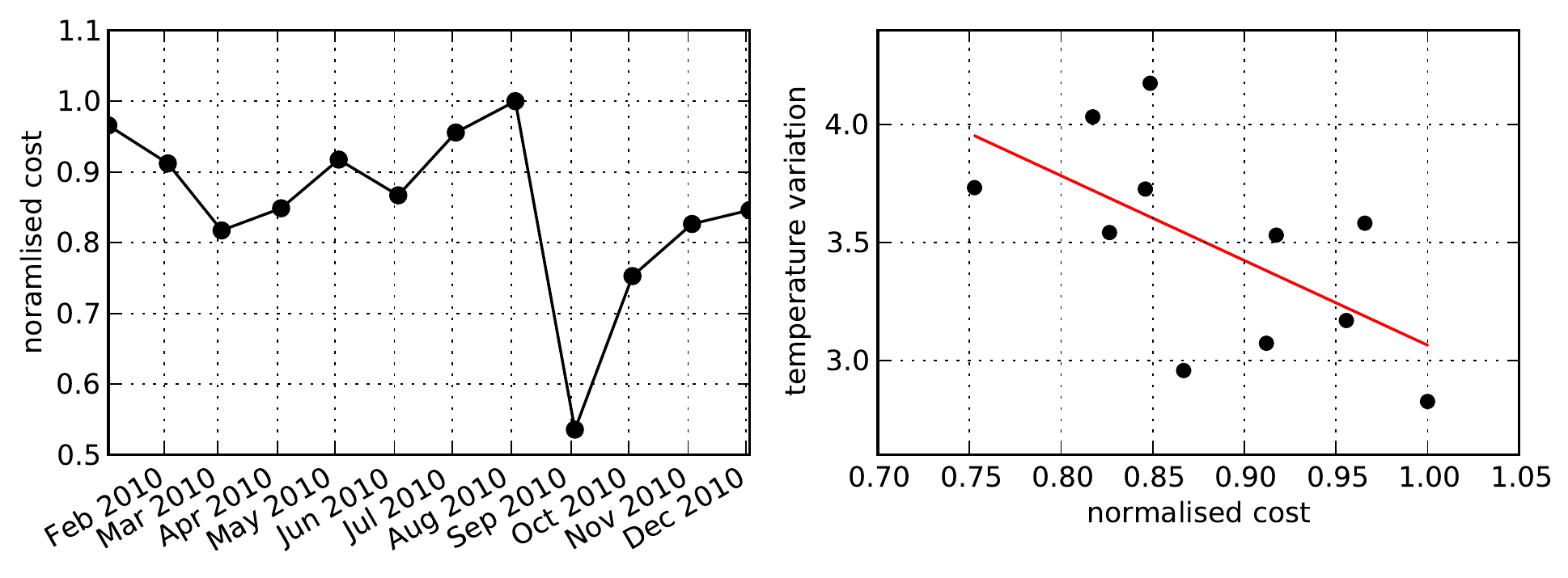}
\vspace{\figcaptionmargin}
\vspace{-0.15cm}
\caption{Energy costs for simulation runs during different months\
of the year (left) and a scatter plot of the same data and the\
temperature variation metric showing linear dependence (right).}
\label{fig:temperature-variation}
\end{figure}

The resulting graphs are shown in Fig.~\ref{fig:temperature-variation}.\
The figure to the left shows the energy cost (normalised as relative\
to the maximum vaule) resulting from the simulation of the pervasive\
cloud controller handling the same \gls{vm} requests, only shifted\
to a different month of the year.\
We can see a gradual trend, with a single sudden drop in October.\
To extract the statistical environment changes between these runs,\
we plotted the same normalised cost as a scatter plot against\
the temperature variation in the figure to the right.\
The temperature variation is calculated as the mean\
of the standard deviations of temperature values\
for individual data center locations.\
Once ordered this way, the trend of the data becomes clearer\
and we calculated a linear correlation between the variables\
with an adjusted $R^2$ of \temperaturerangersquared{} (model shown in red).\
We see that a higher temperature variation\ 
results in lower energy costs.\
This is due to the higher impact avoiding locations with\
unfavourable cooling efficiency conditions has.\

\subsubsection{Data Quality} 

To explore the effect errors in forecasting geotemporal inputs have on\
the pervasive cloud controller's operation, we simulate different\
data quality scenarios.\
%
Additionally, we explore different forecast window sizes\
and their effect on scheduling efficiency.\
A longer forecast window enables the fitness function to\
evaluate the consequences of different management actions\
over a longer interval, reducing the impact of short-term\
geotemporal impact changes, such as electricity price spikes.\

\begin{figure}
\vspace{\figtopmargin}
\includegraphics[width=1.0\columnwidth]{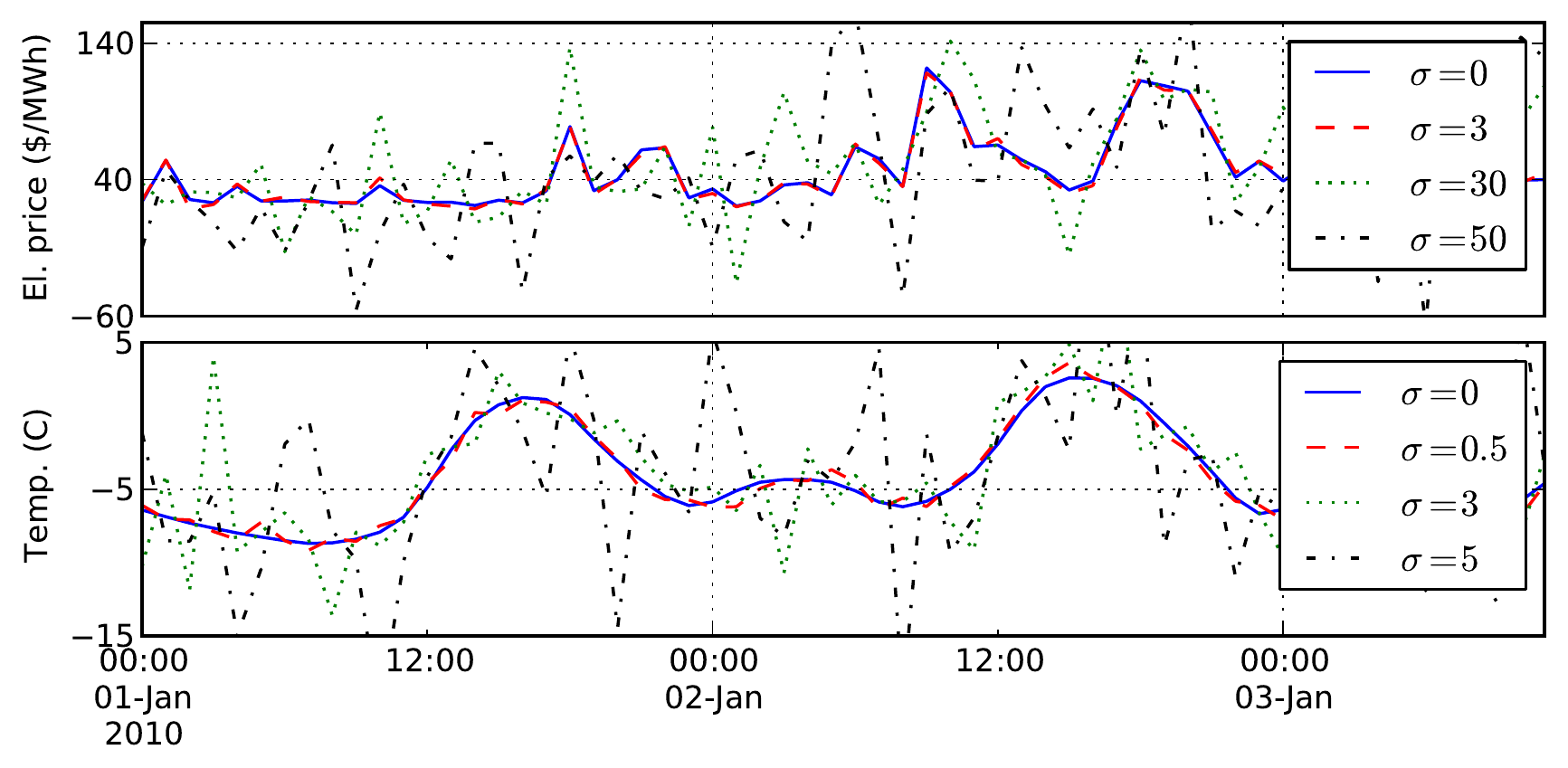}
\vspace{-0.8cm}
\caption{Different levels of forecasting errors applied to\
electricity price and temperature time series (for Mankato, MN).}
\label{fig:forecasting-error-model}
\vspace{-0.6cm}
\end{figure}

The time series provided to the scheduling algorithm\
with different forecasting errors were obtained\
using Eq.~\ref{eq:forecasting_error}\
by selecting different standard deviation ($\sigma_{pred}$) parameters.\
A $\sigma_{pred}$ close to zero represents very accurate forecasting,\
while a higher $\sigma_{pred}$ causes higher signal volatility and\
forecasting errors.\
A segment of the generated time series for one of the cities is\
shown in Fig.~\ref{fig:forecasting-error-model}.\
Both time series are aligned to the same x-axis.\
Each curve represents one $\sigma_{pred}$ scenario.\
It can be seen that smaller error levels\
($\sigma_{pred}$ of 3 \$/MWh and 0.5 C\
for electricity or temperature, respectively)\
still retain the general trend with identifiable peaks and lows.\
Higher error levels\
($\sigma_{pred}$ from 30 to 50 \$/MWh or 3 to 5 C\
for electricity or temperature, respectively)\
start to significantly diverge from the original time series,\
in that peaks are predicted where in reality lows occur and vice versa.\
We simulated forecast window sizes of 4, 12, 24 and 48 hours.\


\begin{figure}
\centering
\vspace{\figtopmargin}
\includegraphics[width=0.6\columnwidth,trim={1.6cm 0 0 0.3cm},clip]{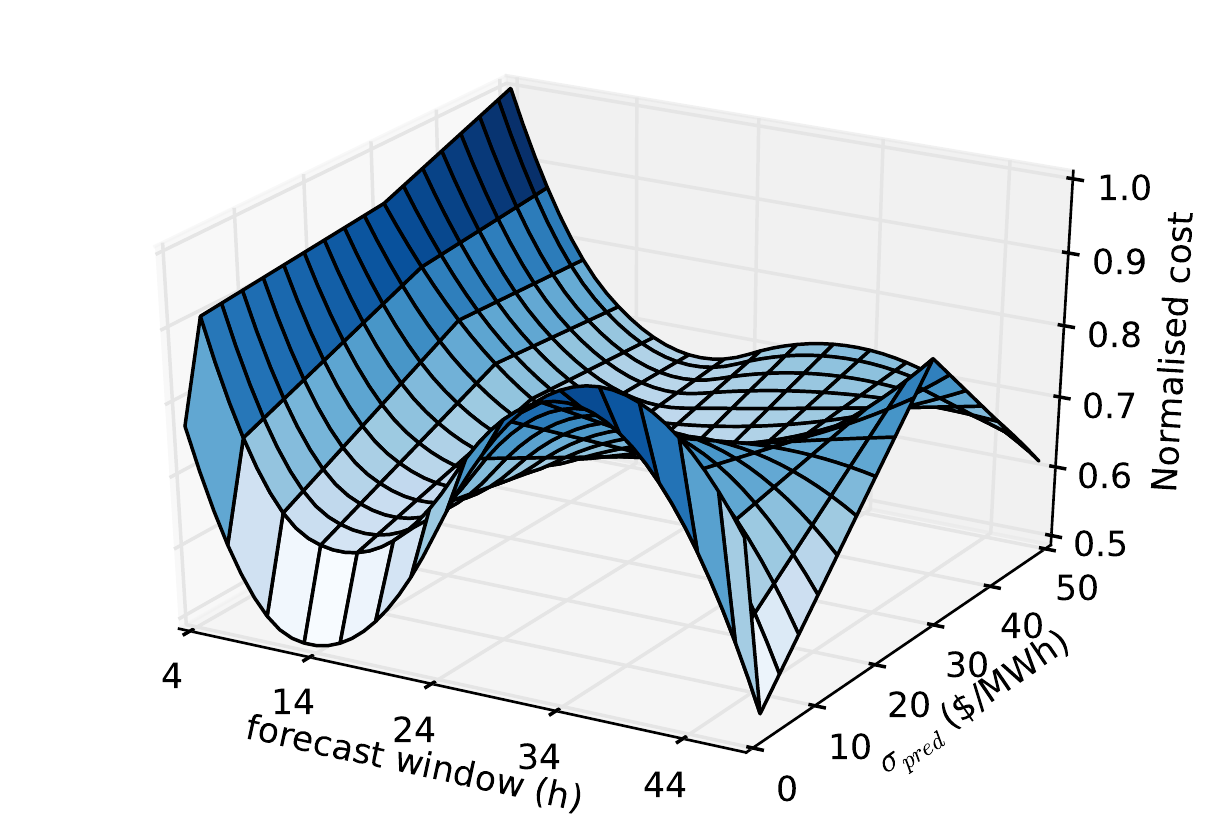}
\caption{Normalised energy costs resulting from different forecasting errors and forecast window sizes.}
\label{fig:results-forecasting-parameters}
\vspace{\figbottommargin}
\end{figure}

The generated time series with forecasting errors were provided\
to the pervasive cloud controller to base its decisions upon.\
The forecast parameter exploration results are shown in\
Fig.~\ref{fig:results-forecasting-parameters}.\
The 3-dimensional visualisation shows the space\
of forecast window sizes and electricity price $\sigma_{pred}$ values\
on the bottom plane. The z-position on the surface (its height) shows\
the total energy cost (including the migration and cooling overhead),\
normalised relative to the worst case ($fw=4\ h,\ \sigma_{pred}=50\ \$/MWh$).\
Missing data points\ 
were interpolated.\ 
The graph only shows $\sigma_{pred}$ used to model electricity price\
forecasting errors, but a matching $\sigma_{pred}$ for temperature\
from Fig.~\ref{fig:forecasting-error-model} was also applied.\

Looking at the forecast window sizes\
in Fig.~\ref{fig:results-forecasting-parameters},\
we can see that initially bigger windows result in lower costs.\
This trend can clearly be seen from 4 h to 20 h for all $\sigma_{pred}$.\
The trend changes, however, for 24 h and bigger windows.\
Higher or lower savings are visible\ 
and the pattern is more randomised.\
The reason is that both electricity prices and temperatures exhibit a daily\
seasonality effect and extending the window further than 24 h does not\
provide much more information to the controller,\
but increases the problem search space.\
We conclude that the rift-like surface shape in\
the forecast window range from 12 to 24 hours\
represents an optimal size for the given geotemporal inputs.

The forecasting error dimension shows\
deviations of around 25\% between large forecasting errors and\
perfect knowledge.\
This can be attributed to the fact that large forecasting errors\
mislead the controller into placing \gls{vm}s in areas where geotemporal\
inputs are in fact worse, so both energy cost losses\
and migration overheads are incurred.\ 
Smaller forecasting errors result in lower energy costs,\
which shows the importance of having accurate forecasting methods\
(or data sources) when managing clouds based on geotemporal inputs.\
Based on our simulation, $\sigma_{pred}$ of 3 \$/MWh\
(\gls{mse} of $\approx 9$) and 0.5 C\
(\gls{mse} of $\approx 0.25$) or less is\
necessary for feasible cost savings.\

\subsection{Cloud Provider Guidelines Case Study}

To show the usage potential of the collected measurements as\ 
guidelines, we performed a case study for several different cloud providers.\
The results are shown in Table~\ref{tab:case_study}.\
The annual electricity cost estimations\
are obtained from \cite{qureshi_cutting_2009}.\
We selected cloud providers of different scale\
(e.g. A and C).\
We compare several environment conditions,\
namely the temperature variation $temp\_var$ and\
data quality metrics $fw$ and $\sigma_{pred}$,\
considering different combinations as hypothetical scenarios that\
cloud providers might be in.\
The cost factor is calculated based on the already analysed\
temperature variation linear model\
from Fig.~\ref{fig:temperature-variation} and\
the forecast data quality impact results\
from Fig.~\ref{fig:results-forecasting-parameters}.\
Finally, we show the order of magnitude of the cost savings\ 
based on the cost factor.\

\vspace{\tabletopmargin}
\begin{table}[H]
\centering
\caption{Cloud provider guidelines case study.}
\vspace{\tablecaptionmargin}
\label{tab:case_study}
\begin{tabular}{llrrrrl}
\toprule
provider & electricity &  $fw$ &  $\sigma_{pred}$ &  $temp\_var$ &  factor & savings\\
\midrule
        A &             \$38M &  14 &     10 &       3.5 &        0.565 &         \$16.5M \\
     B &             \$36M &  48 &     30 &       2.5 &        0.989 &          \$0.4M \\
     C &             \$12M &  24 &     10 &       4.0 &        0.536 &          \$5.6M \\
\bottomrule
\end{tabular}
\end{table}
\vspace{\tablebottommargin}

The results show that significant savings are possible\
using pervasive cloud control with appropriate environmental conditions.\
The B scenario, however, shows that even with high initial costs,\
having bad forecast data quality and a low temperature variation\
can result in lower savings,\
perhaps not enough of an incentive to apply our method.\
On the other hand, even a smaller provider, such as C,\
can achieve promising savings\
in a favourable environment.\ 

\vspace{-0.4cm}
\section{Summary}


In this chapter we presented an approach for pervasive cloud control\
under geotemporal inputs, such as real-time electricity pricing\
and temperature-dependent cooling efficiency.\
The solution is designed for extensibility\
with new geotemporal inputs and cloud regulation mechanisms\
through a modular\
decision support component system and\
a forward-compatible optimisation engine.\
We presented a proof-of-concept controller implementation\
combining forecast-based planning and\
a hybrid genetic algorithm\
with greedy local optimisation.\
The genetic algorithm approach was extended\
with partial population propagation.\

The approach was evaluated in a simulation based on real\
traces of temperatures and electricity prices.\
We estimated energy cost savings\
of up to \gaschedulerEnSavingsNoGeotemp{}\
compared to a baseline cloud control method\
that applies \gls{vm} consolidation without considering geotemporal inputs.\
We analysed per-\gls{vm} migrations\ 
to show that no significant\
\gls{qos} impact is incurred in the process.\
We evaluated different parameters such as geographical data center distributions and\
forecast data quality\ 
as cloud provider guidelines\
to find conditions fit for pervasive cloud control.\



\chapter{Progressive SLA Specification for Energy-Aware Cloud Management}
\label{ch:vmpricing}



The pervasive cloud controller from the previous chapter
dynamically reallocates computation
across geographically distributed data centers to leverage regional electricity
price and temperature differences. As a result, a managed \gls{vm} may suffer
occasional downtimes. Current cloud providers only offer high availability
\gls{vm}s, without enough flexibility to apply such energy-aware management.
In this chapter, we show how to analyse past traces of dynamic cloud management
actions based on electricity prices and temperatures to estimate \gls{vm}
availability and price values. We propose a novel \gls{sla} specification
approach for offering \gls{vm}s with different availability and price values
guaranteed over multiple \gls{sla}s to enable flexible energy-aware cloud
management. We determine the optimal number of such \gls{sla}s as well as their
availability and price guaranteed values. We evaluate our approach
in a user \gls{sla} selection simulation
using Wikipedia and Grid’5000 workloads.
The results show higher customer conversion and \VMPricingEnSavings{} average
energy savings per \gls{vm}.
Section \ref{sec:approach}\
gives an overview of how our progressive \gls{sla} specification can be applied
in energy-aware cloud management based on geotemporal inputs.\ 
The probabilistic technique to build \gls{sla}s\
is presented in Section~\ref{sec:pricing_model}.
The user behaviour model is described in Section~\ref{sec:user_model}.\
The evaluation description along with the results analysis\
are given in Section~\ref{ch:vmpricing:sec:evaluation}.\


\vspace{\sectionendmargin}
\section{Progressive SLA Specification}
\vspace{\sectiontitlemargin}
\label{sec:approach}
\
\



To be able to reason about energy-aware cloud management\
in terms of \gls{sla} specification,\
we analyse two concrete schedulers:\
\
(1) The migration scheduler\
described in Chapter~\ref{ch:gascheduler} -- \
applies a genetic algorithm to dynamically migrate\
\gls{vm}s, such that energy costs based on geotemporal inputs are minimised,
while also minimising the number of migrations per \gls{vm}\
to retain high availability.\
(2) The peak pauser scheduler described in Chapter~\ref{ch:volatility} --\
\
pauses the managed \gls{vm}s for a predefined duration every day,\
choosing the hours of the day that are statistically most likely\
to have the highest energy cost, thus reducing \gls{vm} availability, but also\
the average energy cost.\

With such energy-aware cloud management methods in mind,\
we propose a progressive \gls{sla} specification,\
where services are divided among multiple treatment categories,\
each under a different \gls{sla} with different availability and price values.\
Hence, different schedulers can be used for \gls{vm}s\
in different treatment categories\
(or the same scheduler with different \gls{qos} constraint parameters).\
The goal of this approach is\
to allow different levels of energy-aware cloud management\
and thus achieve higher energy savings on \gls{vm}s\
with lower availability requirements.\
What is given, therefore, are the schedulers for each treatment category,\
and the historical traces of generated schedules.\
What we have to find are the availability and price values\
that can be guaranteed in the SLAs for each treatment category\
and the optimal number of such SLAs\
to balance SLA flexibility and search difficulty for users.

\begin{figure}
\vspace{\figtopmargin}
\centering
\includegraphics[width=0.99\textwidth]{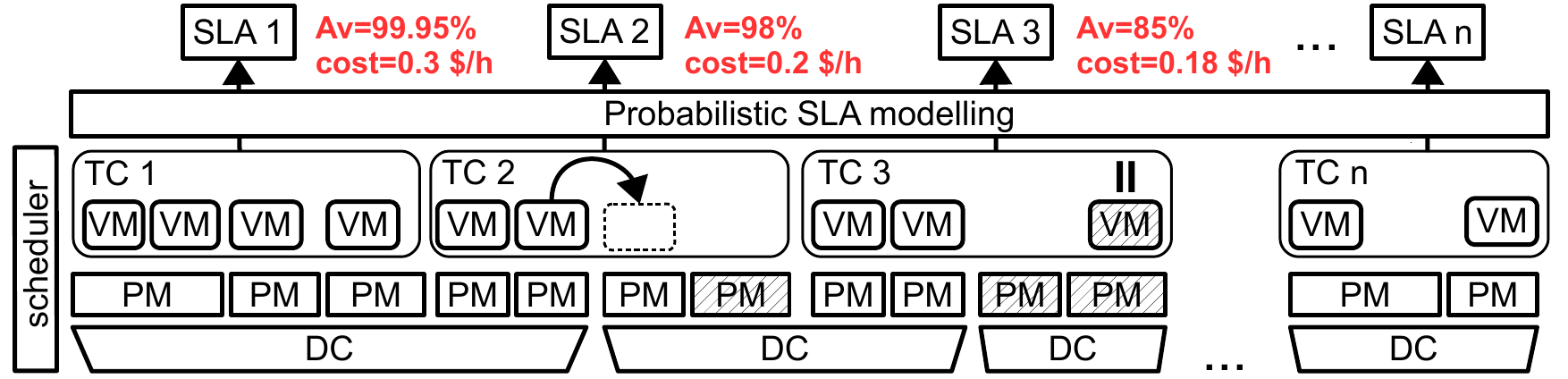}
\caption{Progressive SLA specification}
\label{fig:vmpricing-overview}
\vspace{\figbottommargin}
\end{figure}

We illustrate this approach in Fig.~\ref{fig:vmpricing-overview}.\
A cloud provider operates a number of \gls{vm}s,\
each hosted on a \gls{pm} located\
in one of the geographically distributed data centers (DC).\
Each \gls{vm} belongs to a \gls{tc} that determines the type of\
scheduling that can be applied to it.\
For example, \gls{tc} 1 is a high availability category where no actions\
are applied on running \gls{vm}s.\
\gls{tc} 2 is a category for moderate cloud management actions, such as\
live \gls{vm} migrations (marked by an arrow) which result in short downtimes.\
\gls{tc} 3 is a more aggressive category where \gls{vm}s can be paused\
(marked as hatched with two vertical lines above it)\
for longer downtimes.\
Other \gls{tc}s can be defined using other scheduling algorithms\
or by varying parameters, e.g. the maximum pause duration.\
The optimal number of \gls{tc}s (and therefore also \gls{sla}s) $n$\
is determined by analysing user\
\gls{sla} selection to have enough variety to satisfy most user types,\
yet not make the search too difficult,\
which we explore in Section~\ref{ch:vmpricing:sec:evaluation}.\
Aside from selecting the number of \gls{sla}s, another task is setting\
availability and price values for every \gls{sla}.\
As energy-aware cloud management that depends on geotemporal inputs\
introduces a degree of randomness into the resulting availability and price\
values of a \gls{vm}, we can only estimate the values that can be guaranteed.\
We do this using a \emph{probabilistic \gls{sla} modelling} method for\
analysing historical cloud management action traces\
to calculate the most likely worst-case availability and average\
energy cost for a \gls{vm} in a \gls{tc}.\
For the example in Fig.~\ref{fig:vmpricing-overview}, sample values\
are given for the \gls{sla}s.\
\gls{sla} 1 might have an availability (Av) of $99.95\%$\
and a high cost of $0.3$ \$/h\
due to no energy-aware management,\
\gls{sla} 2 might have slightly lower values\
due to live migrations being applied on \gls{vm}s in \gls{tc} 2,\
\gls{sla} 3 might have even lower values due to longer downtimes\
caused by \gls{vm} pausing etc.\
In the following section,\
we will show how to actually estimate availability and price\
values for the \gls{sla}s using probabilistic \gls{sla} modelling.\

\vspace{\sectionendmargin}

\section{Probabilistic SLA Modelling}
\label{sec:pricing_model}
\vspace{\sectiontitlemargin}
To estimate availability and price values\
that can be guaranteed in an \gls{sla} using probabilistic modelling\
for a certain \gls{tc}, we require historical cloud management traces.\
Cloud management traces\
can be obtained through monitoring,\
but for evaluation purposes we simulate different\
scheduling algorithm behaviour.\ 
\
\gls{vm} price is estimated\
by accounting for the average energy costs.\
To calculate \gls{vm} availability,\ 
we analyse the factors that cause \gls{vm} downtime.\
While the downtime duration of the peak pauser scheduler can be specified\
beforehand, the total downtime caused by the migration scheduler depends\
on \gls{vm} migration duration and rate as dictated by geotemporal inputs,\
so we individually analyse both factors.


\vspace{\subsectionendmargin}

\subsection{Cloud Management Simulation}
\label{ch:vmpricing:sec:cloud-simulation}
\vspace{\subsectiontitlemargin}
Our modelling method can be applied to different\
scheduling algorithms and cloud environments.\
To generate a concrete \gls{sla} offering for evaluation purposes,\
we consider a use case\
%
of a cloud consisting of six geographically distributed data centers
shown in Fig.~\ref{ch:vmpricing:fig:cities}
to represent a deployment similar to large cloud providers such as Google.
We use the world-wide datasets of electricity prices \cite{alfeld_toward_2012}\
and temperatures from the Forecast web service \cite{_forecast_2015}
described in Section~\ref{ch:gascheduler:sec:evaluation_methodology}.
%
The effects of the migration and peak pauser scheduler\
are determined in a simulation\
using the Philharmonic\
cloud simulator also described in Section~\ref{sec:philharmonic}.
%
%
%
The cloud simulation parameters\
are summarised in Table~\ref{ch:vmpricing:tab:simulation-cloud}.\
We illustrate the application of the presented methods with this use\
case as a running example.

\vspace{1cm}
\noindent\begin{minipage}{\textwidth}
  \begin{minipage}{0.6\textwidth}
    \centering

	\includegraphics[width=1.0\textwidth]{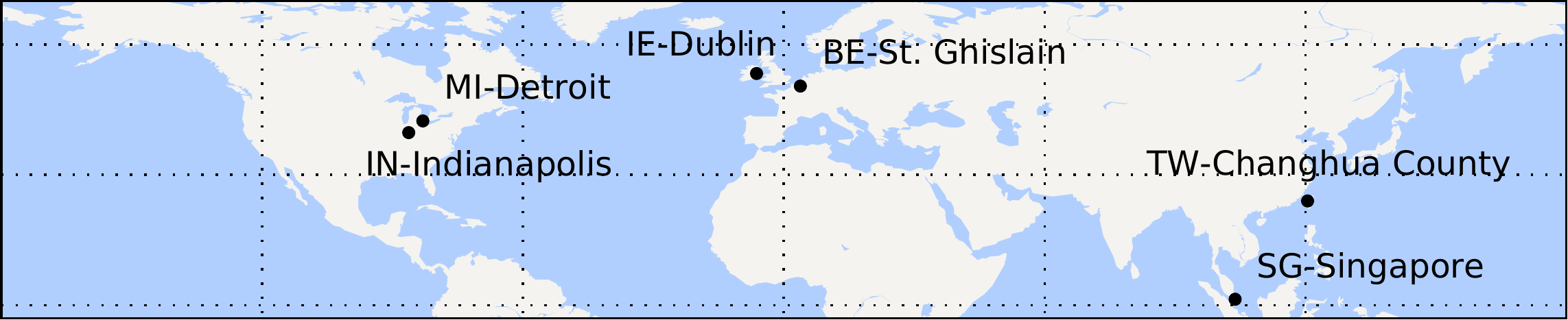}
    \captionof{figure}{Simulated world-wide data centers}
    \label{ch:vmpricing:fig:cities}
  \end{minipage}
  \begin{minipage}{0.4\textwidth}
    \centering
	\begin{tabular}{ c c c c }
	\hline
	duration & DCs & PMs & VMs\\
	\hline
	3 months & 6 & 20 & 80 \\
	\hline
	\end{tabular}
	\captionof{table}{Cloud simulation} 
	\label{ch:vmpricing:tab:simulation-cloud}
  \end{minipage}
\end{minipage}



\vspace{\subsectionendmargin}

\subsection{Migration Duration}
\vspace{\subsectiontitlemargin}
Even a live \gls{vm} migration incurs a temporary downtime,\
in the stop-and-copy phase of \gls{vm} memory transferring.\
A very accurate model, with less than 7\% estimation errors,\
for calculating this downtime overhead\
is presented in \cite{liu_performance_2011}.\
The total \gls{vm} downtime during a single live migration $T_{down}$\
is a function of the \gls{vm}'s memory $V_{mem}$, data transmission rate $R$,\
memory dirtying rate $D$, pre-copying termination threshold $V_{thd}$\
and $T_{resume}$, the time necessary to resume a \gls{vm}.

\vspace{\eqtopmargin}
\begin{eqnarray}
T_{down} = \frac{V_{mem} D^n}{R^{n+1}} + T_{resume} \quad \text{, where} \hskip 1em
n = \ceil[\bigg]{ log_{\frac{D}{R}}\frac{V_{thd}}{V_{mem}} }
\end{eqnarray}
\vspace{\eqbottommargin}

\begin{figure}
\vspace{\figtopmargin}
\centering
\includegraphics[width=0.70\textwidth]{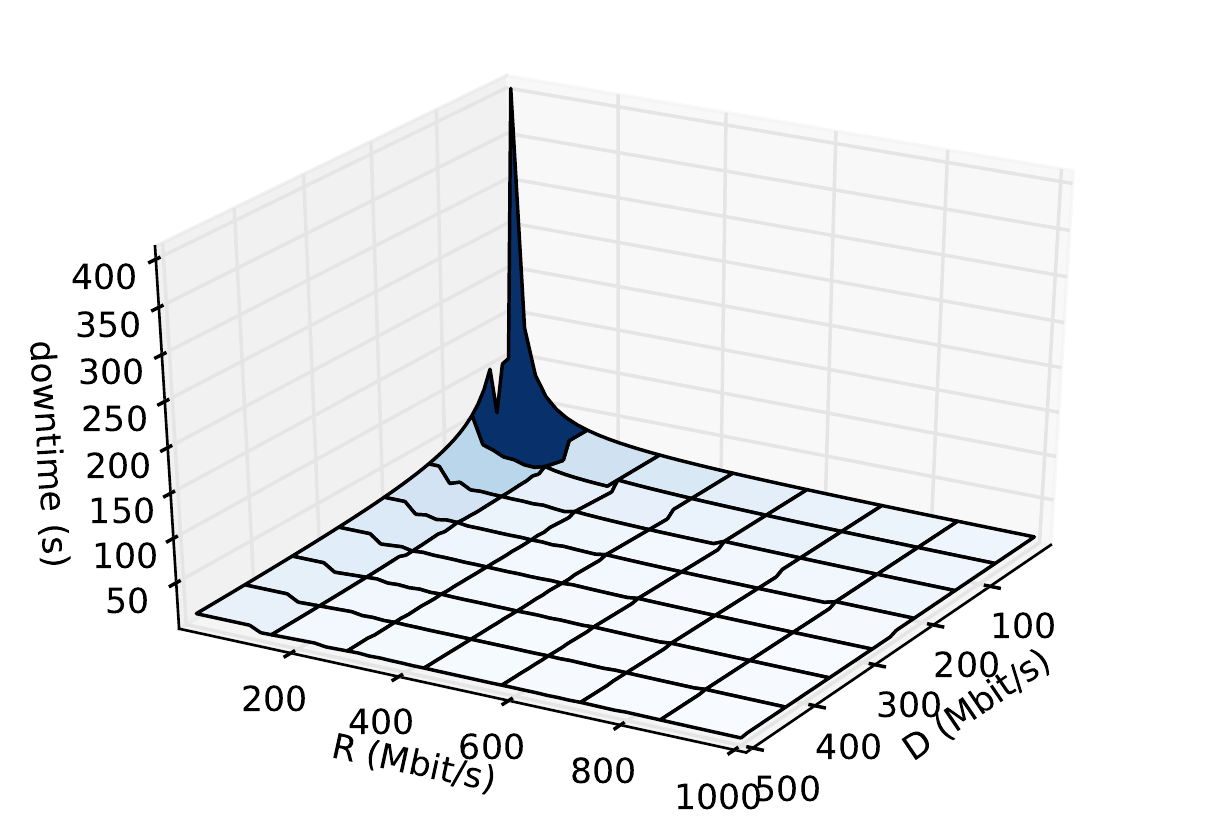}
\caption{VM migration downtime}
\label{fig:migration_duration}
\vspace{\figbottommargin}
\end{figure}

All of the parameters can be determined\
beforehand by the cloud provider, except for $R$ and $D$ which depend\
on the dynamic network conditions and application-specific characteristics.\
Based on historical data, it is possible to reason about the range of these\
variables.\ 
In our running example,\ 
we assume a historical range from low to high values.\
$R$ values in the 10--1000 Mbit/s range were taken based on an\
independent benchmark of Amazon EC2 instance bandwidths.\
$D$ values from 1 kbit/s\
(to represent almost no memory dirtying)\ 
to 1 Gbit/s were taken (the maximum is not important, as will be shown).\
We assumed constant $V_{mem} = 4 GB,\ V_{thd} = 1 GB,\ T_{resume} = 5 s$,\
as these values do not affect the order of magnitude of $T_{down}$.\
We show how $T_{down}$ changes for different $R$ and $D$\ 
in Fig.~\ref{fig:migration_duration}.\
Looking at the graph, we can see that higher $R$ and $D$ values result\
in convergence towards negligible downtime durations and the only area\
of concern is the peak happening when $R$ and $D$ are both very low.\
This happens, because close $R$ and $D$ values lead to a small number\
of pre-copying rounds and copying the whole \gls{vm} under slow speeds\
leads to long downtimes.\
$T_{down}$ is under \VMPricingTdown{} in the worst-case scenario, \ 
which will be our \gls{sla} estimation as\
suggested in \cite{berndt_towards_2013}.\

\vspace{\subsectionendmargin}

\subsection{Migration Rate}
\vspace{\subsectiontitlemargin}

\begin{figure}
\vspace{\figtopmargin}
\centering
\begin{minipage}{0.48\textwidth}
    \centering
    \includegraphics[width=\textwidth]{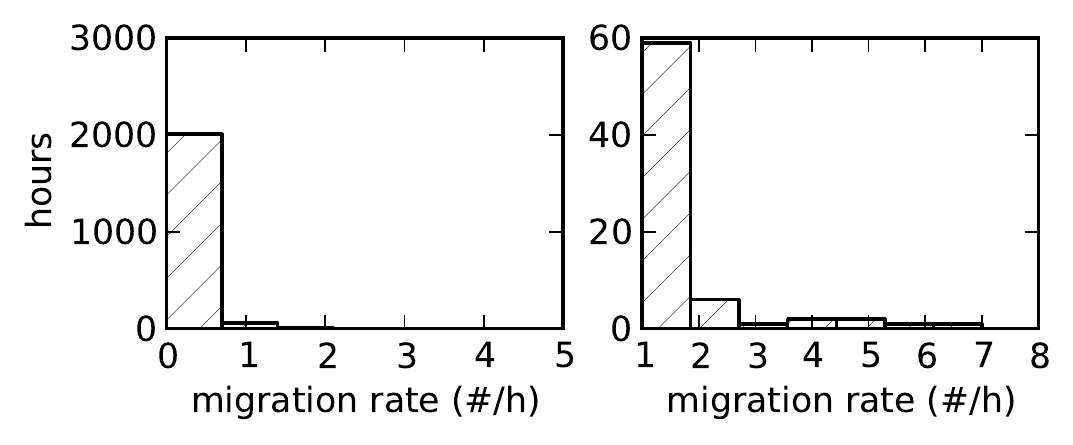}
    \caption{Hourly migration rate histogram}
    \label{ch:vmpricing:fig:migrations-rate}
\end{minipage}
\begin{minipage}{0.48\textwidth}
    \centering
    \includegraphics[width=\textwidth]{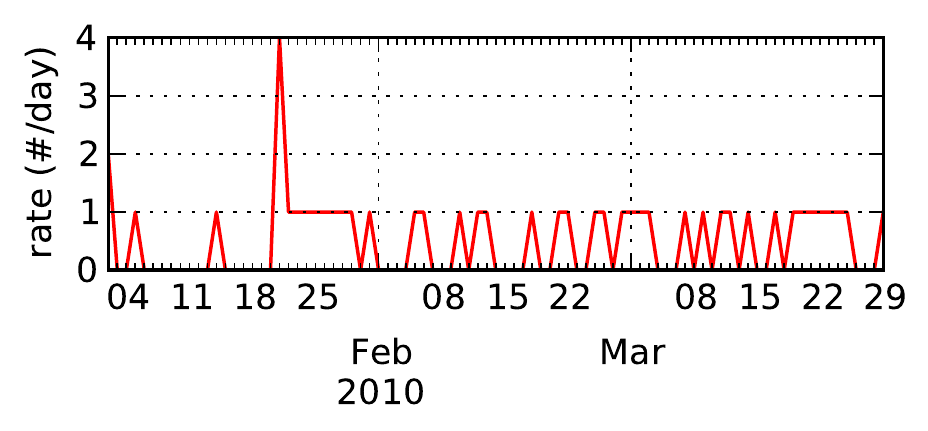}
    \caption{Aggregated worst-case migr. rate}
    \label{ch:vmpricing:fig:migrations-aggregated}
\end{minipage}
\vspace{\figbottommargin}
\end{figure}

Aside from understanding migration effects, we need to analyse how often they\
occur, i.e. the migration rate.\
We presented a method to analyse migration traces\
obtained from the cloud manager's past operation in Chapter~\ref{ch:gascheduler}.\
A histogram of migration rates\ 
for the migration traces from our running example\
described in Section~\ref{ch:vmpricing:sec:cloud-simulation}\
can be seen in Fig.~\ref{ch:vmpricing:fig:migrations-rate}.\
%
%
%
We reuse the \textit{aggregated worst-case} function
presented in Section~\ref{ch:gascheduler:sec:qos_analysis} that\ 
counts the migrations per \gls{vm} per day and selects\
the highest migration count among all the \gls{vm}s in every interval.\
The output time series is shown
in Fig.~\ref{ch:vmpricing:fig:migrations-aggregated}.
There is one or zero migrations per \gls{vm} most of the time,\
with an occasional case with a higher rate, such as the peak in January.\
Such peaks can occur due to more turbulent geotemporal input changes.\
or insufficient schedule optimisation.

Finally, the
bootstrap confidence interval method \cite{efron_introduction_1994}\
was applied
to predict the maximum aggregated migration rate.\ 
For our migration dataset,\ 
the 95\% confidence interval for the worst-case migration rate is
from three to four\
migrations per day.\

\vspace{\subsectionendmargin}

\subsection{\gls{sla} Options}
\vspace{\subsectiontitlemargin}
\label{sec:8slas}

By combining the migration rate and duration analyses,\
we can estimate the upper bound\
for the total \gls{vm} downtime and, therefore, the availability\
that can be warranted in the \gls{sla}.\
We define availability ($Av$) of a \gls{vm} as:
\
\begin{equation}
Av = 1 - \frac{\text{total VM downtime}}{\text{total VM lease time}}
\end{equation}
\
For our migration dataset and the previously discussed migration duration and rate,\
we estimate the total downtime of a \gls{vm}\
controlled by the migration scheduler to be \VMPricinggadown{} per day\
in the worst case, meaning we can guarantee an availability of \VMPricinggaav{}.\
We can precisely control the availability of the \gls{vm}s\
managed by the peak pauser.\

The average energy savings ($en\_savings$) for a \gls{vm} running in\
a treatment category $TC_i$ can be calculated by comparing it to the\
high availability $TC_1$.\
From the already described simulation,\ 
we calculate $en\_cost$, the average cost of energy consumed by a \gls{vm}\
based on real-time electricity prices and temperatures.\
We divide the energy costs equally among \gls{vm}s within a \gls{tc}.\
This is an approximation, but serves as an estimation of the\
energy saving differences between \gls{tc}s.\
We calculate energy savings as:\
\
\begin{equation}
en\_savings(VM_{TC_i}) = 1 - \frac{en\_cost(VM_{TC_i})}{en\_cost(VM_{TC_1})}
\label{eq:energy_model}
\end{equation}
\
where $en\_cost(VM_{TC_1})$ is the average energy cost\
for a \gls{vm} in $TC_1$\ with no actions applied\
and $en\_cost(VM_{TC_i})$ is the average energy cost\
for a \gls{vm} in the target $TC_i$.\

The \gls{vm} cost consists of several components.\
\
Aside from $en\_cost$,
$service\_cost$ groups other \gls{vm} upkeep costs\
(manpower, hardware amortization, profit margin etc.)\
during a charge unit (typically one hour in \gls{iaas} clouds).\
We assume the service component to be charged\
linearly to the \gls{vm}'s availability.\
\
\begin{equation}
cost = en\_cost + Av \cdot service\_cost
\label{eq:vm_cost}
\end{equation}
\vspace{\eqbottommargin}
\
To generate the complete \gls{sla} offering\
we consider an Amazon m3.xlarge instance which costs 0.280 \$/h\
(Table~\ref{tab:base-vm}) as a base \gls{vm} with no energy-aware scheduling.\
Base instances with different resource values (e.g. RAM, number of cores)\
can be used, but this is orthogonal to the \gls{qos} requirements of\
availability that we consider and\
would not influence the energy-aware cloud management potential.\
Similarly, Amazon spot instances were not considered specially as they perform\
exactly the same as normal instances while running,\
as we mentioned in Chapter~\ref{ch:relatedwork}.\
We assumed the service component to be 0.1 \$/h,\
about a third of the \gls{vm}'s price.\
The prices of \gls{vm}s controlled by the two\
energy-efficient schedulers were derived from it,\
applying $en\_savings$ obtained in the cloud simulation.\
The resulting \gls{sla}s are shown in Fig.~\ref{fig:sla}.\ 
SLA 1 is the base \gls{vm}.\ 
SLA 2 is the \gls{vm} controlled 
by the migration scheduler.\
The remaining \gls{sla}s are \gls{vm}s controlled by the peak pauser\
scheduler with downtimes uniformly distributed from \VMPricingDRmin{} to \VMPricingDRmax{}\
to represent a wide spectrum of options.\
We chose \VMPricingslanum{} \gls{sla}s to analyse how \gls{sla} selection changes\
from the user perspective.\
We later show that this number is in the 95\% confidence interval\
for being the optimal number of \gls{sla}s based on our simulation.\
We analyse a wider range of \VMPricingslanummult{} offered \gls{sla}s\
and how they impact customer conversion\
from the cloud provider's perspective\
in Section~\ref{ch:vmpricing:sec:evaluation}.\
The lines in the background\ 
illustrate value progression.\
$Av$ decreases only slightly for the migration \gls{sla}, yet the energy\
savings are significant due to\
dynamic \gls{vm} consolidation and \gls{pm} suspension.\
For peak pauser \gls{sla}s, availability and costs decrease linearly,\
from high to low values.\ 
The $en\_savings$ values are at first lower\
than those attainable with the migration scheduler,\
as the peak pauser scheduler cannot migrate \gls{vm}s\
to a fewer number of \gls{pm}s, but can only pause them for a certain time.\
With lower availability requirements, however,\
the peak pauser can achieve higher $en\_savings$ and lower prices,\
which could not be reached by \gls{vm} migrations alone.

%
%
%


\noindent\begin{minipage}{\textwidth}
  \begin{minipage}{0.75\textwidth}
    \centering

	\includegraphics[width=1.08\textwidth]{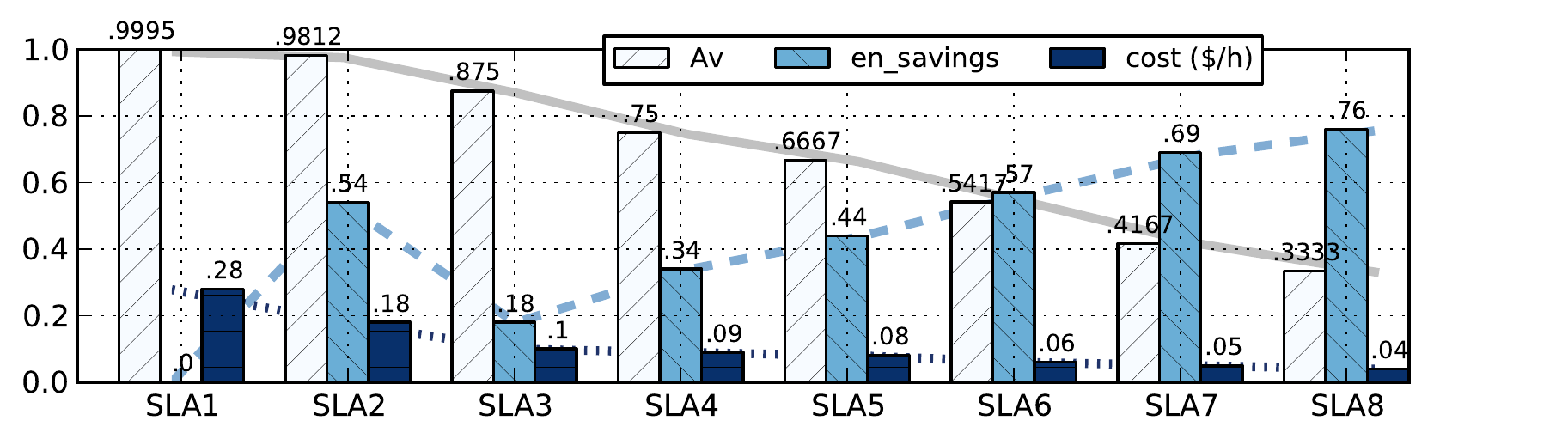}
	\vspace{\figcaptionmargin}
	\captionof{figure}{SLAs generated for different TCs}
	\label{fig:sla}
  \end{minipage}
  \begin{minipage}{0.24\textwidth}
    \centering
	\begin{tabular}{ c c c }
	\hline
	type & m3.xlarge\\
	\hline
	Av & 99.95\%\\
	\hline
	cost & 0.28 \$/h\\
	\hline
	service & 0.1 \$/h\\
	\hline
	\end{tabular}
    \captionof{table}{Base VM} 
	\label{tab:base-vm}
  \end{minipage}
\end{minipage}

\vspace{\sectionendmargin}

\section{User Modelling}
\label{sec:user_model}
\vspace{\sectiontitlemargin}
Knowing the \gls{sla} offering, the next step is\
to model user \gls{sla} selection\
in order to analyse the benefits of our progressive \gls{sla} specification.\
We first describe how we derive user requirements and then the utility\
model used to simulate user \gls{sla} selection based on their requirements.


\vspace{\subsectionendmargin}

\subsection{User Requirements Model}
\vspace{\subsectiontitlemargin}

\begin{figure}
\vspace{\figtopmargin}
\centering
    \begin{minipage}{0.48\textwidth}
        \centering
        \includegraphics[width=\textwidth]{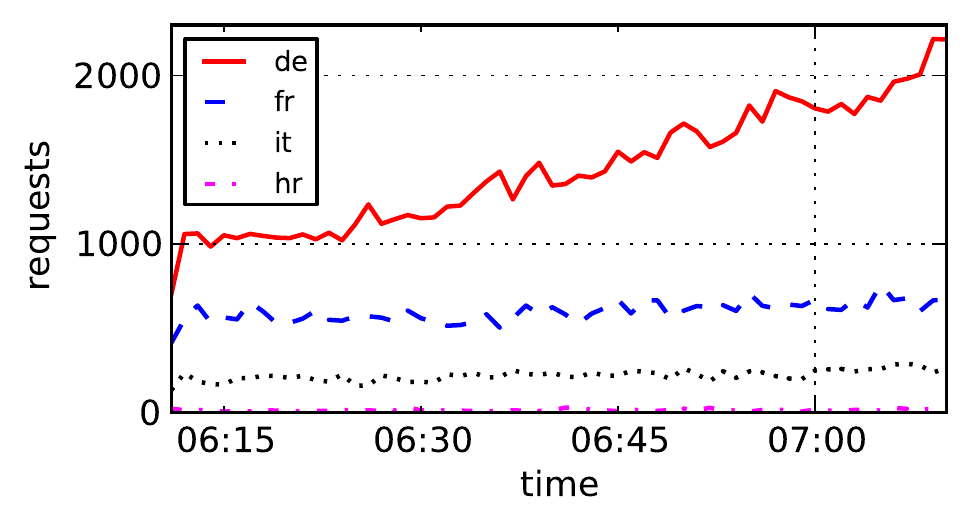}
        \smaller{(a) Wikipedia requests per minute}
    \end{minipage}
    \begin{minipage}{0.48\textwidth}
        \centering
        \includegraphics[width=\textwidth]{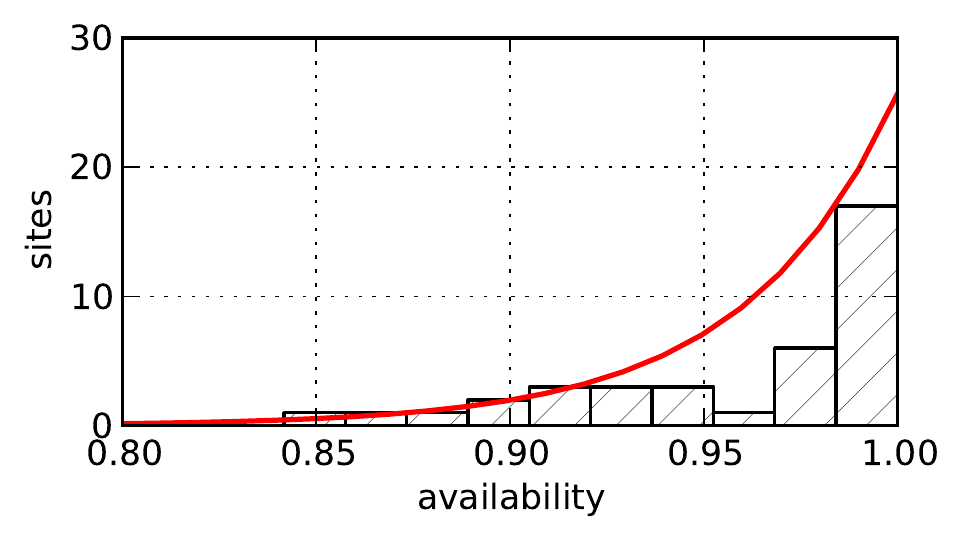}
        \smaller{(b) Wikipedia availability distribution}
    \end{minipage}
\caption{Web user modelling}
\label{fig:user-modelling-web}
\vspace{\figbottommargin}
\end{figure}

To model user requirements, we use real traces\
of web and \gls{hpc} workload,\
since I/O-bound web and CPU-bound \gls{hpc} applications\
represent two major usage patterns of cloud computing.\
As we do not have data on availability requirements of website owners,\
we generate this dataset based on the frequency of end user HTTP requests\
directed at different websites and counting missed requests,\
similarly to how reliability is determined\
from the mean time between failures \cite{oconnor_practical_2011}.\
A public dataset of HTTP requests\
made to Wikipedia \cite{urdaneta_wikipedia_2009} is used.\
To obtain data for different websites, we consider Wikipedia\
in each language as an individual website,\
because of its unique group of end users\
(different in number and usage pattern).\
In this scenario we consider a website owner\
to be the user of an \gls{iaas} service\
(not to be confused with the end user, a website visitor).\
The number of HTTP requests for a small subset of four websites\
(German, French, Italian and Croatian Wikipedia\
denoted by their two-letter country codes)\
is visualised in Fig.~\ref{fig:user-modelling-web} (a)\
for illustration purposes\
(we use the whole dataset with 38 websites for actual requirements modelling).\
The data exemplifies\ 
significant differences in amplitudes.\
Users of the German Wikipedia send between 1k and 2k requests per minute,\
while the Italian and Croatian Wikipedia have less than 300 requests per minute.\
Due to this variability, we assume that different Wikipedia websites\
represent diverse requirements of website owners.\
We model availability requirements by applying a heuristic\
 -- a website's required availability is the minimum\
necessary to keep the number of missed requests\
below a constant threshold (we assume 100 requests per hour).\
Using this heuristic, we built an availability requirements dataset\
for the web user type from 5.6 million requests\
divided among 38 Wikipedia language subdomains.\
The resulting availability requirement histogram\
can be seen in Fig.~\ref{fig:user-modelling-web} (b). It follows an\
exponential distribution (marked in red).\
There is a high concentration of sites that need almost full availability,\
with a long tail of sites that need less (0.85--1.0).

\begin{figure}
\vspace{\figtopmargin}
\centering
    \begin{minipage}{0.48\textwidth}
        \centering
        \includegraphics[width=\textwidth]{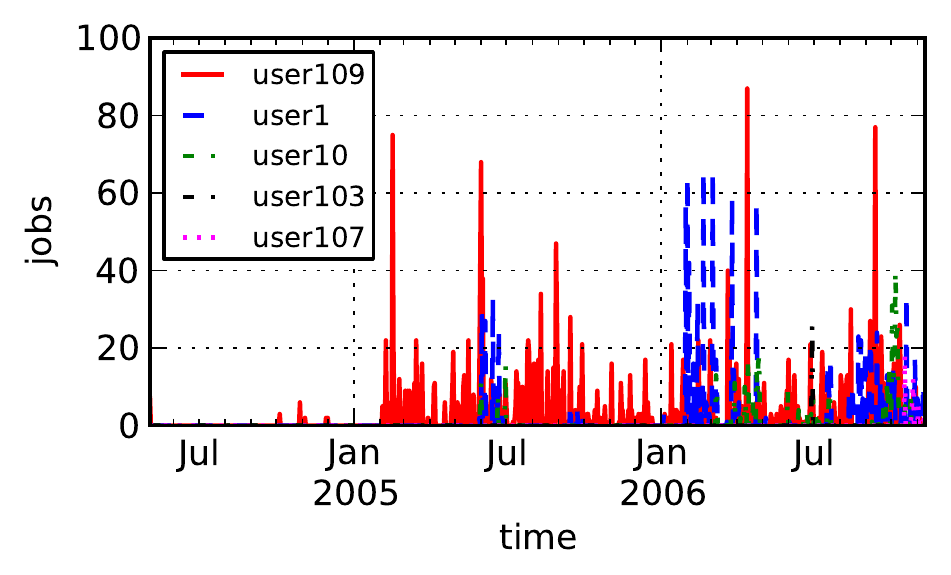}
        \smaller{(a) G5k example job submissions}
    \end{minipage}
    \begin{minipage}{0.48\textwidth}
        \centering
        \includegraphics[width=\textwidth]{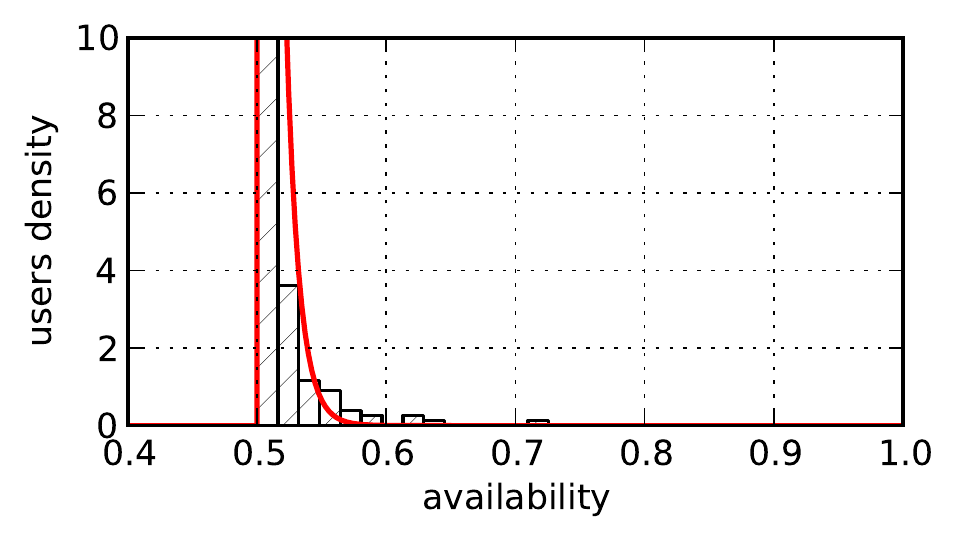}
        \smaller{(b) G5k availability distribution}
    \end{minipage}
\caption{HPC user modelling}
\label{fig:user-modelling-hpc}
\vspace{\figbottommargin}
\end{figure}

For HPC workload, we use\
a dataset\ 
of job submissions made to Grid'5000 (G5k)~\cite{franck_cappello_gwa-t-2_2014},\
a distributed job submission platform spread across 9 locations in France.\
The number of jobs submitted by a small subset of users\
is visualised in Fig.~\ref{fig:user-modelling-hpc} (a).\
While some users submit jobs over a wide period (\emph{user109}), others\
only submit jobs in small bursts (\emph{user1, user107}), but the load is not nearly\
as constant as the web requests from the Wikipedia trace.\
To model HPC users' availability requirements, where jobs have\
variable duration as well as rate (unlike web requests, which typically have\
a very short duration), we use another heuristic.\
Every user's availability requirement is\
mapped between a constant minimum availability (we assume 0.5)\
and full availability using $mean\_duration \cdot mean\_rate$,\
which stands for mean job duration and mean job submission rate per user.\
Using this heuristic, we built a dataset of availability requirements\
for the \gls{hpc} user type from jobs submitted\
over 2.5 years by 481 G5k users.\
The resulting availability requirement distribution\
(normalised such that the area is 1)\
can be seen in Fig.~\ref{fig:user-modelling-hpc} (b).\
The distribution marked red again follows an exponential distribution\
(the first bin, cut off due to the zoom level, shows a density of 100),\ 
but with the tail facing the opposite direction\
than the web requirements.\
\gls{hpc} users submit smaller and less frequent jobs most of the time,\
with a long tail of longer and/or more frequent jobs (from 0.5 to 0.75).\

%
%

Every user's \gls{wtp} is derived by multiplying\
his/her availability requirement with\
the base \gls{vm} price and adding Gaussian noise $\mathcal{N}(0,\,0.05^2)$\
to express subjective value perception.\
We selected the noise standard deviation\
to get positive \gls{wtp} values considering the availability model.\
The resulting \gls{wtp} histogram is shown in Fig.~\ref{fig:wtp}.\
It can be seen that
\gls{hpc} users have lower \gls{wtp} values,
but there is also an overlap area with web users who have similar requirements.


\vspace{\subsectionendmargin}

\subsection{Utility Model}
\vspace{\subsectiontitlemargin}

The utility-based model is used to simulate how users select services\
based on their requirements.\
We use a quasi-linear utility function adopted from\
multi-attribute auction theory \cite{jrad2013}\ 
to quantify the user's preference for a provided \gls{sla}.\
The utility is calculated by multiplying the user's SLA satisfaction score\
with \gls{wtp}\ 
and subtracting the \gls{vm} cost\
charged by the provider.\ 
The utility for user $i$ from selecting a VM instance type $t$\
with availability $Av_t$ is calculated as:
\begin{equation}
	U_{i}(VM_t)=WTP_i \cdot f_i(Av_t)-cost(VM_t)
	\label{eq:utility}
\end{equation}
\vspace{\eqbottommargin}

\begin{figure}
\vspace{\figtopmargin}

    \begin{minipage}{0.48\textwidth}
        \centering
        \includegraphics[width=\textwidth]{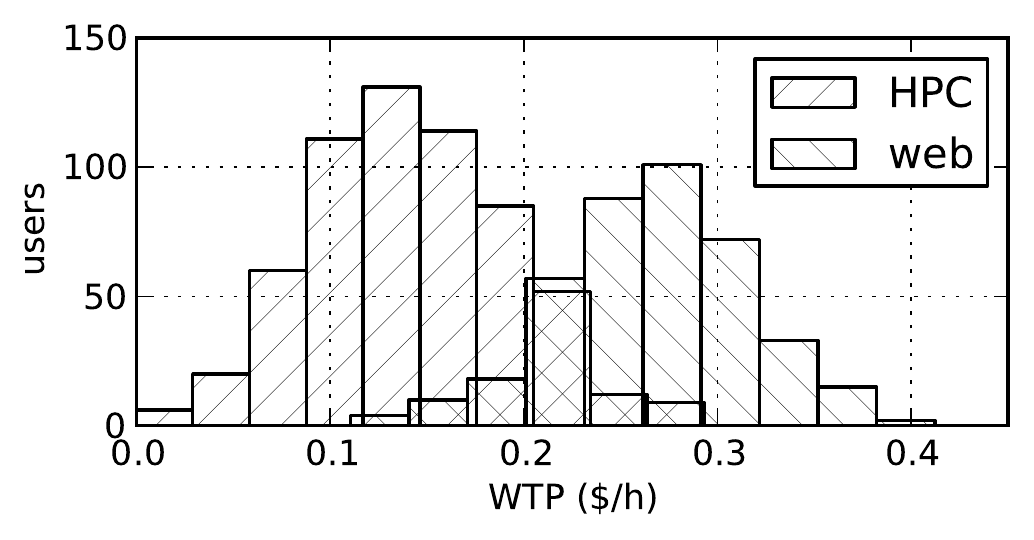}
        \caption{WTP histogram}
        \label{fig:wtp}
    \end{minipage}
    \begin{minipage}{0.48\textwidth}
        \centering
        \includegraphics[width=\textwidth]{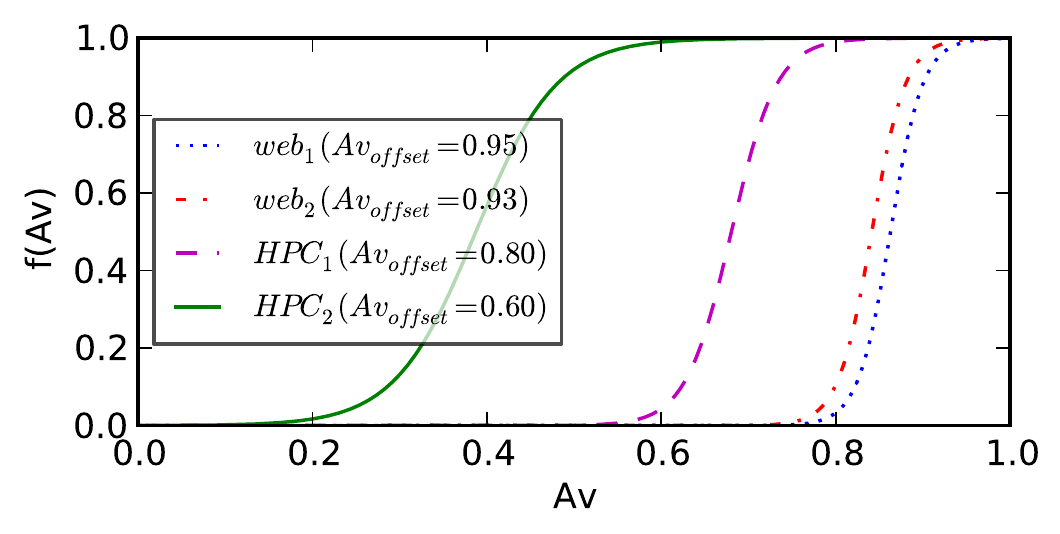}
        \caption{SLA satisfaction function}
        \label{fig:satisfaction}
    \end{minipage}

\vspace{\figbottommargin}
\end{figure}

where 
$f_i(Av_t)$ is the user's satisfaction\
with the offered \gls{vm}'s availability.\
We model it using a mapping function $f_i: [0, 1] \to [0, 1]$,\ 
extended from \cite{jrad2013} with variable slopes:
\
\begin{equation}
f_{i}(Av_t)=\frac{\gamma}{\gamma+\beta e^{Av_{offset_i}^2 \alpha(Av_{offset_i}-Av_t)}}
\label{eq:fitting1}
\end{equation}
\
where $Av_{offset_i}$ is the required availability specific to each user,\
which we select based on the exponential models of Wikipedia and G5k data\
presented earlier.\
$\alpha$, $\beta$ and $\gamma$ are positive constants common for all users,\
which we set to 60, 0.01 and 0.99 respectively.\
These values were chosen for a satisfaction of close to\
1 for the desired availability value\
and a steep descent towards 0 for lower values,\
similar to earlier applications of this satisfaction function \cite{jrad2013}.\
The slope of the function also depends on $Av_{offset_i}$, to model that\
users who require a lower availability have a wider range of acceptable values.\
The mapping function is visualised in Fig.~\ref{fig:satisfaction}\
for a small sample of two \gls{hpc} and two web users.\
We can see that for the two web users,\
the slope is almost the same and very steep\
(at 0.93 their satisfaction is close to 1 and at 0.8 it is almost 0).\
The $HPC_1$ user is similar to the web users,\
only with lower availability requirements and a slightly wider slope.\
The $HPC_2$ user has low requirements and a wide slope\
from an availability of 0.2 to 0.6.\
Later in our evaluation, we generate 1000 users,\
each with different mapping functions, distributed\
according to the user requirements model.\

\
User $i$ chooses a VM instance type $VM_{selected}$\
offering the best utility value:\
\
\begin{equation}
U_{i}(VM_{selected})=\max_{\forall t}\ U_{i}(VM_t)
\end{equation}
\
unless all types result in a negative utility,\
in which case the user selects none.\
Additionally, we model search difficulty by defining $P_{stop}$, a probability\
that a user will give up the search after an \gls{sla} has been examined.\
We model this probability as increasing after every new \gls{sla} check,\
by having $P_{stop_j} = j \cdot check\_cost$,\
where $j$ is the number of checks already performed\
and $check\_cost$ is a constant parameter standing for the probability\
of stopping after the first check.\
Based on every user's requirements and the \gls{sla} offering, $min\_checks$\
is the minimum number of\
checks necessary to reach a \gls{vm} type that yields a positive $U_i(VM)$.\
We define $P_{quit}$ to be the total probability that\
a user will quit the search before reaching a positive-utility \gls{sla}.
By applying the chain probability rule,\
we can calculate $P_{quit}$ as:\
\
\begin{equation}\label{eq:pquit}
P_{quit}=\mathlarger{\sum}_{j=1}^{min\_checks - 1} P_{stop_j} \prod_{k=1}^{j-1}(1 - P_{stop_k})
\end{equation}
\
The outer sum is the joint distribution of all possible\
stop events that may occur for a user\
and the inner product stands for all event outcomes\
when searching continued\
until the $j$-th event was realised as stopping.\ 
We use this expression in the evaluation as a measure of difficulty for\
users to find a matching \gls{sla}.

\vspace{\sectionendmargin}

\section{Evaluation}
\label{ch:vmpricing:sec:evaluation}
\vspace{\sectiontitlemargin}
In this section we describe the simulation of the proposed\
progressive \gls{sla} specification\
using user models based on real data traces\
and analyse the results.


\vspace{\subsectionendmargin}

\subsection{Simulation Environment}
\vspace{\subsectiontitlemargin}

The simulation parameters are summarised in Table~\ref{ch:vmpricing:tab:simulation}.\
The first step of the simulation is to generate\
a population of web and \gls{hpc} users based on the\
requirement models derived from\
the Wikipedia and Grid'5000 datasets, respectively.\
We simulated 1000 users to represent a population with enough variety to\
explore different \gls{wtp} and availability requirements.\
We assume the ratio between web and \gls{hpc} users of 1 : 1.5,\
based on an anlysis of a real system performed in \cite{liu_renewable_2012}.\
We determine each user's \gls{wtp} from the desired availability\
with Gaussian noise, as already explained in the previous section.\
The \gls{sla} offering was derived from the migration and peak pauser\
scheduler using the probabilistic modelling technique\
(Section~\ref{sec:pricing_model}).\
For the examination of \gls{sla} selection from the user's perspective,\
the \VMPricingslanum{} \gls{sla}s\
we already defined in Section~\ref{sec:8slas} were used.\
To examine the cloud provider's perspective,\
we evaluated \VMPricingslanummult{} \gls{sla}s,\
doing 100 simulation runs per offering\
to calculate the most likely optimal number of \gls{sla}s.\
A $check\_cost$ of 0.015 is selected to initially start with a low chance of\
the user quitting and then subsequently increase it for every \gls{sla} check\
per Eq.~\ref{eq:pquit}.\
The same $\alpha,\ \beta,\ \gamma$ values that we already explained\
in the previous section were set that result in\
a utility of 1 for the required availability\
and a gradual decline towards a utility of 0 for lower availabilities.\
The core of the simulation is to determine each user's\
\gls{sla} selection (if any) based\
on the utility model (Section~\ref{sec:user_model}).\

\begin{table}
\vspace{\tabletopmargin}
\centering
\caption{Simulation settings}
\label{ch:vmpricing:tab:simulation}
\vspace{\tablecaptionmargin}

\begin{tabular}{ l | c c c c c c c c c c}
\hline
\textbf{Parameter} & users & web : \gls{hpc} & \#SLAs \ 
& runs & $check\_cost$ & $\alpha$ & $\beta$ & $\gamma$\\
\hline
\textbf{Value} & 1000 & 1 : 1.5 & \VMPricingslanummult{} \ 
& 100 & 0.015 & 60 & 0.01 & 0.99\\
\hline
\end{tabular}

\vspace{\tablebottommargin}
\end{table}

\vspace{\subsectionendmargin}

\subsection{User Benefits}
\vspace{\subsectiontitlemargin}

\begin{figure}
\vspace{\figtopmargin}
\centering
\begin{minipage}{0.48\textwidth}
    \centering
    \includegraphics[width=\textwidth]{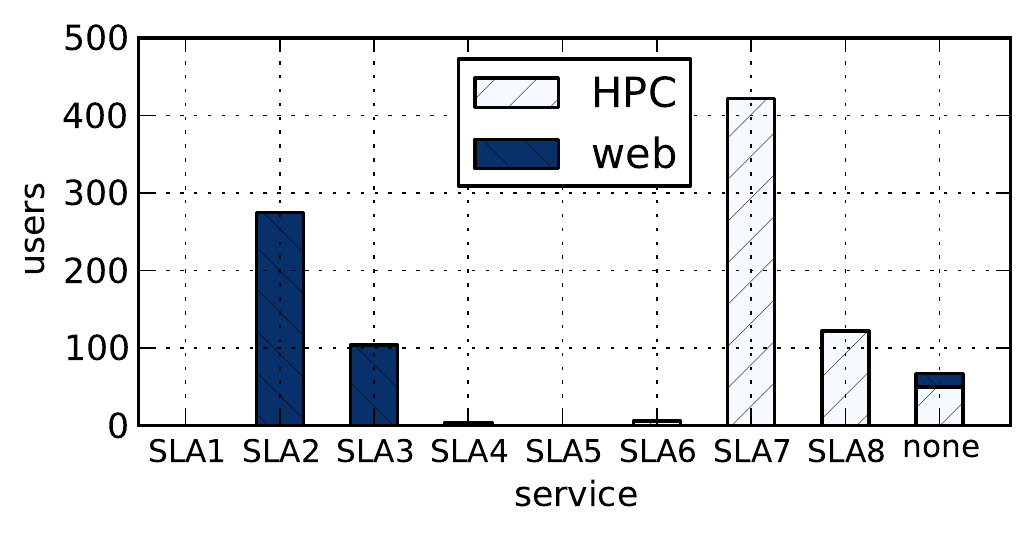}
    \vspace{\figcaptionmargin}
    \caption{Simulated service selection}
    \label{fig:category_distribution}
\end{minipage}
\begin{minipage}{0.48\textwidth}
    \centering
    \includegraphics[width=\textwidth]{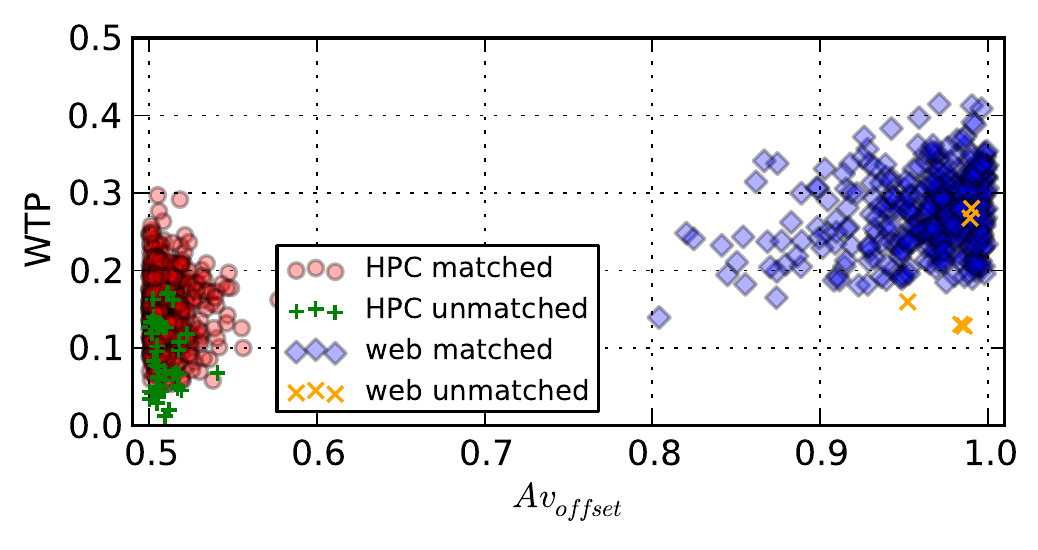}
    \vspace{\figcaptionmargin}
    \caption{Matched and unmatched users}
    \label{fig:unmatched_distribution}
\end{minipage}
\vspace{\figbottommargin}
\end{figure}

Simulation results showing the distribution of users among the\
\VMPricingslanum{} offered \gls{sla}s from Section~\ref{sec:8slas}\
are presented in Fig.~\ref{fig:category_distribution}.\
Different colours are used for web and HPC users types.\
It can be seen that most of the users successfully found a service\
that matches their requirements, with less than 5\% of unmatched requests.\
The majority of HPC users are distributed between SLA 7 and 8 offering\
42\% and 33\% availability, respectively.\
The majority of web users selected SLA 2, the migration scheduler \gls{tc},\
due to its high availability comparable to a full availability\
service, but a more affordable price due to the energy cost savings.\
26\% of web users opted for SLA 3, the\
peak pauser instance which still offers a high availability (87.5\%),\
but at almost half the price of SLA 2.

The distribution of unmatched users\
who did not select any of the offered services\
(where utility was negative for all \gls{sla}s)\
is shown alongside the matched users\
in Fig.~\ref{fig:unmatched_distribution},\
showing their $Av_{offset}$ and \gls{wtp} values.\
We can see that unmatched users have low \gls{wtp} values,\
the cause of them not being able to find a suitable\
service option.

\vspace{\subsectionendmargin}

\subsection{Cloud Provider Benefits}
\begin{figure}
\vspace{\figtopmargin}
\centering
\begin{minipage}{0.48\textwidth}
    \centering
    \includegraphics[width=0.9\textwidth]{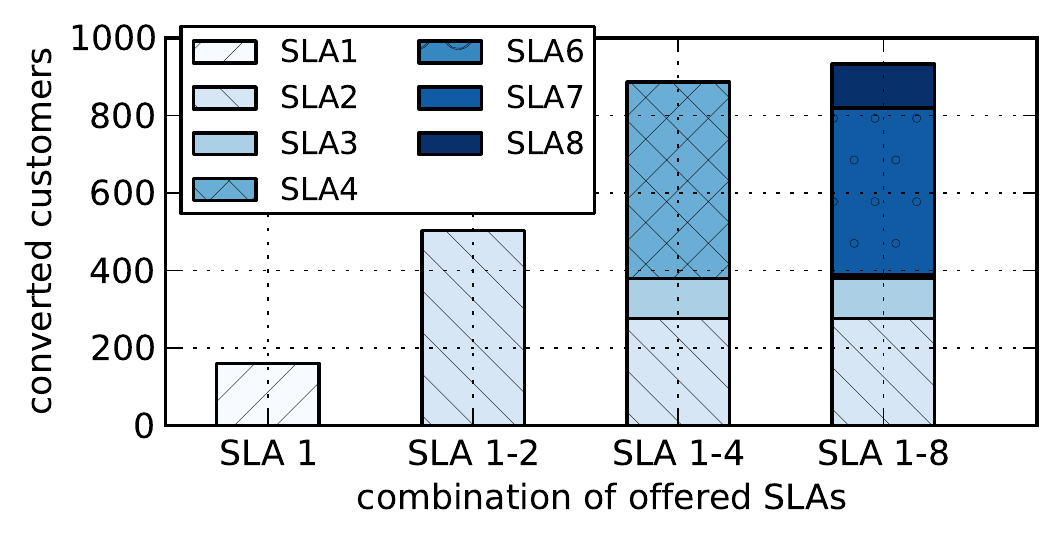}
    \caption{Matched users per SLA combination based on WTP.}
    \label{fig:users_count}
\end{minipage}
\begin{minipage}{0.48\textwidth}
    \centering
    \includegraphics[width=\textwidth]{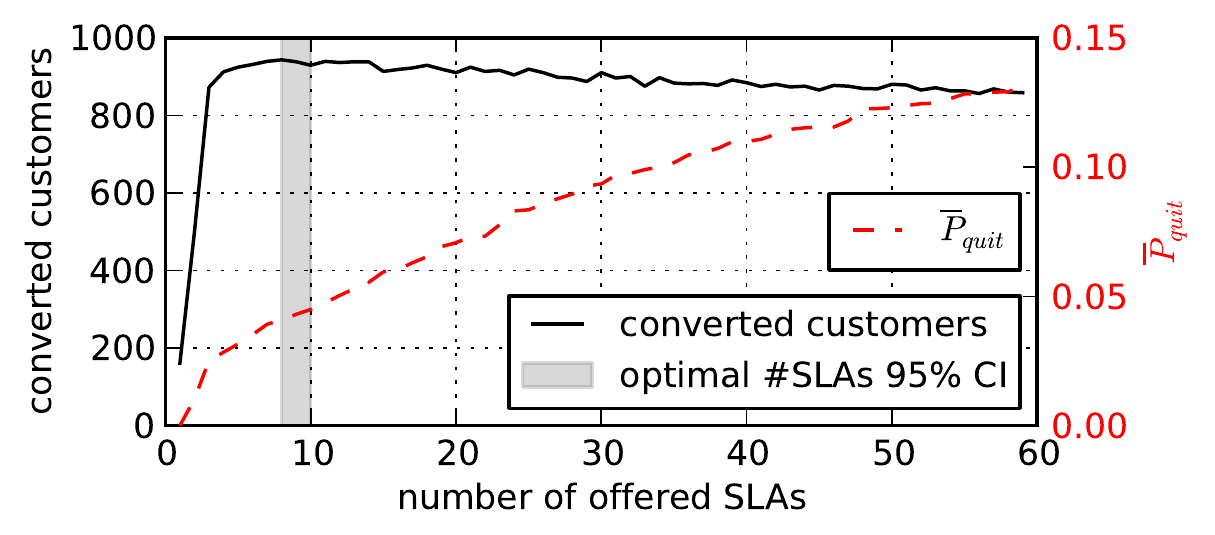}
    \caption{Users and $\overline{P}_{quit}$ per SLA number}
    \label{fig:sla_count}
\end{minipage}
\vspace{\figbottommargin}
\end{figure}

\vspace{\subsectiontitlemargin}
Customer conversion means the number of users\
who looked at the \gls{sla} offering and\
found an \gls{sla} that matches their needs.\
This metric is an indicator of the provider's economic success.\ 
\
To compare the multiple treatment category system with the traditional\
way of only having a full availability option,\
we simulated different \gls{sla} offerings.\
\
Fig.~\ref{fig:users_count} shows customer conversion\ 
with colour indicating the selection distribution\
for different offering combinations\
of the \VMPricingslanum{} previously examined \gls{sla}s.\
Customer conversion growth can be seen with more service types,\
due to users having\
a higher chance of finding a category that matches\
their requirements.\
\
With \gls{sla}s 1--2 offered, only \gls{sla} 2 was selected,\
as it still offers a high-enough availability to satisfy user requirements\
and the price is lower than in \gls{sla} 1.\
As we widen the offering, more \gls{sla}s get selected, but the majority of\
users choose among two \gls{sla}s that best suit\
the two user types that we modelled.\
Still, a small number of users select other \gls{sla}s\
(\gls{sla} 3 and, if offered, \gls{sla} 8)\
which better suit their needs.\ 
\gls{sla} 5 is never selected\
due to user requirements\
and in real clouds such \gls{sla}s should be removed\
to simplify selection.

The introduced service types can be managed in a more energy efficient\
manner. The average energy savings weighted based on the lease time\
per \gls{vm} for the \VMPricingslasall{} offering,\
compared to the current 99.95\% availability Amazon instances represented by\
\gls{sla} 1, are \VMPricingEnSavings{}.\
Full annual lease time was assumed for web users\
(as web applications are typically running all the time)\
and was varied based on job runtime and frequency for \gls{hpc} users\
(we assume that a \gls{vm} is provisioned just to perform the submitted job).\
This shows that more energy efficient management is possible if users\
declare the \gls{qos} levels they require through \gls{sla} selection.
For the \VMPricingslasall{} scenario, where \VMPricingCustomerIncrease{} more users\
can be converted and the annual lease times explained above,\
a \VMPricingRevenueIncrease{} revenue increase is calculated\
from the service component of the selected \gls{vm}s.\
Exact numbers\ 
depend on the user type ratio\
that will vary between cloud providers.\

To find the optimal number of offered \gls{sla}s,\
we performed a simulation where we explore customer conversion\
for a higher number of \gls{sla}s.\
The extra \gls{sla}s were generated for the peak pauser scheduler,\
which allows for arbitrary control of \gls{vm} availability and price.\
The peak pauser \gls{sla}s were uniformly interpolated\
between full and no availability\
to avoid duplicates.\
Fig.~\ref{fig:sla_count} shows how the number of offered \gls{sla}s affects\
the user conversion count and $\overline{P}_{quit}$,\
the mean $P_{quit}$ value over all the users (including unmatched ones).\
After an initial linear growth, we can see that the number of users begins\
to stagnate and slowly decrease.\
Once a sufficient offering to satisfy the majority of users is achieved,\
adding extra \gls{sla} options only increases search difficulty.\
This is seen from the steadily increasing $\overline{P}_{quit}$,\
the probability that a user will quit the search before finding\
a positive-utility \gls{sla}.\
For our scenario, the optimal number of converted customers is achieved between\
6 and 14 \gls{sla}s, depending on the $P_{quit}$ random variable realisations.\
By applying the bootstrap confidence interval method,\ 
we calculate the 95\% confidence interval (CI) for the optimal number of \gls{sla}s\
to be between 8 and 10.


%
%

\vspace{\sectionendmargin}

\section{Summary}
\label{ch:vmpricing:sec:summary}
\vspace{\sectiontitlemargin}
We presented a novel progressive \gls{sla} specification\
suitable for energy-aware cloud management.\ 
We obtained cloud management traces from two schedulers\ 
optimised for real-time electricity prices\
and temperature-dependent cooling efficiency.\
The \gls{sla}s are derived using\
a method for a posteriori probabilistic modelling of cloud management data\
to estimate upper bounds for \gls{vm} availability, energy savings\
and the resulting \gls{vm} prices.\
The \gls{sla} specification is evaluated\
in a utility-based user \gls{sla} selection simulation\
using realistic workload traces from Wikipedia and Grid'5000.\
Results show mean energy savings per \gls{vm} of up to \VMPricingEnSavings{} due to\
applying more aggressive energy preservation\
actions on users with lower \gls{qos} requirements.\
Furthermore, a wider spectrum of user types with requirements\
not matched by the traditional high availability \gls{vm}s can be reached,\
increasing customer conversion.\




\chapter{Performance-Based Pricing in Multi-Core Geo-Distributed \\Cloud Computing}
\label{ch:multicore}
New pricing policies are emerging where cloud providers charge resource
provisioning based on the allocated CPU frequencies.\
As a result, resources are offered to users as combinations of different
performance levels and prices which can be configured at runtime.\
With such new pricing schemes\ 
and the increasing energy costs in data centres, balancing energy savings
with performance and revenue losses
is a challenging problem for cloud providers.\
CPU frequency scaling can be used to reduce power dissipation,
but also impacts \gls{vm} performance and therefore revenue.\
In this chapter, we firstly propose a non-linear power model
that estimates power dissipation of a multi-core \gls{pm} and
secondly a pricing model that adjusts the pricing based on the \gls{vm}'s
CPU-boundedness characteristics.\
Finally, we present a cloud controller that uses these models
to allocate \glspl{vm}\
and scale CPU frequencies of the \glspl{pm}\ 
to achieve energy cost savings that exceed service revenue losses.
We evaluate the proposed approach using simulations with realistic
\gls{vm} workloads, electricity price and temperature traces and estimate
energy savings of up to \Multicoreensavingsmax{}.
This approach is a less invasive approach for reducing power consumption than
\gls{vm} pausing presented in Chapter~\ref{ch:volatility}. Frequency scaling is
orthogonal to the \gls{vm} migration controller
from Chapter~\ref{ch:gascheduler}, meaning that both approaches could be applied
-- at the \gls{pm} and \gls{vm} level, respectively.




We structure the chapter by first introducing in more details the challenges
of frequency scaling in multi-core computers and the inefficiencies of existing
control approaches in Section~\ref{ch:multicore:sec:challenges}. This serves as
a motivation for our detailed power model for multi-core Intel and ARM CPU
architectures in Section~\ref{ch:multicore:sec:power_model}. We then highlight
the economical aspects of frequency scaling in cloud computing by explaining
emerging \gls{vm} pricing schemes and propose
our own perceived-performance pricing scheme
in Section~\ref{ch:multicore:sec:pricing_model}. We combine the power
and pricing models to devise a cloud controller that geographically distributes
\gls{vm}s and applies frequency scaling on \gls{pm}s
in Section~\ref{ch:multicore:sec:scheduler}.
In Section~\ref{ch:multicore:sec:evaluation}, we present the evaluation
methodology and comment on the most significant obtained results.



\section{Challenges of Multi-Core Frequency Scaling}
\label{ch:multicore:sec:challenges}

In this section we introduce the main challenges inherent to multi-core
frequency scaling of \gls{pm}s with multiple CPU cores and motivate our power
model and subsequently frequency control approach. We show that neither
operating system CPU governors, nor traditional race-to-idle approaches
provide optimal energy efficiency because of inaccurate decision making.



\subsection{Limitations of Current Frequency Governors}
The currently used power management system in Linux operating systems is handled by the \textit{frequency governors},
which alter the CPU clock frequency based on a predefined policy.
Even though the intention is to reduce the clock frequency when performance is not needed,
the approach suffers from limitations.

\begin{figure}
\centering
\includegraphics[width=7cm]{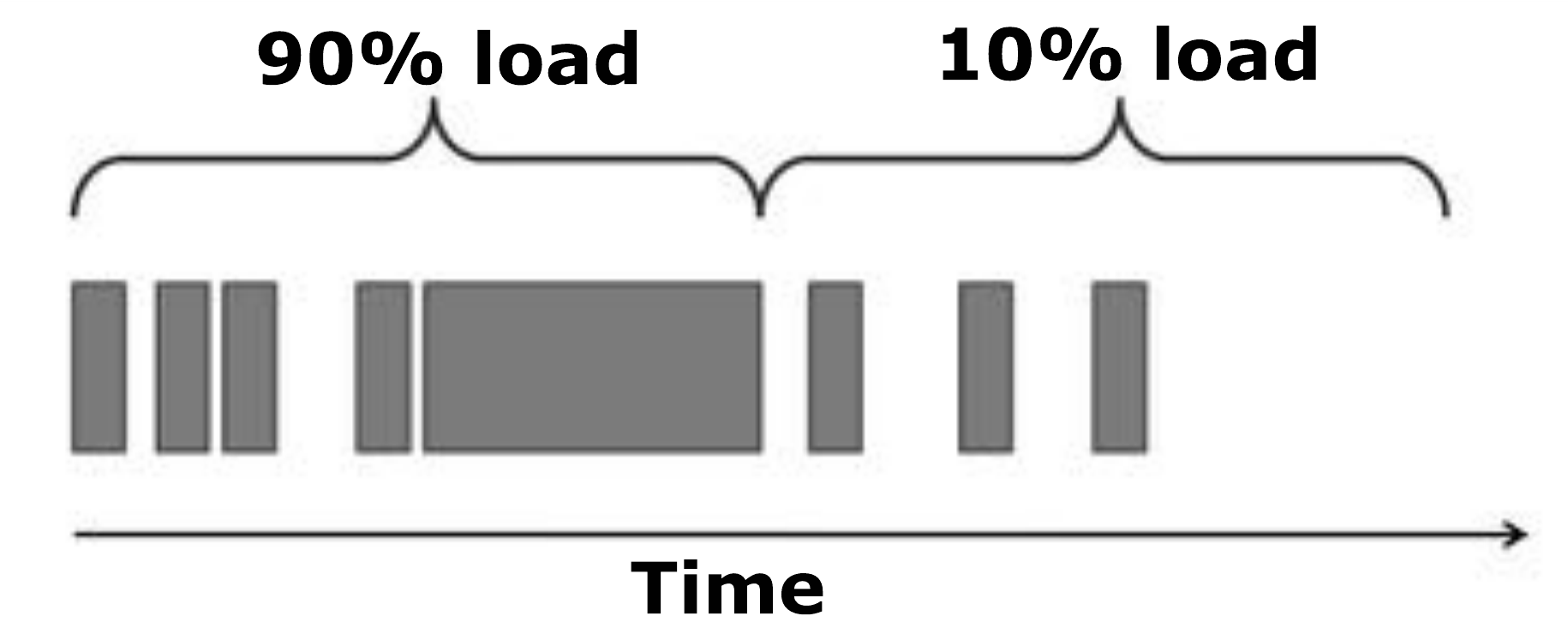}
\caption{Load calculated as a ratio between active and idle CPU for a defined window -- dark squares indicate an active CPU.}
\label{fig:loadwindow}
\end{figure}

The metric used to determine the clock frequency in the governors is the system \textit{workload}.
The workload is expressed as a ratio between an active CPU and an idle CPU over a given time window,
which is illustrated in Fig.~\ref{fig:loadwindow} as two time windows: one with 90\% load and one with 10\% load.

Workload, however, does not represent the performance, or the ``real work'' done
by an application, but mainly the activity level of the CPU.
This means that as long as the CPU is loaded, the performance requirement
is recognised as insufficient and the clock frequency is increased
to the maximum even though the \textit{actual} performance is sufficient
on a moderate clock frequency.

\subsection{Energy Inefficient Execution}

Using workload as the metric for power management decisions often results
in race-to-idle scenarios \cite{sasaki2013model,seeker2014energy},
in which the workload is executed as fast as possible in order to obtain
an idle system. This execution principle was considered
an energy efficient method of executing workload in previous generation
single-core microprocessors, because the minimisation of the execution time
caused minimal energy consumption.

\begin{figure}
\centering
\includegraphics[width=8cm]{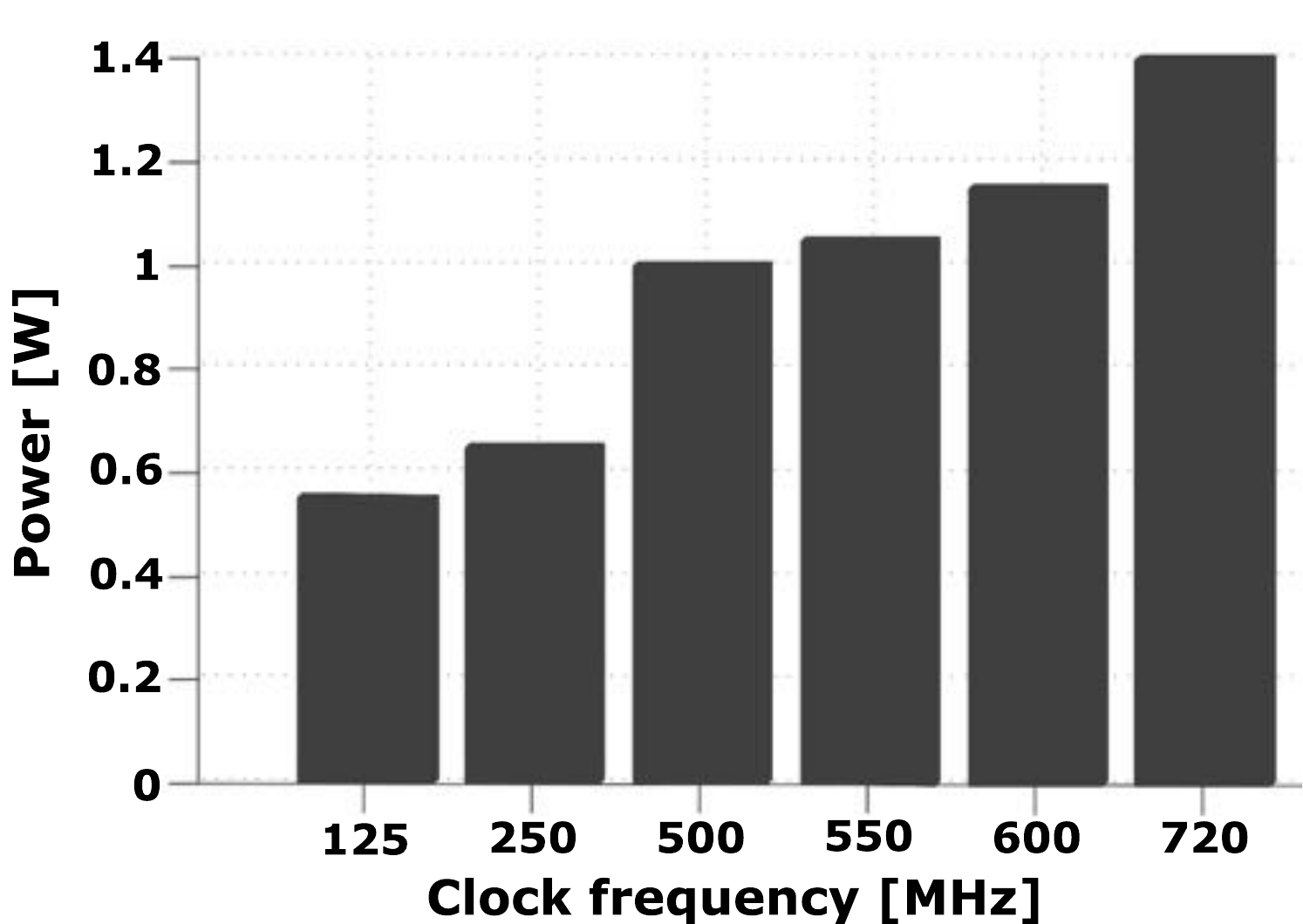}
\caption{Power dissipation for a Cortex-A8 CPU using different clock frequencies.}
\label{fig:cortexa8}
\end{figure}

This is demonstrated in Fig.~\ref{fig:cortexa8}, which shows the total power
dissipation for a single-core ARM Cortex-A8 processor using different clock
frequencies. As seen in the figure, the highest clock frequency (720 MHz)
results in roughly 1.4W of power dissipation.
When scaling down the frequency roughly 3x (250 MHz), the power dissipation
is only reduced by 2x (0.7W), which means that the total energy consumption
may be lower when executing at a higher clock frequency.

\begin{figure}
\centering
\includegraphics[width=0.7\columnwidth]{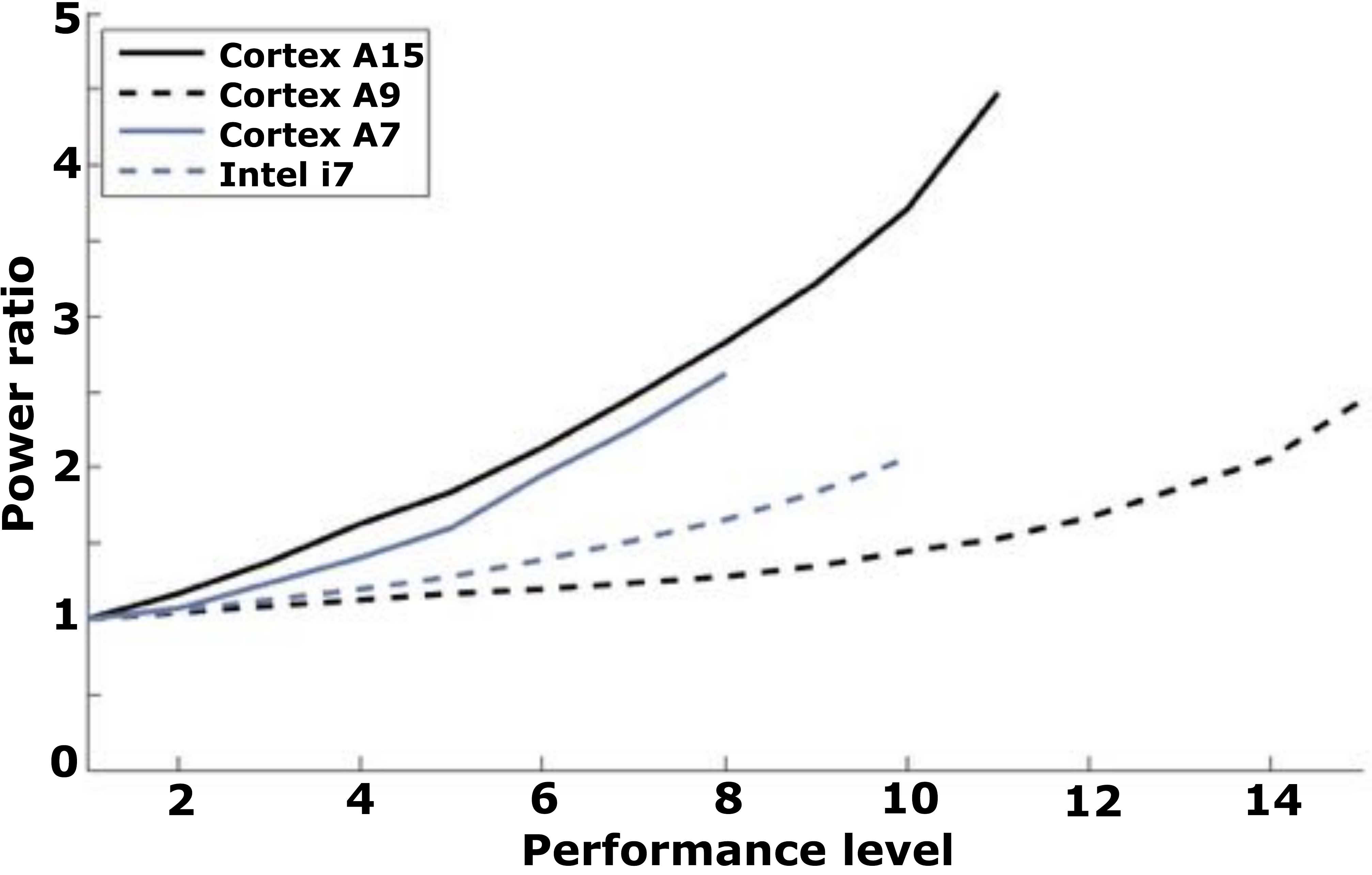}
\caption{Performance to power ratio for different platforms.}
\label{fig:profile}
\end{figure}

However, using more recent microprocessors with higher clock frequencies
and multiple cores, the power dissipation has increased exponentially --
this has reduced the energy efficiency of the race-to-idle principle
because the cost in power is greater than the savings in
execution time~\cite{holmbacka2014energy,holmbacka2015energy,seeker2014energy}.
Fig.~\ref{fig:profile} shows the relative performance-to-power ratio of
four different modern platforms. All of the four platforms show
an exponential profile, which means that the power dissipation required
to operate on the highest clock frequencies is higher
than the relative performance gain of the platform.
The race-to-idle principle should therefore not be used for energy efficient
execution.

\subsection{Energy Efficient Execution}
In order to not race-to-idle, a \textit{performance driven execution}
should be used instead of
a workload driven
execution~\cite{holmbacka2014energy,holmbacka2015energy,seeker2014energy}.
By monitoring the actual performance of an application and adjusting
the clock frequency accordingly,
the energy efficiency can be improved. 

Benchmarks were executed on a quad-core ARM platform with an Exynos 5410 SoC
using the default \textit{ondemand} governor and the modified
performance driven power manager.
By using the ondemand governor, the decoder decodes the frames
as quickly as possible
since the decoding task increases the workload,
and the clock frequency is consequently increased.
As the frame buffer is filled, the decoder is idle until the frame buffer
is emptied by the video display.
By instead decoding at the same frame rate as the video display
is using (25 FPS), the clock frequency can be reduced
to an intermediate clock frequency for the whole execution while still providing
the required video quality.

\begin{figure}
\centering
\includegraphics[width=0.7\columnwidth]{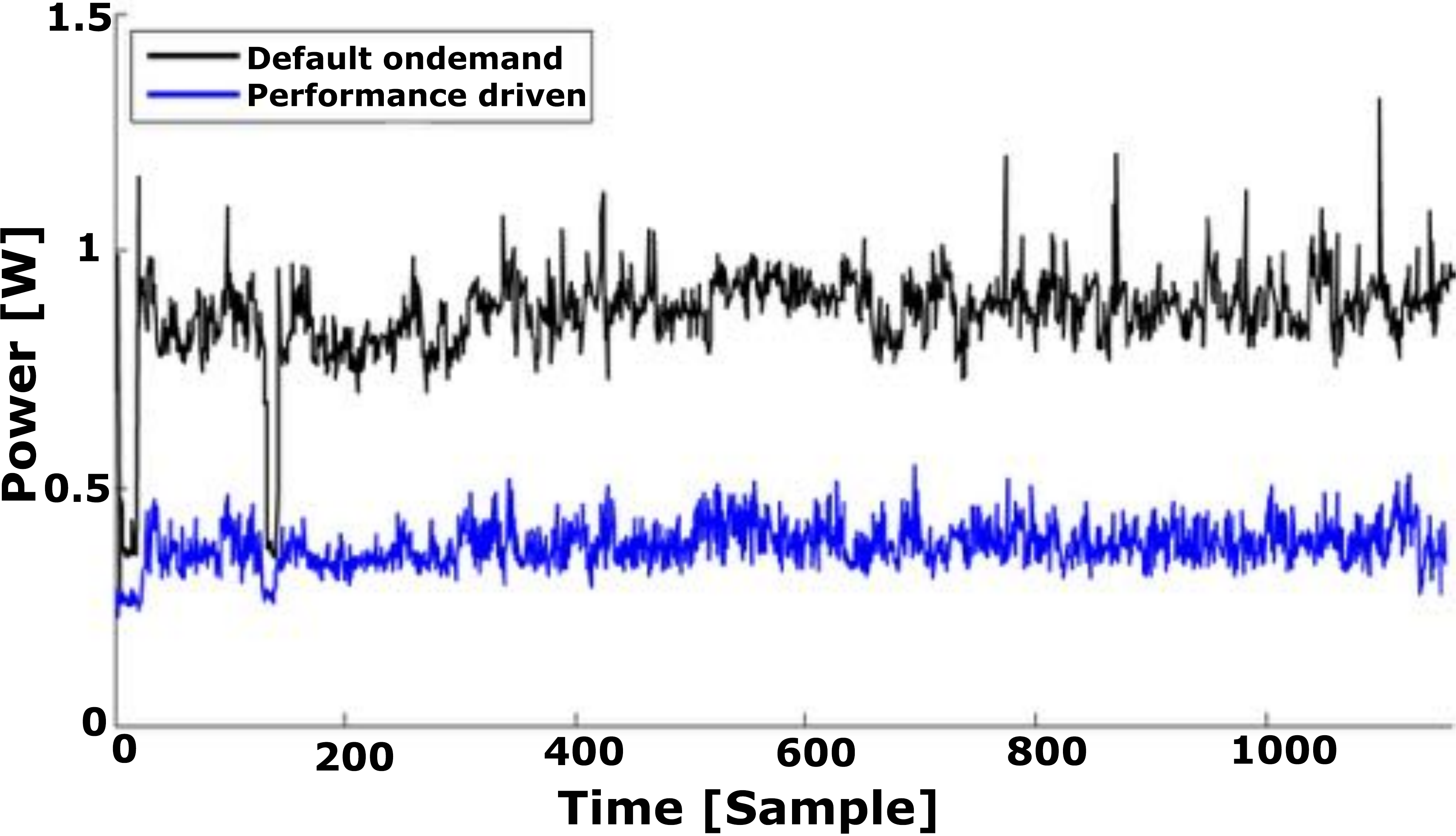}
\caption{Experiments with video decoding using the ondemand governor and performance driven clock frequency scaling on an Exynos 5410 platform \cite{holmbacka2015energy}.}
\label{fig:mplayer}
\end{figure}

The power dissipation was measured by internal power sensors
for both power managers and the result is shown in Fig.~\ref{fig:mplayer}
(more details can be found in \cite{holmbacka2015energy}).
The power dissipation of executing the video decoder using the ondemand governor
is shown as the upper, black line and the performance driven power manager
is shown as the lower, blue line.
Since the standard ondemand governor increases the clock frequency
while workload is present, most of the execution demands the highest
clock frequencies, which causes excess power dissipation
as seen in Fig.~\ref{fig:mplayer} (upper, black line).
By matching the decoding frame rate to the output frame rate (of 25 \gls{fps}),
the lower clock frequencies are providing enough performance to decode the
frames at the intended phase of 25 \gls{fps},
and the power is significantly reduced as seen in Fig.~\ref{fig:mplayer}
(lower, blue line).

We intend to bring the performance driven clock frequency scaling using
multi-core hardware from CPU level to the cloud level consisting
of many parallel machines.
The difference is a much more diverse execution platform with additional
parameters such as \gls{vm} migration, network I/O
and variable electrical cost models.
A power model capable of reflecting such details is needed
to get an accurate cost model of the cloud system.


\section{Multi-Core Power Model}
\label{ch:multicore:sec:power_model}


In this section we show how we modelled the behaviour of multi-core CPU
power dissipation, accounting for both the CPU frequency and the number of
active cores for generic multi-core systems.
Such a power model allows us to determine what performance level to execute at,
depending on the performance requirements, which is one of the key parts
of our cloud controller.

As the power characteristics of modern multi-core CPUs are
highly non-linear~\cite{holmbacka2014energy}, a non-linear model should be
created to accommodate as accurately as possible to the real-world power
dissipation. The model should also not be computationally
heavy to introduce unnecessary overhead.

\subsection{ARM and Intel Architectures in the Cloud}
\label{sec:mont_blanc}

Aside from the popular Intel architecture used as a typical server platform,
the architecture based on ARM processors made popular through wide usage
in smartphones is currently also being investigated for use in servers.
ARM processors are much more energy efficient than Intel processors,
though their maximum CPU frequency capacity is lower, potentially increasing
the necessary number of servers and therefore the communication overhead.
The Mont Blanc EU project~\cite{rajovic2013supercomputing,francesquini2015benchmark}
was devoted to determine whether this approach is valid for large scale
cloud platforms. Companies like Calxeda already ship ARM based server machines
and Lenovo is pushing its NextScale \cite{shah2015platform} platform
with the motivation to increase the performance-per-watt ratio by focusing on
possibly more energy efficient architectures.

We therefore included the ARM architecture in our evaluation
as a viable candidate for investigating the effects of performance-based
frequency scaling in order to provide a comparison to the Intel architecture.

\subsection{Power and Energy Consumption Model}
\label{ch:multicore:sec:power_model_arm}

\begin{figure}
\centering
\includegraphics[width=0.7\columnwidth]{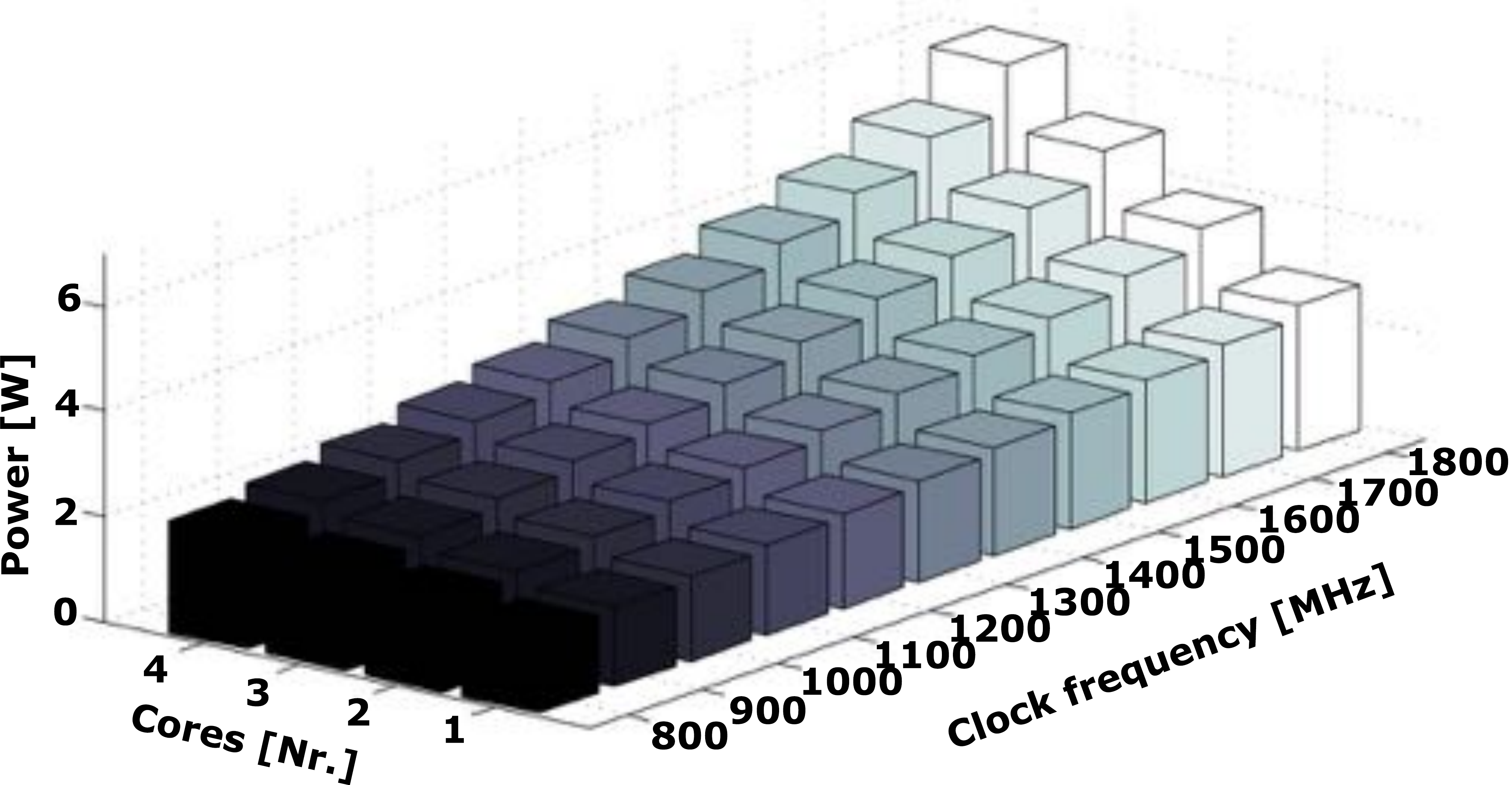}
\caption{A top-down model of a quad-core ARM Cortex-A15 CPU.}
\label{fig:odroidpower}
\end{figure}

For our cloud controller, we created an ARM and an Intel power
consumption model.\ 
The ARM model was created by reading internal power sensors on an
Exynos 5410 board,
and the Intel model by using an external measurement device connected
directly to the ATX socket on the motherboard.
Both models were used in the evaluation, but we describe
our modelling procedure with a higher focus on the ARM model for brevity.
The same procedure we describe in this section is also applicable
to the Intel model (or any other derived model).

We used a similar methodology as the work in~\cite{holmbacka2014energy}
to derive the power model, where the model was created by stressing the system
to full capacity using all combinations of the clock frequencies
and all number of cores. Our power measurements for stressing the ARM board is
shown in Fig.~\ref{fig:odroidpower}.
The figure shows power dissipation of the CPU during full load
using different configurations of the clock frequency and the number of cores.
Naturally, more cores and a higher clock
frequency cause a higher power dissipation.

\begin{figure}
\centering
\includegraphics[width=0.7\columnwidth]{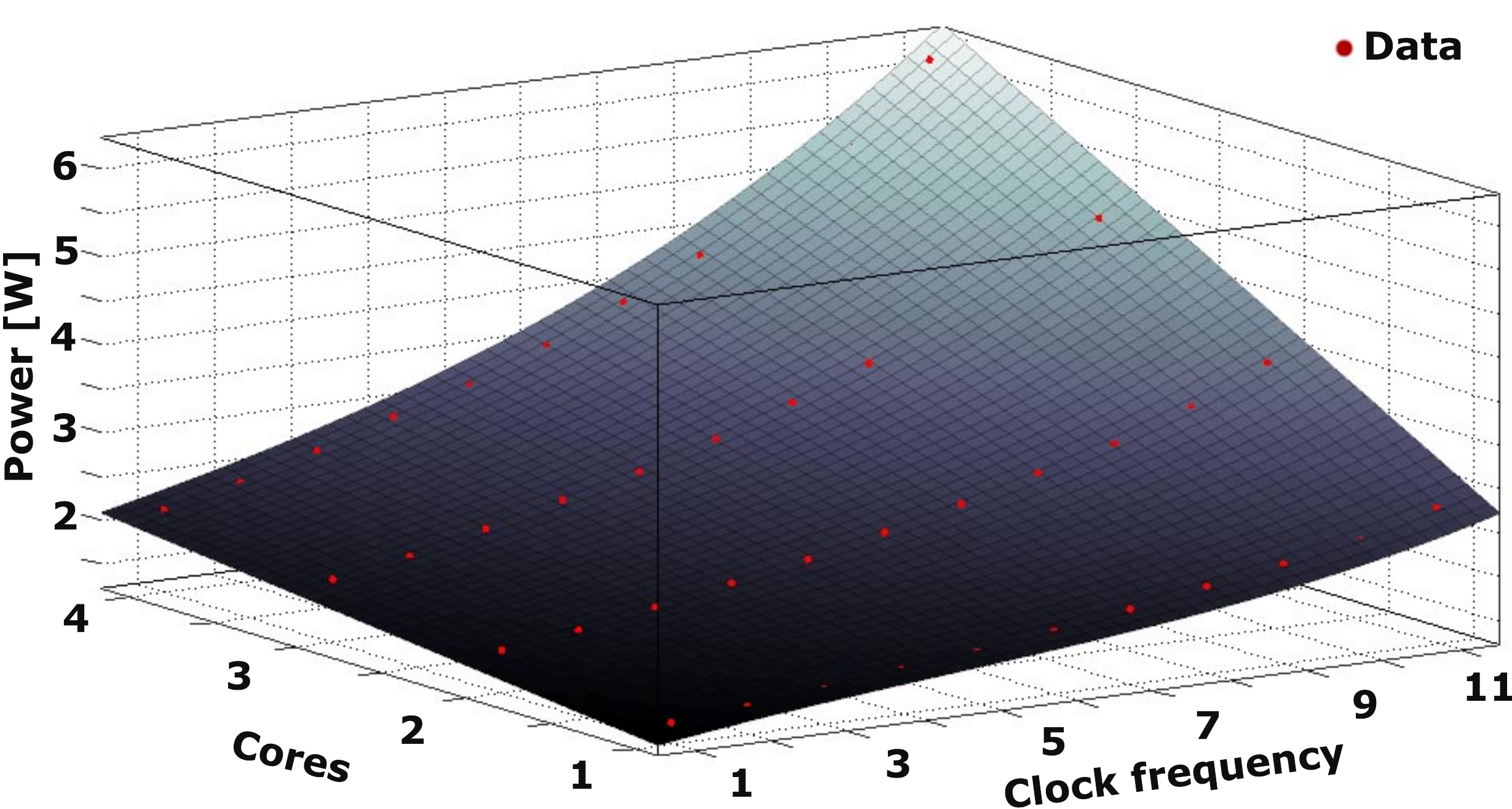}
\caption{A mathematical representation of the power values in Fig.~\ref{fig:odroidpower}
obtained using surface fitting methods.}
\label{fig:planefitting}
\end{figure}

The power measurements were then used as basis for a two dimensional plane
fitting algorithm,
in order to build a mathematical expression of the multi-core system and its
power dissipation.
We used a least-squares algorithm~\cite{lawson1987optimization} provided
in Matlab to obtain the polynomial of the form:
\begin{align}\label{eq:ppeak}
P_f(q,c)=p_{00}+p_{10}q+p_{01}c+p_{20}q^2+p_{11}qc+p_{30}q^3+p_{21}q^2c
\end{align}
which is a function of the clock frequency ($q$) and number of cores ($c$) used.
Fig.~\ref{fig:planefitting} shows the analytical representation of the power
dissipation and the data points obtained from Fig.~\ref{fig:odroidpower}
for the ARM platform. The clock frequency and the number of cores used are
represented as discrete steps from 1 to 11 and from 1 to 4 respectively.
The plane shown in Fig.~\ref{fig:planefitting} was fitted to the data values
using the obtained parameters shown in Table~\ref{tab:parameters}.
The same method was used for the Intel platform,
and other parameters were then obtained.


\begin{table}[!b]
 \caption{Power model coefficients.}
  \begin{center}
  \begin{tabular}{ | c | c | c | c | c | c | c|}
  \hline
  $p_{00}$  & $p_{01}$ & $p_{10}$ & $p_{11}$ & $p_{20}$ & $p_{21}$ & $p_{30}$   \\ \hline
  \footnotesize 1.318  &\footnotesize 0.03559 &\footnotesize 0.2243 &\footnotesize -0.00318 &\footnotesize 0.03137 &\footnotesize 0.000438 &\footnotesize 0.00711  \\ \hline
  \end{tabular}
\label{tab:parameters}
\end{center}
\end{table}

By comparing the model to the measurements, a maximum deviation of 18.7\% was
obtained and an average deviation of 6.4\% compared to the experimental data,
which we considered feasible for our cloud controller evaluation.

The idle power was modelled similarly to Eq.~\ref{eq:ppeak},
but without the core component $c$ as:
\begin{align}\label{eq:pfixed}
P_{idle}=p_{00}+p_{10}q+p_{20}q^2+p_{30}q^3
\end{align}
where $q$ is the clock frequency step and the p-parameters are identical to the
fitted parameters in Table~\ref{tab:parameters}.

\subsection{I/O-based Power Dissipation} \label{section:berserk}






The power dissipation of a \gls{pm} varies depending on the CPU utilisation
of the machine, which is dependent of the I/O usage of the workload.
A \emph{CPU-boundedness parameter ($\beta$)} is therefore used to model
the portion of the execution which consists of low intensity I/O operations.
The parameter may range between 0 and 1 to represent workloads of different
CPU-boundedness properties with values close to 0 corresponding to I/O-intensive
workloads and values close to 1 corresponding to CPU-bound
workloads~\cite{etinski2010optimizing}. 
The value of $\beta$ normalises the ratio between low intensity I/O bound
and high intensity CPU bound instructions in the workload.

The power model was therefore extended to account for the CPU utilisation based
on the \gls{vm} CPU-boundedness of the executing workload. To do so,
the CPU utilisation $u$ is expressed as:
 \begin{align}\label{eq:utilisation}
 u=\sum\limits_{core}\frac{\gamma_{core}}{cores_{active}}
 \end{align}
where $\gamma_{core}$ is a power ratio depending on the \gls{vm} CPU-boundedness
$\beta$ and $cores_{active}$ represents the number of the currently used cores
of the \gls{pm}.


Similarly to the basic power model (Eq.~\ref{eq:ppeak}), the power dissipation
was evaluated on the same platform with different $\beta$
in order to train the model.
The experiments were run using the \textit{Berserk} benchmark on a single active
core (to avoid any lack of scalability from using multiple cores and get the
pure effect of only the I/O).
The Berserk benchmark framework we also used in Chapter~\ref{ch:volatility} was
extended~\cite{_berserk_2015} for stressing the CPU cores
at various CPU-boundedness ratios $\beta$.
The workload itself is again a CPU-intensive task -- repeated recursive
calculation of Fibonacci numbers,
only parallelised to execute on all available CPU cores.
By passing different $\beta$ parameters to the benchmark, proportional ratios of
the workload are deferred to a remote server, making the work less or more
CPU-bound for monitoring purposes.
For example for a value of $\beta$ equal to 0, all the work is sent to and
received from a remote server via the network, making the task fully I/O-bound.
For a value of $\beta$ equal to 1, all the work is executed locally,
resulting in a CPU-bound task.

\begin{figure}
\centering
\includegraphics[width=0.7\columnwidth]{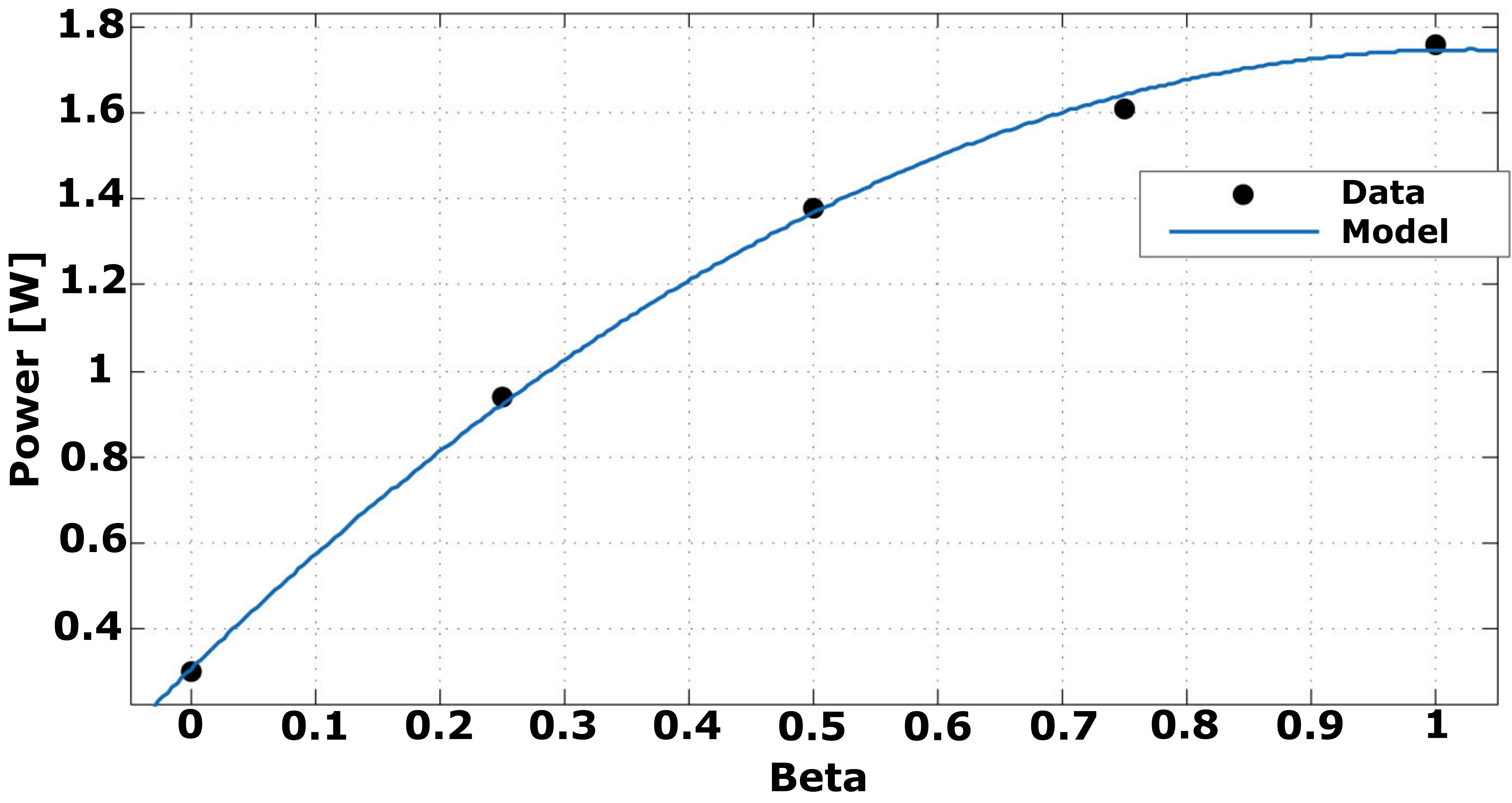}
\caption{Measurements of power dissipation based on I/O expressed as a ratio $\beta$. The model is expressed as a second degree polynomial.}
\label{fig:gammamodel}
\end{figure}

The explored I/O ratios ($\beta$ values) were selected in the range
[0.0, 0.25, 0.5, 0.75, 1.0] where 0.0 indicates total I/O blocking and 1.0
indicates no I/O (a fully CPU-intensive workload).
The CPU (in this case the ARM architecture) was executing at 1600 MHz for all
measurements (we show the behaviour of the model at different CPU frequencies in
the following experiment), and the measurement results are shown
in Fig.~\ref{fig:gammamodel} as the data values.

A one-dimensional curve fitting technique was used to model the power ratio
$\gamma_{core}$ as a second degree polynomial:
 \begin{align}
 \gamma_{core} = \frac{p_o \beta^2 + p_1 \beta + p_2}{P_{max}}
 \end{align}
where $\beta$ is the CPU-boundedness of the core, $P_{max}$ is the maximum power
dissipation of a core and the obtained function parameters are listed in
Table~\ref{tab:gammaparameters}. The curve in Fig.~\ref{fig:gammamodel} shows
the model of the $\gamma_{core}$ function.

\begin{table}[!b]
 \caption{Coefficients for the power ratio $\gamma_{core}$ model.}
  \begin{center}
  \begin{tabular}{ | c | c | c |}
  \hline
  $p_{0}$  & $p_{1}$ & $p_{2}$ \\ \hline
  \footnotesize -1.362   &\footnotesize 2.798  &\footnotesize 1.31  \\ \hline
  \end{tabular}
\label{tab:gammaparameters}
\end{center}
\end{table}

\begin{figure}
\centering
\includegraphics[width=0.7\columnwidth]{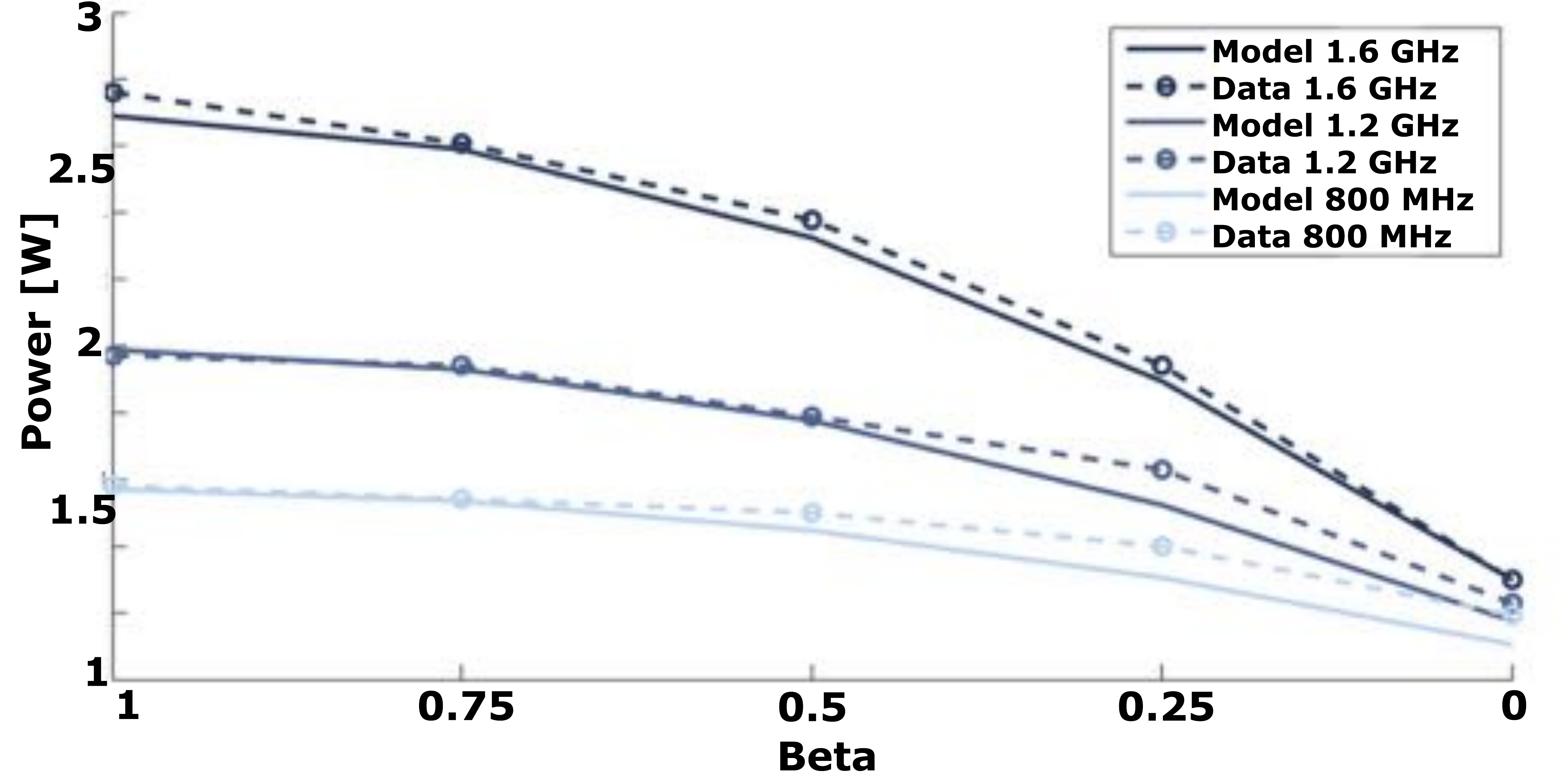}
\caption{Verification of the I/O-based power model from Fig.~\ref{fig:gammamodel}. Verification performed with three different frequency levels.}
\label{fig:gammaverify}
\end{figure}

The accuracy of the $\gamma_{cores}$ function at different CPU frequencies was
evaluated in another experiment. We executed the Berserk benchmark with the same
$\beta$ parameters at clock frequencies 1600MHz, 1200MHz and 800MHz.
Both the measurement data and the model for each experiment is shown in
Fig.~\ref{fig:gammaverify}, in which the circles are measurement points.
The maximum difference between the data and the model was 10.59\% and the
average difference was 4.32\%, which we considered as acceptable for our
cloud controller evaluation.

To combine all of the presented model components, the power dissipation $P$ of a \gls{pm} in the cloud system was modelled as:
\begin{align}\label{eq:putil}
P = P_{idle} + (P_f - P_{idle}) u
\end{align}
where $P_{idle}$ is the idle power expressed in Eq.~\ref{eq:pfixed},
$P_{f}$ is the active power expressed in Eq.~\ref{eq:ppeak}
and $u$ is the utilisation modelled in Eq.~\ref{eq:utilisation}
based on the I/O activity parameter $\beta$. The model to compute the
power dissipation of a \gls{pm} in the cloud system was based on the
approach proposed in \cite{liu_renewable_2012}, where \gls{pm}'s power
dissipation is linearly related with its CPU utilisation. The model was extended
to take into account the active power dissipation of the currently used cores of
the \gls{pm} at the operating frequency and the CPU-utilisation of the cores
based on the \gls{vm} CPU-boundedness $\beta$.

\subsection{Energy Calculation}
\label{ch:multicore:sec:energy_calculation}

With the multi-core, frequency-dependent power model, we now
present the details of electricity cost calculation based on geotemporal inputs.
This conversion of power dissipation to a monetary value is crucial for
comparing potential energy savings with revenue losses under performance-based
\gls{vm} pricing that we explore in the next section. This part of the model
was described in more detail
in Section~\ref{ch:gascheduler:sec:cost_components},
so we just summarise the important expressions necessary to understand
the integration with other model components from this chapter.

The power model so far was expressed for instantaneous values,
but to express the energy costs for the cloud provider we need to add a time
dimension to account for geotemporal inputs and actions like frequency scaling.
Therefore, we time-stamp the expressions that change
as time progresses, so for example $P_t$ is the power dissipation for a \gls{pm}
at time $t$ (as it depends on the current CPU frequency $f$).
We define an observed time period of $N$ equidistant time stamps\
in the range from $t_{0}$ to $t_{N}$, denoted $[t_{0}, t_{N}]$.\

Cooling overhead based on local temperatures\ 
is derived from the power signal of each \gls{pm} at its corresponding data
center location. For this purpose, the model for computer room air conditioning
using outside air economisers from \cite{xu_temperature_2013} was applied.
Cooling efficiency is expressed as partial \gls{pue} -- $pPUE_{dc,t}$\
at data center $dc$ at time $t$,\
which affects the power overhead based on the following formula:
\begin{align}\label{eq:P_tot}
P_{tot,t} = P_t + P_{cool,t} = pPUE_{dc,t} \cdot P_t(pm)
\end{align}
where $P_{cool,t}$ is the power necessary to cool the physical machine, and
$P_{tot,t}$ stands for the combined cooling and computation power.
For each data center location $dc$, there is a time series of temperature values
$\{T_{dc,t}:\ t \in [t_{0}, t_{N}]\}$.\
The dynamic value of $pPUE_{dc,t}$ is modelled as a function of
temperature $T$ to match hardware specifics as:
\begin{align}\label{eq:pPUE}
pPUE_{dc,t} = 7.1705 \cdot 10^{-5}T_{dc,t}^2 + 0.0041T_{dc,t} + 1.0743
\end{align}

These power signals are then integrated over time and combined with fixed
or real-time electricity prices (both models are explored in the evaluation)
for the corresponding data center location to compute the total
energy cost $C_{en}$ of every individual \gls{pm}.
\vspace{\eqtopmargin}
\begin{align}\label{eq:price}
C_{en} = \frac{t_N-t_0}{N} \mathlarger{\sum}_{t=t_0}^{t_N-1} P_{tot,t}e_{dc,t}
\end{align}
The integration is approximated using the rectangle numerical integration
method.
At each $dc$ location, there is a time series of electricity prices\
$\{e_{dc,t}:\ t \in [t_{0}, t_{N}]\}$.\




\section{Performance-based VM Pricing}
\label{ch:multicore:sec:pricing_model}

After covering the detailed components that make up energy costs in modern
multi-core physical machines, in this section we continue analysing
the economical side of cloud computing by looking at \gls{vm} pricing.
Concretely, we cover state of the art performance-based \gls{vm} pricing schemes
used by providers such as ElasticHosts and CloudSigma where the user pays
for the \gls{vm} proportionally to the allocated CPU frequency.
We then show on a practical experiment that these schemes do not account
for properties like the CPU-boundedness of the \gls{vm}'s workload
and its effect on \gls{qos} in the price calculation.
To address such drawbacks, we present our own perceived-performance \gls{vm}
pricing scheme as a next step in performance-based pricing, adapted
for both Intel and ARM architectures. Having models for both the energy costs
and \gls{vm} revenue accounting for frequency scaling
on multi-core \gls{pm}s will allow us to explain our cloud controller
in the next section. 



\subsection{Emerging Performance-based Pricing Cloud Providers}

In the performance-based pricing model used by several cloud providers, each user is charged on a per-time-unit basis (e.g. hourly) depending on the provisioned resources and their characteristics. The overall cost includes the cost for CPU provisioning, the allocated amount of RAM and the use of other resources, e.g. storage. Such a pricing scheme is offered by several providers, such as ElasticHosts \cite{elastichosts} and CloudSigma \cite{cloudsigma}, that allow the provisioning of different CPU frequency and core quantities, calculating the total CPU capacity allocated for the final invoice. This enables users to choose between equivalent combinations of frequencies and number of virtual CPU cores that incur same CPU provisioning costs \cite{elastichosts}.

Based on the performance-based pricing scheme offered by ElasticHosts and CloudSigma, we fitted a pricing model that describes the behaviour of both schemes, similarly to the work in \cite{pietri2015cost} where the ElasticHosts pricing scheme was initially modelled and analysed. In our obtained model, the price charged for CPU provisioning changes linearly with the total requested CPU capacity, as CPU capacity can be customised for different selected CPU frequencies and number of cores. Also, we extended the model to describe the RAM allocation. As \gls{vm}s may have different RAM capacity requirements, the price varies according to the selected RAM size. Finally, the cost for other resources used is considered to be fixed in the model as it is not the focus in this work.
Hence, the price $C_{vm}$ of each \gls{vm} at frequency $f_{CPU}$ is computed as:
\begin{align}\label{eq:simpleprice}
C_{vm} = C_{base} + C_{CPU} \sum \limits_{cpu \in vm} {(\frac{f_{cpu} - f_{base}}{f_{base}})} + C_{RAM} \frac{RAMsize}{RAMsize_{base}},
\end{align}
where $C_{base}$ is the price of the \gls{vm} at minimum capacity, i.e. a CPU at minimum frequency $f_{base}$. $C_{CPU}$ and $C_{RAM}$ are cost weights used to generate the \gls{vm} price for different CPU and RAM capacities. By replacing these variables with actual constants (presented later in the evaluation), the prices for configurations offered by ElasticHosts or CloudSigma can be approximated.


\subsection{Workload Heterogeneity Implications}

While the performance-based pricing model offered by ElasticHosts and CloudSigma
enables the cloud provider to balance energy savings with the revenue losses
from actions such as CPU frequency scaling, it ignores
the impact of \gls{vm} workload characteristics.
We illustrate this in an empirical experiment we have performed
to show how operating CPU frequency may affect workload performance
in a different way depending on the application's CPU boundedness ($\beta$)
characteristics.

We executed the Berserk benchmark (already explained in
Section~\ref{section:berserk}) on one local server with a \emph{remote\_ratio}
parameter indicating the portion of the work to offload to a different,
remote server. The rest of the tasks were executed locally. Both servers had the
same Quad-core 2.6 GHz AMD Opteron 4130 processor. The \emph{remote\_ratio}
parameter therefore controlled the workload's CPU boundedness,
as we could control if the task was more CPU-bound
(i.e. a low \emph{remote\_ratio}) or more I/O-bound where they would wait on the
results to arrive from a network resource (i.e. a high \emph{remote\_ratio}).
We calculated the CPU boundedness parameter $\beta$ as inversely proportional
to \emph{remote\_ratio}.\
This approach enabled us to set arbitrary workload CPU boundedness.
The experiment was run for six equidistant $\beta$ values between 0 and 1.

To also measure the effects CPU frequency scaling has in this context,
we executed each of the workload characteristics
on all five CPU frequency levels (applied both locally and to the remote server)
that our server offered using the \emph{cpufrequtils} tool
(the scripts we developed are included together with
the Berserk benchmark~\cite{_berserk_2015}).
We collected the duration of the benchmark under each of the workload
CPU boundedness $\beta$ and server CPU frequency combination.
The energy consumption was collected through
an Eaton Wattmeter\footnote{EATON ePDU PW104MA0UC34} using its SNMP API
(the energy measurement source code
is part of our Philharmonic simulator~\cite{drazen_lucanin_philharmonic_2014}).


 \begin{figure}
 \centering
 \includegraphics[trim=2.5cm 1cm 1.2cm 1.5cm, clip=true, width=0.7\columnwidth]{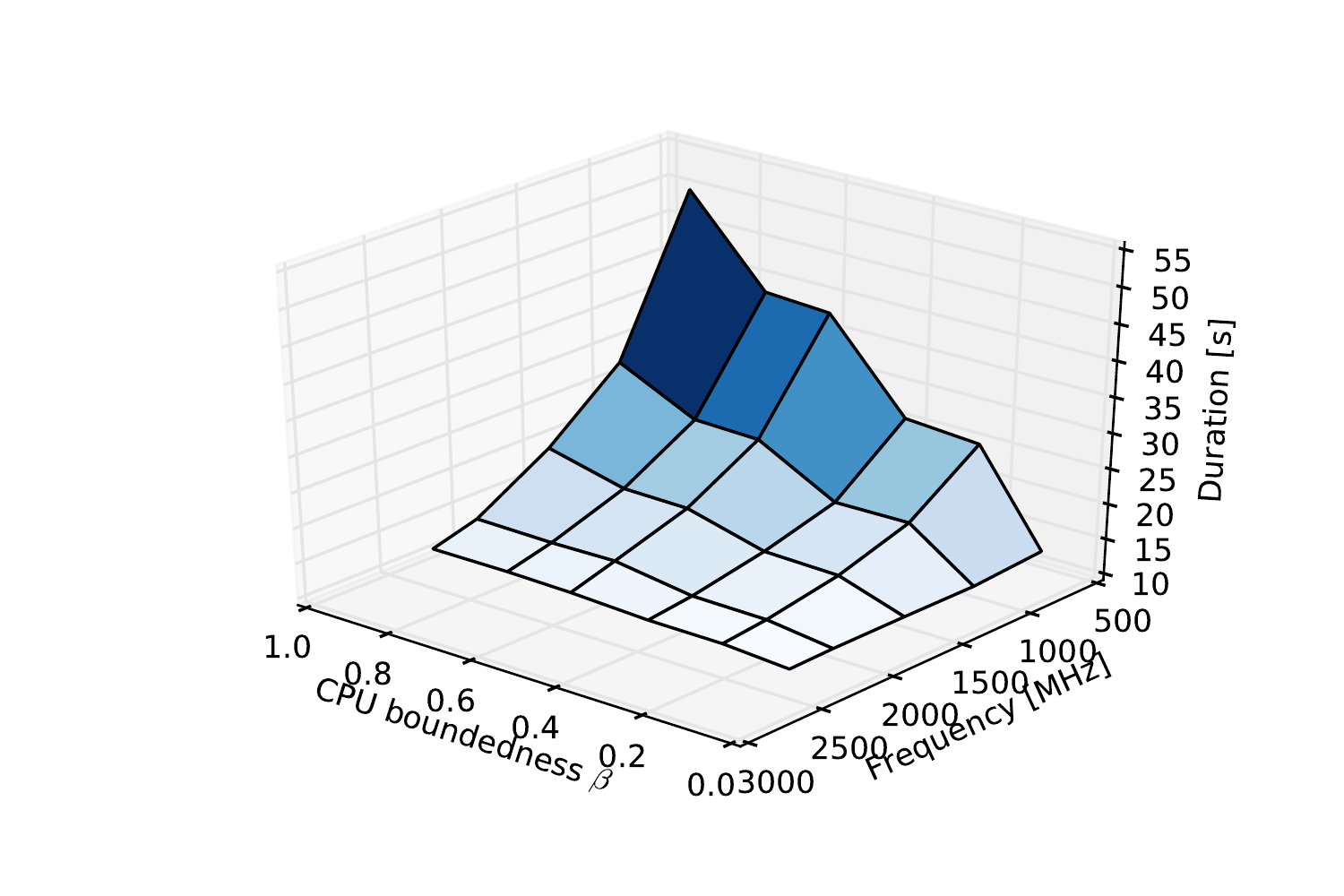}
 \caption{Benchmark duration for different workload CPU-boundedness ($\beta$) parameters and server CPU frequencies.}
 \label{fig:motExample}
 \end{figure}



The results are shown in Fig.~\ref{fig:motExample}. As can be seen, the execution time of an application with high CPU-boundedness ($\beta$) increases significantly when using a lower frequency and remains mostly unaffected for application with a lower CPU-boundedness ($\beta$ close to 0). Using the current performance-based pricing for CPU provisioning, like ElasticHosts and CloudSigma, a low frequency for jobs with low CPU boundedness would result in significantly lower revenue for the provider, even though the application performance would not be greatly affected. This was the main motivation for our perceived-performance pricing scheme that we present in the following section.


\subsection{Perceived-Performance Pricing} \label{section:perc_pricing}

\begin{figure}
\centering
\includegraphics[width=0.7\columnwidth]{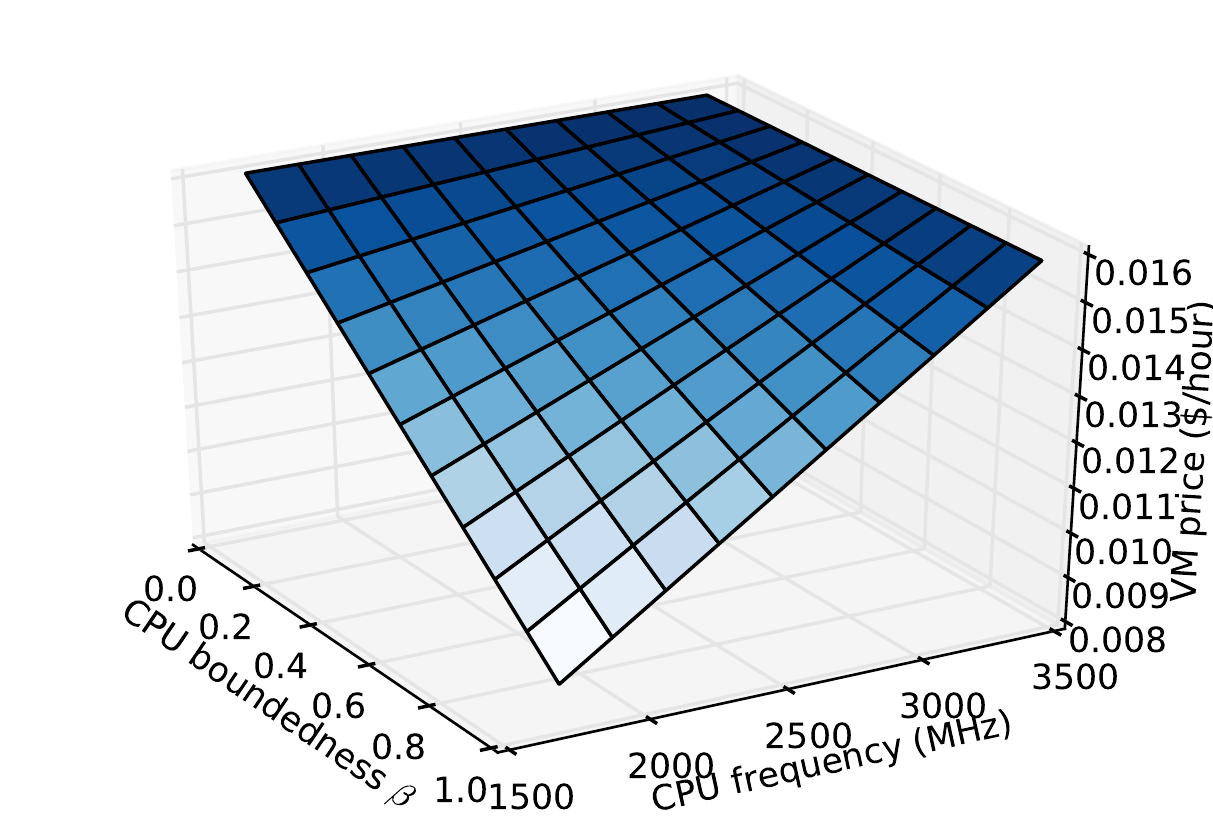}
\caption{Perceived-performance pricing model.}
\label{fig:pricing_model}
\end{figure}

As mentioned earlier, performance-based pricing, used by cloud providers like
ElasticHosts and CloudSigma, does not consider the impact of frequency reduction
on \gls{vm} performance. To do so, Eq.~\ref{eq:simpleprice} is extended
for pricing based on the performance perceived by the hosted application based
on the approach we proposed in \cite{lucanin_cloud_2015}
by defining $f_{cpu}$ as:
\
\begin{align}\label{eq:perc_pricing}
f_{cpu} = \beta f + (1 - \beta) f_{max},
\end{align}
where $f_{max}$ is the maximum operating frequency of the host
so that CPU-bound applications incur lower CPU provisioning cost when using
lower frequencies that may result in lower performance. On the other hand,
I/O-bound applications that are not affected by the change in frequency do not
receive significant decrease in cost. It is assumed that the impact of frequency
on application performance is same for each core of the \gls{vm} ($\beta$).

The pricing model is presented in Fig.~\ref{fig:pricing_model}, where the axes
show the provisioned CPU capacity, the CPU boundedness of the \gls{vm}
(parameter $\beta$) and the respective \gls{vm} price. The CloudSigma model
constants were used. The amount of RAM was not varied. CPU frequency values
for the Intel platform were used. We assumed four active CPU cores where
the CPU frequency is scaled linearly (so the sum of assigned CPU frequencies
of all the cores was used in the price calculation).

The linear curve for $\beta=1$ corresponds to the case of performance-based
pricing where the CPU provisioning cost changes linearly with the actual CPU
frequency ($f_{cpu} = f$). As the value of $\beta$ is decreased, the price
is less affected by CPU scaling (becoming constant for $\beta=0$), which
matches the workload behaviour from the experiment described earlier.

It is assumed that resources are charged on an hourly basis. The total service
revenue is computed by adding the per-hour provisioning cost of
Eq.~\ref{eq:simpleprice} for each \gls{vm} served, allowing us to compare
it with the frequency-based energy costs
from Section~\ref{ch:multicore:sec:power_model}.

\subsection{Prices for Different Architectures}
\label{ch:multicore:sec:pricing_model_arm}

As mentioned in Section~\ref{sec:mont_blanc}, besides using Intel architectures with ElasticHosts and CloudSigma pricing, in our work we are also interested in analysing the new pricing models on ARM architectures. Since the performance of an ARM-based CPU is significantly lower than the Intel, the price was scaled according to this factor.
The authors in \cite{chi2015benchmark} evaluated the ARM Cortex-A15 against a Haswell i5 executing \gls{hevc} decoding tasks and obtained a 6x performance advantage for the Intel.
Similarly, the authors in \cite{francesquini2015benchmark} evaluated a Sandy-Bridge based Xeon and a Cortex-A15 using various benchmarks, with roughly an 8x performance advantage for the Intel.
Finally, a set of scientific benchmarks was used in \cite{padoin2015benchmark} with both a Cortex-A15 CPU and a Sandy-Bridge based high-end i7 CPU.
The results indicated that the i7 performed roughly 16x better than the ARM.

Based on these numbers, and by using one ARM Cortex-A15 CPU and one Sandy-Bridge based high-end i7,
we normalised the performance values to the clock frequency of both platforms in order to match with the cost model of the cloud provider.
We assumed a 11x performance advantage for the Intel architecture, and therefore we assumed a 11x cost reduction in the cloud service when using the ARM platform.


\section{Cloud Controller Description}
\label{ch:multicore:sec:scheduler}

Having described both the multi-core, geographically-dependent energy
consumption model used to compute the energy costs from operating the cloud
and the perceived-performance pricing scheme used to compute the revenue from
\gls{vm} provisioning, in this section we explain our cloud controller.
The cloud controller balances both of these cost-related components
to obtain a quantitative comparison of energy saving
and revenue loss trade-offs,
which can be addressed as an optimisation problem.\ 
In other words, both cost-related components addressed in this chapter
are used to determine the actions invoked by the cloud controller.

We describe the \Multicoreschedulerfreq{} cloud controller that determines
the \gls{vm} migration and frequency scaling actions to be triggered
in order to achieve energy cost savings that exceed the revenue losses.
As these two control actions -- \gls{vm} migration and CPU frequency scaling
-- are mutually orthogonal, they are considered as two complementary actions
in order to optimise the allocation and configuration of \gls{vm}s to \gls{pm}s.
Hence, the two actions are examined in the algorithm separately as two stages,
firstly migrating the \gls{vm}s to \gls{pm}s and then scaling
the CPU frequencies of the \gls{pm}s to achieve further energy cost savings.
During the \gls{vm} migration stage, the controller migrates \gls{vm}s
to \gls{pm}s so that resource utilisation is maximised while preferring
more economical locations in terms of electricity and cooling costs.
Then, the controller reduces the CPU frequencies of the \gls{pm}s iteratively
as long as the energy cost savings exceed the service revenue losses.
The algorithm is invoked periodically to trigger appropriate actions.
In a real cloud system the algorithm could be automatically invoked
by new \gls{vm} arrivals or threshold violations of geotemporal inputs,
e.g. a temperature increase of 1 C. Next, the two stages of the algorithm
are described in more detail.


\subsection{VM Migration Stage}

During the first stage, the controller allocates newly requested \gls{vm}s
or reallocates \gls{vm}s from underutilised hosts using migration based on the
power overhead and the geotemporal input parameters of the \gls{pm}s.
As the underlying bin packing problem of \gls{vm} allocation is NP-hard,
for this stage we propose the heuristic polynomial time \gls{bcf} algorithm
we presented as Alg.~\ref{alg:bcf} in Chapter~\ref{ch:gascheduler}.

To briefly summarise the logic of the algorithm once again,
first all the newly requested \gls{vm}s or \gls{vm}s that run on underutilised
hosts (as discussed in Chapter~\ref{ch:gascheduler}) are selected
for allocation.
The selected \gls{vm}s are then migrated, prioritising \gls{vm}s larger
in their resource requirements (e.g. more required RAM, CPU cores).
The host \gls{pm} of each \gls{vm} is selected by preferring active
(already hosting another \gls{vm}), smaller machines
(in order to minimise the idle power overhead) and data centers with
a lower combined electricity price and cooling overhead cost based
on the geotemporal input prices model
outlined in Section~\ref{ch:multicore:sec:energy_calculation}.

\subsection{Frequency Scaling Stage}

\begin{algorithm}[!t]
\caption{Frequency Scaling Stage.}
\label{alg:frequency_scaling}
{
\begin{algorithmic}[1]

\Ensure Reduce CPU frequencies while energy savings exceed revenue losses.
\Procedure{Frequency Scaling Stage}{}

\State $decrease\_feasible \gets False$
\State reset frequency of $\forall pm \in active$ to $f_{max}$\
    \label{alg:fs:reset}

\For {$pm \in active$} \label{alg:fs:pmiterate} 
    \State $f \gets f_{max}$ \Comment{Start the loop at max frequency}
    \State $revenue\_cur \gets get\_revenue(pm, f_{to\_apply})$
    \Comment Service revenue, $\forall vm \in pm$
    \State $en\_cost\_cur \gets get\_en\_cost(pm, f_{to\_apply})$
    \Comment Energy cost of the $pm$

    \While{$f>f_{min}$} 

        \State $f \gets f-f_{step}$ \
            \label{alg:fs:freqStep}
        \State $revenue\_new \gets get\_revenue(pm, f)$
        \Comment {Revenue for the new frequency}
        \label{alg:fs:get_revenue}
        \State $en\_cost\_new \gets get\_en\_cost(pm, f)$
        \Comment {New energy cost}
        \label{alg:fs:get_cost}
        \State $revenue\_loss \gets revenue\_cur - revenue\_new$
        \State $en\_savings \gets en\_cost\_cur - en\_cost\_new$ \
        \If {$en\_savings > revenue\_loss$} \
            \label{alg:fs:comparison}
            \State $revenue\_cur \gets revenue\_new$
                \Comment{Update current service revenue}
            \State $en\_cost\_cur \gets en\_cost\_new$
                \Comment{Update current energy cost}
            \State $decrease\_feasible \gets True$ \
            \State $f_{to\_apply} \gets f$
            \Comment{Update currently selected frequency} \
            \label{alg:fs:apply}
        \Else
            \State break \label{alg:fs:break}
        \EndIf
    \EndWhile

    \If {$decrease\_feasible$} \
        \State apply $f_{to\_apply}$ to $pm$
    \Else
        \State remove from $active$: $\forall \hat{pm} \in PMs$ s.t. $\hat{pm}$
        has higher mean $\beta_{vm}$ and lower electricity price
        and temperature than $pm$ \
        \label{alg:fs:remove_worse}
    \EndIf
\EndFor
\EndProcedure

\end{algorithmic}
}
\end{algorithm}


Having allocated the \gls{vm}s to \gls{pm}s, the CPU frequencies
of the \gls{pm}s are adjusted in the next stage.
We assume that each host can operate between a minimum and maximum frequency,
$f_{min}$ and $f_{max}$ respectively. The appropriate CPU frequencies
are selected based on both the geotemporal inputs
and the workload characteristics, by considering their overall impact
on the cost components presented in Sections~\ref{ch:multicore:sec:power_model}
and~\ref{ch:multicore:sec:pricing_model}. To do so, a \gls{pm}'s CPU frequency
is reduced only when energy savings from the reduction in the CPU frequency
exceed the revenue losses under perceived-performance pricing.
The algorithm is described in Alg.~\ref{alg:frequency_scaling}.
From a high level, the CPU frequency of each \gls{pm} is initially set
to its maximum frequency $f_{max}$ (line~\ref{alg:fs:reset}).
Then, the algorithm iterates through the list of $active$ \gls{pm}s
(line~\ref{alg:fs:pmiterate}) to determine the most efficient CPU frequency
for each one, analysing the range of the available CPU frequencies
(line~\ref{alg:fs:freqStep}). Note that the actions determined in each step
do not have to be executed physically before the procedure halts,
after which the final \gls{pm} frequencies can be determined and applied.

To pick the best CPU frequency, the cost-related components
for the current \gls{pm} are calculated for the previously determined
and the next lower frequency
(lines~\ref{alg:fs:get_revenue}--\ref{alg:fs:get_cost}).
The components include the service revenue from the \gls{vm}s allocated
to the current \gls{pm} and the \gls{pm}'s energy cost based on the
multi-core power model and energy cost calculation presented
in Section~\ref{ch:multicore:sec:power_model}. Whenever the consideration
of the lower CPU frequency results in energy cost savings
which exceed the subsequent revenue loss, the new frequency is chosen
for the current \gls{pm} (line~\ref{alg:fs:comparison}) and the algorithm
continues to the next lower available frequency (line~\ref{alg:fs:freqStep}).
The procedure in the inner loop terminates when the revenue losses
exceed the respective energy savings (line~\ref{alg:fs:break}).
If no frequency reduction occurred for the \gls{pm} ($decrease\_feasible$
stays $False$), the procedure will remove \gls{pm}s with
higher average $\beta$ and lower electricity price and temperature
(line~\ref{alg:fs:remove_worse}) before continuing. The idea is that
such \gls{pm}s may incur even lower energy savings and higher revenue losses,
hence they can be omitted from the analysis to prune the search space.


\section{Evaluation}
\label{ch:multicore:sec:evaluation}

In this section we evaluate the presented cloud controller in a simulation
that we first describe as part of the evaluation methodology.
We then proceed with presenting the simulation results showing the impact of our
cloud controller with a focus on different environment factors.

\subsection{Methodology}

The \Multicoreschedulerfreq{} method is evaluated in a\
simulation of \Multicorevmnumsimulationmulticore{} \gls{vm}s based
on real traces of geotemporal inputs and \gls{vm} CPU-boundedness values.
The goal of the evaluation is to show the cost savings attainable
using our approach, the impact on service revenue and to analyse the dependence
on external factors, such as electricity prices and \gls{vm} workloads.
The simulations were executed on the Philharmonic simulator described
in Section~\ref{sec:philharmonic}.
The simulated controller is called to determine cloud control actions,
such as \gls{vm} migrations or \gls{pm} frequency scaling.

To compute the energy costs of the simulated geographically distributed cloud,
we consider the same use case of six data centers as
in Section~\ref{ch:vmpricing:sec:cloud-simulation}.
A dataset of real-time electricity prices described in \cite{alfeld_toward_2012}\
and temperatures from the Forecast~\cite{_forecast_2015} 
web service were again used (with synthetically generated data for cities where
we had no \gls{rtep} data).
The data center locations used in the simulation
(Fig.~\ref{ch:vmpricing:fig:cities}) were selected to resemble Google's
deployment.
Additionally, a scenario with fixed electricity prices over time is considered
in the evaluation, using the mean values for each location.

\begin{table}[!t]
\centering
\caption{Infrastructure parameters.}
\label{tab:simulation}
\begin{tabular}{cccccc}
\hline
 Architecture & \gls{pm}s & \gls{vm}s & \
 $f_{min}$ & $f_{max}$ & $f_{step}$ \\
\hline
 ARM & \Multicorepmnumsimulation{} & \Multicorevmnumsimulationmulticore{}& \
 0.8 GHz & 1.8 GHz & 100 MHz \\
\hline
 Intel & \Multicorepmnumsimulation{} & \Multicorevmnumsimulationmulticore{}& \
 2.6 GHz & 3.4 GHz & 200 MHz \\
\hline
\end{tabular}
\end{table}

\begin{table}[!h]
\centering
\caption{Pricing model parameters.}
\label{tab:pricing_simulation}
\begin{tabular}{ccccccccccccc}
\hline
 Pricing Model & $C_{base}$ & $C_{CPU}$ & $C_{RAM}$ &\
 $RAMsize_{base}$  \\
\hline
ElasticHosts & 0.027 \$/h  & 0.018 \$/h  & 0.025 \$/h & 1 GB\\
\hline
CloudSigma & 0.0045 \$/h  & 0.0017 \$/h  & 0.004 \$/h & 1 GB\\
\hline
\end{tabular}
\end{table}

The simulator was set up using the infrastructure parameters shown
in Table~\ref{tab:simulation}.\
The table shows two architecture types: ARM and Intel. Their respective
performance characteristics were derived from the real specifications,
such as minimum CPU frequency $f_{min}$, maximum CPU frequency $f_{max}$
and the absolute frequency increase or decrease step size $f_{step}$.\
\
\
The parameters we fitted for the pricing models
in Section~\ref{ch:multicore:sec:pricing_model} to calculate hourly \gls{vm}
prices based on the pricing schemes offered by ElasticHosts \cite{elastichosts}
and CloudSigma \cite{cloudsigma} are shown
in Table~\ref{tab:pricing_simulation}. The cost of other resources which is not
the focus in this work, e.g. disk, was considered to be fixed.
\
We show results for both CPU architectures
and both pricing schemes only in sections where we compare the effects of these
respective factors on the attainable energy and cost savings. Other presented
results are limited to the ARM architecture and the CloudSigma pricing scheme,
which proved to be more promising for the application of our method,
as will be shown later.\

Each run simulated the cloud system for seven days of operation (168 h) with
an hourly step size (1 h). The step size was chosen based on the available
datasets of geotemporal inputs. However, note that different time intervals
and triggering events, e.g. thresholds in geotemporal input changes
or new \gls{vm} arrivals could invoke the cloud controller
in production environments.\
The characteristics of each resource considered in this work, namely the number
of CPU cores and the amount of RAM, were uniformly distributed.\
Heterogeneous \gls{vm}s were assumed with 1 or 2 CPU cores and RAM capacity
ranging between 8 and 16 GB RAM in order to model different \gls{vm} requests
and prices. Each \gls{pm} consists of 1--4 CPU cores and 16--32 GB RAM to model
specification diversity. For each \gls{vm}, the boot time and duration were
varied using a uniform distribution to generate random values within
the simulation time and distribute delete events over the simulation period
and range the utilisation of the resources.

\begin{figure}
\centering
\includegraphics[width=0.7\columnwidth]{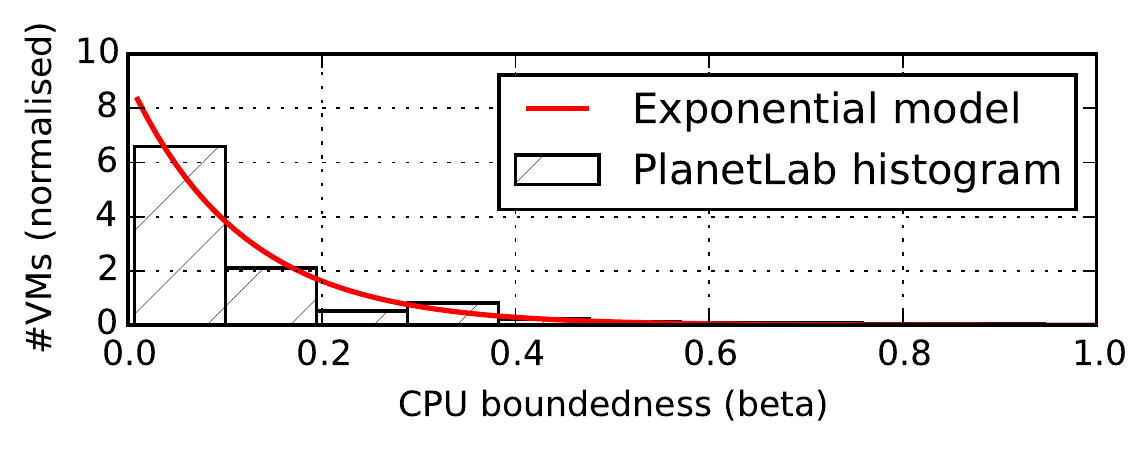}
\caption{VM CPU-boundedness distribution from PlanetLab traces.} 
\label{fig:beta_model}
\end{figure}

The CPU-boundedness of each \gls{vm} was modelled based on the CPU usage traces
from the PlanetLab dataset \cite{planetlab}.\
The dataset includes CPU usage traces of 1024 \gls{vm}s. The data was collected
every five minutes throughout a day. To generate realistic \gls{vm}
CPU-boundedness values in the simulation, the average CPU usage of each \gls{vm}
in the dataset was calculated and mapped to a $\beta$ value. From the generated
dataset of $\beta$ values, an exponential distribution was fitted.
The distribution is shown in Fig.~\ref{fig:beta_model}. The figure also includes
the empirical histogram of the traces normalised to an area of 1. The $\beta$
values of the \gls{vm}s used in the simulation were generated based
on this model.




We consider two baseline controllers for results comparison.\
The \Multicoreschedulerbase{} algorithm developed
in~\cite{beloglazov_energy-aware_2012}
(also used as the baseline in Chapter~\ref{ch:gascheduler})
is a cloud controller that migrates
\gls{vm}s, dynamically adapting to user requests.
The second baseline controller, \Multicoreschedulermigr{},
is a variant of the \Multicoreschedulerfreq{} controller that only applies the
\gls{vm} migration stage based on geotemporal inputs, but does not consider
frequency scaling.
The \Multicoreschedulerfreq{} controller allows us\
to quantify the improvement brought by CPU frequency scaling in isolation. 

The remainder of the section presents individual results for the different
simulation scenarios we performed to compare the energy and cost savings
and the performance implications from applying the proposed cloud controller
approach. The parameters specified earlier are used in all of the experiments,
unless otherwise stated.


\begin{figure}
\vspace{\figtopmargin}
\centering
\includegraphics[width=0.7\columnwidth]{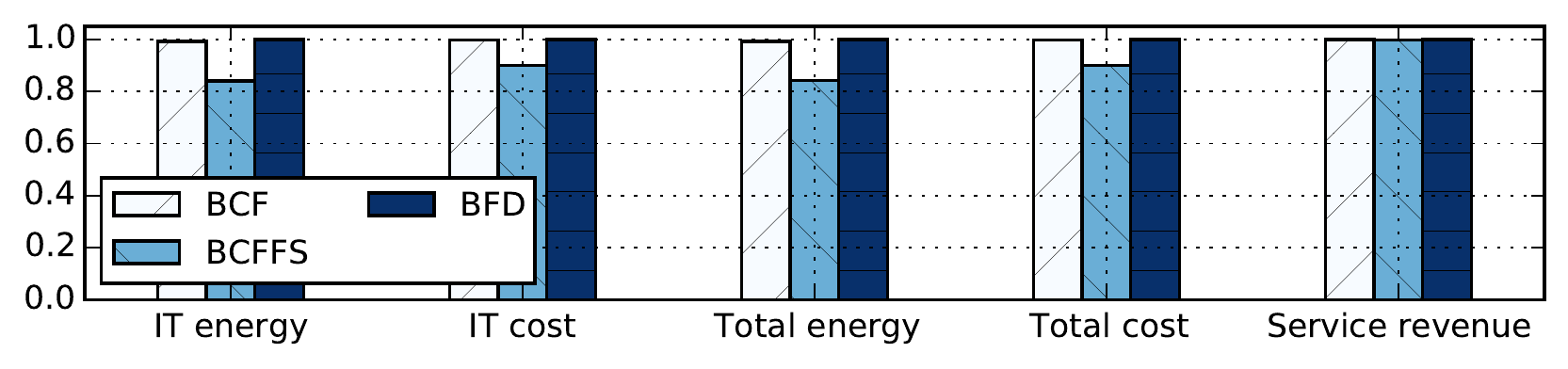}
\vspace{\figcaptionmargin}
\caption{Aggregated results for a \Multicorepmnumsimulation{} Intel PM simulation.} 
\label{fig:aggregated_results-single_simulation}
\vspace{\figbottommargin}
\end{figure}

\subsection{Cloud Controller Evaluation}

We begin by showing the cloud controller evaluation and comparison
to the baselines for the Intel architecture.
Fig.~\ref{fig:aggregated_results-single_simulation} includes the aggregated
results for the achieved energy costs and service revenue -- the energy
and cost used by the IT equipment, total energy and cost which include
the cooling overhead taking into account the outside temperatures
and the service revenue from hosting \gls{vm}s considering
the perceived-performance pricing model described
in Section~\ref{section:perc_pricing}.
The values are normalised as a relative value of the results obtained
using the baseline \Multicoreschedulerbase{} controller,
while the absolute values
can be found in Table~\ref{table:aggregated_results-single_simulation}.
It can be seen that the \Multicoreschedulerfreq{} controller
achieves \Multicoreensavingsmaxintel{} energy savings
compared to the baseline controller \Multicoreschedulerbase{},
out of which \Multicoreensavingsfreqintel{} are the additional energy savings
achieved by using frequency scaling
(the savings compared to the \Multicoreschedulermigr{} controller).
The service revenue losses from using perceived-performance pricing were not
significant with a drop of less than 0.3\%,
compared to the \Multicoreschedulerbase{} baseline.
This is because the frequency scaling algorithm presented in the previous
section does not scale frequencies if the revenue loss exceeds
the energy cost savings.



\begin{table}[!b]
\centering
\caption{Absolute aggregated Intel simulation results.}
\begin{tabular}{lrrr}
\toprule
{} &        BCF &      BCFFS &        BFD \\
\midrule
IT energy (kWh)     &   18793.73 &   15933.10 &   18943.00 \\
IT cost (\$)         &     974.60 &     878.50 &     977.00 \\
Total energy (kWh)  &   22501.25 &   19095.69 &   22678.65 \\
Total cost (\$)      &    1161.18 &    1046.75 &    1163.89 \\
Service revenue (\$) &    6543.78 &    6524.54 &    6543.78 \\
\bottomrule
\end{tabular}
\label{table:aggregated_results-single_simulation}
\end{table}


\subsection{Architecture Impact}

Having shown the results for the Intel architecture, we now show results for the same simulation, only this time using the ARM power model, presented in Section~\ref{ch:multicore:sec:power_model_arm}. This allows us to compare the architecture impact on attainable energy cost savings. As we previously mentioned, ARM processors are increasingly popular due to their good computation-per-watt ratio and are being explored for use in data centers as part of the Mont Blanc project \cite{francesquini2015benchmark}. The \gls{vm} pricing is also adapted for the lower ARM performance compared to Intel, as detailed in Section~\ref{ch:multicore:sec:pricing_model_arm}.

\begin{figure}
\centering
\includegraphics[width=0.7\columnwidth]{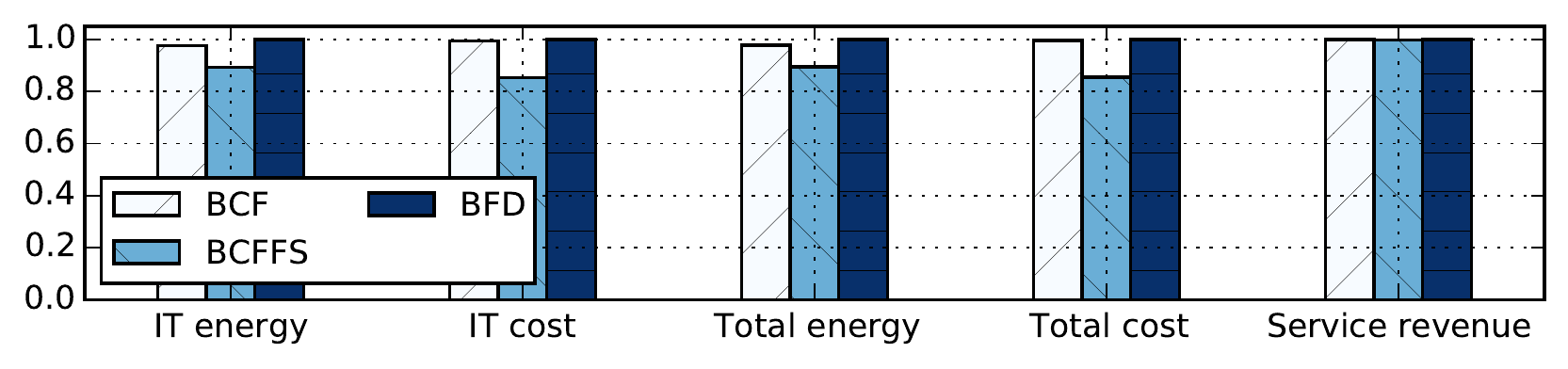}
\caption{Aggregated results for the ARM power and pricing model.}
\label{fig:aggregated_results-ARM}
\end{figure}

The results for the ARM architecture are shown in Fig.~\ref{fig:aggregated_results-ARM}. Even higher savings are achieved than for the Intel architecture -- the \Multicoreschedulerfreq{} controller achieves \Multicoreensavingsmax{} energy cost savings compared to the \Multicoreschedulerbase{} baseline and \Multicoreensavingsfreq{} compared to the \Multicoreschedulermigr{} baseline. Even lower service revenue losses (less than 0.25\% drop compared to \Multicoreschedulerbase{}) again indicate that the impact on \gls{vm} performance is not significant. Given the better applicability of our controller method to ARM architectures, we limit the remainder of the results to this architecture. 
Absolute values can be found in Table~\ref{table:aggregated_results-single_simulation-arm}.


\begin{table}[!b]
\centering
\caption{Absolute aggregated ARM simulation results.}
\begin{tabular}{lrrr}
\toprule
{} &        BCF &      BCFFS &        BFD \\
\midrule
IT energy (kWh)     &     681.79 &     623.50 &     698.19 \\
IT cost (\$)         &      39.23 &      33.68 &      39.48 \\
Total energy (kWh)  &     817.12 &     747.47 &     835.61 \\
Total cost (\$)      &      46.78 &      40.17 &      47.02 \\
Service revenue (\$) &     588.81 &     587.33 &     588.81 \\
\bottomrule
\end{tabular}
\label{table:aggregated_results-single_simulation-arm}
\end{table}


\subsection{Dynamic CPU Frequency Analysis}

\begin{figure}
\centering
\includegraphics[width=0.65\columnwidth]{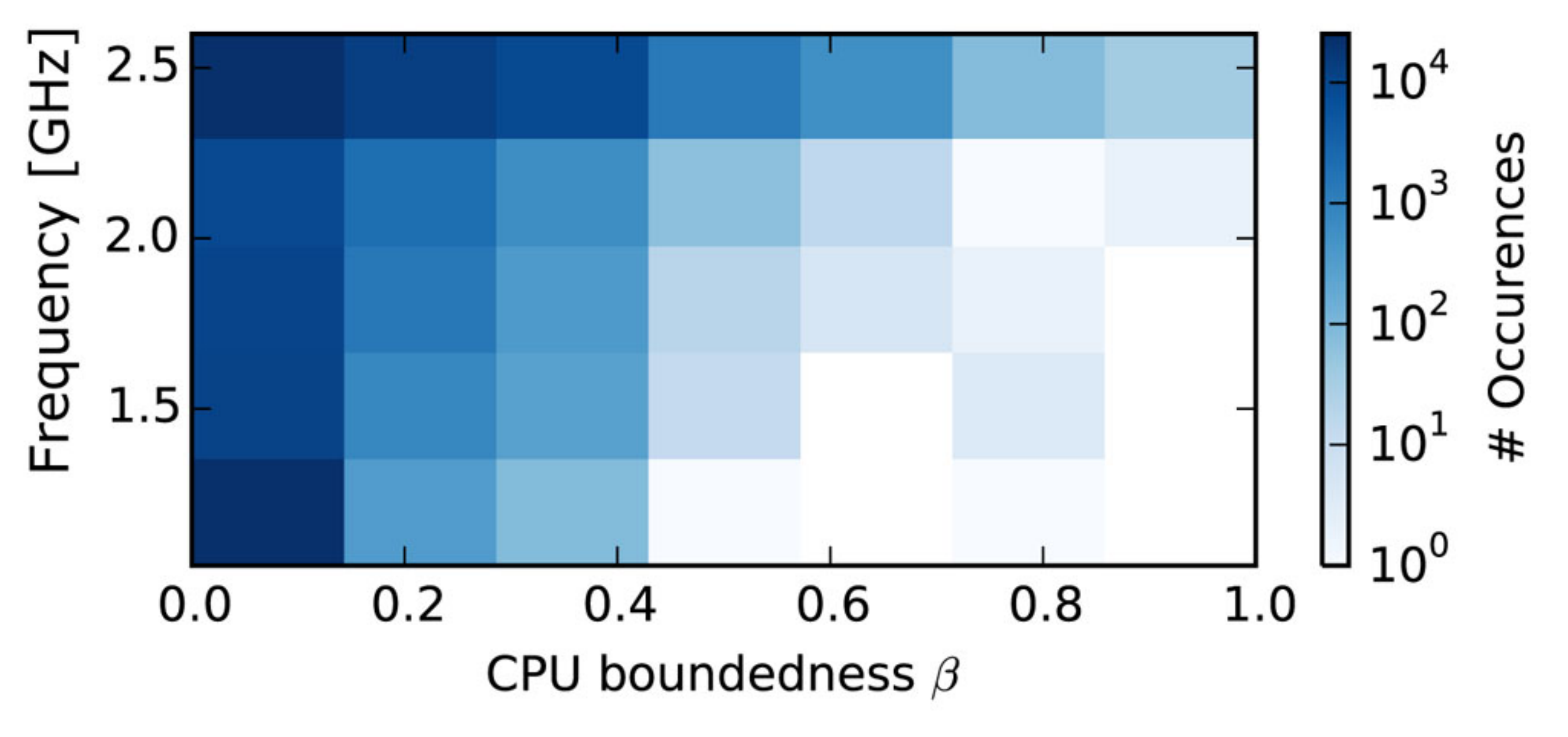}
\caption{Occurrences of $(\beta, f)$ combinations among the controlled VMs for the ARM architecture.}
\label{fig:beta_freq_histogram}
\end{figure}

To explore the frequencies $f$ assigned to \gls{vm}s dynamically during the simulation and compare them with the \gls{vm}s' CPU boundedness $\beta$, we counted the number of occurrences of each $(\beta, f)$ combination for every \gls{vm} and time slot. This data is illustrated\
as a bivariate histogram in Fig.~\ref{fig:beta_freq_histogram} with the number of occurrences shown on a logarithmic scale.\
Darker areas indicate a higher number of frequency occurrences for the respective $(\beta, f)$ combination.\
It can be seen that the occurrences of CPU frequencies assigned based on each \gls{vm}'s CPU boundedness match the areas where \gls{vm} prices are high, based on the perceived-performance pricing model from Fig.~\ref{fig:pricing_model}.
The area with high $\beta$ and low $f$, where prices would be the lowest, contains no occurrences.\
The darkest areas of the graph with a high number of occurrences\
represent the balance between energy savings and profit losses,\ 
which is\
in line with the controller requirements\ 
that energy cost savings should be maximised, but not exceeded by revenue losses.

\subsection{Provider Pricing Impact}



\begin{figure}
\centering
\includegraphics[width=0.7\columnwidth]{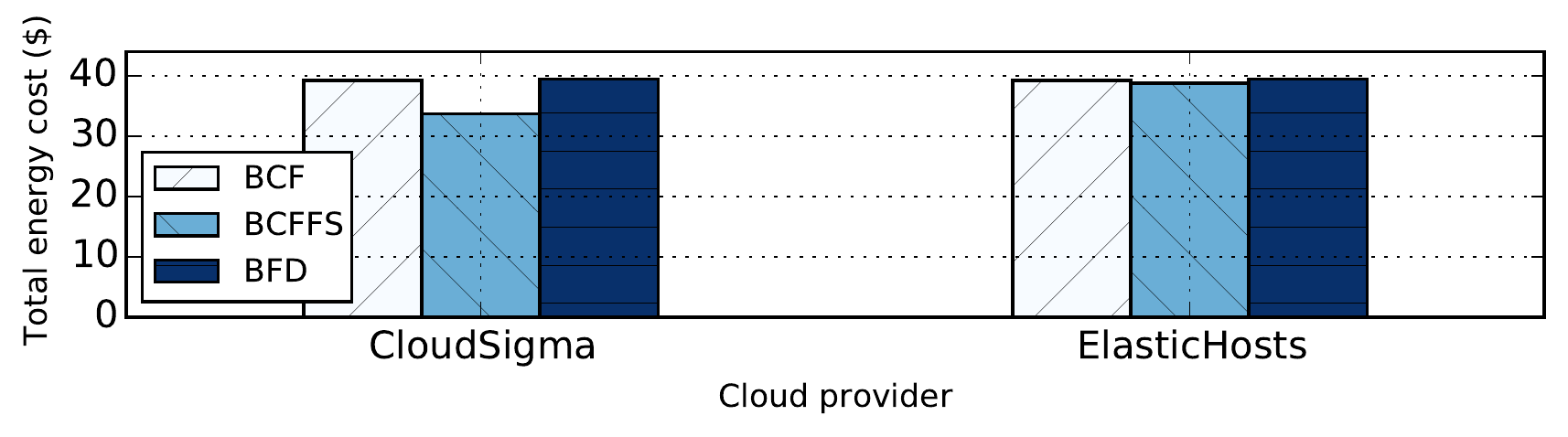}
\caption{Energy costs for the pricing models used by different cloud providers.}
\label{fig:results_provider_pricing}
\end{figure}

In this set of experiments we evaluated and compared the performance of the algorithms for different pricing models in order to investigate the impact of the pricing model on the savings from using the proposed approach. Fig.~\ref{fig:results_provider_pricing} presents the results for the CloudSigma and ElasticHosts cloud providers.\ 
As can be seen, higher energy cost savings are possible for the CloudSigma pricing scheme ($14\%$) than for the pricing offered by ElasticHosts ($2\%$). This is because CPU provisioning offered by CloudSigma is charged at a lower price resulting in service revenue being closer to the energy costs. As a result, energy savings gain comparably more weight in the revenue-energy balancing performed by the cloud controller. 
Since our method applies better to cloud providers like CloudSigma, we used their pricing scheme in all the other simulation scenarios.


\begin{figure}
\centering
\includegraphics[width=0.7\columnwidth]{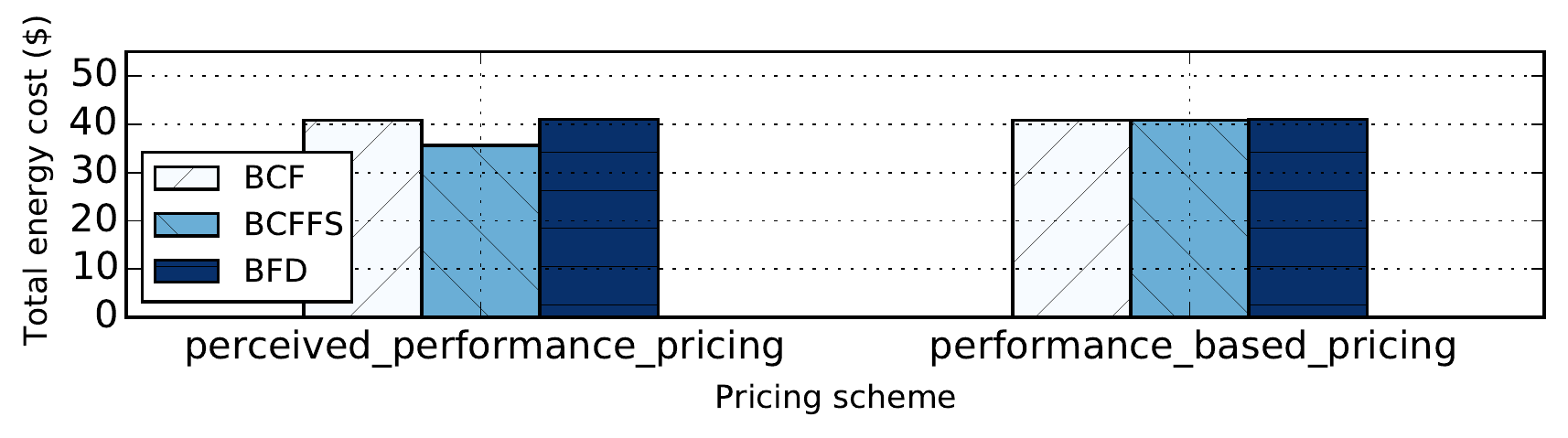}
\caption{Energy cost savings for perceived-performance and performance-based pricing.}
\label{fig:pricing_impact_arm}
\end{figure}

\subsection{Pricing Model Impact}

In this experiment we compared the savings obtained by using different pricing models. These include the perceived-performance pricing model proposed in Section~\ref{section:perc_pricing} and performance-based pricing offered by the current providers. The results\ 
are presented in Fig.~\ref{fig:pricing_impact_arm}.\
It can be seen that using performance-based pricing does not lead to energy savings, as the reduction in prices is high compared with the energy costs. As a result, CPU frequency scaling is not feasible. On the other hand, using perceived-performance pricing, savings are possible as CPU frequency reduction does not lead to substantially reduced service revenues.

\subsection{Electricity Cost Variation}

\begin{figure}
\centering
\includegraphics[width=0.7\columnwidth]{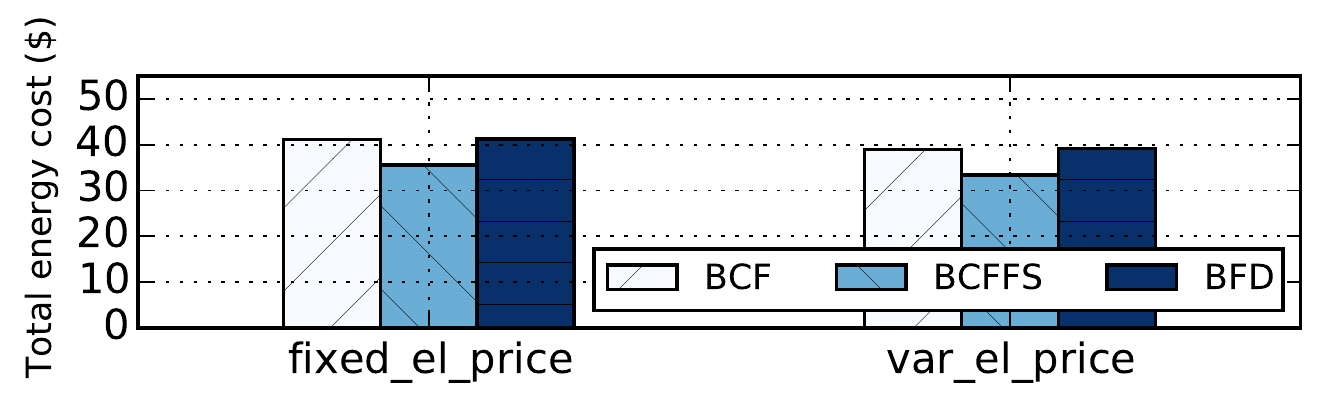}
\caption{Energy cost savings for fixed and variable electricity prices.}
\label{fig:fixed_variable_el_price}
\end{figure}

As not all cloud providers may have access to real-time electricity pricing, in this set of experiments we evaluate the performance of the proposed controller under fixed electricity pricing. In Fig.~\ref{fig:fixed_variable_el_price} scenarios for fixed and variable electricity prices are compared to investigate the impact of electricity pricing on the energy savings obtained using the \Multicoreschedulerfreq{} controller.\
The \Multicoreschedulerfreq{} controller achieves better performance under variable electricity pricing reducing the energy costs by exploiting runtime information and adapting the cloud configuration according to the electricity price changes within the day. However, cost savings of $10\%$ (compared to the \Multicoreschedulerbase{} baseline) that are still significant are achieved for the fixed electricity cost scenario.


\subsection{Variation of Parameter $\beta$}

\begin{figure}
\centering
\includegraphics[width=0.86\columnwidth]{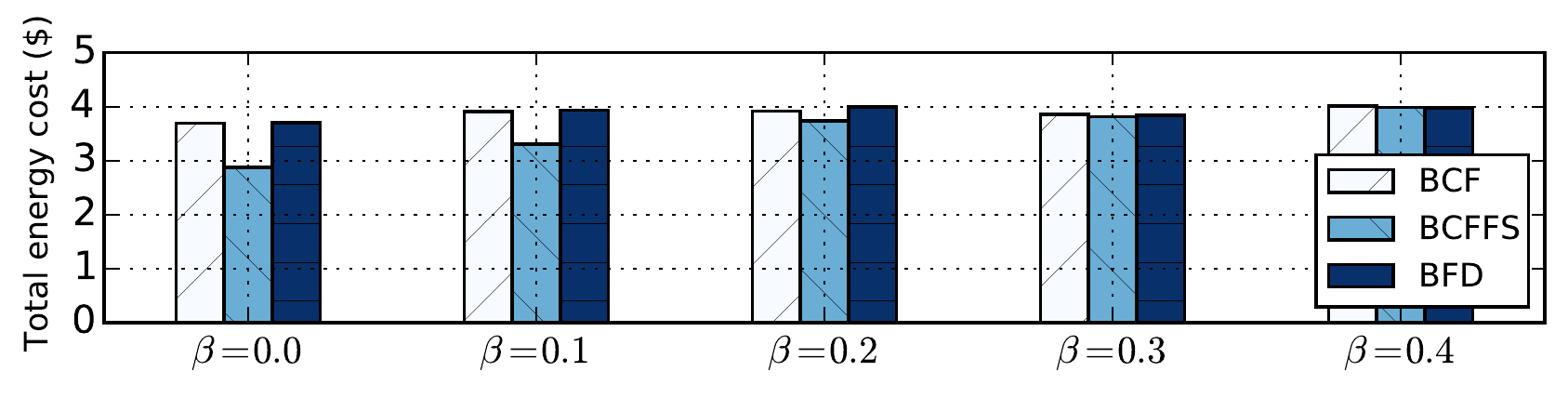}
\caption{Energy cost savings for VMs with different fixed CPU-boundedness.}
\label{fig:beta_variation}
\end{figure}

Fig.~\ref{fig:beta_variation} shows the results for \gls{vm}s with fixed CPU-boundedness properties. The aim is to evaluate the impact of different workloads on energy cost savings under the proposed controller and identify workload types where our approach is more beneficial.
To do so, simulations using the same set of \gls{pm}s and \gls{vm} requests were used, while \gls{vm} CPU-boundedness properties were varied between $0.0$ to $0.4$. 
The results are omitted for larger values of $\beta$, where savings are limited due to the impact on application performance.
The energy savings achieved by the \Multicoreschedulerfreq{} controller decrease gradually while approaching higher values of CPU-boundedness ($\beta$). Between a $\beta$ of 0.0 and 0.2 there is a substantial increase in energy cost as even a small reduction in frequency results in high energy cost savings that exceed the revenue losses. For higher values of $\beta$, the savings are limited due to the impact on application performance. As a result, the \Multicoreschedulerfreq{} controller achieves the best results for I/O-bound workloads where application performance is not greatly affected by the reduction in frequency.

\section{Summary}
\label{sec:conclusion}

As the demand for cloud platforms increases and as the workload becomes
more diverse, a one-fit-all pricing policy does not only provide
poor flexibility to the user, but is also not energy efficient.
To keep up with the rapid evolution in information infrastructure,
a more flexible way of controlling cloud systems must be provided
to both satisfy the user and minimize the energy costs.

We have presented a flexible cloud control approach capable
of system-level resource management to fit the performance guarantees
requested by the user and minimise energy waste
by scaling CPU resources on demand.
Our cloud controller is driven by a
model which covers realistic aspects of real-world cloud platforms.
Geotemporal inputs such as real-time electricity pricing and
temperature-dependent cooling affecting a geographically distributed
cloud provider have been modelled together with a multi-architecture,
multi-core power model based on real experiments and used in the Philharmonic
cloud simulator to estimate operational costs and the \gls{vm} service revenue.
Several scenarios were examined and two baseline methods were used resulting
in energy cost savings of up to \Multicoreensavingsmax{} for ARM and
up to \Multicoreensavingsmaxintel{} for Intel architectures.

The lessons learned from our research can be applied to cut costs in data
centers. For example, a cloud provider with an \$12M annual electricity bill
providing \gls{vm}s on ARM infrastructure at prices similar to CloudSigma
for mostly I/O intensive workloads can save around \$1.7M, assuming no frequency
scaling was previously used.\
Even if not all ideal factors are satisfied, e.g. the \gls{vm} prices are
in the ElasticHosts range, around \$750k savings can be achieved
for a larger cloud provider with a \$38M annual energy bill
(estimated in~\cite{qureshi_cutting_2009}).

Our results show that energy costs can be significantly reduced
without noticeably impacting the service revenue
by scaling the CPU frequencies of the \gls{pm}s
according to the hosted \gls{vm} characteristics.
We have shown that this method applies better to some cloud providers
like CloudSigma, where service revenue is closer to energy costs.
Savings can be achieved for fixed electricity pricing,
but \gls{rtep} pricing allows higher energy savings.
For our method, ARM architectures are more suitable than Intel,
and more I/O-intensive workloads allow for higher savings
than CPU-intensive workloads.


\chapter{Conclusion}
\label{ch:conclusion}
In this chapter we give the concluding remarks on the contributions
of this thesis, list the limitations of our research and outline possible
future work directions.

\section{Summary}
\label{ch:conclusion:sec:summary}

A major challenge of the popular cloud computing paradigm is the
rising energy consumption of data centers hosting its computational resources.
Furthermore, modern cloud providers frequently operate multiple geographically
distributed data centers which result in dynamically-changing environments,
constituting geotemporal inputs such as real-time eletricity prices
and temperatures that affect cooling efficiency.
Existing research explores cloud management actions like \gls{vm} migration and
CPU frequency scaling, but disregarding the additional challenges of geotemporal
inputs.
We argue that a pervasive cloud control approach is necessary that considers
a wider model -- from the low-level infrastructure details, like geotemporal
inputs affecting power consumption and electricity price -- to the high-level
aspects of \gls{sla} definition of \gls{vm} \gls{qos} properties
and pricing details.
In this thesis we have presented a collection of management models
and cloud control methods that allow
geographically distributed cloud providers to address geotemporal inputs
with the goal of reducing energy consumption and costs, while satisfying
the \gls{qos} requirements of end users.

We firstly provided background information on geotemporal inputs, their origins
in deregulated electricity markets and temperature-dependent cooling technology.
We surveyed the forecasting methods that analyse historical time series
to predict their future behaviour. We evaluated two representative forecasting
methods -- Theta and \gls{aam} on geotemporal inputs relevant
in the cloud domain as well as presented their performance on a wider range of
time series data from the M3 competition.
We also presented a concept that would allow for \coe\ emissions to also
be treated as an energy cost, considering the Kyoto protocol emission trading
scheme. Under this model, \gls{coe} emissions can be treated as another
geotemporal input. We consider that involving data centers
in the trade of credits for emission reduction would help encourage
more environmentally friendly computational resource allocation.

As our main contributions, we explored methods
for dynamically controlling cloud resources based on geotemporal inputs
to improve energy and cost efficiency
in geographically distributed data centers. We started with
a straightforward method for finding peaks in real-time electricity pricing
and pausing managed \gls{vm}s which was evaluated in a realistic OpenStack
deployment to show measurable energy savings.
We then presented a broader concept of a pervasive cloud controller
where a number of geotemporal inputs,
such as \gls{rtep} and temperature-dependent cooling
can be considered in a forward-compatible optimisation engine
to reduce energy costs and preserve \gls{qos}.
We evaluated the approach in a simulation for which we developed
a proof-of-concept controller combining forecast-based planning and
a hybrid genetic algorithm with greedy local optimisation.
Real traces of temperatures and electricity prices were used as input datasets
to show a significant energy cost reduction
without noticeable \gls{qos} implications.
A number of parameters such as geographical data center distributions and
forecast data quality were explored to serve as cloud provider guidelines
to find conditions fit for pervasive cloud control.
Similar promising results were shown on the \gls{pm} layer where we presented
methods for determining the best runtime CPU frequency on multi-core
Intel and ARM architectures considering both
geotemporal inputs and potential revenue losses in performance-based pricing.

To tackle the side effects of energy-aware cloud control methods where \gls{vm}
properties like availability are affected, we proposed a progressive \gls{sla}
specification system.
Traces from cloud control simulations optimised for geotemporal inputs
were used to derive corresponding \gls{sla}s with precise \gls{vm} availability
and cost estimation.
A separate utility-based user \gls{sla} selection simulation
using real-world datasets was used to validate
the energy saving and customer satisfaction potential.
Our results show that diverse cloud providers complement each other.
For users that require high performance, like web applications,
less invasive methods can be used. Scientific computing use cases,
on the other hand, can benefit from lower costs of energy-aware \gls{vm}s.

Finally, we analysed new pricing policies in cloud computing where
\gls{vm} costs are based on the runtime allocated CPU frequencies.
We considered ways to adapt cloud control for such performance-based pricing
in the context of workloads with variable degrees of CPU boundedness.
We developed a perceived-performance pricing scheme that combines both the
CPU frequency and the workload's CPU boundedness.
This new pricing scheme supports energy-aware methods that consider
geotemporal inputs like our CPU frequency scaling cloud controller.
Our results show the importance of considering both the computational
and the economical side of cloud computing to successfully reduce
energy consumption and cost in the underlying geographically distributed
data center infrastructure.

\section{Limitations}

In this section we list the limitations and constraints of the contributions
presented in this thesis. The topics listed here were out of scope
for our research, but can be considered as possible future work directions:

\begin{itemize}
\item As mentioned in the introduction, our work focused on \gls{iaas} public
  clouds and while some portions of our work could also apply
  to other paradigms, certain intricacies of \gls{paas} clouds,
  different computational resources like cloud storage etc.
  can be explored separately in the context of geotemporal inputs
  for further energy efficiency improvements.
\item A question that remains open is how to extend the pervasive cloud
  control approach\
  for integrated forecasting of arbitrary time series data,\
  e.g. application-level\
  load predictions or local renewable energy availability\
  without using external information sources\
  such as weather forecast services.\
  We have analysed forecasting methods such as \gls{aam} and Theta,
  but we have not integrated them into our proof-of-concept cloud controller,
  since domain-specific forecast services can provide better accuracy
  which proved to be critical for planning \gls{vm} migrations
  that result in energy savings.
\item The \gls{aam} method for forecasting geotemporal inputs performed better
  in our evaluation, but related work shows that other methods, such as Theta
  perform better with a shorter history available.
  Additional tests are necessary to find the history
  length breakpoints that are decisive for the choice of
  the most accurate forecasting method.
\item \gls{coe} emission prices in markets such as the EU ETS are currently
  not economically comparable to electricity prices and
  even less so to \gls{vm} prices. As a result, it has no effect on cloud
  control strategies from a financial sense, as any revenue loss from
  performance-based pricing outweighs savings in CERs.
  This is why \gls{coe} was only theoretically studied
  in Chapter~\ref{ch:background} with no simulated evaluation. In the future,
  however, if environmental aspects of energy consumption become more important,
  it is possible that a market with higher \gls{coe} emission prices will arise.
  The benefits of accounting for \gls{coe} emissions in cloud control
  could then become more significant in practice.
\item We were mostly focusing on CPU-bound and I/O-bound workload
  use cases in our work, often in a black box manner, without a deeper
  analysis of in-app performance metrics.
  The details of what types of applications are
  running inside the \gls{vm}s, whether they are single-threaded, parallel,
  memory-intensive or disk-intensive,
  what caching approaches they have etc. can make a big difference
  on the allocation of \gls{vm}s.
  Analysing a wider variety of applications could give more realistic
  performance and energy consumption models and allow for
  higher energy savings in cloud control.
\item We were mostly using simulations for cloud controller evaluation.
  Although we used real data traces to make them more relevant, every simulation
  model must make certain simplifications of real-life phenomena. Only
  Chapter~\ref{ch:volatility} evaluates the proposed cloud controller on a real
  OpenStack cloud deployment. The Philharmonic simulator and other data analysis
  methods used to evaluate the remaining contributions were based on the
  experience and data obtained from working with real systems,
  but it is possible that
  adjustments would be necessary for
  practical applications of the presented methods.
\end{itemize}

\section{Future Work}

Finally, in this section we conclude this thesis with a number of possible
future work directions beyond the challenges we addressed:

\begin{itemize}
\item A next steps in our progressive \gls{sla} specification research could be
  to examine variable \gls{sla} metrics,
  where the probabilistic model would reflect only the most recent data and
  explore how users might react to such volatile pricing options.
  As predictions change based on day-night and seasonal changes,\
  exploring time-changing \gls{sla}s\
  in the manner of stocks and bonds\
  to match the volatile geotemporal inputs would be feasible. 
\item \gls{sla} violation detection could also be explored
  as something to integrate into our progressive \gls{sla} specification.
  Perhaps a combination of performance-based pricing, only considering \gls{vm}
  availability and not only CPU performance could be used for this purpose.
\item A possible future work direction would also be to investigate approaches
  where the \gls{vm} migration cloud controller stage of our \gls{bcffs}
  cloud controller also considers
  the workload CPU-boundedness characteristics\
  in order to maximise the energy savings from using perceived-performance
  pricing.
\item In addition to \gls{vm}s, Docker containers are emerging as a popular
  computation encapsulation method in public \gls{iaas} clouds.
  Docker containers are often used in an architecture of immutable application
  containers and stateful database containers. Such architectures would enable
  much more efficient computation migration by simply starting
  application containers at different data center locations with no downtime.
  It would be interesting to explore our pervasive cloud control approach
  in such an environment where network latencies between application
  and database containers would pose a new challenge to address.
\end{itemize}

All in all, as cloud computing and future large-scale computing paradigms
evolve, especially alongside other complex systems we group under the umbrella
of geotemporal inputs, methods for making them more energy efficient
and environmentally sustainable will continue to be
an interesting and challenging field of research.



\bibliographystyle{plainurl}
\bibliography{references,references-foued,references-ilia-simon}

\appendix{}\renewcommand\chaptername{Appendix}
\setcounter{chapter}{0}
\renewcommand{\thechapter}{\Alph{chapter}}%

\chapter{Curriculum Vit\ae}
\label{ch:cv}

\togglefalse{full}
\togglefalse{publicationURLs}

{\huge \textbf{ \name }}


\bigskip

\hrule


\section*{Education}
\label{sec:education}

\begin{tabular}{lp{\textwidth}}
2011--2016 & Computer Science PhD,\newline
Faculty of Informatics, Vienna University of Technology \\
2009--2011 & Computer Science MSc,\newline
Faculty of Electrical Engineering and Computing, University of Zagreb\\
2006--2009 & Computer Science BSc,\newline
Faculty of Electrical Engineering and Computing, University of Zagreb

\end{tabular}
%
%

\section*{Employment}
\label{sec:employment}

\begin{tabular}{lp{\textwidth}}
2015-- & Software developer, 	Dražen Lučanin e.U. \\
2011--2015 & Research assistant,\newline
Distributed Systems Group,
Vienna University of Technology
\\
2011--2012 & External associate,\newline
Department of Electronics, Ruđer Bošković Institute, Zagreb\\
2009--2010 & Software developer,\newline
Department of Distributed and Multimedia Systems, University of Vienna
\label{sec:employment.amasl}
\end{tabular}

%
%
%

\section*{Research}
\label{sec:research}
\subsection*{Research Projects}
\begin{itemize}
  \item \href{http://www.infosys.tuwien.ac.at/linksites/haley}{HALEY} -- Holistic
  Energy Efficient Approach for the Management of Hybrid Clouds, Vienna University of Technology research award, 2011--2015
  \item
  \href{http://www.focproject.net/}{FOC-II} -- Forecasting Financial Crises,
  \href{http://cordis.europa.eu/projects/rcn/95753_en.html}{FP7}, 2011--2012
  \item
  \href{http://www.univie.ac.at/AMASL}{AMASL} -- Ambient Assisted Shared Living,
  \href{http://cs.univie.ac.at/research/projects/projekt/infproj/809/}{University
  of Vienna}, 2009--2010
\end{itemize}

\subsection*{Publications}


\begin{itemize}
\subsubsection*{Refereed Publications in Journals}

   \item Dražen Lučanin, Ivona Brandić.\
   \emph{Pervasive Cloud Controller for Geotemporal Inputs.}\
   IEEE Transactions on Cloud Computing, 2016.
   \href{http://dx.doi.org/10.1109/TCC.2015.2464794}{doi:10.1109/TCC.2015.2464794}
   \iftoggle{publicationURLs}{(\href{http://ieeexplore.ieee.org/xpl/articleDetails.jsp?arnumber=7180314}{IEEE Xplore})}

   \item Dražen Lučanin*, Ilia Pietri*, Simon Holmbacka*, Ivona Brandic, Johan Lilius, Rizos Sakellariou.\
   \emph{Performance-Based Pricing in Multi-Core Geo-Distributed Cloud Computing.}\
   IEEE Transactions on Cloud Computing (under review, *equal contribution).

\subsubsection*{Refereed Publications in Conference Proceedings}
   \item Soodeh Farokhi, Pooyan Jamshidi, Dražen Lučanin, Ivona Brandić.\
   \emph{Performance-based Vertical Memory Elasticity.}\
   \nth{12} IEEE International Conference on Autonomic Computing (ICAC 2015),\
   7--10 July, 2015, Gronoble, France.
   \iftoggle{publicationURLs}{(\href{http://www.infosys.tuwien.ac.at/staff/sfarokhi/soodeh/papers/Soodeh-Farokhi_CameraReady_ICAC-2015.pdf}{preprint})}
   \href{http://dx.doi.org/10.1109/ICAC.2015.51}{\path{doi:10.1109/ICAC.2015.51}}

   \item Dražen Lučanin, Ilia Pietri, Ivona Brandić, Rizos Sakellariou.\
   \emph{A Cloud Controller for Performance-Based Pricing.}\
   \nth{8} IEEE International Conference on Cloud Computing (CLOUD 2015),\
   27 June -- 2 July, 2015, New York, USA.\ 
   \iftoggle{publicationURLs}{(\href{http://ieeexplore.ieee.org/xpl/articleDetails.jsp?arnumber=7214040}{IEEE Xplore})}
   \href{http://dx.doi.org/10.1109/CLOUD.2015.30}{\path{doi:10.1109/CLOUD.2015.30}}

    \item Dražen Lučanin, Foued Jrad, Ivona Brandić, and Achim Streit.\
      \emph{Energy-Aware Cloud Management through Progressive\
        SLA Specification.}\
      \nth{11} International Conference
      Economics of Grids, Clouds, Systems, and Services
      (GECON 2014).\
      16--18 September, 2014, Cardiff, UK.
      \iftoggle{publicationURLs}{(\href{http://arxiv.org/abs/1409.0325}{arXiv})}
      \href{http://dx.doi.org/10.1007/978-3-319-14609-6_6}{\path{doi:10.1007/978-3-319-14609-6_6}}

      \item Dražen Lučanin, Ivona Brandić.\
  	\emph{Take a break: cloud scheduling optimized for\
  	  real-time electricity pricing.}\
  	Proceedings of the \nth{3} International Conference\
  	on Cloud and Green Computing (CGC),\
  	30 September -- 2 October, 2013, Karlsruhe, Germany.\
  	\iftoggle{publicationURLs}{(\href{http://arxiv.org/abs/1307.7037}{arXiv},
  	\href{http://ieeexplore.ieee.org/xpls/abs_all.jsp?arnumber=6686017}{IEEE Xplore})}
      \href{http://dx.doi.org/10.1109/CGC.2013.25}{\path{doi:10.1109/CGC.2013.25}}

    \item Dragan Gamberer, Dražen Lučanin, Tomislav Šmuc.\
      \emph{Analysis of World Bank Indicators for Countries with Banking Crises\
        by Subgroup Discovery Induction.}\
      Proceedings of the \nth{36} International\
      Convention on Information and Communication Technology,\
      Electronics and Microelectronics -- MIPRO.\
      20--24 May, 2013, Opatija, Croatia.\
      \iftoggle{publicationURLs}{(\href{http://bib.irb.hr/prikazi-rad?&lang=EN&rad=631311}{CROSBI})}
      \href{http://dx.doi.org/10.1007/978-3-642-33492-4_8}{\path{doi:10.1007/978-3-642-33492-4_8}}

      \item Toni Mastelić, Dražen Lučanin, Andreas Ipp, Ivona Brandić.\
  	\emph{Methodology for trade-off analysis when moving scientific\
  	  applications to the Cloud.} CloudCom 2012, \nth{4} IEEE International\
  	Conference on Cloud Computing Technology and Science.\
        3--6 December, 2012, Tapei, Taiwan.
  	\iftoggle{publicationURLs}{(\href{http://ieeexplore.ieee.org/xpl/articleDetails.jsp?arnumber=6427575}{IEEE Xplore})}
        \href{http://dx.doi.org/10.1109/CloudCom.2012.6427575}{\path{doi:10.1109/CloudCom.2012.6427575}}

      \item Dragan Gamberer, Dražen Lučanin, Tomislav Šmuc. \emph{Descriptive\
	modeling of systemic banking crises.} The \nth{15} International\
	Conference on Discovery Science (DS 2012), 29--31 October, 2012, Lyon,\
	France. \iftoggle{publicationURLs}{(\href{http://lis.irb.hr/~gambi/DS2012paper/ds2012-for-web.pdf}{preprint},\
	\href{http://bib.irb.hr/prikazi-rad?&lang=EN&rad=601608}{CROSBI})}
      \href{http://dx.doi.org/10.1007/978-3-642-33492-4_8}{\path{doi:10.1007/978-3-642-33492-4_8}}

      \item Matija Gulić, Dražen Lučanin, Nina Skorin-Kapov. \emph{A Two-Phase\
	Vehicle based Decomposition Algorithm for Large-Scale Capacitated Vehicle\
	Routing with Time Windows.} Proceedings of the \nth{35} International\
	Convention on Information and Communication Technology, Electronics\
	and Microelectronics -- MIPRO, 21--25 May, 2012, Opatija, Croatia.\
	\iftoggle{publicationURLs}{(\href{http://ieeexplore.ieee.org/xpl/articleDetails.jsp?tp=&arnumber=6240808}{IEEE Xplore})}

	      \item Dražen Lučanin, Ivan Fabek, Domagoj Jakobović.\
	\emph{A visual programming language for drawing and executing flowcharts.}\
	Proceedings of the \nth{34} International Convention on Information\
	and Communication Technology, Electronics and Microelectronics -- MIPRO.\
	23--27 May, 2011, Opatija, Croatia. \textbf{Best student paper award.}\
	\iftoggle{publicationURLs}{(\href{http://arxiv.org/abs/1202.2284}{arXiv},\
	\href{http://ieeexplore.ieee.org/search/srchabstract.jsp?tp=&arnumber=5967331}{IEEE Xplore})}

      \item Matija Gulić, Dražen Lučanin, Ante Šimić, Šandor Dembitz.\
	\emph{A digit and spelling speech recognition system for the Croatian language.}\
	Proceedings of the \nth{34} International Convention on Information and\
	Communication Technology, Electronics and Microelectronics -- MIPRO. 23--27\
	May, 2011, Opatija, Croatia.\
	\iftoggle{publicationURLs}{(\href{http://ieeexplore.ieee.org/search/srchabstract.jsp?tp=&arnumber=5967330}{IEEE Xplore})}

\subsubsection*{Refereed Publications in Workshop Proceedings}

      \item  Dražen Lučanin, Michael Maurer, Toni Mastelić, Ivona Brandić.\
	\emph{Energy Efficient Service Delivery in Clouds in Compliance with the Kyoto\
	  Protocol.}\
	\nth{1} International Workshop on Energy-Efficient Data Centers, \nth{8} May,\
	2012, Madrid, Spain.\
	\iftoggle{publicationURLs}{(\href{http://arxiv.org/abs/1204.6691}{arXiv},\
	\href{http://www.springerlink.com/content/f1r83222335t6671/}{SpringerLink})}

\subsubsection*{Theses}

    \item  Dražen Lučanin. \emph{Energy Efficient Cloud Control and Pricing in Geographically Distributed Data Centers.}
	PhD thesis, 
	Faculty of Informatics, Vienna University of Technology,
	June, 2016.

    \item  Dražen Lučanin. \emph{Visual definition of procedures for automatic\
	virtual scene generation.}
	Master's thesis, 
	Faculty of Electrical Engineering and Computing, University of Zagreb,
	June, 2011.
	\iftoggle{publicationURLs}{(\href{http://arxiv.org/abs/1202.2868}{arXiv})}

	\item Dražen Lučanin.
	\emph{Graphical user interface for a video-conferencing application accessible to the elderly.}
	Bachelor's thesis,
	Faculty of Electrical Engineering and Computing, University of Zagreb,
	June, 2009.

\end{itemize}


\subsection*{Activities}

\subsubsection*{Research Visits}
\begin{itemize}
  \item Short Term Scientific Mission (STSM),\
    \emph{Cloud Control for Performance-Based Pricing.}\
    \href{http://www.nesus.eu/}{COST Action IC1305\
      Network for Sustainable Ultrascale Computing (NESUS)},\
    carried out at the University of Manchester, Manchester, United Kingdom;\
    5--20 December, 2014.
  \item \nth{2} COST IC804 Training School on\
  \href{http://cost804trainingschool.uib.es/}{Energy Efficiency\
    in Large Scale Distributed Systems},\
  University of Balearic Islands (UIB), Palma de\
  Mallorca, Spain; 24--27 April, 2012.
\end{itemize}

\subsubsection*{Scientific Talks}
\begin{itemize}
  \item \emph{Energy-Aware Cloud Management\
    through Progressive SLA Specification.}\
    \href{http://www.nesus.eu/ai1ec_event/paris-2014-working-groups-meeting}{\
      COST IC1305 NESUS Meeting},\
    Paris, France;\
    1--2 December, 2014.
  \item \emph{Energy-aware Cloud Management through\
    Progressive SLA Specification.}\
    \href{http://cloudresearch.org/workshops/5th/}{\nth{5} Cloud Control\
      Workshop}, Moelle, Sweden;\ 
    20--22 August, 2014.
  \item \emph{Collaborations with Univ. Toulouse.}\
    \href{http://www.irit.fr/~Georges.Da-Costa/ee-lsds2013/}{Energy Efficiency\
      in Large Scale Distributed Systems conference},\
    Vienna, Austria; 22--24 April 2013.

  \item \emph{Kyoto Protocol Compliant Management of Data Centers.} COST 804 focus group
  \href{http://www.irit.fr/cost804/index.php/focus-groups/energy-and-qos-aware-workload-management-in-clouds}{``Energy
  and QoS-aware Workload Manangent in Clouds''} meeting, INRIA Rennes -- Bretagne
  Atlantique, France; \nth{27} March, 2012.
\end{itemize}

\emph{Additionally, personally gave talks for all the conference and workshop
publications with first authorship.}


\subsubsection*{Other Activities}

\begin{itemize}
  \item Track chair for
    \href{http://www.gecon-conference.org/gecon2015/committees}{GECON 2015},\
    \nth{12} International Conference on Economics of Grids, Clouds, Systems
    and Services.
  \item Program comitee member for
  	\href{http://cloudcomp.eu/2012/show/home}{CloudComp 2012},\
  	\nth{3} International Conference on Cloud Computing, Vienna, Austria;\
  	24--26 September, 2012.
  \item Reviewer for
    \textit{IEEE Transactions on Parallel and Distributed Systems}, 
  	\textit{IEEE Transactions on Cloud Computing}, 
  	\textit{SuperComputing}, 
  	\textit{CCGrid}, 
  	\textit{(EC)2}, 
  	\textit{Euro-Par}, 
  	\textit{PDP}, 
  	\textit{International Conference on Runtime Verification}, 
  	\textit{WETICE}, 
  	\textit{WORKS}, 
  	\textit{BISE / WIRTSCHAFTSINFORMATIK}. 
  \item Member, IEEE, 2012--2015.
\end{itemize}


\section*{Honours, Awards \& Scholarships}
\begin{itemize}
	\item \href{https://ep2013.europython.eu}{EuroPython} \href{https://ep2013.europython.eu/grants}{attendance grant}, 2013.
	\item \href{http://www.google.com/intl/en/jobs/students/}{Google student grant} for attending the EuroPython conference, 2012.
	\item Best Student Paper award -- \emph{A visual programming language for drawing and executing flowcharts.} MIPRO 2011. (see section \nameref{sec:research})
	\item Best Student Computer Program award -- \emph{RoboDJ} \iftoggle{full}{(see
	section \nameref{sec:projects})}, Faculty of Electrical Engineering and Computing, 2010.
	\item Two times Croatian Ministry of Education student scholarship,
          2006--2009 and 2009--2011.
	\item Two times Erasmus student exchange scholarship,
          2009 and 2011 at the University of Vienna.
\end{itemize}

\medskip

\begin{center}
  \begin{small}
    Curriculum vit\ae\ last updated: \today
  \end{small}
\end{center}

\label{appendix:cv}

\end{document}